\documentclass[12pt,nofootinbib,floats,floatfix,oneside]{book}
\usepackage[utf8]{inputenc}

\usepackage{graphicx,mathtools,amssymb,amsmath,amsthm,amsfonts,epsfig,epsf}
\usepackage[a4paper,top=3cm,bottom=3cm,left=4.5cm,right=3cm]{geometry}
\usepackage[outdir=./]{epstopdf}
\usepackage{hyperref}
\hypersetup{hidelinks}
\usepackage[usenames]{color}		
\usepackage{tensor}
\usepackage{setspace}
\usepackage{appendix}
\usepackage{cite}
\usepackage{csquotes}
\usepackage{mathrsfs}
\hypersetup{pdfstartview=}
\usepackage[caption=false]{subfig}
\usepackage{enumitem}
\usepackage{floatrow}

\usepackage{tocbasic}[2016/05/10]
\DeclareTOCStyleEntry[numwidth=2.5em]{tocline}{chapter}
\DeclareTOCStyleEntry[indent=2.5em,numwidth=3em]{tocline}{section}
\DeclareTOCStyleEntry[indent=5.6em,numwidth=3em]{tocline}{subsection}
\DeclareTOCStyleEntry[dynnumwidth]{tocline}{figure}

\renewcommand{\thechapter}{\Roman{chapter}}

\newcommand{\note}[1]{\text{\tiny{#1}}}

\newcommand{\dd}{\mathrm{d}}
\newcommand*\DAlembert{\mathop{}\!\mathbin\Box}

\newcommand{\gb}{\mathscr{G}}

\begin{document}


\begin{titlepage}
    \begin{center}
        
        \large
        \textsc{University of Nottingham}\\
        \vspace{.4cm}
        \includegraphics[width=0.5\textwidth]{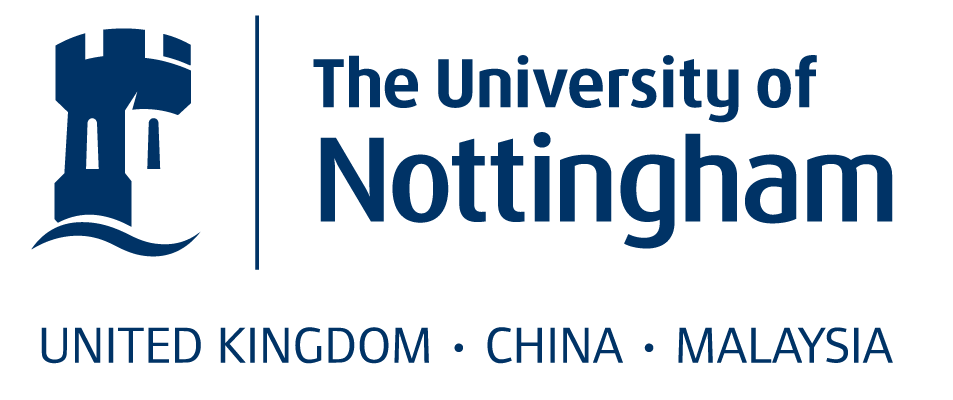}
            
        \large
        \textsc{School of Mathematical Sciences}\\
        
         \vspace{3cm}
            
        \Huge
        \textbf{New Perspectives on Spontaneous Scalarization in Black Holes and Neutron Stars}
            
        \vspace{0.5cm}
        \Large

        \vspace{1.5cm}
            
        \textbf{Giulia Ventagli}
            
        \vfill
        
        \Large    
        A thesis submitted to the University of Nottingham for the degree of\\
        \textsc{Doctor of Philosophy}
            
        \vspace{0.8cm}
        
        \Large
        April 2022        
            
    \end{center}
\end{titlepage}

\pagenumbering{roman}

\cleardoublepage
\thispagestyle{empty}
\vspace*{\stretch{1}}
\begin{flushright}
\itshape A mia nonna Iole\\
e Leonardo
\end{flushright}
\vspace{\stretch{2}}
\cleardoublepage

\chapter*{Abstract}

 Although general relativity passes all precision tests to date, there are several reasons to go beyond the current model of gravitation and search for new fundamental physics. This means looking for new, so far undetected, fields. Scalars are the easiest fields to consider and they are ubiquitous in both extensions of the Standard Model and in alternative theories of gravity. That is why a lot of attention has been drawn towards the investigation of scalar-tensor theories, where gravity is described by both the metric tensor and a scalar field.
 
 A particularly interesting phenomenon that has recently gained increasing interest is spontaneous scalarization. In gravity theories that exhibit this mechanism, astrophysical objects are identical to their general relativistic counterpart until they reach a specific threshold, usually either in compactness, curvature or, as recently shown, in spin. Beyond this threshold, they acquire a nontrivial scalar configuration, which also affects their structure.
 
 In this thesis, we focus on the study of this mechanism in generalized scalar-tensor theories. We identify a minimal action that contains all of the terms that can potentially trigger spontaneous scalarization. We first focus on the onset of scalarization in this specific theory and determine the relevant thresholds in terms of the contributing coupling constants and the properties of the compact object. Finally, we study the effect of this model on the properties of both scalarized black holes and neutron stars, such as affecting their domain of existence or the amount of scalar charge they carry.

\chapter*{Acknowledgements}

 First of all, I would like to thank my supervisor, Thomas Sotiriou, for all he taught me in the last four years, for all the time he dedicated me and the guidance he gave me. I would also like to thank all the past and present members of our research group, who directly or indirectly helped me get through my PhD. In particular, I would like to thank Antoine Lehébel who was always there when I needed help, and had the patience to mentor me even from the other side of Europe, and Georgios Antoniou, with whom I shared a bit of the loneliness of doing research during a pandemic.

 Next, I have to thank my parents. Thank you for your love, your support, your patience and your understanding. You both have taught me precious lessons and the dearest values I have in life. I would also like to thank my grandmother Iole, that always supported me, especially during my five years of university in Rome. I am sure I was not the easiest housemate to live with. I also wish to thank my (not so little anymore) cousin, Leonardo, who made me become a child again in more than one occasion. This thesis is dedicated to both of you.

 I have had many great teachers and professors, in schools and university, but I would have never even imagined of studying physics if it were not for my high school teacher, Professor Nadia Bonaldo. She has taught me what it means to do scientific research, and I probably only realized- it almost a decade later. 
 
 I would like to thank all my friends that were there along my studies. Some of you have been there for ages: Sara, Gioia, Gianluca and Nicolò, thank you for all the time we spent together, all the times you listened to me and all the fun we had through the years.
 To my amazing fellow band (and university) companions: Matteo, Francesco, Mattia, Lorenzo and Luigi, the music maybe was not that good, and we are now all scattered around the world, but we are always there for each other.
 To the friends with whom I shared that small room while we were all writing our master thesis: Filippo, Giulia, Davide and Angelo, we stayed sane together in challenging times and we still support each other.
 To my friends in Nottingham: Lucía, Marco, Elisabetta, Maria and Luisa, thank you for all the time spent together especially in the last year.
 To all my West Bridgford rugby teammates: in the last months you all taught me so much about sport and life.
 Last but certainly not least, I would like to thank Andrea, who was always there, even in the darkest moments, always ready for a phone call, even if it was in the middle of the night.
 
 A special thanks to Francesca, my therapist, who in the last year helped me through a lot and taught me to see things about myself that I never even dared.
 
 I wish to thank, again, Luigi. I think that neither of us ever thought of finding us where we are now. But, believe me, I am incredibly grateful. Thank you for your almost infinite patience and your dedication in understanding me.
 
 Finally, a bit unconventionally, I would like to acknowledge my own achievements. It is a thing that I seldom do, and maybe it is time I learn to give myself a bit more credit. These last four years have been at times incredibly amazing, at times undoubtedly painful. I am grateful for the help I received from all the people around me, and I am also definitely grateful that I decided to believe in myself.

\chapter*{Conventions}

 Throughout this thesis, the metric signature is chosen to be $(-,+,+,+)$ and spacetime indices are denoted with Greek letters.
 What follows is a list detailing the conventions used throughout the text.
 \begin{description}[labelindent=.5cm, labelwidth=2cm,leftmargin=!,font=\normalfont]
 \item[$g_{\mu\nu}$] spacetime metric
 \item[$g$] determinant of the metric
 \item[$\Gamma^\alpha_{\mu\nu}$] Christoffel symbol of the spacetime metric $g_{\mu\nu}$
 \item[$\nabla_\mu$] covariant derivative corresponding to the Christoffel symbol $\Gamma^\alpha_{\mu\nu}$
 \item[$R^\alpha_{\mu\kappa\nu}$] Riemann tensor
 \item[$R_{\mu\nu}$] $R^\alpha_{\mu\alpha\nu}$, Ricci tensor
 \item[$R$] $g^{\mu\nu}R_{\mu\nu}$, Ricci scalar
 \item[$\mathscr{G}$] $R^2-4\,R_{\mu\nu}R^{\mu\nu}+R_{\mu\nu\rho\sigma}R^{\mu\nu\rho\sigma}$, Gauss-Bonnet invariant
 \item[$G_{\mu\nu}$] Einstein tensor
 \item[$G$] Newtonian gravitational constant
 \item[$c$] speed of light, we used units such as $c=1$
 \item[$M_\odot$] mass of the Sun
 \item[$\kappa$] $8\pi G/c^4$
 \item[$X$] $-\partial_\mu\phi\partial^\mu\phi/2$
 \item[$S_\text{M}$] matter action
 \item[$\psi_\text{M}$] matter fields
 \item[$T_{\mu\nu}$] matter stress-energy tensor
 \item[$\epsilon$] energy density of a perfect fluid
 \item[$p$] pressure of a perfect fluid
 \item[$u_\mu$] 4-velocity of the perfect fluid
 \item[$T^\text{PF}_{\mu\nu}$] $(\epsilon+p)u_\mu u_\nu+p g_{\mu\nu}$, matter stress-energy tensor for a perfect fluid
 \item[$M$] ADM mass of the compact object
 \item[$Q$] scalar charge of the compact object
 \item[$\text{EOS}$] equation of state
 \end{description}

\chapter*{Publications}

The original work in this thesis is based on the following works produced during my studies:
\begin{itemize}
    \item Chapter~\ref{Chap:scalarizationHordneski}: N. Andreou, N. Franchini, G Ventagli, and T. P. Sotiriou, ``Spontaneous scalarization in generalized scalar-tensor theor'', \textit{Phys. Rev. D}, vol. 99, no. 12, p. 124022, 2019,
    \item Chapter~\ref{Chap:Threshold}: G. Ventagli, A. Lehébel, and T. P. Sotiriou, ``Onset of spontaneous scalarization in generalized scalar-tensor theories'', \textit{Phys. Rev. D}, vol 102, no. 2, p.024050, 2020,
    \item Chapter~\ref{Chap:blackholes}: G. Antoniou, A. Lehébel, G. Ventagli, and T. P. Sotiriou, ``Black hole scalarization with Gauss-Bonnet and Ricci couplings'', \textit{Phys. Rev. D}, vol. 104, no. 4, p. 044002, 2021,
    \item Chapter~\ref{Chap:neutronstars}: G. Ventagli, G. Antoniou, A. Lehébel, and T. P. Sotiriou, ``Neutron star scalarization with Gauss-Bonnet and Ricci couplings'', \textit{Phys. Rev. D}, vol. 104, no. 12, p. 124078, 2021.
\end{itemize}

\tableofcontents
\listoffigures
\clearpage
\pagenumbering{arabic}


\chapter{Introduction}\label{Chap:Introduction}

\section{Gravitational waves: a laboratory for new physics}\label{Sec:gw}

 In 1915, Einstein formulated his theory of general relativity. The revolutionary concept behind it is that the geometry of spacetime is a new physical entity, with degrees of freedom and its own dynamics~\cite{Misner:1973prb}. In other words, the spacetime is curved and its intrinsic properties are described by a metric. Thus, instead of introducing a gravitational field as a new field, we can think of it as a deviation of the spacetime geometry from the flat geometry of special relativity~\cite{Wald:1984rg}. The equivalence principle of the Newtonian theory of gravitation is then generalized into the \textit{Einstein equivalence principle} that states that the motion of freely falling test particles follow geodesics of the metric\footnote{Note that this is the formulation for the so-called \textit{weak equivalence principle}, which only refers to test-body whose self-gravity is negligible. If we generalize the statement to self-gravitating bodies, we obtain the \textit{strong equivalence principle}. This distinction is non-trivial in the context of scalar-tensor theories of gravity, as we will see in Chapter~\ref{Chap:sttheories}.}. In general relativity, the motion of matter in the universe is influenced by the geometry of the spacetime. At the same time, the curvature of spacetime is related to the stress-energy momentum tensor of matter via the Einstein field equations. Thus, the structure of spacetime is itself influenced by the presence of matter in the universe~\cite{Wald:1984rg}.

 General relativity has been widely accepted as a fundamental theory of gravity and it successfully predicted various phenomena such as the gravitational redshift, the bending of light by the sun, the precession of the perihelion of Mercury, the existence of neutron stars and black holes. One of the most remarkable validations arrived exactly a century after the formulation of Einstein's theory: in 2015 the Advanced LIGO detectors in Livingston and Hanford observed for the first time the emission of gravitational waves from a black-hole binary merger~\cite{LIGOScientific:2016aoc,LIGOScientific:2016sjg}. Gravitational waves are predicted by general relativity and they consist of ripples of the gravitational field caused by accelerated masses that propagate across the spacetime. They are commonly produced by cataclysmic events such as colliding black holes and neutron stars. Since the first detection, in the span of three observing runs, of which the last one ended on March 2020, more than ninety events were detected with an increasing precision thanks also to the addition to the collaboration of the Advanced Virgo detector in Italy in August 2017~\cite{LIGOScientific:2018jsj,LIGOScientific:2020ibl,LIGOScientific:2021djp}. Most excitingly, the joint collaboration was able to detect not only black-hole binary coalescences, but also more rare events such as gravitational waves from the collision of two neutron stars~\cite{LIGOScientific:2017vwq,LIGOScientific:2020aai}, and from neutron star-black hole mergers~\cite{LIGOScientific:2021qlt}. Yet, the number of events detected is drastically destined to increase. On one side, the collaboration is currently preparing for a fourth observing run, possibly starting in the second half of 2022, where the LIGO and Virgo detectors will be joined by the KAGRA interferometers based in Japan, forming a network of ground based detectors that will certainly enhance the precision of the observations. On the other, in the next decades we will witness the advent of third generation ground detectors, like the Einstein Telescope and the Cosmic Explorer, and new detectors in space, like the LISA mission. Hence, in the future years we will inevitably find ourselves with an incredible amount of data from coalescing compact objects.

 Such a thriving scenario allows us to have multiple direct channels to areas of our universe characterised by so far unexplored energy scales and spacetime curvature. For the first time, we are equipped with the tools to probe the \textit{strong gravity regime}, where the gravitational interaction is predominant over the other fundamental forces. Such experimental conditions are impossible to recreate in any laboratory on Earth. Even at the scale of our Solar System, we are only able to perform tests on the gravitational fields in the \textit{weak gravity regime}. On the other hand, compact objects inspirals, where the curvature of spacetime is high, constitute the perfect natural laboratory where the conditions for strong gravity are satisfied.
 
 We can better understand this by giving a more formal classification of the weak and the strong field regime. Let us assign to a gravitational system a characteristic length $L$, a characteristic total mass $M$ and a characteristic velocity $V$. We can then define the compactness of the system, a quantity which expresses the strength of the gravitational field, as
 \begin{equation}
    C=\frac{GM}{L},
 \end{equation}
 where $G$ is the Newtonian gravitational constant, and where we set $c=1$. We can refer to weak-field gravity when the following conditions are both satisfied
 \begin{equation}\label{eq:weakfield}
    C\ll1,\quad V\ll 1.
 \end{equation}
 This ensures that the gravitational field is weak relative to the mass-energy of the system, the characteristic velocities of gravitating bodies are small relative to the speed of light, and the gravitational field is stationary or quasi-stationary relative to the characteristic size of the system~\cite{Yunes:2013dva}. Conversely, the strong-field regime is the region of the spacetime where both conditions on Eq.~\eqref{eq:weakfield} are not satisfied. It is then clear why, when conducting experiments in the Solar System, we are still bound in the weak field regime: considering the gravitational field at the surface of the Sun, we find $C\approx 2 \times 10^{-6}$, whereas for the Earth-Sun system, this further reduces to $C\approx 9.8 \times 10^{-9}$, with a characteristic velocity of $V\approx9.9\times10^{-5}$. Even in the case of binary pulsars, conditions~\eqref{eq:weakfield} are still not violated: for the double binary pulsar J07373039, $C\approx 6 \times 10^{-6}$ and $V\approx2\times10^{-3}$. Despite neutron stars being sources of strong gravity, indeed their compactness is typically $C\approx O(10^{-1})$, binary pulsars are most sensitive to the quasi-static part of the post-Newtonian effective potential or to the leading-order of the radiation-reaction force~\cite{Yunes:2013dva}. If we take into consideration compact binary coalescence, however, the situation drastically changes: both the gravitational compactness and the characteristic velocity can reach values of $O(1)$. Hence, the gravitational information produced during these events provides the strongest gravity field tested to date.

 Gravitational waves, therefore, represent the perfect way not only to perform null tests of general relativity, but also, more intriguingly, to challenge the actual state of physics, from the current theory of gravitation to the Standard Model. There is a general shared hope between physicists involved in diverse fields such as cosmology, astrophysics, modified gravity, quantum gravity and particle physics, that testing the strong-field highly dynamical regime will lead to the discovery of new unexpected phenomena, giving some answers to the big open questions in physics.

 Nevertheless, this research field is still at its beginning. While compact objects are generally well understood in general relativity, this is not always the case in other theories of gravity. However, if one intends to obtain information from gravitational-wave observations, understanding black holes and compact stars in these alternative models is crucial. Indeed, these detections consist in extremely noisy data. In order to extract the gravitational wave signal from the observations, one has to \textit{a priori} model the source, given a specific theory. Thus, while observations probe the object, its structure is determined by the theoretical model. It is then understandable why, in the recent years, so much attention has been drawn to the study of compact objects in different scenarios that contain new physics. Just as Einstein's theory attracted a lot of interest because it was rich in applications~\cite{Misner:1973prb}, the study of modified theories of gravity, as well as particle physics, is gaining more and more attention thanks to the rapid development and growth of gravitational-wave observations.
 
\section{Beyond General Relativity and the Standard Model}\label{Sec:overcomegr}

 We discussed how the rapid expansion of the gravitational waves research field is encouraging the development of new alternative theories that challenge general relativity and the Standard Model and, more in general, it represents a promising tool for the discovery of new fundamental physics. It is, however, natural to ponder the question: why do we expect new physics? Why do we suppose that general relativity and the Standard Model might fail in the strong gravity regime? The answer to these questions is not easy and is not unique. There are several reasons, each of which stems from different considerations. In this Section, we give a concise review of the main motivations for questioning the current state of gravity and particle physics.

 First of all, general relativity is a classical theory, thus all quantum physics contributions are neglected. Furthermore, in the attempt of treating the gravitational field from a standard quantum field theory perspective, one is inevitably faced with the fact that general relativity is non renormalizable. This is a signal that the theory must fail to describe gravity at some energy scale usually associated with the Planck scale. However, it is not possible to assign a precise value to this cutoff energy scale, since there is no way for current experiments to probe quantum effects in gravitational physics, or vice versa. It has been shown that allowing the action to contain higher order curvature terms, one finds a theory which is indeed renormalizable, but affected by the presence of ghosts, i.e. massive quantum states with negative norm, therefore violating unitarity~\cite{Stelle:1976gc}. To overcome this problem, several candidate theories of quantum gravity have been proposed. For instance, a well-received theory is loop quantum gravity, which is an effort toward the quantization of general relativity through the postulate that spacetime itself is discrete. Remarkably, some of the proposed quantum-gravity theories also try to tackle the problem of unification, addressing the question of whether there exists a unified description of all known forces, possibly including gravity. As a result, the Standard Model would be seen as a low-energy limit\footnote{Note that low energy physics is not synonymous of weak-field gravity.} of a more fundamental theory. For example, among the most well-studied proposals, we certainly find string theory, which replaces the point-like particles of particle physics by one-dimensional objects, the so-called strings, and studies how their dynamics evolve in the spacetime. Interestingly, one could look at classical modified theories of gravity as limit of quantum gravity models at low energy scales, or in other words, they can be seen as effective field theories with a cutoff energy scale that characterizes the underlying fundamental quantum gravity theory.

 Another theoretical conundrum regarding general relativity is that it predicts the appearance of spacetime singularities. These can either be cosmological singularities or points of the spacetime that are located in the centre of black holes, as a result of the gravitational collapse of massive stars. The \textit{cosmic censorship conjecture} claims that the appearance of naked singularities is forbidden, and each singularity is clothed in an event horizon~\cite{Penrose:1969pc}. Nonetheless, this conjecture is not precise and it is not proven. Moreover, one of the most urgent open problems in theoretical physics is the so-called
\textit{information-loss paradox}~\cite{Mathur:2009hf}, which is related to loss of unitarity at the end of
the black holes evaporation due to Hawking’s radiation~\cite{Hawking:1974rv}. Once again, these singularity problems can be avoided by high-energy corrections or resolved by a consistent quantum gravity theory.

 The objections to general relativity and the Standard Model are not merely theoretical. For instance, there is rich and overwhelming astronomical evidence for the existence of a large amount of unseen dark matter. The first hints of dark matter was given in 1933 by F. Zwicky~\cite{Zwicky:1933gu}, who estimated that the Coma cluster of galaxies had a mass-to-light ratio of two orders of magnitude larger that in the solar neighborhood. The arguments for dark matter accumulated over the decades (see~\cite{Bertone:2004pz,Turner:1999kz} for detailed reviews). However, the identity and properties of dark matter remain unknown. Scalar fields have been seen as potential candidate for a long time. The most well motivated examples are the axion particles from quantum chromodynamics or axion-like-particles from string theory, we return on the subject in Section~\ref{Sec:scalarnewdof}. Furthermore, evidence from type Ia supernovae suggests that the universe has evolved from a past decelerated phase to a present accelerated period. To account for this accelerated phase, the universe must be composed by ordinary baryonic matter for the $4\%$, by dark matter for the $20\%$ and by dark energy for the $76\%$~\cite{SDSS:2005xqv, SupernovaSearchTeam:2004lze,WMAP:2006bqn}. The latter is an unknown form of energy which not only has not been detected directly, but also does not satisfy the Strong Energy Condition. The presence of dark energy can be explained by adding a cosmological constant $\Lambda$ in the Einstein’s equations, interpreting it as the energy density associated with the quantum vacuum. However, this approach leads to a discrepancy of several order of magnitude between the theoretical and observational value of the cosmological constant itself, signaling another conflict between gravity and quantum physics. A way to resolve this issue is to either include a novel matter field with new fundamental properties or to interpret dark energy as some gravitational effect, thus believing that its origin must come from beyond general relativity (see~\cite{Carroll:2000fy} for a review on the cosmological constant problem). It is noteworthy that both the issue of dark matter and the cosmological constant problem suggest that the standard effective field theory argument, stating that quantum gravity corrections are unimportant in the infrared regime might be wrong, motivating the speculation that such corrections might persist at low energies. We reflect on the consequences of these low energy imprints in Section~\ref{Sec:scalarimprints}.

 Both dark matter and dark energy cannot be explained in the context of the Standard Model. Moreover, further theoretical and experimental arguments are accumulating, pointing to the existence of particle physics beyond the Standard Model. On the experimental side, for instance, there is robust evidence for oscillations of atmospheric neutrinos, which can be explained by assuming that neutrinos have mass, in contrast to the current hypothesis of zero mass in the Standard Model~\cite{Maltoni:2003da}. On the theoretical side, the \textit{hierarchy problem} is of particular note~\cite{deGouvea:2014xba}. This problem is linked to the huge discrepancy between the electroweak and the Planck scale, and arises in the size of the mass of the Higgs boson. The Higgs field has an observed mass of about $125$ GeV, however the radiative corrections to the Higgs mass lead us to expect that the Higgs mass would be much higher, comparable to the Planck scale. Reconciling these inconsistencies would require a much higher accurate fine-tuning of parameters or assuming that there is some other fundamental reason not explained by the Standard Model. For example, postulating the existence of new particles with similar masses but with spin different by one half would solve the hierarchy problem~\cite{Bertone:2004pz}. These types of particles emerge naturally in the framework of supersymmetric theory. However, this is still a rather open topic of discussion.

 We tried to give a condensed summary of the various currently open questions in physics. Even though we did not provide a complete and detailed review, this still suffices to convey the idea of why there is such an active and pressing research for new physics beyond both general relativity and the Standard Model.

\section{Scalar fields as new degrees of freedom}\label{Sec:scalarnewdof}

 So far, we highlighted the main motivations for challenging our current knowledge of gravitational and particle theories. It is then natural to ask how we can potentially describe this promising new physics. To overcome general relativity and the Standard Model one necessarily needs to relax one or more of the assumptions on which such theories rely. Let us explicitly examine the case of general relativity. The Lovelock's theorem~\cite{Lovelock:1971yv,Lovelock:1972vz} assures that general relativity is the only theory that satisfies all of the following assumptions:
 \begin{itemize}
    \item its action is invariant under diffeomorphism;
    \item it has second order field equations for the metric;
    \item the theory is restricted to 4 dimensions;
    \item only the metric field is involved in the gravitational action.
 \end{itemize}
 To go beyond general relativity one needs to evade the Lovelock's theorem, hence any alternative model of gravitation must break at least one of the assumptions above. This inevitably introduces more degrees of freedom within the theory (see~\cite{Sotiriou:2014yhm}).

 A possible way to include new degrees of freedom is to allow alternative models to contain new, so far undetected, fundamental fields either in the matter or in the gravitational sector of the theory. Scalar fields are the easiest field to consider, and they can act as basic probes for novel models of gravity. The work of this thesis is entirely focused on the study of generic scalar-tensor theories, where a scalar field comes into play through a non-minimal coupling with gravity. The first prototype of such theories was first conceived by Jordan~\cite{Jordan}. He embedded a four-dimensional curved manifold into a five-dimensional flat space and showed that a four-dimensional scalar field can be a constraint in formulating projective geometry while enabling a gravitational constant dependent on the spacetime. We dedicate Chapter~\ref{Chap:sttheories} to a thorough review of scalar-tensor theories. These models have drawn vast interest since they \textit{``appear to provide a small window through which one can look into phenomenological aspects of more fundamental theories to which one is still denied any direct access otherwise''}~\cite{fujii_maeda_2003}. We have already mentioned how scalar fields have been proposed as major candidate to resolve the issues with dark matter. However, they are not only employed in this context, they are ubiquitous in several proposed alternative theories of gravitation and particle physics, as we now discuss.

 A first scalar field candidate emerged in the Kaluza-Klein theory in the 1920s. The theory proposed a five-dimensional spacetime to which general relativity was applied, and one of the spatial dimension was compactified to a small circle. The size of the compactified space dimension can be recast as a four-dimensional scalar field. This theory gained new interest in the 1970s~\cite{Cho:1975sf}, when it was realized that string theory requires higher dimensional spacetime. Additionally, scalar fields are also provided in string theory in the form of dilatons. Indeed, the zero mode of a closed string describes a symmetric tensor, the graviton, that in the low energy limit reduces to the spacetime metric, as well as a scalar field that emerges from the trace of a symmetric second rank tensor and an antisymmetric one. The field equations for the graviton and the scalar field can be shown to be derived from a scalar-tensor theory-like lagrangian. This prototype of scalar-tensor theory emerges due to the requirement of the finiteness of the theory coming from the invariance under conformal transformation in the two-dimensional spacetime where the strings propagate. This conformal invariance generalizes in D-dimensions as dilatation invariance and it is implemented via the scalar field, hence the origin of the name dilaton.

 Scalar fields arise in the framework of yet another quantum gravity focus, non-commutative geometry. The basic idea is to quantize the spacetime coordinate replacing them by generators of a noncommutative algebra~\cite{Madore:1998hu}. This approach provides scalar fields as gauge fields introduced in addition to ordinary continuous spacetime~\cite{fujii_maeda_2003}. For example, in the electroweak unified theory, these scalar fields can be associated with the Higgs field. It was shown that a natural extension to the theory of gravity yields a scalar-tensor theory~\cite{Kokado:1996ua,Kokado:1998uw}. Interestingly, unlike the case of Kaluza-Klein theory or string theory, in noncommutative geometry the scalar field is a non-ghost field, i.e. they have a correct negative sign in front of the kinetic term in the action, ensuring unitarity\footnote{Note that in the case of Kaluza-Klein theory, even if the scalar field is a ghost, this does not lead to physical inconsistencies since the energy of the whole system remains positive. In the case of string theory, for $D>2$ the diagonalized scalar field is also a non-ghost.}.

 As we have already mentioned in Section~\ref{Sec:overcomegr}, quantum chromodynamics predicts the appearance of pseudoscalar fields called axions, i.e. scalar fields which are odd under parity transformations. In particular, they emerge as a consequence of the Peccei and Quinn's solution to the problem of the absence of CP violation in quantum chromodynamics~\cite{Peccei:1977hh,Peccei:1977ur}. Such particle can be interpreted as a pseudo Goldstone-boson of a new spontaneously broken global U(1) symmetry. It can couple to other fields, however, this interactions are suppressed by a scale $f_a$ which can be arbitrarily large. Hence, the axion can be engineered to be weakly coupled to the Standard Model. This weakly interacting aspect is the reason why the axion is an extremely viable candidate for dark matter. Besides the case of quantum chromodynamics, axion-like-particles are often predicted by embeddings of the Standard Model in string theory~\cite{Ringwald:2014vqa}. They also arise from the breaking of accidental global U(1) symmetries that appear as low energy remnants of exact discrete symmetries. Axion and axion-like particles are very well motivated both theoretically and on cosmological and astrophysical grounds, rendering them attractive candidates to test.

 Lastly, let us discuss the case of the cosmological inflaton. In 1977, it was discovered that the entropy of the universe is extremely large, with an estimate of $S\approx 10^{88}$~\cite{Barrow:1977}, even though one would expect this value to be of order of unity, since it is a dimensionless constant. In order to explain this large amount of entropy, the universe must have been close to spatial flatness at early times. Even if such balance is in principle possible, it still seems an unlikely case of fine-tuning. This dilemma is known as the \textit{flatness problem}. A possible way to solve this issue was first proposed by Guth in 1981~\cite{Guth:1980zm}. He postulated that the universe went through a non-adiabatic cosmic expansion for a finite time interval in its early history, creating a process of entropy production called \textit{inflation}. A necessary condition for this to happen is to assume that the universe underwent an accelerated expansion. But what drives this expansion? A simple answer is to assume the presence of a scalar field with a self-interaction potential, which takes the name of inflaton. In the modern concept of inflation, what drives it is a single scalar field slow-rolling from a regime of high potential energy. However, the precise identity of this field is not known and several different inflationary scenarios have been developed. Possible candidates are the Higgs boson, the dilaton field of string theory, which we mentioned earlier, the extras degrees of freedom associated with higher metric derivatives in modified gravity and the time-varying gravitational coupling of scalar-tensor theories (for a review on inflation see~\cite{Lidsey:1995np,Bassett:2005xm,Pajer:2013fsa,Amin:2014eta}). There are several other scalar fields proposed in cosmology, other than the inflaton. We briefly discuss two other examples, the galileon and the chameleon scalar, in the next Section.
 
 This review of the diverse employments of scalar fields in gravitational theories and particle physics is certainly not exhaustive. Besides, we do not expect that introducing a scalar field to a theory can possibly solve all of the several problematics to which we referred in Section~\ref{Sec:overcomegr}. However, our aim was to give a solid motivation for the study of scalar fields coupled to gravity and their phenomenology. Understanding if these fields leave an imprint on compact objects is more relevant than ever, since we finally have the tools to explore and test this possibility thanks to gravitational-wave observations.

\section{Scalars' imprints at low energy scales}\label{Sec:scalarimprints}

 We gave several motivations for the presence of scalar field in regimes of strong gravity, however we still have to debate if it is possible to find their imprints at scales where energy is low, and consequently detectable by current experiments. At first, corrections deriving from quantum gravity, or more generally from unified theories, were expected to be detectable only at high energy scales, typically at the Planck scale. It was thus believed that there was no need to have scalar fields imprints at low energy. However, this speculation lost its appeal when the issues of dark matter and the cosmological constant problem emerged. Indeed, they both pose experimental contradictions that appear in the infrared regime. Solving these open questions would necessarily require a more complete theory able to introduce corrections also at low energy scales.

 Nonetheless, this is a delicate matter. On one side, general relativity passed all weak gravity tests performed so far. On the other, we have several reason to expect to find deviations to the current model of gravitation in the strong gravity regime. Hence, introducing new scalar fields poses an important dilemma: we need to introduce a theory able to provide detectable modifications to the current gravitational theory in regimes where gravity is strong, and at the same time explain why there are no signs of scalar forces in laboratory or solar system observations. An effective way to achieve this is to resort to a \textit{screening mechanism}, which is able to yield scalar imprints in specific regions, while concealing them in the rest of the spacetime.
 
 An example of the employment of such process can be found in the framework of Galileon cosmology, originally proposed as a possible candidate of inflation theory~\cite{Nicolis:2008in}. The Galileon model is a scalar effective field theory that contains higher derivative terms in the Lagrangian, which is invariant under the Galileon-shift symmetry $\partial_\mu\phi\to\partial_\mu\phi+b_\mu$, while its field equations are still second order in derivatives, hence it is free of ghost-like instabilities. The scalar field is required to produce interesting cosmological effects in sparse cosmic environment, while it decouples locally. This is achieved thanks to the so-called \textit{Vainshtein screening effect}. In regions where the density is small, the galileon field perturbation acts like a gravitational potential, thus the scalar mediates a sizeable fifth force. On the other hand, in areas of high density, near massive bodies, the scalar field decouples and the theory reduces to Newtonian gravity~\cite{Chow:2009fm,Brax:2011sv,Barreira:2014zza}.
 
 Another example of screening mechanism can be found in chameleon cosmology~\cite{Khoury:2003aq}, which proposes an alternative explanation to the expansion of the universe. In this theory the scalar field can couple directly to baryons with gravitational strength while acquiring a mass which is dependent on the local matter density. This effectively enables a screening process: in regions of high density, the mass of the field is large, exponentially suppressing the scalar field itself; in areas where the density is much lower, such as the solar system, the wavelength of the field can be larger than the size of the solar system; however, on cosmological scales, where the density is considerably smaller, the mass can be of order of the present Hubble parameter and the chameleon field can potentially cause the acceleration of the universe~\cite{Khoury:2003rn}. Both these mechanisms are qualitative similar, with the difference that the Vainshtein effect for galileons relies on derivative iterations while the chameleon mechanism depends on the scalar potential.

 In this work, we focus on yet another screening process, originally formulated in the 1990s, that has recently gained renewed and growing attention. The outcome of this phenomenon is a scalar field that can potentially be non-vanishing close to compact objects, where gravity is strong. Nevertheless, far away from the source the scalar field is suppressed and general relativity is retrieved in accord with solar system experiments. This mechanism can potentially leave an imprint, such as a \textit{scalar charge}, a quantity that measures how much scalar field is present in the object, which could be detectable in gravitational-wave observations. In its first formulation proposed by Damour and Esposito-Far\`{e}se~\cite{Damour:1993hw}, this phenomenon applied only to neutron stars in the context of scalar-tensor theories. It consisted of a phase transition which occurs to the scalar field when the compactness of the star exceeds a certain threshold, it has thus been dubbed \textit{spontaneous scalarization}, to recall the phase transition mechanism of spontaneous magnetization. Compared to other theories where all compact objects have a scalar configuration, the peculiarity of this phenomenon is that general relativity solutions are still admissible, but in specific regions of the spacetime, in particular inside or close to compact objects, they become unstable, while scalarized solutions are preferred.
 In recent years, the study of this screening mechanism has gained new interest, since it was understood that it can be generalized to other theories while also affecting black holes~\cite{Silva:2017uqg,Doneva:2017bvd,Antoniou:2017acq}. Even though spontaneous scalarization is fundamentally a non-perturbative effect, one can understand its generic properties  employing perturbative techniques. At the linear level, this mechanism manifests as a tachyonic instability. Such negative effective mass instability triggers an exponential growth of the scalar field, which can then be quenched by non-linearities of the theory, creating a stable configuration of the scalar around black holes or neutron stars.

 As we mentioned, even though this mechanism was originally proposed for scalar-tensor theories, it is not unique to this model alone and it can be applied to other theories as well as to different field contents, ({\em e.g.}~\cite{Herdeiro:2018wub,Ramazanoglu:2017xbl,Ramazanoglu:2018hwk}). The work of this thesis strictly focuses on the study of spontaneous scalarization oi generalized scalar-tensor theories. In Chapter~\ref{Chap:sttheories}, we review scalar-tensor theories. We first discuss the easiest example of a scalar field included in a gravity theory, that is a scalar field minimally coupled to the Einstein-Hilbert action. We then concentrate on scalar-tensor theory, where we explicitly examine the case of Brans-Dicke theory, and its generalization to Horndeski gravity. We conclude with a discussion on scalar Gauss-Bonnet gravity.
 
 We return on the concept of spontaneous scalarization in Chapter~\ref{Chap:spontaneousscalarization}.
 We first provide a detailed description of the process as a tachyonic instability at the linear level, and then analyze the contribution of non-linearities. 
 We then review the original formulation developed by Damour and Esposito-Far\`{e}se, and we present the more recent case of spontaneous scalarization in scalar Gauss-Bonnet gravity.

 In Chapter~\ref{Chap:scalarizationHordneski}, we present the generalization of this mechanism to the case of Horndeski gravity. We identify all of the terms in the action that can contribute to the tachyonic instability, and hence, can trigger scalarization.
 
 Having determined the minimal action that contains all of these terms, in Chapter~\ref{Chap:Threshold} we then investigate the threshold of this phenomenon in terms of the contributing coupling constants and the properties of the compact object.

 Chapter~\ref{Chap:blackholes} and~\ref{Chap:neutronstars} are dedicated to the study of the properties and the domain of existence of scalarized black holes and neutron stars, respectively, for a theory that has a coupling to both the Ricci and the Gauss-Bonnet scalar.

 Finally, we draw the conclusion of this thesis in Chapter~\ref{Chap:neutronstars}.


\chapter{Scalar-tensor theories of gravity}\label{Chap:sttheories}

 In Chapter~\ref{Chap:Introduction}, we have discussed the importance that scalar fields have in the framework of gravitational physics, while we also reviewed some of the most well-known applications of this type of field in several proposed theories. 
 
 In this Chapter, we present more accurately scalar-tensor theories and their generalization to Horndeski theory. This type of theory not only includes two kinds of fields, a tensor and a scalar, but the latter specifically enters the theory through a nontrivial term, a \textit{nonminimal coupling term}, as we will illustrate.

 \section{Minimally coupled scalar field}\label{Sec:minimallycoupled}
 
 Let us first concisely discuss the easiest example of a scalar field included into a gravitational theory: a scalar field $\varphi$ minimally coupled to the Einstein-Hilbert action. The theory takes the form
 \begin{equation}\label{eq:EinstHilb}
 S= \frac{1}{2\kappa} \int d^4x \sqrt{-g} \left[ R - \frac{1}{2} g^{\mu\nu} \nabla_\mu\varphi\,\nabla_\nu\varphi \right].
 \end{equation}
 Note that, for simplicity, we are neglecting any contribution from the matter sector. The negative sign of the kinetic term for $\varphi$ is important, since it guarantees that the field has a positive energy and hence it is not a ghost, that is a field whose negative norm cause lack of unitarity.
 
 Let us now clarify why the scalar field in action~\eqref{eq:EinstHilb} is said to be minimally coupled to gravity. The Einstein equivalence principle, which we mentioned in Chapter~\ref{Chap:Introduction}, has as its ultimate formulation that the space tangent to any point of the curved spacetime should be Minkowskian. Following the approach of~\cite{fujii_maeda_2003}, we call this the \textit{covariant equivalence principle}, to identify it as a theoretical development of the \textit{weak equivalence principle}, which states that test particles in a gravitational field, when no other forces are present, fall locally with a common acceleration. The ultimate equivalence principle provides a useful tool to derive a physical law in the presence of gravity from the one that holds in its absence. Starting from a theory defined in flat Minkowski spacetime, one applies the following substitution rule
 \begin{equation}
    \eta_{\mu\nu}\to g_{\mu\nu}, \qquad \partial_\mu\to\nabla_\mu.
 \end{equation}
 Thus, the kinetic term in action~\eqref{eq:EinstHilb} is obtained from $- \frac{1}{2}\eta^{\mu\nu} \partial_\mu\varphi\,\partial_\nu\varphi$, and the scalar field $\varphi$ couples to gravity only through $\sqrt{-g}\,g_{\mu\nu}$. A gravitational coupling obtained by applying this ``minimum'' rule is called a minimal coupling. Conversely, a term is said to be ``nonminimally'' coupled when it cannot be retrieved by applying this rule, so that in Minkowski spacetime such term simply vanishes, as we will see in the next Section.

 Lastly, the field equations for the theory described by action~\eqref{eq:EinstHilb} are
 \begin{subequations}
 \begin{equation}\label{eq:einstScal}
    G_{\mu\nu}= \nabla_\mu\varphi\,\nabla_\nu\varphi-\frac{1}{2}g_{\mu\nu}\nabla_\lambda\varphi\,\nabla^\lambda\varphi,
 \end{equation}
 \begin{equation}
    \DAlembert{\varphi}=0.
 \end{equation}
 \end{subequations}
 Focusing on Eq.~\eqref{eq:einstScal}, if we identify the contribution from the scalar field in a single tensor, that is we define $T^\varphi_{\mu\nu}\equiv\nabla_\mu\varphi\,\nabla_\nu\varphi-\frac{1}{2}g_{\mu\nu}\nabla_\lambda\varphi\,\nabla^\lambda\varphi$, then Eq.~\eqref{eq:einstScal} simply becomes the Einstein field equation where $T^\varphi_{\mu\nu}$ plays the role of the scalar field stress-energy tensor. Thus, we can effectively interpret the contribution of a minimally coupled scalar field as coming from the matter sector, and rewrite action~\eqref{eq:EinstHilb} as the Einstein-Hilbert action plus the massless Klein-Gordon action for the field $\varphi$ as the matter action. 
 
\section{Scalar-tensor theories}\label{Sec:scalartensor}
 \subsection{Scalar-tensor theories as effective field theories}\label{Sub:steffective}
 
 In Section~\ref{Sec:minimallycoupled}, we presented a theory with a minimal coupling between the metric tensor and the scalar field; in the context of scalar-tensor theories, however, the scalar field is introduced through a nonminimal coupling term. An easy way to obtain such theory is to start from action~\eqref{eq:EinstHilb} and promote the coefficients of each terms to generic functions of the scalar field $\varphi$. This approach can be motivated from an effective field theory perspective. Typically, in such theories one performs a dimensional analysis and treats the action as an expansion in momenta. Action~\eqref{eq:EinstHilb}, with the addition of a scalar potential $V(\varphi)$, can then be interpreted as the lowest order of a generic effective field theory, in which terms with higher derivatives are suppressed by some mass scale. One can then generalize the action by turning the coefficients to each terms into functions of the scalar field itself, the result is
 \begin{multline}\label{eq:ScalTens}
 S= \frac{1}{2\kappa} \int d^4x \sqrt{-g} \left[F(\varphi)R - \frac{1}{2}G(\varphi)g^{\mu\nu} \partial_\mu\varphi\,\partial_\nu\varphi+V(\varphi) \right]\\+S_\text{M}(g_{\mu\nu},\psi_\text{M}).
 \end{multline}
 Note that we also simplified our notation by substituting the covariant derivative of a scalar field with a partial one. Action~\eqref{eq:ScalTens} represents the most generic way to write a scalar-tensor theory and it allows us to emphasize the unique feature of these models encapsulated in the particular type of coupling between the scalar field and gravity. Indeed, following the prescription we gave in Section~\ref{Sec:minimallycoupled}, the first term on the right-hand side of~\eqref{eq:ScalTens} cannot be obtained by the ``minimum'' rule, and in flat Minkowski spacetime it just vanishes. Hence, the scalar field is nonminimal coupled to gravity.
 
 Historically, the nonminimal coupling term between the scalar field and the Ricci scalar was first introduced in the 1950s by Jordan~\cite{Jordan}, de facto marking the birth of scalar-tensor theories. This precursory model can be retrieved from action~\eqref{eq:ScalTens} by fixing $F(\varphi)=\varphi^\gamma$, $G(\varphi)=2\,\omega_J/\varphi^{2-\gamma}$ and $V(\varphi)=0$, where $\gamma$ and $\omega_J$ are some constants.
 
 In the next Section, we focus on a particular class of such theories, that is Brans-Dicke theory. In Section~\ref{Sub:conformal}, we return to the broader framework of scalar-tensor theories and we provide some further considerations on action~\eqref{eq:ScalTens}.
 
 \subsection{Brans-Dicke theory}\label{Sub:bransdicke}
 
 A specific class of theories that stems from action~\eqref{eq:ScalTens} is the so-called Brans-Dicke model. It is generally viewed as a prototype, since it is based on certain assumptions made for the sake of simplicity and, being oversimplified, it is generally not welcomed as entirely realistic. Inspired by the work done by Jordan~\cite{Jordan}, Brans and Dicke developed their theory in 1961~\cite{Brans:1961sx}, simplifying the choice for the coupling function $F(\varphi)$ by simply identifying it with the scalar field itself. The action proposed in their work is
 \begin{equation}\label{eq:BransDicke}
    S_\text{BD}= \frac{1}{2\kappa}\int d^4x\sqrt{-g} \left[ \varphi R - \frac{\omega_0}{\varphi}g^{\mu\nu} \partial_\mu\varphi\,\partial_\nu\varphi \right]+S_\text{M}(g_{\mu\nu},\psi_\text{M}),
 \end{equation}
 where $\omega_0$ is the Brans-Dicke parameter. Note that the Brans-Dicke action can be retrieved from~\eqref{eq:ScalTens} by choosing $F(\varphi)=\varphi$, $G(\varphi)=2\,\omega_0/\varphi$ and a vanishing potential, that is it corresponds to the Jordan model with $\gamma=1$.
 
 By comparing the first right-hand side term of action~\eqref{eq:BransDicke} with the Einstein-Hilbert term, one can notice how, instead of being characterized by a gravitational ``constant'', the Brans-Dicke model has an effective gravitational constant defined by $G_\text{eff}=\varphi/2\kappa$ as long as the dynamical field $\varphi$ varies slowly. In particular, in a cosmological framework, the scalar field is a function of the cosmic time to a first approximation, allowing us to consider $\varphi$ to be spatially uniform. Nevertheless, if there is in fact a time dependence of $G_\text{eff}$, it should be of order of $10^{-10}$ or less in a year~\cite{fujii_maeda_2003}, thus the time variability can be significant only at cosmological time scales.

 Focusing on the second term on the right-hand side of the Brans-Dicke action, we emphasize that this kinetic term has an unusual form. Indeed, on one side the presence of a pole at $\varphi=0$ suggest the appearance of a singularity, on the other there is a multiplying parameter $\omega_0$. These peculiarities, however, can both be reabsorbed by a redefinition of the scalar field and the kinetic term can be cast into the standard canonical form. Indeed, performing the transformation 
 \begin{equation}
 \varphi\to\varphi=\frac{1}{2} \xi \Phi^2,
 \end{equation}
 with $\xi^{-1}=4\omega_0$, then the kinetic term of action~\eqref{eq:BransDicke} can be rewritten as
 \begin{equation}
 -\frac{1}{2}g^{\mu\nu} \partial_\mu\Phi\,\partial_\nu\Phi,
 \end{equation}
 at the price of now having a quadratic coupling between the scalar field and the Ricci scalar~\cite{fujii_maeda_2003}.
 
 We now highlight an important feature of the Brans-Dicke action. Let us start by considering the weak equivalence principle. This law is based on the empirical fact $m_I=m_G$, where $m_I$ and $m_G$ are respectively the inertial and the gravitational mass\footnote{The proper relation between the inertial and the gravitational masses is $m_I/m_G=\text{const}$, however by choosing adequate units of measurement this ratio can be set equal to unity.}, nowadays tested with a precision of $1$ part in $10^{13}$~\cite{Schlamminger:2007ht}. It is then natural to demand that this equivalence must hold for the Brans-Dicke model as well. This is indeed guaranteed by the fact that the scalar field is not contained in the matter action, as it is manifest from Eq.~\eqref{eq:BransDicke}. This can be explained in a non-rigorous way by considering a point particle whose field equation, in general relativity, can be derived from the action
 \begin{equation}
     S_\text{M}=-m\int d\tau,
 \end{equation}
 where $\tau$ is the proper time and $m$ is the inertial mass of the point particle itself. The latter only appears as an overall factor, thus it does not affect the trajectory in spacetime of the point particle. As a consequence, the universal free-fall stated in weak equivalence principle is satisfied. If the scalar field appeared in the matter action, however, the inertial mass would have a dependence on $\varphi$, i.e. $m(\varphi)$, and it would be impossible to factor it out of the integral, thus violating the weak equivalence principle. Note, however, that, due the presence of a dynamical scalar field, the Brans-Dicke theory, and more in general scalar-tensor theories, violates the \textit{strong equivalence principle}, a generalization of the weak equivalence principle to self-gravitating bodies as well as test bodies. The reason for this lies in the dependency of the effective gravitational constant on the scalar field. In fact, when a self-gravitating object moves in a region of the spacetime where $\varphi$ is not constant, its internal gravitational energy, and consequently its total mass-energy, varies~\cite{Berti:2015itd}.

 After these considerations, we derive the field equations from action~\eqref{eq:BransDicke}. The result is
 \begin{subequations}
 \begin{equation}\label{eq:BDgrav}
    G_{\mu\nu}=\frac{\kappa}{\varphi}T_{\mu\nu}+\frac{\omega_0}{\varphi^2}\left( \partial_\mu\varphi\partial_\nu\varphi-\frac{1}{2}g_{\mu\nu}\partial_\lambda\varphi\partial^\lambda\varphi \right)-\frac{1}{\varphi}(g_{\mu\nu}\Box\varphi-\nabla_\mu\partial_\nu\varphi),
 \end{equation}
 \begin{equation}
    \DAlembert{\varphi}=\frac{1}{2\varphi}\partial_\lambda\varphi\partial^\lambda\varphi-\frac{\varphi}{2\,\omega_0}R,
 \end{equation}
 \begin{equation}
    \nabla_\mu T^{\mu\nu}=0.
 \end{equation}
 \end{subequations}
 Taking the trace of Eq.~\eqref{eq:BDgrav}, we can write the Ricci scalar in terms of $\varphi$ and the stress energy tensor, which allows us to rewrite the scalar field equation as
 \begin{equation}\label{eq:BDscal}
     \DAlembert{\varphi}=\zeta \kappa T,
 \end{equation}
 where we have defined $\zeta^{-1}=3+2\,\omega_0$. Eq.~\eqref{eq:BDscal} is particularly insightful. First of all, it shows that the scalar field is sourced from the stress-energy tensor. This might seem contradictory, given the assumption that the scalar field is not present in the matter action. However, the underlying reason for this ``mixing mechanism'' lies in the nonminimal coupling term, since it provides an interaction between the scalar field and the metric tensor. Therefore, the decoupling in the matter action imposes a restriction on how $\varphi$ couples to matter, rather than forcing $\varphi$ to not be related to the matter fields.
 
 Secondly, Eq.~\eqref{eq:BDscal} indicates that the scalar field mediates a long-range force between massive objects, similarly to the Newtonian force in the weak field limit of general relativity. However, imposing that the theory is in accordance with experimental result leads to a heavy constraint on the Brans-Dicke parameter. Indeed, from the Cassini experiment, today we have $\omega_0 > 4 \times 10^4$~\cite{Scharer:2014kya}. This was an initial problem for the validity of the theory. However, considering a scalar field potential, {\em e.g.} introducing a mass term, one can bypass the issue of the value of the Brans-Dicke parameter. Indeed, if the field acquires a nonzero mass, so that the corresponding force-range of the scalar force turns out to be smaller than the size of the solar system, it no longer affects physical phenomena well explained by general relativity, such as the perihelion advance of Mercury. In this case, the constraint mentioned before can be evaded.
 
 Lastly, note that the coupling in Eq.~\eqref{eq:BDscal} vanishes if $\omega\to\infty$. Thus, in this limit, the theory reduces to general relativity with a constant scalar field playing the role of a cosmological constant.
 
 \subsection{Conformal transformations and field redefinitions}\label{Sub:conformal}
 
 We conclude this Section with a further discussion on action~\eqref{eq:ScalTens}. In defining the action for scalar-tensor theories, we intentionally wrote it in the most generic way possible. However, in the literature, the coupling term between the Ricci scalar and the scalar field is typically identified as being simply $\varphi R$. This can always be achieved starting from action~\eqref{eq:ScalTens} by performing a redefinition of the scalar field, without losing generality. Indeed, let us consider the following redefinition
 \begin{equation}\label{eq:fieldRescale}
 F(\varphi)=\phi.
 \end{equation}
 Note that we are here improperly using the connotation ``field redefinition'', since the mapping in Eq.~\eqref{eq:fieldRescale} is not a diffeomorphism. Indeed, we are effectively narrowing the space of solutions for the scalar field to that of only positive values of $\phi$. Thus, a mapping like the one in Eq.~\eqref{eq:fieldRescale} becomes an actual field redefinition, provided that ones suitably restrict the domain and codomain of the map. Throughout the work of this thesis, to streamline the presentation, we do not explicitly specify the relevant domains and codomains of each field redefinitions, which is the standard practice in the literature.\footnote{Note that the generalized theory studied in this thesis, defined in Eq.~\eqref{eq:ACI}, is invariant under sign redefinition of the scalar field. Thus, even though performing a field redefinition on the theory one ``selects'' only half the space of solutions, we are not neglecting any possible scenario.}

 Using Eq.~\eqref{eq:fieldRescale}, we can rewrite action~\eqref{eq:ScalTens} as
 \begin{equation}\label{eq:Jordanframe}
 S= \frac{1}{2\kappa} \int d^4x \sqrt{-g} \left[\phi R - \frac{\omega(\phi)}{\phi}g^{\mu\nu} \partial_\mu\phi\,\partial_\nu\phi+V(\phi) \right]+S_\text{M}(g_{\mu\nu},\psi_\text{M}),
 \end{equation}
 where $V(\phi)$ is the potential rewritten in terms of the new scalar field $\phi$ and we have also defined
 \begin{equation}
     \omega(\phi)=\frac{\phi\,G(\varphi)}{2F'(\varphi)^2},
 \end{equation}
 so that action~\eqref{eq:Jordanframe} can be seen as a generalization of the Brans-Dicke action, namely Eq.~\eqref{eq:BransDicke}, where $\omega_0\to\omega(\phi)$. Eq~\eqref{eq:Jordanframe} is the conventional way to write the action of scalar-tensor theory in the so-called \textit{Jordan frame} and $g_{\mu\nu}$ is identified as the ``Jordan frame metric''. Such frame is defined as that where the coupling between the scalar field and gravity enters through the $\phi R$ term and where the scalar field does not appear in the matter action, which we remind is a necessary assumption to preserve the weak equivalence principle. 
 
 The field equations for action~\eqref{eq:Jordanframe} are
 \begin{subequations}
 \allowdisplaybreaks
 \begin{align}
    G_{\mu\nu}=&\frac{\kappa}{\phi}T_{\mu\nu}+\frac{\omega(\phi)}{\phi^2}\left( \partial_\mu\phi\partial_\nu\phi-\frac{1}{2}g_{\mu\nu}\partial_\lambda\phi\partial^\lambda\phi \right)\\
    \notag
    &-\frac{1}{\phi}(g_{\mu\nu}\Box\phi-\nabla_\mu\partial_\nu\phi) +\frac{1}{2\,\phi}g_{\mu\nu}V(\phi),\\
    \DAlembert{\phi}=& \frac{1}{2\phi}\partial_\lambda\phi\partial^\lambda\phi-\frac{\phi\,V'(\phi)}{2\omega(\phi)}-\frac{\phi\,R}{2\,\omega(\phi)}-\frac{\omega'(\phi)}{2\,\omega(\phi)}\partial_\lambda\phi\partial^\lambda\phi.
 \end{align}
 \end{subequations}
 
 It is often convenient to rewrite action~\eqref{eq:Jordanframe} in a different frame. In order to do so one has to perform what is called a \textit{conformal transformation}. Let us first introduce the concept of conformal transformation. Given a smooth $D$-dimensional manifold $M$, we consider two Lorentzian metrics $g_{\mu\nu}$ and $\tilde{g}_{\mu\nu}$, which define two spacetimes $(M, g_{\mu\nu})$ and $(M, \tilde{g}_{\mu\nu})$. A conformal transformation from the metric $g_{\mu\nu}$ to $\tilde{g}_{\mu\nu}$ is defined as
 \begin{equation}\label{eq:confTrasf}
 \tilde{g}_{\mu\nu}(x)=\Omega^2(x)g_{\mu\nu},
 \end{equation}
 where $\Omega$ is called the \textit{conformal factor} and it is a smooth, non vanishing spacetime function. Performing a conformal transformation on the metric tensor is de facto a rescaling of the metric itself. It allows to switch from a reference frame to another by shrinking or stretching the distances between two points described by the same coordinate system on the manifold, but preserving the angles between vectors, thus preserving the global causal structure of the manifold. Indeed, the definition in Eq.~\eqref{eq:confTrasf} is equivalent to apply the transformation directly to the line element, that is
 \begin{equation}\label{eq:lineElementconf}
     d\tilde{s}^2=\Omega^2 ds^2.
 \end{equation}
 Note that a conformal transformation changes distances by a rate that differs from point to point on the space-time manifold and this is done isotropically in a four-dimensional space, i.e. it changes spatial distances and time interval at the same rate. Moreover, Eq.~\eqref{eq:lineElementconf} shows that a conformal transformation is different from a coordinate transformation, since the latter leaves the line element invariant. Naturally, a conformal transformation affects also the geometric quantities related to the metric and the matter terms, such as the stress-energy tensor. The corresponding transformation rules are given in Appendix~\ref{App:conformal}.
 
 The two metrics $g_{\mu\nu}$ and $\tilde{g}_{\mu\nu}$ are called conformally equivalent. Nevertheless, it has to be stressed that the field equations that describe the same physical phenomena in the two conformal frames are not formally equivalent. A simple proof of this statement is that the conservation law does not generally hold under a conformal transformation. Indeed, let us assume that in the frame described by the metric $g_{\mu\nu}$ the conservation law stands, that is
 \begin{equation}\label{eq:consLaw}
    \nabla_\mu T^{\mu\nu}=0.
 \end{equation}
 Making use of the relations provided in Appendix~\ref{App:conformal}, Eq.~\eqref{eq:consLaw} can be rewritten in the conformal frame as
 \begin{equation}\label{eq:consLawTransf}
    \tilde{\nabla}_\mu \tilde{T}^{\mu\nu}=-\frac{\partial^\nu\Omega}{\Omega}\tilde{T}.
 \end{equation}
 As previously anticipated, from Eq.~\eqref{eq:consLawTransf} it is clear that, if the stress-energy tensor is conserved in the frame described by the metric $g_{\mu\nu}$, it is still conserved in the frame with the metric $\tilde{g}_{\mu\nu}$ only if its trace is null, a condition that is not valid for all kind of matter.
 
 Let us now perform the following conformal transformation on the Jordan metric of action~\eqref{eq:Jordanframe}
 \begin{equation}
     g_{\mu\nu}\to \tilde{g}_{\mu\nu}=\phi g_{\mu\nu},
 \end{equation}
 where we have defined the conformal factor as $\Omega=\sqrt{\phi}$. The Jordan frame action can thus be rewritten as
 \begin{equation}\label{eq:Intermediate}
 \begin{split}
    S= & \frac{1}{2\kappa}\int d^4x \sqrt{-\tilde{g}} \left[ \tilde{R}- \frac{3}{2}\frac{\partial_\lambda\phi\partial^\lambda\phi}{\phi^2}-\frac{\omega(\phi)}{2}\frac{\partial_\lambda\phi\partial^\lambda\phi}{\phi^2} +\frac{V(\phi)}{\phi^2} \right]\\
    &+S_\text{M}(\phi^{-1}g_{\mu\nu},\psi_\text{M})
 \end{split}
 \end{equation}
 If we further perform a field redefinition $\phi\to\tilde{\phi}$ defined as~\cite{Palenzuela:2013hsa}
 \begin{equation}\label{eq:fieldRedef}
 \left(\frac{ \text{d log}\,\phi}{\text{d}\tilde{\phi}}\right)^2=\frac{2\kappa}{3+2\,\omega(\phi)},
 \end{equation}
 up to integration constants, action~\eqref{eq:Intermediate} can be rewritten as 
 \begin{equation}\label{eq:EinsteinFrame}
    S= \int d^4x \sqrt{-\tilde{g}} \left[ \frac{\tilde{R}}{2\kappa}- \frac{1}{2}\partial_\lambda\tilde{\phi}\partial^\lambda\tilde{\phi}+U(\tilde{\phi}) \right]
    +S_\text{M}\left(\frac{\tilde{g}_{\mu\nu}}{\phi(\tilde{\phi})},\psi_\text{M}\right),
 \end{equation}
 where we have defined the new potential as
 \begin{equation}
     U(\tilde{\phi})=\frac{V(\phi)}{\phi^2}.
 \end{equation}
 Action~\eqref{eq:EinsteinFrame} represents scalar-tensor theory in the so-called \textit{Einstein frame}. It is often useful to perform calculation in such frame since the gravity action is formally equivalent to that of a scalar field minimally coupled to gravity, i.e. Eq.~\eqref{eq:EinstHilb}, with the addition of a scalar field potential. However, the ``price'' to pay is the introduction of a nonminimal coupling between the scalar field and the matter fields. Indeed, as a consequence of the conformal transformation, now the scalar field appears in the matter action through the conformal factor. Thus, in the Einstein frame the conservation law of the stress-energy tensor does not hold, which means that the weak equivalence principle does not hold in this frame. Matter, then, does not follow the geodesics of the Einstein metric $\tilde{g}_{\mu\nu}$, but instead those of the Jordan metric or a conformally equivalent metric that does not introduce a coupling between matter and the scalar field.
 
 The field equations in the Einstein frame read as
 \begin{subequations}
 \begin{align}
    &\tilde{G}_{\mu\nu}=\kappa\tilde{T}_{\mu\nu}+\kappa\left( \partial_\mu\tilde{\phi}\partial_\nu\tilde{\phi} -\frac{1}{2}\partial_\lambda\tilde{\phi}\partial^\lambda\tilde{\phi}\right)+\kappa\,\tilde{g}_{\mu\nu}U(\tilde{\phi}),\\
    \label{eq:ScalEinstein}
    &\tilde{\Box}\tilde{\phi}=\frac{1}{2}\frac{ \text{d log}\,\phi}{\text{d}\tilde{\phi}}\tilde{T}-U'(\tilde{\phi}).
 \end{align}
 \end{subequations}
 
 We conclude this Section with one last final remark. Actions~\eqref{eq:Jordanframe} and~\eqref{eq:EinsteinFrame} are simply different representations of the same theory. There is nothing exceptional about the Jordan or the Einstein frame, and one can actually find infinitely many conformal frames~\cite{Flanagan:2004bz,Sotiriou:2007zu}.

\section{Horndeski gravity}\label{Sec:Horndeski}

 In Section~\ref{Sec:scalartensor}, we derived the action for scalar-tensor theories by considering only the lowest order terms of an expansion in the scalar field derivatives of a generic effective field theory, i.e. suppressing higher order derivatives of the scalar field. However, we can relax this assumption, and still be able to construct an effective action while controlling the number of degrees of freedom of the theory by requiring that the theory yields second order field equations.
 
 This approach has been used in~\cite{Deffayet:2011gz}, where they obtained the most general scalar field theory with an action that depends on second order derivatives or less, but still has second order field equations for both the metric and the scalar field. They thus provided the most general extension to Galileons theories, mentioned in Chapter~\ref{Chap:Introduction}. It was shown in~\cite{Kobayashi:2011nu} that the action for generalized Galileon can be mapped to that of Horndeski action, proving that the two theories are equivalent. Horndeski theory was first formulated in 1974~\cite{Horndeski:1974wa}, under the requirements of a four-dimension theory constructed from the metric $g_{\mu\nu}$ and a scalar field $\phi$, and their derivatives $\partial g_{\mu\nu}$, $\partial^2 g_{\mu\nu}$, $\partial^3 g_{\mu\nu}$, $...$, $\partial \phi$, $\partial^2 \phi$, $\partial^3 \phi$, $...$ but still having second-order field equations. Note that the assumptions made by Horndeski are weaker than those of~\cite{Deffayet:2011gz}. Indeed, he allows for the action to contain higher order derivatives, this, however, still leads to the same result as~\cite{Deffayet:2011gz}.
 
 \subsection{Action and field equations}\label{Sub:2.3.1}
 
 Let us now present Horndeski theory. As we have already mentioned, the action of the theory is obtained by considering all of the possible operators constructed with derivatives of the metric tensor and the scalar field which still leads to second order field equations. Each of these operators is coupled to generic functions $G_i$ of the scalar field $\phi$, with $i=2,...5$, and a dynamical term defined as
 \begin{equation}\label{eq:X}
    X= -\frac{1}{2}\partial_\mu\phi\partial^\mu\phi. 
 \end{equation}
 Following the notation of~\cite{Kobayashi:2011nu}, the action for the theory can then be written as
 \begin{equation}\label{Horndeski}
 S=\frac{1}{2\kappa}\sum_{i=2}^{5}\int \dd^4x\,\sqrt{-g}\mathcal{L}_i+S_{\note{M}},
 \end{equation}
 where we have defined
 \begin{subequations}\label{eq:LagrHorn}
 \begin{align}
 \label{L2}
 \mathcal{L}_2 = & \, G_2(\phi,X),\\
 \label{L3}
 \mathcal{L}_3 = & \, -G_3(\phi,X)\DAlembert\phi,\\
 \label{L4}
 \mathcal{L}_4 = & \, G_4(\phi,X)R+G_{4X}[(\DAlembert\phi)^2-(\nabla_\mu\nabla_\nu\phi)^2],\\
 \notag
 \mathcal{L}_5 = & \, G_5(\phi,X)G_{\mu\nu}\nabla^\mu\nabla^\nu\phi \\
 \label{L5}
 - & \frac{G_{5X}}{6}\left[\left(\DAlembert\phi\right)^3 -3\DAlembert\phi(\nabla_\mu\nabla_\nu\phi)^2+2(\nabla_\mu\nabla_\nu\phi)^3\right],
 \end{align}
 \end{subequations}
 and $G_{iX}=\partial G_{i}/\partial X$. Note that we have used the shorthanded forms $(\nabla_\mu\nabla_\nu\phi)^2=\nabla_\mu\nabla_\nu\phi\nabla^\mu\nabla^\nu\phi$, $(\nabla_\mu\nabla_\nu\phi)^3= \nabla_\mu\nabla_\nu\phi\nabla^\nu\nabla^\lambda\phi\nabla_\lambda\nabla^\mu\phi$, in the expressions for $\mathcal{L}_i$. In principle, the second derivative terms in the action would lead to third order derivative in the equations. However, the action is constructed in such a way that these terms would cancel identically, leaving ghost-free second order equations. Furthermore, the nonminimal coupling to gravity in $\mathcal{L}_4$ and $\mathcal{L}_5$ are essential to eliminate higher order derivatives that would otherwise appear in the field equations. 
 
 We stress that we are presenting the theory in the so-called Jordan frame, since in action~\eqref{Horndeski} matter is assumed to couple minimally to the metric only.
 
 Scalar-tensor theory studied in Section~\ref{Sec:scalartensor} can clearly be retrieved as a particular case of Horndeski theory by choosing
 \begin{equation}
    G_2=2\frac{\omega(\phi)}{\phi}X+V(\phi), \qquad G_4=\phi, \qquad G_3=G_5=0.
 \end{equation}
 
 Lastly, varying the action with respect to the metric $g^{\mu\nu}$ and the scalar field $\phi$ yields respectively
 \begin{align}
 \label{HornEq1}
 &\sum_{i=2}^{5}\mathcal{G}^{i}_{\mu\nu} =\kappa T_{\mu\nu}, \\
 \label{HornEq2}
 &\sum_{i=2}^{5}\left( P^i_\phi -\nabla^\mu J^i_\mu \right) =0.
 \end{align}
 See Appendix~\ref{App:Horndeski} for the definition of $\mathcal{G}^{i}_{\mu\nu}$, $P^i_\phi$ and $J^i_\nu$ and for the explicit form of Eq.~\eqref{HornEq2}.

 \subsection{Disformal transformations}\label{Sub:2.3.2}
 
 It is natural to ask whether one can generalize the concept of conformal transformation to the case of Horndeski theory and if this can offer more insights into this type of models. In fact, it has been shown that action~\eqref{Horndeski} is formally invariant under the following transformation~\cite{Bettoni:2013diz}
 \begin{equation}\label{eq:disformal}
 g_{\mu\nu} \rightarrow C(\phi)\left[g_{\mu\nu}+D(\phi)\nabla_\mu\phi\nabla_\nu\phi\right].
 \end{equation}
 This type of transformation is known as \textit{special disformal transformation}. The functions $C(\phi)$ and $D(\phi)$ are free functions of the scalar field. Note that, when $D=0$, Eq.~\eqref{eq:disformal} reduces to a conformal transformation, whereas for $C=1$ one has a so-called \textit{purely disformal} transformation.
 
 Disformal transformations were first introduced by Bekenstein in a more general form in which $C$ and $D$ are also allowed to depend on $X$~\cite{Bekenstein:1992pj}, such that
 \begin{equation}\label{eq:disformalGeneral}
 \tilde{g}_{\mu\nu}=C(X,\phi)\left[g_{\mu\nu}+D(X,\phi)\nabla_\mu\phi\nabla_\nu\phi\right].
 \end{equation}
 While a conformal transformation can be seen as a change of local units of length, the disformal transformation can be interpreted as a change of local units of length for which the units for intervals along the gradient of $\phi$ are different than those for intervals orthogonal to it~\cite{Bekenstein:1992pj}. Indeed, disformal transformations can distort the causal structure of the spacetime. The line elements are related by $d\tilde{s}^2=C ds^2+D(\nabla_\mu\phi\, dx^\mu)^2$, so that, assuming $C>0$, a 4-vector which is null with respect to $g_{\mu\nu}$ will be space-like or time-like with respect to $\tilde{g}_{\mu\nu}$ depending on whether $D$ is positive or negative locally~\cite{Zumalacarregui:2013pma}. 
 
 In the already mentioned work of~\cite{Bettoni:2013diz}, it was shown that performing a general disformal transformation of the type of~\eqref{eq:disformalGeneral} on the Horndeski action introduces terms that cannot be expressed in the form~\eqref{eq:LagrHorn}. The action is formally invariant only under the special transformation, i.e. Eq.~\eqref{eq:disformal}. Nevertheless, the transformation of Eq.~\eqref{eq:disformalGeneral} still provides a powerful tool to construct new gravitational theories based on pairs of disformally related geometries. Indeed, a particular class of theories, the so-called \textit{degenerate higher-order scalar-tensor theories}, or DHOST theories, is invariant under the general disformal transformation~\cite{BenAchour:2016cay}. They were originally identified in~\cite{Langlois:2015cwa}, as a broad class of theories that due to the degeneracy of their Lagrangian are able to avoid the Ostrogradski's theorem, for which nondegenerate Lagrangians with higher order derivatives lead to ghost-like instabilities, also known as Ostrogradski instabilities~\cite{Ostrogradsky:1850fid,Woodard:2006nt,Woodard:2015zca}. A degenerate theory is identified by a Lagrangian that, after introducing auxiliary variables to replace the second order time derivatives by first order time derivatives, has a degenerate kinetic matrix, which is composed by the coefficients of the kinetic terms. They have later been systematically classified up to cubic order in~\cite{BenAchour:2016fzp}.
 
 Finally, we stress that DHOST theories are an extension of the Horndeski theory, as well as Beyond Horndeski theory, a generalization of Horndeski theories that leads to higher order field equations. Horndeski theory can thus be seen as a particular subclass of DHOST. 
 
\section{Scalar Gauss-Bonnet gravity}\label{Sec:GaussBonnet}
 
 We conclude this Chapter by considering a specific class of scalar-tensor theories: scalar Gauss-Bonnet gravity. This model is characterized by a nonminimal coupling between the scalar field and gravity through the Gauss-Bonnet term
 \begin{equation}\label{eq:GaussBonnet}
    \mathscr{G} =R^2 -4R_{\mu\nu} R^{\mu\nu} +R_{\mu\nu\rho\sigma}R^{\mu\nu\rho\sigma}
 \end{equation}
 The action then contains quadratic contractions of the Riemann and Ricci tensors.
 Note that in four-dimension $\mathscr{G}$ is a topological invariant, thus including this term in the Einstein-Hilbert action can introduce modification to general relativity only when the Gauss-Bonnet invariant is coupled to a nonzero scalar field. This quadratic gravity model can be seen as a low energy effective string theory~\cite{Campbell:1991kz,Boulware:1985wk,Green:2012oqa,Green:2012pqa,Burgess:2003jk}.

 \subsection{Coupling with the Gauss-Bonnet invariant}\label{Sub:2.4.1}
 
 Let us present the action for scalar Gauss-Bonnet gravity
 \begin{equation}\label{eq:sGBaction}
 S= \frac{1}{2\kappa} \int d^4x \sqrt{-g} \left[ R - \frac{1}{2}g^{\mu\nu} \partial_\mu\phi\,\partial_\nu\phi+f(\phi) \mathscr{G}\right]+S_\text{M}(g_{\mu\nu},\psi_\text{M}),
 \end{equation}
 where $f(\phi)$ is a generic coupling function. In literature, typical choices for the coupling functions are
 \begin{itemize}
    \item $f=\alpha e^\phi$, with $\alpha$ as a coupling constant, corresponding to Einstein-dilaton-Gauss-Bonnet gravity~\cite{Mignemi:1992nt,Kanti:1995vq},
    \item $f=\sigma \phi$, with $\sigma$ as a coupling constant, referred to as shift-symmetric scalar Gauss-Bonnet gravity~\cite{Sotiriou:2013qea},
    \item $f=\lambda \phi^2/2$, with $\lambda$ as a coupling constant, labeled as quadratic scalar Gauss-Bonnet gravity~\cite{Silva:2017uqg,Doneva:2017bvd,Antoniou:2017acq}.
 \end{itemize}
 
 Maintaining an implicit expression for the coupling function, the field equations for action~\eqref{eq:sGBaction} are
 \begin{align}
    \label{eq:sGBgrav}
    & G_{\mu\nu}=\kappa T_{\mu\nu}+ T^\phi_{\mu\nu} \\
    \label{eq:sGBscal}
    & \DAlembert{\phi}=-f'(\phi)\,\mathscr{G},
 \end{align}
 where we have identified a stress-energy tensor for the scalar field defined as~\cite{Kanti:1995vq}
 \begin{equation}
    T^\phi_{\mu\nu}=\frac{1}{2}\partial_\mu\phi\,\partial_\nu\phi-\frac{1}{4}g_{\mu\nu}\partial_\lambda\phi\,\partial_\lambda\phi-\frac{1}{g} g_{\mu(\rho}g_{\lambda)\nu}\epsilon^{\kappa\lambda\alpha\beta}\epsilon^{\rho\gamma\sigma\tau}R_{\sigma\tau\alpha\beta}\nabla_\gamma\partial_\kappa f,
 \end{equation}
 where $\epsilon^{\kappa\lambda\alpha\beta}$ is the Levi-Civita tensor.

 \subsection{The Gauss-Bonnet term in Horndeski gravity}\label{Sub:2.4.2}
 
 From Eqs.~\eqref{eq:sGBgrav}-\eqref{eq:sGBscal}, one can clearly see that scalar Gauss-Bonnet gravity leads to second order field equations. It must then belong to the Horndeski class. In fact, even if action~\eqref{eq:sGBaction} is not manifestly in the Horndeski form presented in Section~\ref{Sec:Horndeski}, it has been shown that the nonminimal coupling between the Gauss-Bonnet invariant and $f(\phi)$ can be reproduce in the formalism of action~\eqref{eq:LagrHorn} by taking~\cite{Kobayashi:2011nu}
 \begin{equation}
 \begin{aligned}\label{G_GB}
 G^{\note{GB}}_2= &8f^{(4)}X^2(3-\text{ln} X), \\
 G^{\note{GB}}_3= &4f^{(3)}X(7-3\text{ln} X), \\
 G^{\note{GB}}_4= &4f^{(2)}X(2-\text{ln} X), \\
 G^{\note{GB}}_5= &-4f^{(1)}\text{ln} X,
 \end{aligned}
 \end{equation}
 where, for simplicity, we defined $f^{(n)}\equiv\partial^n f/\partial\phi^n$. Note that the $G^{\note{GB}}_i$ functions of the Horndeski representation are nonanalytic in $X$; nevertheless, there is an analytic representation of the action, namely Eq.~\eqref{eq:LagrHorn}, and the field equations are analytic at $X=0$. This observation is crucial for the analysis that we perform in Chapter~\ref{Chap:scalarizationHordneski}.
 


\chapter[\texorpdfstring{Spontaneous scalarization of compact objects}{Spontaneous scalarization of compact objects}
]{\chaptermark{Spontaneous scalarization} Spontaneous scalarization of compact objects}
\chaptermark{Spontaneous scalarization}
\label{Chap:spontaneousscalarization}

 In Chapter~\ref{Chap:Introduction}, we outlined the issue of scalar imprints at low energy. If they are in fact present as an additional degree of freedom, they must in some way be screened in regimes of weak gravity, where one expects to retrieve general relativity as the correct gravitational theory in accordance with observations. We then showed how screening mechanisms have been used in the literature in different context.
 
 Here, and throughout the thesis, we focus on a specific phenomenon, the so-called spontaneous scalarization. This mechanism provides an elegant way to dress a compact object with a scalar hair, while still admitting general relativity solutions far away from the object itself.
 
 We first provide a detailed analysis of the phenomenon in Section~\ref{Sec:spmechanism}. We then proceed to present two well-known examples of this mechanism, the first one for scalar-tensor theories, and the second one for scalar Gauss-Bonnet gravity.

\section{The spontaneous scalarization mechanism}\label{Sec:spmechanism}

 In Chapter~\ref{Chap:Introduction}, we briefly presented the phenomenon of spontaneous scalarization as a phase transition mechanism, where, when a specific threshold is crossed, general relativity solutions are no longer stable and scalarized solution are instead favoured. In this Section we aim to give a more thorough description of the phenomenon itself.

 \subsection{Tachyonic instability}\label{Sub:tachyonicinst}
 
 Let us first discuss the onset of the mechanism. We have mentioned that, at a linear level, spontaneous scalarization is triggered by a tachyonic instability. A tachyon is a wave degree of freedom whose frequency becomes imaginary due to a negative mass square. Indeed, given the Klein-Gordon field equation
 \begin{equation}\label{eq:KG}
    \DAlembert{\phi}-m^2\,\phi=0,
 \end{equation}
 and assuming $\phi\propto e^{-i\omega t+i k\cdot x}$, one can retrieve the dispersion relation for the scalar field, that is
 \begin{equation}
    \omega^2=k^2+m^2.
 \end{equation}
 For sufficiently low wave numbers $k$ and negative mass square, the frequency $\omega$ becomes imaginary. This leads to an exponential growth of the field and consequently to an instability. 
 
 For a generic theory to present this kind of instability, and hence being affected by spontaneous scalarization, the scalar field equation must then have the following behavior 
 \begin{equation}\label{eq:scalarGeneric}
    \DAlembert{\phi}+\mathcal{I}(g_{\mu\nu})\phi+\text{h.p.t.}=0,
 \end{equation}
 where $\mathcal{I}(g_{\mu\nu})$ is a generic spacetime invariant and in the last term we include all higher power terms that do not contribute at the linear level. Note that a crucial aspect of spontaneous scalarization is that it consists of a phase transition from general relativity branches of solutions to scalarized ones. Thus, the theory should first of all admits the former, that is Eq.~\eqref{eq:scalarGeneric} should be satisfied for a constant $\phi=\phi_0$. This requirement puts some model-dependent conditions on the theory itself.
 
 If one performs a linear analysis of Eq.~\eqref{eq:scalarGeneric} around a constant scalar field background, i.e. a general relativity background, the result is
 \begin{equation}\label{eq:tachyon}
    \DAlembert{\delta\phi}-m_\text{eff}^2\,\delta\phi=0,
 \end{equation}
 which is exactly Eq.~\eqref{eq:KG} for the scalar perturbation $\delta\phi$ and where we identified the spacetime invariant as an effective mass square, that is $m_\text{eff}^2=-\mathcal{I}(g_{\mu\nu})$. It is then clear that if in Eq.~\eqref{eq:tachyon} the sign of the mass like term has the same sign of the principal term, that is $m^2_\text{eff}<0$, then the scalar field is indeed affected by a tachyonic instability. 
 Note that we did not make any assumption on the nature of $\mathcal{I}(g_{\mu\nu})$ that constitutes the effective mass square, apart from requiring that it has the right sign to trigger the instability.
 
 As long as the theory presents the behavior of Eq.~\eqref{eq:tachyon}, it can develop spontaneous scalarization. In Section~\ref{Sec:def} and~\ref{Sec:sGB} we will present two specific cases. For the case of scalar Gauss-Bonnet it is easy to see that the invariant is in fact the Gauss-Bonnet term. For the Damour and Esposito-Far\`{e}se model it is less straightforward, since the theory is studied in the Einstein frame. Once one conformally transforms it back to the Jordan frame, linearizing the scalar field equation, it is clear that the Ricci scalar is the term that plays the role of the effective mass square.
 
 We point out that the instability is a purely linear effect and, furthermore, it is constrained to a specific region of the spacetime. Indeed, the invariant coupled to the scalar field appears as a constant only when one zooms in a small patch, realistically it is a scalar function of the spacetime itself. Thus, there might be some region of the spacetime where the effective mass square is non-negative and the instability is not developed. This is what makes spontaneous scalarization a screening mechanism. It allows the compact object to acquire a non-trivial scalar configuration only in certain parts of the spacetime, typically inside or close to the compact object, while the solutions will reduce to those of general relativity where experimental observations require it, {\em e.g.} far away from a black hole or a neutron star. Thanks to the existence of a scalarization threshold, the theory thus has two branches of solutions.
 
 As a last remark, we want to stress that in curved spacetime, the effective mass square being negative is a necessary condition for spontaneous scalarization, but not sufficient. The effect of the curvature can in fact stabilize the spacetime, and the instability is triggered only if a certain threshold is exceeded. This is not the case of Minkowski spacetime, where a negative mass square is already sufficient to trigger the instability.

 \subsection{The contribution of non-linearities}\label{Sub:nonlinearities}
 
 The spontaneous scalarization process is completed once one takes into consideration the nonlinearities of the system, which are all included in the third term of Eq.~\eqref{eq:scalarGeneric}. Indeed, if the nonlinearities are strong enough, they will eventually suppress the instability, leaving a stable configuration of the scalar field around the compact object.
 
 \begin{figure}[h!]
    \centering
	\includegraphics[width=.8\linewidth]{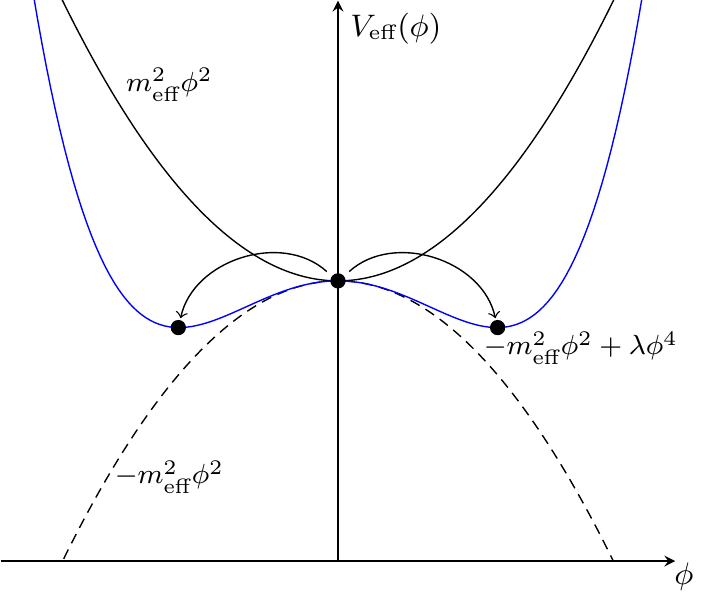}%
	\caption[Effective potential for a scalar field with both a positive and negative mass and non-linear contributions]{Effective potential for the scalar field in a local patch in the case of a positive mass square (solid black line) or a negative one (dashed black line). The black dot in the middle is a stable position for the $m^2 \phi^2$ potential, whereas it is an unstable maximum for the $-m^2 \phi^2$ potential. Once nonlinearities switch on (blue solid line), the tachyonic potential becomes bounded from below. The $\phi=0$ is still an unstable maximum, but two new minima appear, favouring the scalar field phase transition.}
	\label{fig:Potential}
 \end{figure}
 
 This behavior is illustrated in Fig.~\ref{fig:Potential}, where we present the effective potential felt by the scalar field in a local patch. The solid black line represents the case of a positive effective mass square, where it is clear that the potential has a stable minimum. However, when the effective mass square is negative, shown in a dashed black line, this stable minimum becomes an unstable maximum and the potential is not bounded from below. The field is then subject to an instability. In Fig.~\ref{fig:Potential} we also show the effect of nonlinearities. The solid blue line represent the potential with the addition of a quartic contribution. While a local unstable maximum in the origin is still present, the potential is now bounded from below, acquiring two new stable minima.
 
 This oversimplified example aims to simulate the phase transition of spontaneous scalarization. When the potential is quadratic and well-behaved general relativity solutions, i.e. solutions for which $\phi=0$, are stable. In the case of the quartic potential, general relativity solutions exist, but are unstable. The theory, however, still admits stable solutions which are scalarized, corresponding to the two stable minima in Fig.~\ref{fig:Potential}.
 
 Note that in the oversimplified example of Fig.~\ref{fig:Potential} we neglected the contribution of the curvature of the spacetime. As we mentioned in Section~\ref{Sub:tachyonicinst}, the curvature itself can stabilize the scalar field instability at the nonlinear level. However, these gravitational effects are not always enough to suppress other types of instabilities, such radial ones, and nonlinearities needs to be present in the scalar field potential in order to get fully stable solutions\cite{Macedo:2019sem,Silva:2018qhn,Minamitsuji:2018xde}.

\section{The Damour-Esposito-Far\`ese model}\label{Sec:def}

 The phenomenon of spontaneous scalarization was originally investigated by Damour and Esposito-Far\`{e}se in~\cite{Damour:1993hw}. They 
 discovered that, in the context of scalar-tensor theory, neutron star underwent a phase transition. They studied the theory in the Einstein frame in the case of a vanishing potential, that is Eq.~\eqref{eq:EinsteinFrame} with $U=0$. For convenience, we report here the scalar field equation for the theory, i.e. Eq~\eqref{eq:ScalEinstein},
 \begin{equation}\label{eq:DEF}
    \tilde{\Box}\tilde{\phi}=\alpha(\tilde{\phi})\tilde{T},
 \end{equation}
 where we have redefined
 \begin{equation}
    \alpha(\tilde{\phi})\equiv\frac{1}{2}\frac{ \text{d log}\,\phi}{\text{d}\tilde{\phi}}=\frac{1}{2}\sqrt{\frac{2\kappa}{3+2\,\omega(\phi)}}.
 \end{equation}
 We remind that a tilde identifies a quantity in the Einstein frame and that $\phi$ and $\tilde{\phi}$ are related through the field redefinition defined in Eq.~\eqref{eq:fieldRedef}. The specific choice taken in~\cite{Damour:1993hw} for $\omega(\phi)$ is 
 \begin{equation}\label{eq:omegaDEF}
    \omega_\text{DEF}(\phi)=-\frac{3}{2}-\frac{1}{4\,\beta\,\text{Log}\phi},
 \end{equation}
 which corresponds to the following field redefinition
 \begin{equation}\label{eq:DEFfield}
    \phi=e^{-\frac{\beta\kappa}{2}\tilde{\phi}^2}, 
 \end{equation}
 where $\beta$ is dimensionless parameter. Note that we rescaled the quantities of~\cite{Damour:1993hw} to our formalism. As a consequence of Eq.~\eqref{eq:DEFfield}, we can rewrite $\alpha(\tilde{\phi})=-\beta\kappa\,\tilde{\phi}/2$.
 
 The theory is covered by a no-hair theorem~\cite{Hawking:1972qk,Bekenstein:1971hc,Sotiriou:2011dz}, thus all scalar configurations for black holes are trivial. Indeed, observing Eq.~\eqref{eq:DEF}, it is clear that in vacuum, where $T=0$, the scalar field is not excited and all solutions are general relativity solutions with $\tilde{\phi}=\tilde{\phi}_0=\text{const}$. 
  
 In the presence of matter, however, the Einstein frame scalar field $\tilde{\phi}$ is coupled to the matter stress-energy tensor, while $\alpha(\tilde{\phi})$ plays the role of the effective coupling constant between the two. Hence, one can possibly retrieve other solutions different to those of general relativity. The stability of the latter depends on the behavior of the scalar perturbations $\delta\tilde{\phi}$, governed by the equations
 \begin{equation}
    \tilde{\Box}\delta\tilde{\phi}+\frac{1}{2}\beta\kappa\,\tilde{T}\delta\tilde{\phi}=0,
 \end{equation}
 The term $-\kappa\beta\tilde{T}/2$ plays the role of an effective mass $m^2_{eff}$ for the scalar perturbation. If we assume matter to be described by a perfect fluid, then we have $T^\textit{PF}_{\mu\nu}=(\epsilon+p)u_\mu u_\nu+p\, g_{\mu\nu}$, where $\epsilon$, $p$ and $u_\mu$ are respectively the energy density, the pressure and the 4-velocity of the fluid. We then have $T^\textit{PF}=3p-\epsilon$. For a highly compact neutron star, if $\epsilon>3p$, the trace of the stress-energy tensor becomes negative. Thus, if a negative $\beta$ yields a sufficiently negative effective mass square, it triggers a tachyonic instability for the scalar field. Therefore, a threshold exists below which scalarized solution are favoured with respect to those of general relativity, which are indeed unstable. Notably, a curved spacetime is destabilized only if such threshold is exceeded, while for a Minkowski spacetime any negative effective mass squared will cause an instability.

 As we have already emphasized in Section~\ref{Sec:spmechanism}, this instability, however, is a linear phenomenon. Solving the full system of equations, one can find a new stable solution with a non-trivial configuration of the scalar field surrounding the neutron star. This is possible thanks to the nonlinearities of the system, which can quench the instability, yielding a new branch of solutions. In Ref.~\cite{Damour:1993hw}, they found that scalarized solutions exist when $\beta\le -4.5$. This result was later improved by Ref.~\cite{Harada:1997mr}, where it was found $\beta\le -4.35$. Note that the exact value of $\beta$ associated with the threshold for the tachyonic instability depends on the choice of equation of state one uses for describing the neutron star model and on the initial value of the central energy density of the star.

\section{Scalarization with scalar Gauss-Bonnet gravity}\label{Sec:sGB}

 The limit of the mechanism originally proposed by Damour and Esposito-Far\`{e}se is its application only to neutron stars. As we mentioned in Section~\ref{Sec:def}, scalar-tensor theory is covered by a no-hair theorem~\cite{Hawking:1972qk,Bekenstein:1971hc,Sotiriou:2011dz}, thus all solutions in vacuum are those of general relativity. 
 
 It was recently shown, however, that scalar Gauss-Bonnet gravity, i.e. action~\eqref{eq:sGBaction}, can exhibit spontaneous scalarization for both black holes and neutron stars~\cite{Silva:2017uqg,Doneva:2017bvd,Antoniou:2017acq}. We report here the scalar field equation, that is Eq.\eqref{eq:sGBscal}, for the reader's convenience,
 \begin{equation}\label{eq:sGBscal2}
    \DAlembert{\phi}=-f'(\phi)\,\mathscr{G}.
 \end{equation}
 The theory does not admit $\phi=\text{const}$ solutions unless
 \begin{equation}\label{eq:sGBgrCond}
    f'(\phi_0)=0,
 \end{equation}
 for some constant $\phi_0$. Eq.~\eqref{eq:sGBgrCond} can be interpret as a condition for general relativity solutions to exist, we thus focus on theories that satisfy it. Note that such requirement discards two of the subclasses we presented in Chapter~\ref{Chap:sttheories}, namely Einstein-dilaton-Gauss-Bonnet gravity with $f\sim e^\phi$ and shift-symmetric scalar Gauss-Bonnet with $f\sim \phi$. 
 
 As for the previous model, studying the behavior of the scalar perturbation provides insights on the stability of general relativity solutions. Linearizing Eq.\eqref{eq:sGBscal2} around a fixed background solution to general relativity, namely $\phi=\phi_0$, gives
 \begin{equation}\label{eq:GBlin}
  \DAlembert\delta\phi+f_{,\phi\phi}(\phi_0)\mathscr{G}\delta\phi=0.
 \end{equation}
 Once again, we can interpret $-f_{,\phi\phi}\mathscr{G}$ as an effective mass square for the scalar perturbation. When such effective mass square is negative, i.e. when $f_{,\phi\phi}\mathscr{G}>0$, the theory can be affected by a tachyonic instability and the scalar field can grow exponentially. Remarkably, it was shown in Refs.~\cite{Silva:2017uqg,Antoniou:2017acq} that, in the case where the effective mass is positive, the general relativity configurations for this theory are unique, that is under such condition the theory is covered by a no-hair theorem. Hence, the condition on the effective mass arises naturally in the context of a no-hair theorem. Note that, in contrast with the Damour and Esposito-Far\`{e}se model, where spontaneous scalarization is induced by the presence of matter, in this case the scalarization is caused by the curvature of the spacetime itself.

 In Ref.~\cite{Silva:2017uqg}, they focused on quadratic scalar Gauss-Bonnet gravity choosing the specific coupling function $f=\eta\phi^2/8$. They first solved~\eqref{eq:GBlin} on a fixed background, thus neglecting the backreaction from the metric. As a result, they found that the equation admits a non-trivial solution for a discrete spectrum of values of the coupling parameter $\eta$. By solving the full set of equations, they showed that scalarized solutions can be found for both black holes and neutron stars. Interestingly, for black holes they exist when the coupling parameter belongs to a set of ``scalarization bands'', whereas for neutron stars they can exist for both positive and negative values of $\eta$.
 
 In Ref.~\cite{Doneva:2017bvd} they considered a different coupling function $f=\lambda^2(1-e^{-3\phi^2/2})/12$, rescaled to our formalism, and they showed that the theory exhibits spontaneous scalarization for black holes. We point out that for what concerns the onset of the instability, the model studied in Ref.~\cite{Doneva:2017bvd} reduces to the one considered in Ref.~\cite{Silva:2017uqg}, upon suitable rescaling. Indeed, Taylor-expanding the exponential coupling functions, the only contribution that renders the effective mass square non-vanishing comes from the quadratic term. Nevertheless, higher power terms will come into play when fully studying the set of equations, and will affect the final properties of the scalarized compact objects.

 We conclude this Chapter with a final remark. Throughout the work of this thesis, we only work in the framework of spontaneous scalarization as first identified by the work of Damour-Esposito-Far\`ese: a phase transition triggered at linear level by a tachyonic instability and stabilized thanks to the contribution of nonlinearities. It is, however, conceivable that other mechanisms can trigger a similar phase transition or that the concept of spontaneous scalarization can be extended to other field contents. We further address this topic in the Conclusion of the thesis.


\chapter[\texorpdfstring{Spontaneous scalarization in Horndeski}{Spontaneous scalarization in Horndeski}
]{\chaptermark{Scalarization in Horndeski} Spontaneous scalarization in Horndeski}
\chaptermark{Scalarization in Horndeski}
\label{Chap:scalarizationHordneski}

 The recent discovery of new models of scalarization in scalar Gauss-Bonnet gravity have clearly demonstrated that the Damour and Esposito-Far\`{e}se model is not unique in this respect. This suggests that there might be more, yet to be discovered, theories that exhibit spontaneous scalarization. In this Chapter, we present the work done in Ref.~\cite{Andreou:2019ikc}, where we address this question for a scalar field that belongs to the Horndeski class.
 
 We restrict our attention to the onset of scalarization. In particular, we focus on the conditions that a theory must satisfy for scalarization to be triggered by a tachyonic instability. As discussed already in Chapter~\ref{Chap:spontaneousscalarization}, even though nonlinearities are essential for determining the fate of the instability and pinning down the end state~\cite{Silva:2018qhn,Macedo:2019sem}, the onset of the instability can be captured in the linear regime already. This implies that one can obtain necessary conditions for spontaneous scalarization simply by inspecting the linearized field equations and the contributions to the effective mass term for the perturbation of the scalar. As a final result, we identify the minimal action that contains all the terms that can potentially trigger spontaneous scalarization.

\section{Tachyonic instability in Horndeski gravity}\label{Sec:instabilityHorndeski}

 \subsection{General relativity as a solution}\label{Sub:solutionGR}
 
 Our aim is to identify a subclass of Horndeski gravity affected by a tachyonic instability around solutions of general relativity. Hence we need to impose that the theory actually admits as a solution any spacetime of general relativity with $\phi=\phi_0=\text{const}$. This requires imposing certain conditions on the $G_{i}$ functions of the Horndeski action in Eq.~\eqref{eq:LagrHorn}. These conditions have been fully worked out for shift-symmetric classes~\cite{Saravani:2019xwx} but not for theories that do not respect shift symmetry (and hence, can have a bare or effective mass).

 The obvious thing one can do to do away with theories that do not admit solutions $\phi=\phi_0=\text{const}$, or $X=0$,  is to require that the $G_i$ functions be analytic around $X=0$. In this case one can expand them in a power series in terms of $X$,
 \begin{equation}
 \label{G2analytical}
 G_i=g_{i0}(\phi)+g_{i1}(\phi)X+\dots.
 \end{equation}
 However, imposing analyticity for the $G_i$ function is too restrictive\footnote{Another possibility is to extrapolate theories within Horndeski which admit general relativity as a solution only at leading order~\cite{McManus:2016kxu}. However this goes beyond the scope of this thesis.}. As discussed in Chapter~\ref{Chap:spontaneousscalarization}, scalar Gauss-Bonnet gravity is already known to exhibit spontaneous scalarization, and it can be represented in the Horndeski framework through nonanalytic $G_i$ functions at $X=0$, as shown in Chapter~\ref{Chap:sttheories}.
 Hence, we should certainly relax our analyticity assumption on the $G_i$ functions in order to accommodate it. To this end, we rewrite the $G_i$ functions as a sum of an explicitly analytic part, which we label as $\tilde{G}_i$, and a nonanalytic part, coming from Eqs.~\eqref{G_GB}. Explicitly we have
 \begin{gather}
 \label{Gfunctions}
 G_i(\phi,X)=\tilde{G}_i(\phi,X) + G^{\note{GB}}_i(\phi,X), \\
 \label{eq:GtildeExp}
 \tilde{G}_i(\phi,X)=g_{i0}(\phi)+g_{i1}(\phi)X+\dots
 \end{gather}
 where in Eq.~\eqref{eq:GtildeExp} we expanded $\tilde{G}_i$ as in Eq.~\eqref{G2analytical}.

 The results and classification of Ref.~\cite{Saravani:2019xwx} for shift-symmetric theories suggest that $G_i$ functions that contain $\sqrt{|X|}$ might be another form of mild nonanalyticity that is compatible with general relativity solutions. However, we do not explore this possibility further. Moreover, in principle, there could be another type of non shift-symmetric theories described by nonanalytic $G_i$ functions that admit all of the solutions of general relativity. This deserves further investigation, but we do not pursue it in this thesis.

 Once one has imposed the above conditions on the $G_i$ functions, the terms $-g_{30}(\phi)\DAlembert\phi$ and $g_{21}(\phi)X$ appear in the action and they coincide up to total derivative. Thus, without loss of generality we can set $g_{30}(\phi)=0$, which is equivalent to the redefinition $g_{21}(\phi)\rightarrow g_{21}(\phi)+2g_{30\phi}(\phi)$, where henceforth we use as convention that a subscript $\phi$ denotes a derivative with respect to $\phi$. Moreover, \textit{at the level of the linearized equations}, which is our interest here, the terms $g_{41}(\phi)$ and $-g_{50\phi}(\phi)$ give the same contribution. Hence, we similarly redefine $g_{41}(\phi)\rightarrow g_{41}(\phi)+g_{50\phi}(\phi)$.

 Let us now look explicitly at the equations of motion. The metric satisfies Einstein equations~\eqref{HornEq1}, which for any constant scalar field $\phi=\phi_0$ read
 \begin{equation}
 \label{EinstEq}
 R_{\mu\nu}-\frac{1}{2}g_{\mu\nu}R+\Lambda g_{\mu\nu}=\tilde{\kappa}T_{\mu\nu},
 \end{equation}
 where
 \begin{equation}\label{eq:Lambda,Kappa}
 \Lambda=-g^0_{20}/2g^0_{40}, \qquad \tilde{\kappa}=\kappa/g^0_{40},
 \end{equation}
 provided that $g_{40}^0\neq0$. The superscript $0$ in $g^0_{20}$, $g^0_{40}$, etc., means that the function is evaluated at $\phi=\phi_0$. The equations above imply clearly that the metric is a solution of general relativity Einstein equations and that all solutions of Einstein's equation are admissible.

 Let us now take the scalar field equation~\eqref{HornEq2}, with the choice of functions of Eq.~\eqref{Gfunctions}. One can show that only terms that contain up to one derivative operator (which, in this case, can be only a second order operator) acting on $\phi$ will contribute to the linearization of the equation around the constant value $\phi_0$ made in the next paragraph. Hence, we keep only these terms. We stress that first order derivatives do not contribute to the linearized equations. Indeed, these terms appear at least in the form $\nabla\phi\nabla\phi$, which, upon linearization, vanishes when the background field is constant. With this prescription, the scalar field equation takes the form
 \begin{equation}
 \label{ScalEq}
 \tilde{g}^{\mu\nu}\nabla_\mu\nabla_\nu\phi +\frac{g_{20\phi}+g_{40\phi}R+f^{(1)}\mathscr{G}}{A(\phi)}=0,
 \end{equation}
 where
 \begin{equation}\label{eq:Acoeff}
 A(\phi)=g_{21}+g_{41}R,
 \end{equation}
 and the effective metric reads\footnote{Note that the effective metric~\eqref{effMetr1} must have a Lorentzian signature in order for the linearized equation to be hyperbolic and hence describe the time evolution of the system. This imposes some further conditions on $g_{21}$ and $g_{41}$. In this work, we implicitly assume that such conditions are  satisfied.}
 \begin{equation}
 \label{effMetr1}
 \tilde{g}^{\mu\nu}=g^{\mu\nu} - \frac{2g_{41}R^{\mu\nu}}{A(\phi)}.
 \end{equation}

 We now impose that $\phi=\phi_0$ is a solution of Eq.~\eqref{ScalEq}. There are two distinct cases for which this happens,
 \begin{align}
 \text{case I:}  \qquad & g^0_{20\phi}+g^0_{40\phi}R+f^{(1)}_0\mathscr{G} =0, \notag \\
 \label{GRcond1}
 & A_0\,\, \text{finite} ; \\
 \text{case II:} \qquad & g^0_{20\phi}+g^0_{40\phi}R+f^{(1)}_0\mathscr{G} \neq0, \notag \\
 \label{GRcond2}
 & A_0\rightarrow \infty,
 \end{align}
 where
 \begin{equation}\label{eq:A0}
 A_0\equiv A(\phi_0)=g_{21}^0+g_{41}^0 R.
 \end{equation}
 Case II is rather interesting, as it provides a way to have a general relativity solution even when the term $g_{20\phi}+g_{40\phi}R+f^{(1)}\mathscr{G}$ does not depend on $\phi$ at all (or equivalently when $g^0_{20\phi\phi}=g^0_{40\phi\phi}=f^{(2)}_0=0$) and would otherwise act as a source term for the scalar field.
 For example, as we see in more detail in the next Section, standard scalar-tensor theories belong to case II, as they correspond to $g_{40}=\phi$, $g_{41}=0$. They admit general relativity solutions only when $g_{21}(\phi)=2\omega(\phi)/\phi\rightarrow\infty$ for $\phi\rightarrow\phi_0$. Another interesting term in this context is that with $f=\phi$. As already mentioned, this choice leads to the $\phi\,\mathscr{G}$ term, which is shift symmetric, and the Gauss-Bonnet invariant would appear in the scalar field equation as a pure source for the scalar field. Thus, only theories that satisfy condition~\eqref{GRcond2} can afford to include this term and still admit general relativity solution. This possibility is absent in shift-symmetric theories~\cite{Saravani:2019xwx}.

 Note that an analysis similar to the one presented here has been conducted in Ref.~\cite{Motohashi:2018wdq} for multiscalar-tensor theories, but with more restrictive assumptions that appear to exclude case II.
 
 \subsection{Linearized scalar field equations}\label{Sub:linearization}
 
 We now proceed to perform a linear analysis of Horndeski gravity in order to determine all of the terms that can contribute to the linear tachyonic instability that will eventually trigger the process of spontaneous scalarization.
 
 Linearizing Eq.~\eqref{HornEq2} divided by $A(\phi)$ [or equivalently Eq.~\eqref{ScalEq}] for small $\delta\phi=\phi-\phi_0$ yields
 \begin{equation}
 \label{LinEq1}
 \tilde{g}^{\mu\nu}\nabla_\mu\nabla_\nu\delta\phi-m_\note{I}^2\delta\phi-m_\note{II}^2\delta\phi=0,
 \end{equation}
 where
 \begin{align}
 \label{massI}
 m^2_\note{I} &=-\frac{g^0_{20\phi\phi}+g^0_{40\phi\phi}R+f^{(2)}_0\mathscr{G}}{A_0}, \\
 \label{massII}
 m^2_\note{II} &=\frac{g^0_{20\phi}+g^0_{40\phi}R+f^{(1)}_0\mathscr{G}}{A_0^2}\frac{\partial A}{\partial\phi}\bigg|_{\phi_0}
 \end{align}
 are the effective masses obtained in the two separate cases. We notice that the two cases give mutually exclusive contributions to the mass. Indeed, if relation~\eqref{GRcond1} holds, then $m_\note{II}=0$; and when the condition~\eqref{GRcond2} holds, $m_\note{I}$ vanishes. Note that in the latter case,  $A_0\to \infty$ and  having a nonzero effective mass $m_\note{II}$ requires that $\frac{\partial A}{\partial\phi}\big|_{\phi_0} \to \infty$ is such that
 \begin{equation}\label{eq:A0primecondition}
 \frac{1}{A_0^2}\frac{\partial A}{\partial\phi}\bigg|_{\phi_0}\neq 0\quad \text{and finite}.
 \end{equation}
 Hence, around $\phi=\phi_0$ it must be $A(\phi)\sim (\phi-\phi_0)^{-1}$.

 We can now single out the theories which can exhibit a tachyonic instability around a general relativity background. They either satisfy condition~\eqref{GRcond1} and have $m^2_\note{I}<0$  or they satisfy condition~\eqref{GRcond2} and have $m^2_\note{II}<0$.
 We stress that our perturbative analysis is done around  a general relativity background and we perturb only the scalar without taking into account its backreaction to the metric. This approximation (decoupling) offers drastic simplification. In Chapter~\ref{Chap:blackholes} and Chapter~\ref{Chap:neutronstars}, we provide a complete study of the full set of field equations.
 
\section{The minimal actions}\label{Sec:minimalaction}

 We now analyze what the theories are that belong in one of the categories we identified above. At first, we write down for each case the minimal action that consists of all the terms that contribute to the linearized equation and admits general relativity solutions when $\phi=\phi_0$. Let us redefine the scalar field such that $\phi_0=0$. The minimal action for case I is
 \begin{multline}\label{eq:ActionCaseI}
 S_\note{I}=\int\dd^4x\frac{\sqrt{-g}}{2\kappa}\bigg[R-2\Lambda+(a_{21}+a_{41}R)X +a_{41}R_{\mu\nu}\nabla^\mu\phi\nabla^\nu\phi
 \\-\left( m_\phi^2+\frac{\beta}{2} R-\alpha\,\mathscr{G}\right) \frac{\phi^2}{2}\bigg]+S_\note{M},
 \end{multline}
 whereas for case II we have
 \begin{multline}\label{eq:ActionCaseII}
 S_\note{II}=\int\dd^4x\frac{\sqrt{-g}}{2\kappa}\bigg[R-2\Lambda+\frac{b_{21}+b_{41}R}{\phi}X +\frac{b_{41}}{\phi}R_{\mu\nu}\nabla^\mu\phi\nabla^\nu\phi
 \\+\left(\tau+\eta\, R+\lambda\,\mathscr{G}\right)\phi\bigg] 
 + S_\note{M}.
 \end{multline}
 We normalized the actions~\eqref{eq:ActionCaseI} and~\eqref{eq:ActionCaseII} by the constant multiplying $R$, which is equivalent to setting $g_{40}^0=1$. Moreover, we can identify the constants written in the actions~\eqref{eq:ActionCaseI} and~\eqref{eq:ActionCaseII} in terms of the function $g_{ij}$ evaluated at $\phi=0$,
 \begin{equation}\label{eq:constantsGs}
 \begin{aligned}
 \Lambda=-\frac{g_{20}^0}{2}, & \qquad \tau=g_{20\phi}^0, \qquad m_\phi^2=-g_{20\phi\phi}^0, \\
 a_{21}=(\phi\,g_{21})_\phi^0, & \qquad b_{21}=\left(\phi^2\,g_{21}\right)_\phi^0, \\
    & \qquad \eta=g_{40\phi}^0, \qquad \beta=-2\,g_{40\phi\phi}^0, \\
 a_{41}=(\phi\,g_{41})_\phi^0, & \qquad b_{41}=\left(\phi^2\,g_{41}\right)_\phi^0, \\
    & \qquad \lambda=f^{(1)}_0, \qquad \alpha=f^{(2)}_0.
 \end{aligned}
 \end{equation}
 The actions above could be supplemented with any term that does not contribute to the linearized equations without affecting the onset of the tachyonic instability. However, such nonlinear terms are crucial for determining the end state of the instability and the properties of scalarized solutions~\cite{Silva:2018qhn,Macedo:2019sem}. Hence, one can start from the minimal models above and bootstrap their way to theories that exhibit scalarization but differ quantitatively thanks to terms that introduce different nonlinear corrections.
 
 So far we have treated case I and case II separately because they lead to distinct contributions to the effective mass and, naively, they appear to be qualitatively different. Actually, they are equivalent as different representations of the same physics. Indeed, one can start from action~\eqref{eq:ActionCaseII}, perform the scalar field redefinition
 \begin{equation}\label{eq:FieldRescalingLinear}
 \phi \rightarrow \phi^2\,,
 \end{equation}
 and obtain action \eqref{eq:ActionCaseI} with the correspondence of parameters
 \begin{equation}\label{eq:ConstantsRedefinition}
 \begin{split}
 a_{21}=4\,b_{21}, \qquad a_{41}=4\,b_{41}, \\
 m_\phi^2=-2\,\tau, \qquad \beta=-4\,\eta, \qquad \alpha=2\,\lambda.
 \end{split}
 \end{equation}
 Hence, any theory in the minimal action of case II can be mapped onto an equivalent case I theory, at least in what regards their linear behavior and the onset of the tachyonic instability. This observation simplifies our analysis and reduces significantly the different scenarios of scalarization.

 Having shown that the two cases are equivalent, we now focus on the action outlined in Eq.~\eqref{eq:ActionCaseI} and  consider each term that contributes to the mass separately. This helps us identify its relation with known models of scalarization.  The term that contains $X$ in the action~\eqref{eq:ActionCaseI} contributes to the effective mass only as a multiplicative constant on a general relativity background. The parameter $a_{21}$ can be set to $1$ through a constant rescaling of the scalar and we do so implicitly in what follows. The $a_{41}$ is rather distinct from the rest so, for the time being, let us set $a_{41}=0$ and reduce the $X$-dependent term  to the canonical kinetic term. We relax this assumption in the next Section.

 The first term that contributes to the effective mass is the bare mass of the scalar field $m_\phi^2$. If the mass square is negative, it could lead to a tachyonic instability that would persist in flat space. So, we disregard this possibility. If it is positive, it needs to be sufficiently small not to prohibit the other terms from inducing a tachyonic instability. A small bare mass can actually be beneficial, as it can help suppress the non-general relativity effects away from the compact object. One can generalize the bare mass term to a full-fledged potential and this would introduce nonlinearities that could affect the end point of scalarization~\cite{Macedo:2019sem}. However, it is rather clear that a bare mass term or a potential cannot lead to scalarization by itself. Nevertheless, Refs.~\cite{Sperhake:2017itk,Rosca-Mead:2019seq,Rosca-Mead:2020ehn} dynamically studied spontaneous scalarization during the core collapse process for massive scalar-tensor theories and showed that, if the scalar field is endowed with a mass, a wider range of the parameter space is compatible with current gravitational observations. Moreover, the scalar mass causes the gravitational wave signal to disperse as it propagates. The signal thus carries a highly characteristic imprint of the massive scalar-tensor theory, implying potential detection through observations or, in the case of nondetection, more stringent constraints on the parameter space.

 Next we consider the coupling term between $\phi$ and the Gauss-Bonnet invariant. For the choice $m_\phi=\beta=0$ (and $a_{21}=1$, $a_{41}=0$) one has the action
 \begin{equation}\label{eq:CaseI_GBTerm_Action_0}
 S=\int \dd^4x \frac{\sqrt{-g}}{2\kappa}\left[R-\frac{1}{2}\nabla^\mu \phi \nabla_\mu \phi+\frac{1}{2}\alpha \phi^2\mathscr{G}\right]+S_{\note{M}},
 \end{equation}
 This is the quadratic coupling scalarization model considered in Ref.~\cite{Silva:2017uqg}. Allowing for a more general coupling function one gets the action considered in Refs.~\cite{Doneva:2017bvd,Silva:2017uqg},
 \begin{equation}\label{eq:CaseI_GBTerm_Action}
 S=\int \dd^4x \frac{\sqrt{-g}}{2\kappa}\left[R-\frac{1}{2}\nabla^\mu \phi \nabla_\mu \phi+f(\phi)\mathscr{G}\right]+S_{\note{M}},
 \end{equation}
 where, from the requirement that general relativity is a solution of the field equations, that is the theory admits a constant scalar field solution, one can infer that $f_\phi(0)=0$. This condition guarantees that the leading term in $f(\phi)$ is indeed $\phi^2$.

 Finally, if we set $m_\phi=\alpha=0$, we have
 \begin{equation}\label{eq:CaseI_RTerm_Action_0}
 S=\int \dd^4x \frac{\sqrt{-g}}{2\kappa}\left[\left(1-\frac{\beta\phi^2}{4}\right)R-\frac{1}{2}\nabla^\mu \phi \nabla_\mu \phi \right]+S_{\note{M}}.
 \end{equation}
 We can generalize this theory in a similar fashion as above and write
 \begin{equation}\label{eq:CaseI_RTerm_Action}
 S=\int \dd^4x \frac{\sqrt{-g}}{2\kappa}\left[F(\phi)R-\frac{1}{2}\nabla^\mu \phi \nabla_\mu \phi\right]+S_{\note{M}},
 \end{equation}
 where we assume $F(0)\neq0$. The condition~\eqref{GRcond1} implies $F_\phi(0)=0$, and $F_{\phi\phi}(0)<0$ is the requirement for a tachyonic instability of the theory.

 One may be tempted to think that this is a new model. However, we recall that we can always perform a redefinition of the scalar field, as we did to relate the minimal actions of case I and case II. Indeed, consider the redefinition
 \begin{equation}\label{eq:FieldRescaling}
 \Phi=F(\phi)\,.
 \end{equation}
 Action \eqref{eq:CaseI_RTerm_Action} can be rewritten as
 \begin{equation}\label{eq:CaseII_STT}
 S=\int \dd^4 x \frac{\sqrt{-g}}{2\kappa}\left[\Phi R-\frac{\omega(\Phi)}{\Phi} \nabla^\mu \Phi \nabla_\mu \Phi\right] + S_\note{M}\,,
 \end{equation}
 if we just introduce the definition
 \begin{equation}\label{eq:RescalingSTT}
 \omega(\Phi)\equiv\frac{\Phi}{2F'^2(\phi)}.
 \end{equation}
 Action~\eqref{eq:CaseII_STT} is that of scalar-tensor theories for a scalar field $\Phi$ written in the so-called Jordan frame, i.e Eq.~\eqref{eq:Jordanframe}. The condition $F_\phi(0)=0$ translates into $\omega(\Phi_0)\to \infty$, where $\Phi_0=F(0)$. This picks a specific subclass of scalar-tensor theories, which is precisely that originally considered by Damour and Esposito-Far\`{e}se.

 Indeed, the minimal model in action~\eqref{eq:CaseI_RTerm_Action_0} corresponds to $F(\phi)=1-\beta \phi^2/4$ and hence $\Phi=1-\beta\phi^2/4$,
 \begin{equation}
 \omega(\Phi)=\frac{\Phi}{2\beta(1-\Phi)}=-\frac{1}{2\beta}+\frac{1}{2\beta(1-\Phi)}\,,
 \end{equation}
 and $\Phi_0=1$. In Chapter~\ref{Chap:spontaneousscalarization}, we showed how the most commonly studied Damour and Esposito-Far\`{e}se model corresponds in the Jordan frame to a choice for the $\omega$ function as in Eq.~\eqref{eq:omegaDEF}, which we report here for convenience,
 \begin{equation}\label{eq:omega}
 \omega_\note{DEF}(\Phi)=-\frac{3}{2}-\frac{1}{4\,\beta_\note{DEF} \text{Log}\Phi}\,,
 \end{equation}
 where we have used the subscript DEF to distinguish the commonly used $\beta$ parameter from our notation above. As $\Phi\to \Phi_0=1$ one has
 \begin{equation}
 \omega_\note{DEF}(\Phi)\to -\frac{1}{4\beta_\note{DEF}(\Phi-1)}\,,
 \end{equation}
 which is precisely the same behavior as our minimal model up to a redefinition of constants. The two models are indistinguishable at the linear level.

 We stress that the scalar field redefinition that related the $F(\phi)R$ model with the Damour and Esposito-Far\`{e}se class was basically mapping a case I theory onto a case II theory. Indeed, one can straightforwardly identify the Damour and Esposito-Far\`{e}se class as a subcase of the action~\eqref{eq:ActionCaseII}, with the constant coefficients generalized to functions of $\phi$. Furthermore, these results clearly show that some models that might appear as new are simply combinations of known models rewritten after a scalar field redefinition. For instance, the action
 \begin{equation}\label{eq:Action_R_GB}
 S=\int\dd^4 x \frac{\sqrt{-g}}{2\kappa}\left[\phi R +2\frac{\omega(\phi)}{\phi}X+\eta\phi\,\mathscr{G}\right]+S_\note{M},
 \end{equation}
 with the condition $\omega(\phi_0)\to \infty$ for some $\phi_0$ would yield a seemingly intriguing case II model upon linearization, but it can straightforwardly be mapped onto a combination of actions~\eqref{eq:CaseI_GBTerm_Action} and~\eqref{eq:CaseI_RTerm_Action}.

  Our analysis allowed us to determine a minimal action that contains all of the terms that contribute to the effective mass at linearized level, i.e. action~\eqref{eq:ActionCaseI}. This can be thought of as containing four distinct terms that contribute to scalarization. Through suitable field redefinitions, one of them can be directly linked to the known Damour and Esposito-Far\`{e}se model and another to the scalar Gauss-Bonnet scalarization models. The third term comes from a potential for a scalar and, although it cannot trigger spontaneous scalarization on its own, it affects the onset of the tachyonic instability in all other models. In this Section we neglected the contribution from the fourth term, parametrized by $a_{41}$ in action~\eqref{eq:ActionCaseI}, consisting in a coupling between the kinetic term and the Ricci scalar. We explore the role of this term in the next Section.
 
\section{Disformal transformations and matter coupling}\label{Sec:disformal}

 Hitherto, we have assumed that the matter couples minimally to the metric only. Moreover, in the previous Section we set the coefficients $a_{41}$ of the action~\eqref{eq:ActionCaseI} to the specific value $a_{41}=0$. At linear level (which is our main interest throughout this Chapter), it turns out that one can always do so without loss of generality by relaxing the matter coupling assumption.

 To show this, let us start with action~\eqref{eq:ActionCaseI}  and elevate all of the constants to generic functions of $\phi$ (retaining the minimal coupling to matter, described by some generic fields $\psi_\text{M}$),
 \begin{multline}\label{eq:GeneralAction}
 S=\int\dd^4 x\frac{\sqrt{-g}}{2\kappa}\Big[(g_{40}(\phi)+g_{41}(\phi)X)R +g_{21}(\phi)X \\
 +g_{41}(\phi)R_{\mu\nu}\nabla^\mu\phi\nabla^\nu\phi+g_{20}(\phi)+f(\phi)\mathscr{G}\Big] +S_\note{M}\left[g_{\mu\nu},\psi_\text{M}\right].
 \end{multline}
 We stress that the unknown functions of $\phi$ are assumed to be such that linearizing this action around $\phi=0$ must yield~\eqref{eq:ActionCaseI}, with the identification of the constants~\eqref{eq:constantsGs}. Consider now a special disformal transformation as defined in Eq.~\eqref{eq:disformal}, i.e. $ g_{\mu\nu} \rightarrow C(\phi)\left[g_{\mu\nu}+D(\phi)\nabla_\mu\phi\nabla_\nu\phi\right]$.
 As seen in Chapter~\ref{Chap:sttheories}, this transformation leaves the Horndeski action~\eqref{Horndeski} formally invariant~\cite{Bettoni:2013diz,Zumalacarregui:2013pma}. Applying this transformation to \eqref{eq:GeneralAction} and keeping only the terms which contribute to the linear level in the equations yields
 \begin{multline}\label{eq:DisfGeneralAction}
 S=\int \dd^4 x \frac{\sqrt{-g}}{2\kappa} \Big[(\bar{g}_{40}(\phi)+\bar{g}_{41}(\phi)X)R \\+\bar{g}_{41}(\phi)R_{\mu\nu}\nabla^\mu\phi\nabla^\nu\phi 
 +\bar{g}_{21}(\phi)X+\bar{g}_{20}(\phi)+f(\phi)\,\mathscr{G} \Big] \\ +S_{\note{M}}\left[C(\phi)\left(g_{\mu\nu}+D(\phi)\nabla_\mu\phi\nabla_\nu\phi\right), \psi_\text{M}\right],
 \end{multline}
 where we made explicit the disformal coupling in the matter sector, and the new functions are defined as follows,~\footnote{We derived independently the effect of the disformal transformation~\eqref{eq:disformal} on the Horndeski Lagrangian~\eqref{Horndeski}. However, there is a mismatch with the results of~\cite{Bettoni:2013diz}. See Appendix~\ref{App:Disftransf}.}
 \begin{align}\label{eq:FunctionsDisf}
  \bar{g}_{20} = & \, C^2 g_{20}\\ \label{eq:g21}
  \bar{g}_{21} = & \, C g_{21}-C^2D g_{20}-3g_{40}\frac{C_\phi^2}{C}-6g_{40\phi}C_{\phi} \\ \label{eq:g40}
  \bar{g}_{40} = & \, Cg_{40}, \\ \label{eq:g41}
  \bar{g}_{41} = & \, g_{41}-CDg_{40}-4\frac{C_\phi}{C}f^{(1)},
 \end{align}
 whereas $f(\phi)$ remains invariant. Here we are using again the same convention that a subscript $\phi$ denotes a derivative with respect to $\phi$. Hence, the action~\eqref{eq:DisfGeneralAction} yields field equations whose linear perturbation is formally invariant under the transformation~\eqref{eq:disformal}.

 One notices that two out of the five functions $g_{40}$, $g_{41}$, $g_{21}$, $C$ and $D$ are redundant. That is, one can always perform a disformal transformation and choose $C$ and $D$ in order to redefine two of $g_{40}$, $g_{41}$, $g_{21}$. For example, from Eq.~\eqref{eq:g41} one can set $\bar{g}_{41}=0$, by choosing
 \begin{equation}\label{eq:DisfChoice}
 D=\frac{g_{41}}{Cg_{40}}-4\frac{C_{\phi}}{C^2g_{40}}f^{(1)}.
 \end{equation}
 This choice fixes uniquely the disformal function $D$. This implies that the condition $g_{41}=0$ imposed throughout the previous Section is equivalent to a specific type of disformal coupling. In other words, though having a nonzero $a_{41}$ does lead to a new theory, this theory is simply one of the known scalarization models, or a combination thereof, disformally coupled to matter (see~\cite{Minamitsuji:2016hkk} for a discussion of Damour and Esposito-Farèse spontaneous scalarization plus a disformal coupling).

 For example, let us indeed impose Eq.~\eqref{eq:DisfChoice} in order to set $\bar{g}_{41}=0$ and we further choose
 \begin{equation}\label{eq:ConfChoice}
 C(\phi)=\frac{1}{g_{40}(\phi)},
 \end{equation}
 and redefine the scalar field as
 \begin{equation}\label{eq:RescChoice}
 \varphi = \varphi(\phi), \qquad \varphi'(\phi) = \frac{\sqrt{\bar{g}_{21}(\phi)}}{2},
 \end{equation}
 where $\bar{g}_{21}(\phi)$ is defined in Eq.~\eqref{eq:g21}. With these choices,  action~\eqref{eq:DisfGeneralAction} takes the form
 \begin{multline}
 \label{eq:DisfGeneralActionEF}
 S_\note{E}=\frac{1}{2\kappa}\int \dd^4 x \sqrt{-g} \left[R+V(\varphi)-2\partial_\mu\varphi\,\partial^\mu\varphi+F(\varphi)\mathscr{G} \right] \\ +S_{\note{M}}\left[G(\varphi)\left( g_{\mu\nu}+H(\varphi)\nabla_\mu\varphi\nabla_\nu\varphi \right), \psi_\text{M}\right],
 \end{multline}
 where we defined the new functions
 \begin{equation}
 \begin{split}
 V(\varphi)=\bar{g}_{20}(\phi(\varphi)), \qquad F(\varphi)=f(\phi(\varphi)), \\
 G(\varphi)=C(\phi(\varphi)), \qquad H(\varphi) = \frac{4D(\phi(\varphi))}{\bar{g}_{21}(\phi(\varphi))}.
 \end{split}
 \end{equation}
 For $f(\phi)=0$, this action reduces to the spontaneous scalarization model with disformal coupling studied for the first time in~\cite{Minamitsuji:2016hkk}.
 
 In this Section, we investigated the role of the term parametrized by $a_{41}$ in action~\eqref{eq:ActionCaseI}, i.e. a coupling between the Ricci scalar and the kinetic term $X$. We showed that this term can be thought of as a disformal coupling to matter. Indeed, while including this term leads to a new theory, such model can be seen as one of the scalarization models presented in Section~\ref{Sec:minimalaction} disformally coupled to matter.

\chapter{The threshold of scalarization}\label{Chap:Threshold}

 Having identified a subclass of Horndeski theory that contains all the terms that can trigger spontaneous scalarization, we now determine quantitatively the bounds on the couplings that allow scalarization to take place. We combine the effect of all relevant couplings simultaneously and also examine how the structure of the compact object affects the threshold of the instability.

 In Chapter~\ref{Chap:scalarizationHordneski}, we showed how the two minimal actions of case I and case II obtained after a linearization of Horndeski gravity are in fact equivalent upon a field redefinition. We thus simply focus on the former, namely Eq.~\eqref{eq:ActionCaseI}, where we set $a_{21}=1$ to obtain a canonical kinetic term and we relabel $a_{41}=\gamma$ for convenience. The action then reads
 \begin{multline}\label{eq:ACI}
 S=\frac{1}{2\,\kappa}\int\dd^4x\sqrt{-g}\bigg\{R+X+ \gamma\, G^{\mu\nu}\nabla_\mu\phi\,\nabla_\nu\phi
  \\-\left(m_\phi^2+\dfrac{\beta}{2} R-\alpha\mathscr{G}\right)\dfrac{\phi^2}{2}\bigg\}
 +S_\mathrm{M}.
 \end{multline}
 Note that we used a slightly different notation for identifying the coupling between gravity and the kinetic term parametrized by $\gamma$. However, the two actions in Eq.~\eqref{eq:ActionCaseI} and Eq.~\eqref{eq:ACI} are in fact equivalent. From a dimensional analysis of Eq.~\eqref{eq:ACI}, one can clearly see that $\beta$ is dimensionless, while $\alpha$ and $\gamma$ have the dimension of a length squared. Throughout this Chapter we assume that matter couples minimally to the metric only: we are working in the Jordan frame.

 Our goal is to investigate whether general relativity solutions with $\phi=0$ are stable or not, by studying the perturbations of the scalar field on a fixed general relativity background. Thus, we solely focus on the scalar field equation associated with action~\eqref{eq:ACI}, which reads
 \begin{equation}
 \label{LinEq}
 \tilde{g}^{\mu\nu}\nabla_\mu\nabla_\nu\phi-m^2\phi=0,
 \end{equation}
 where the effective metric and the mass term are respectively 
 \begin{align}
 \label{effMetr}
 \tilde{g}^{\mu\nu}&=g^{\mu\nu} - \gamma G^{\mu\nu},
 \\
 \label{eq:massI}
 m^2&=m_\phi^2+\frac{\beta}{2} R-\alpha\mathscr{G}.
 \end{align}
 It is clear from these equations that $\beta$ and $\alpha$ generate an effective mass for $\phi$ in a curved background. On the other hand, $\gamma$ only determines the effective metric that defines the d'Alembertian which acts on the scalar field perturbations. We stress that Eq.~\eqref{LinEq} is linear by construction, since we kept in the action only the terms that contribute linearly. This approach is valid when focusing on the onset of the scalarization, when linear terms dominate. In order to determine the final state of the process, one needs to include non-linear terms as well. Additionally, we work in the decoupling limit where we only perturb the scalar field. These perturbations will eventually back-react onto the metric, and a consistent analysis (beyond the onset of the instability) should thus include metric perturbations. We perform a full calculation for both cases of black holes and neutron stars in Chapters~\ref{Chap:blackholes} and~\ref{Chap:neutronstars} respectively.
 
 The results presented in this Chapter have been published in Ref.~\cite{Ventagli:2020rnx}.

\section{The case of black holes}\label{Sec:bhthreshold}
 
 
 As we have shown in Chapter~\ref{Chap:spontaneousscalarization}, in linearized theory, scalarization manifests itself as a tachyonic instability around a general relativity solution. Hence, we first consider the case of a Schwarzschild background with the line element defined as
 \begin{equation}\label{eq:lineElement}
 {\dd s}^2=-h(r){\dd t}^2+f(r)^{-1} {\dd r}^2
 +r^2{\dd \Omega}^2.
 \end{equation}
 Thanks to spherical symmetry, the scalar field is decomposed into spherical harmonics as
 \begin{equation}\label{eq:sphersymm}
 \phi=\sum_{\ell,m}\hat\phi_{\ell m}(t,r)Y_{\ell m}(\theta,\phi),
 \end{equation}
 and we focus on the breathing mode, $\ell=m=0$, which is the first one to exhibit instability when it is present. We can then rescale this mode according to $\hat\phi_{00}(t,r) =\sigma(t,r)/r$ and we can recast the scalar equation~\eqref{LinEq} into the following form:
 \begin{equation}\label{eq:sigma}
   -\frac{1}{c^2}\frac{\partial^2 \sigma}{\partial t^2}+\frac{\partial^2 \sigma}{\partial r_*^2}= V_{\text{eff}}\,\sigma,
 \end{equation}
 where we have defined a new radial coordinate $r_*$ such that $\dd r=\dd r_* \sqrt{h f}$, and
 \begin{equation}\label{eq:bhpotential}
   V_{\text{eff}}=\frac{h\,f'+f\,h'}{2\,r}+h\left( m_\phi^2-\alpha\gb \right),
 \end{equation}
 From Eq.~\eqref{eq:bhpotential}, it is clear that the $\beta$ and $\gamma$ terms do not influence the effective potential, since a Schwarzschild background is Ricci flat. It is also straightforward to see that the effect of a positive scalar field bare mass is simply that of shifting the threshold scalarization.
 If we further neglect this term, then the effective potential (and hence the onset of scalarization) is controlled only by the rescaled mass $\hat M =M/\sqrt{\alpha}$, where $M$ is the ADM mass of the black hole. General relativity solutions correspond to $\alpha=0$, thus, they will become unstable for small values of $\hat M$, which correspond to larger curvatures or larger $\alpha$ couplings.
 
 We further focus on exponentially growing perturbations: $\sigma(t,r_\ast)=\hat\sigma(r_\ast) e^{\omega t}$, with $\omega>0$.\footnote{One could also look for the quasi-normal modes associated with $V_\text{eff}$, allowing complex values of $\omega$. However, this requires a much wider set-up, which is not needed to establish the presence of an instability.} Equation~\eqref{eq:sigma} then boils down to a Schr\"odinger equation:
 \begin{equation}\label{scalarEqPot2}
 \frac{\mathrm{d}^2\hat\sigma}{\mathrm{d} r_\ast^2}=\left[V_{\text{eff}}(r_\ast)+\left(\dfrac{\omega}{c}\right)^2\right]\hat\sigma,
 \end{equation}
 where $V_\mathrm{eff}$ is clearly an effective potential, and $-(\omega/c)^2$ plays the role of the energy of the perturbation. The existence of a bound state for $V_\mathrm{eff}$ with `energy' $E_0<0$ implies the existence of an instability, with characteristic growth rate $\omega=c\sqrt{-E_0}$. Our strategy throughout this Chapter, will thus be the following: we start from values of the parameters for which the theory reduces to general relativity with a minimally coupled scalar field and hence there cannot be any instability. We gradually increase the parameters governing the effective potential, thus progressively deforming it. Whenever a bound state appears for $V_\mathrm{eff}$, we identify it with a new unstable mode for $\phi$. By continuity, when deforming the potential, a new bound state will appear with a vanishing energy, $E_0=0$. Note that we identify as a bound state those with $\mathrm{d}\hat\sigma/\mathrm{d}r_\ast\to0$ for $r_\ast\to+\infty$.
 
\begin{figure}[t]
    \centering
    \includegraphics[width=0.7\textwidth]{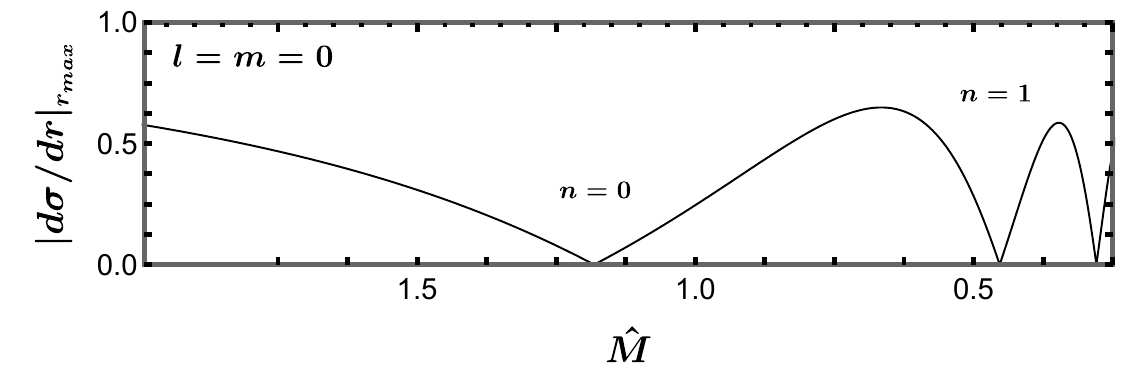}
    \caption[Numerical solution of the decoupled scalar equation on a general relativity background.]{Numerical solution of the decoupled scalar equation on a general relativity background. The points where the line touches the horizontal axis correspond to the scalarization thresholds $\hat{M}^{(n)}_{\text{th}}$: for $\hat M<\hat{M}^{(n)}_{\text{th}}$, general relativity black holes are unstable under scalar perturbations with $n$ nodes. Note that the horizontal axis is inverted.}
    \label{fig:fig_Mth}
\end{figure}

 We then search for such bound states while varying $\hat M$ and identify the threshold rescaled masses, $\hat{M}^{(n)}_{\text{th}}$, ($n=0,1,2,$ etc). This is shown in Fig.~\ref{fig:fig_Mth}.
 The mode associated with a threshold mass $\hat{M}^{(n)}_{\text{th}}$ has $n$ nodes. Hence, whenever $\hat M<\hat{M}^{(n)}_{\text{th}}$, general relativity black holes become unstable to a perturbation with $n$ nodes. Numerically, these thresholds are $(n,\hat{M}^{(n)}_{\text{th}})\approx (0,1.18),\, (1,0.45),\, (2,0.28)$, etc. They are in agreement with the results of Ref.~\cite{Silva:2017uqg} under an appropriate rescaling. 
 
 We investigated the threshold of scalarization for the minimal action~\eqref{eq:ACI}, in the context of black holes. We showed that the scalar field equation can be recast as a Schr\"odinger equation with an effective potential. The bound states of the latter identify unstable modes for the scalar field. Crucially, only the Gauss-Bonnet coupling parameter $\alpha$ and the scalar field bare mass $m^2_\phi$ can influence the effective potential. Our results agree with previous results obtained for black hole scalarization.
 
\section{The case of neutron stars}\label{Sec:nsthreshold}

 It is clear that considering black hole solutions does not allow us to fully explore the parameters of the theory~\eqref{eq:ACI}, since we are not able to determine the contribution from $\gamma$ or $\beta$. Considering that we are interested in the combined effect of all the parameters of the theory, henceforth we will focus on neutron stars solutions, where matter is present under the form of a perfect fluid with $T^\text{PF}_{\mu\nu}=(\epsilon+p)u_\mu u_\nu+p\,g_{\mu\nu}$, where $\epsilon$, $p$ and $u_\mu$ are respectively the energy density, the pressure and the 4-velocity of the fluid. The pressure is directly related to the energy density through the equation of state. We further consider a static and spherically symmetric background spacetime defined as in Eq.~\eqref{eq:lineElement}.
 The metric functions $h$ and $f$ are determined as solutions of an equivalent to the Tolman-Oppenheimer-Volkoff system of equations~\cite{Tolman:1939jz,Oppenheimer:1939ne}, together with the pressure $p$ and energy density $\epsilon$ of the perfect fluid that composes the star. These equations can be found in Appendix~\ref{App:TOV}. In this framework, one has to specify some equation of state $p(\epsilon)$. We use two equations of state, SLy and MPA1~\cite{Gungor:2011vq}, both favored by LIGO-Virgo tidal measurements~\cite{TheLIGOScientific:2017qsa}, which seem to prefer soft equations of state. We work here in units where $c=1$, $G=1$ and $M_\odot=1$.

 As for the case of black holes, we can decompose the scalar perturbation in the basis of spherical harmonics as in Eq.~\eqref{eq:sphersymm}, and, once again, we focus solely on the breather mode, $\ell=m=0$. 
 In order to make the scalar field equation more transparent, we rescale this mode according to $\hat\phi_{00}(t,r) =K(r)\sigma(t,r)$ with 
 \begin{equation}
 \begin{split}
 K(r)&=\left\{\left\{ r^2 - 2\gamma\left[-1+ \dfrac{f}{h} (rh)'\right]\right\}
\{r^2 -2 \gamma[ (rf)'-1]\}\right\}^{-1/4},
 \end{split}
 \end{equation}
 and we trade off the radial coordinate $r$ for a new one, $r_\ast$, defined through
 \begin{equation}
 \frac{dr_\star}{dr}=\frac{\sqrt{h}}{\sqrt{f} K^2 \left[ 2 \gamma r f h'-(2 \gamma+r^2-2 \gamma f)h \right]}\,.
 \label{eq:drast}
 \end{equation}
 Equation~\eqref{LinEq} then can be recast into Eq.~\eqref{eq:sigma}
 where the full expression of $V_\text{eff}$ can be found in Appendix~\ref{App:Veff}. Finally, focusing on exponentially growing perturbations, that is $\sigma(t,r_\ast)=\hat\sigma(r_\ast) e^{\omega t}$, with $\omega>0$, the scalar equation can be written as a Schr\"odinger equation as in Eq.~\eqref{scalarEqPot2}. We proceed on the study of the existence of bound states as illustrated in Section~\ref{Sec:bhthreshold}.
 Therefore, we solve Eq.~\eqref{scalarEqPot2} for $\omega=0$, while scanning the parameters $\beta$, $\alpha$ and $\gamma$. This is less intuitive than choosing a set of parameters and scanning $\omega$, but the final result is equivalent, and the procedure is easier to implement. 

 Equation~\eqref{scalarEqPot2} will admit a solution for any set of values of the parameters. Among these solutions, as in the case of black holes, we identify as a bound state those with $\mathrm{d}\hat\sigma/\mathrm{d}r_\ast\to0$ for $r_\ast\to+\infty$.\footnote{On a more technical level, we perform a numerical integration of eq.~\eqref{LinEq} expressed in terms of $r$, rather than in terms of $r_\ast$ as in eq.~\eqref{scalarEqPot2}; we extract $\mathrm{d}\sigma/\mathrm{d}r$ at a radius $r_\mathrm{max}$ equal to 200 times the Schwarzschild radius of the star.} Note that, physically, this is necessary for the scalar perturbation to be localized in space. In terms of quantum mechanics, this corresponds to the fact that $K(r_\ast)\hat\sigma(r_\ast)$ has to be square integrable. Note that, contrary to the naive expectation from eq.~\eqref{scalarEqPot2}, it is $K(r_\ast)\hat\sigma(r_\ast)$ that should be interpreted as the wave function, rather than $\hat\sigma(r_\ast)$ itself; this is very similar to the $1/r$ rescaling of the wave functions that allows to solve for the bound states of 3D spherically symmetric quantum wells (see {\em e.g.} secs.~14-16 of~\cite{schiff1955quantum}).

\subsection{Changing the effective mass}\label{Sub:effectivemass}

 The main terms that can contribute to the effective mass $m^2$ are the bare mass term of the scalar field, the coupling between $\phi$ and the Ricci scalar, and the coupling between $\phi$ and the Gauss-Bonnet invariant, parametrized respectively by $m_\phi$, $\beta$ and $\alpha$.
 Although the terms proportional to $\gamma$ can affect the instability threshold, they contribute solely as a multiplicative constant. Therefore, we will set $\gamma=0$ for the time being, and explore the role of this parameter in full detail in Section~\ref{Sub:effectivemetric}.
 Several works already investigated the influence of each of the parameters $m_\phi$, $\beta$ and $\alpha$ separately ({\em e.g.},~\cite{Damour:1993hw,Harada:1997mr,Silva:2017uqg,Doneva:2017bvd,Antoniou:2017acq,Blazquez-Salcedo:2018jnn,Minamitsuji:2018xde,Silva:2018qhn,Macedo:2019sem,Doneva:2019vuh}). We study how varying several parameters simultaneously affects the threshold; in this way, we explore much wider regions of the parameter space. Most of our results are presented as 2D plots, where we freeze all parameters but two, and show the stable/unstable regions.

\begin{figure}
 \subfloat{\includegraphics[height=0.54\textwidth, width=0.2\textwidth]{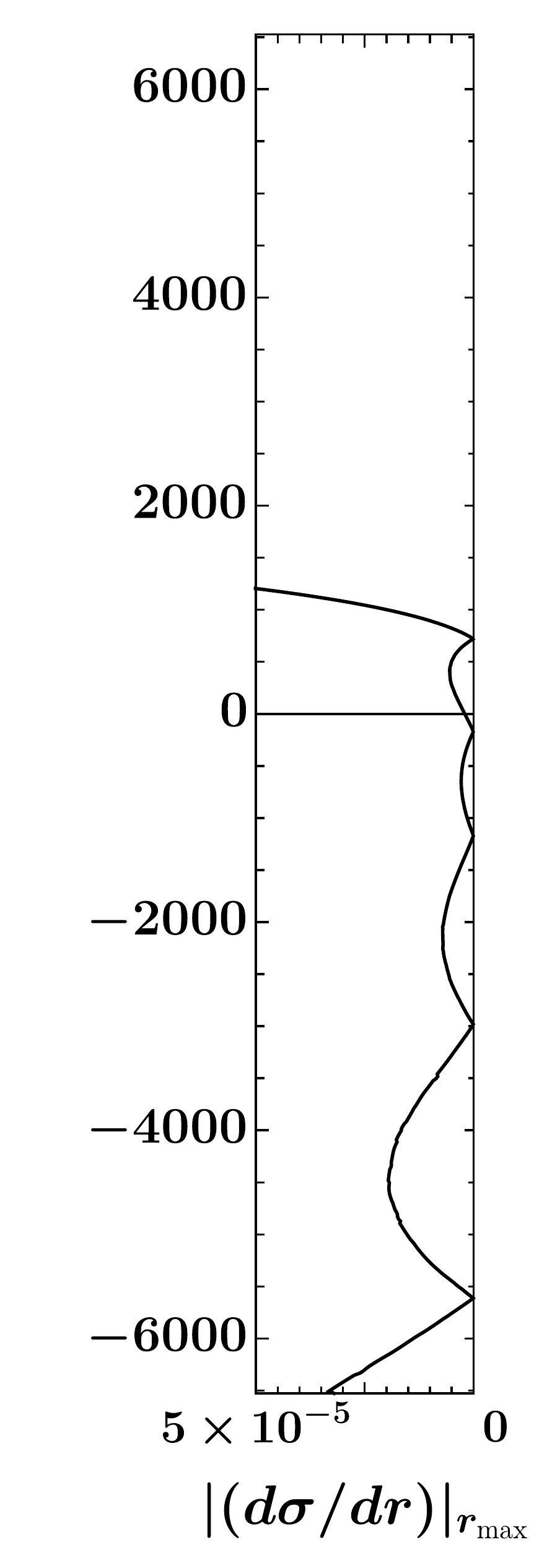}%
 \quad\includegraphics[height=0.557\textwidth, width=0.557\textwidth]{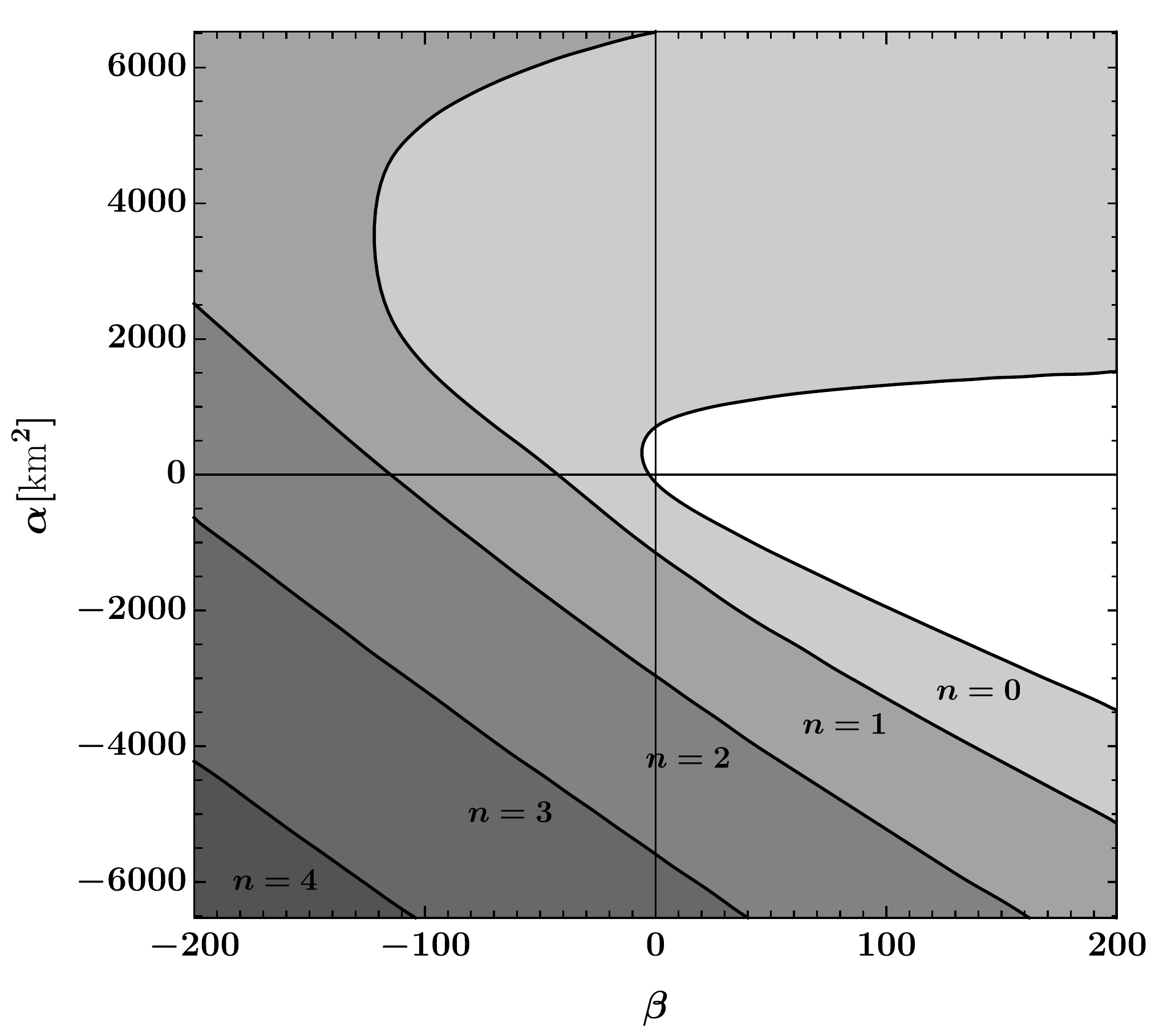}}
 \\
 \qquad\qquad\qquad\quad
 \subfloat{\includegraphics[height=0.19\textwidth, width=0.58\textwidth]{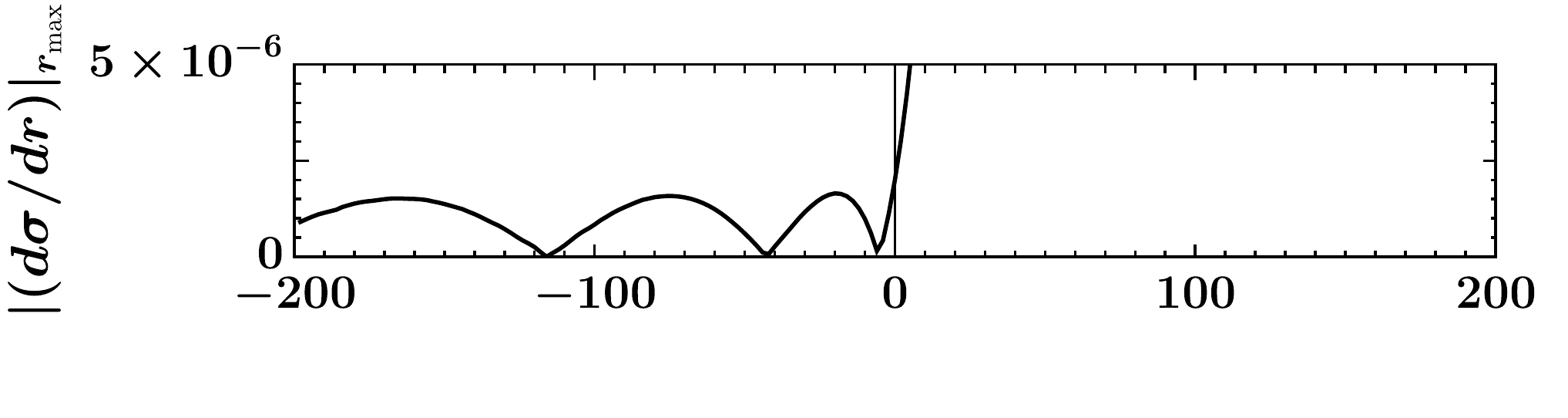}}
 \caption[Stable and unstable regions in the parameter space for a light neutron star]{Stable and unstable regions in the $(\beta,\alpha)$ space for a light star ($M=1.12~M_\odot$, SLy equation of state). In the 2D plot, each line is labeled according to the number of nodes $n$ of the corresponding unstable mode. Inside the white region, where the point $(\beta,\alpha)=(0,0)$ lies, the general relativity solution is stable. Every line crossed while moving away from the origin corresponds to the appearance of a new unstable mode; any point in parameter space that lies within a grey region corresponds to an unstable solution. The lower panel shows $|\mathrm{d}\sigma/\mathrm{d}r|$ at $r_\mathrm{max}$ when varying $\beta$ in the same range as the 2D plot, with $\alpha=0$; it can be understood as a cut in the $(\beta,\alpha)$ plane along the $\beta$ axis (each cusp corresponding to a line-crossing in the 2D plot). Similarly, the left panel shows a cut along the $\alpha$ axis.}
 \label{fig:alphabeta}
 \end{figure}

 We first consider a vanishing bare mass, $m_\phi=0$. The model is then parametrized by $\beta$ and $\alpha$ only. The background we consider is a neutron star described by the SLy equation of state. We choose its central energy density to be $\epsilon_0=8.1\times10^{17}$~kg/m$^3$, so that its gravitational mass is $M=1.12~M_\odot$, which corresponds to the bottom of the mass range for observed neutron stars~\cite{Martinez:2015mya,Suwa:2018uni}. The radius of the star is then $R_\mathrm{s}=11.7$~km. The results are summarized in Fig.~\ref{fig:alphabeta}.
 The white area corresponds to the region of the parameter space where the background solution is stable. A new unstable mode appears when crossing each line while moving away from the origin. The lines are labeled with the number of nodes $n$ of the associated mode, ranging from 0 to infinity. The $n=0$ line is the boundary of the stable region. Any choice of parameters beyond this line will make the general relativity solution unstable. The left and bottom panels show cuts in the $(\beta,\alpha)$ plane, along the $\alpha$ and $\beta$ axes respectively; these panels actually reproduce known results, {\em e.g.} of~\cite{Harada:1997mr} and~\cite{Silva:2017uqg}. Notably, the scalarization threshold when only $\beta$ is presents takes the well-known order of magnitude, $\beta=-5.42$.

\begin{figure}
 \subfloat[]{%
  \includegraphics[width=0.45\linewidth]{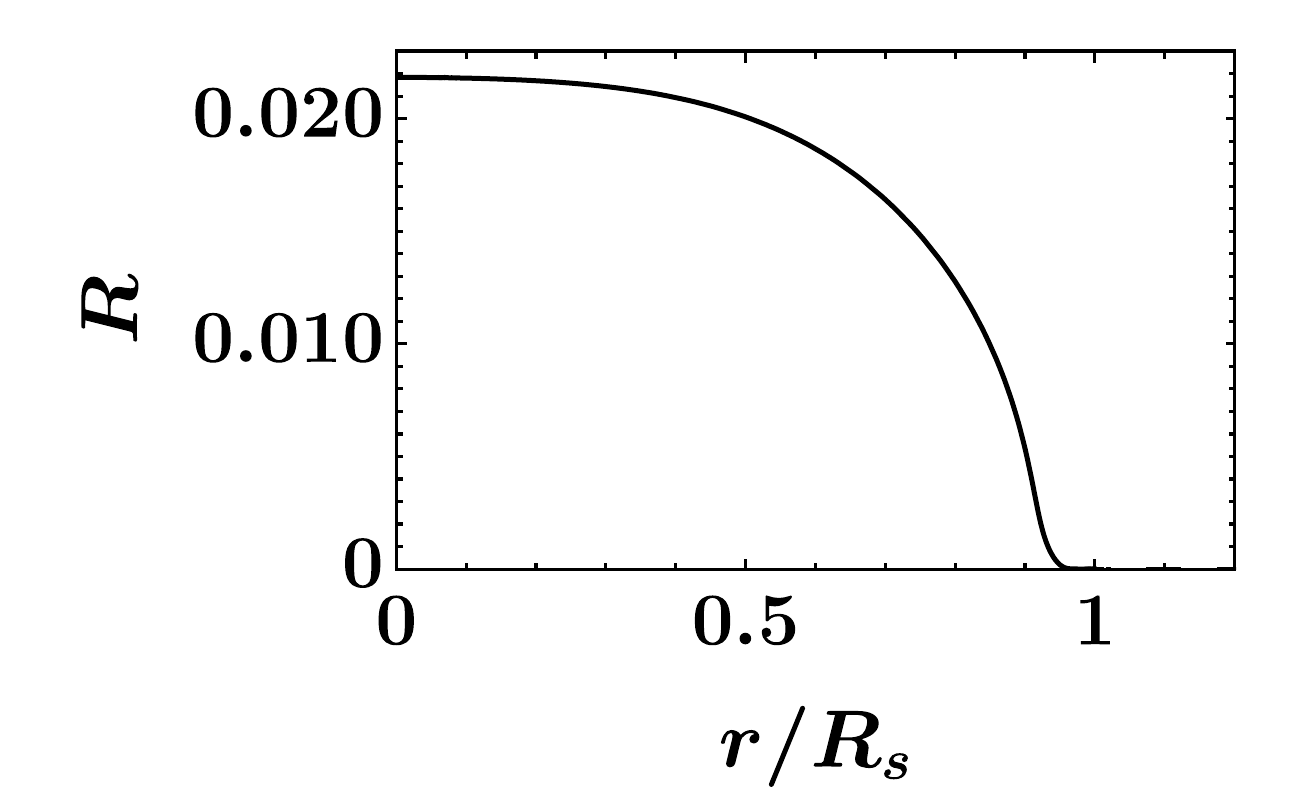}%
 }\hfill
 \subfloat[]{%
  \includegraphics[width=.45\linewidth]{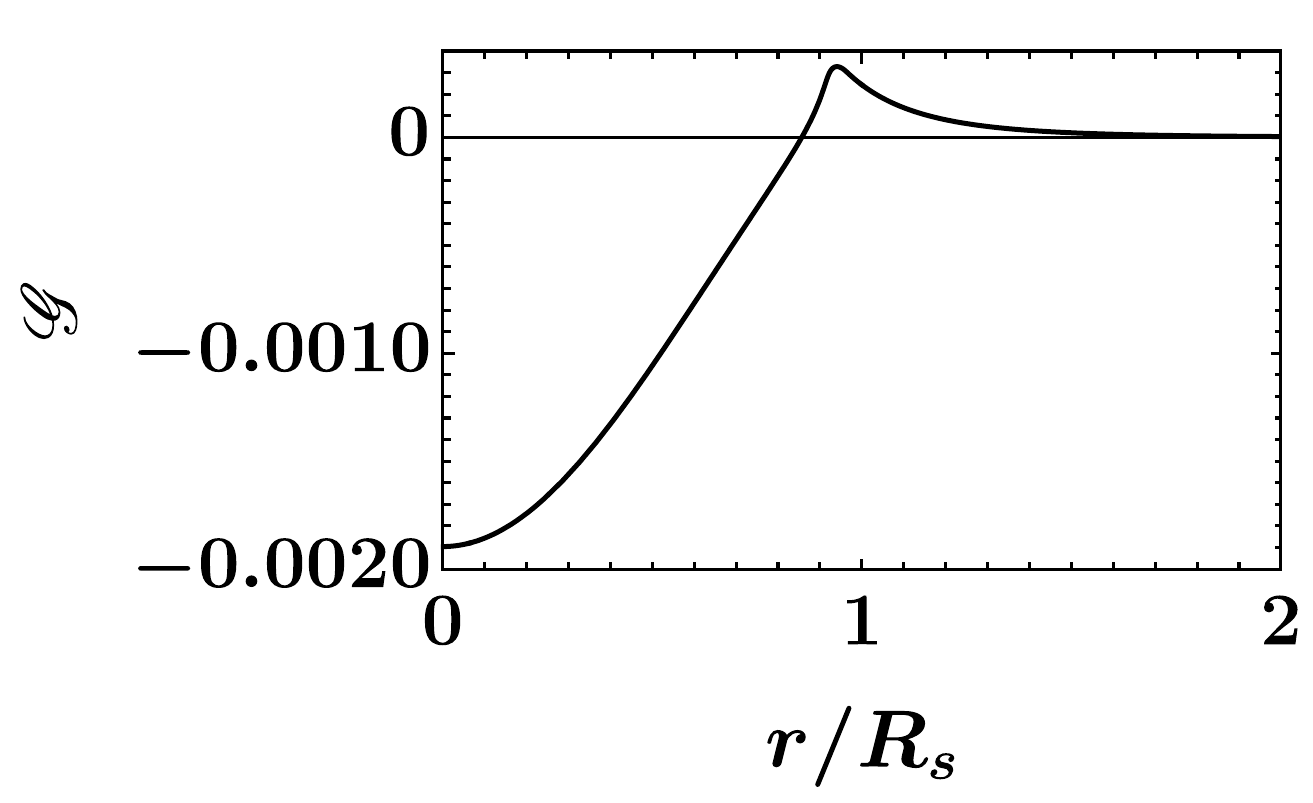}%
 }
 \caption[Ricci scalar and Gauss-Bonnet invariant for a light neutron star]{Ricci scalar and Gauss-Bonnet invariant for a light star ($M=1.12~M_\odot$, SLy equation of state). The radial coordinate is rescaled by the radius of the star $R_\mathrm{s}=11.7$~km. The left panel shows that the Ricci scalar is non-negative everywhere; correspondingly, only $\beta<0$ can lead to an instability. On the other hand, the Gauss-Bonnet invariant, shown in the right panel, is negative in the core of the star and positive towards its surface, leading to instabilities both for $\alpha<0$ and $\alpha>0$.}
 \label{fig:RGB112}
 \end{figure}

 To understand better the shape of the plot, especially along the $\beta$ and $\alpha$ axes, we plot in Fig.~\ref{fig:RGB112} the Ricci scalar and the Gauss-Bonnet invariant.
 The Ricci scalar is always positive on this background. This is due to the fact that we consider a relatively light neutron star, and due to the following relation:
 \begin{equation}
 R=\kappa(\epsilon-3p).
 \label{eq:ricci}
 \end{equation}
 When the medium is not too dense, $\epsilon\gg p$ and $R>0$. We will see how this changes for a very dense star in Section~\ref{Sub:starproperties}. As a consequence, only negative $\beta$ can generate a negative effective mass. On the other hand, as shown in Fig.~\ref{fig:RGB112}, the Gauss-Bonnet invariant is  positive in some regions and negative in others. This is enough to trigger an instability when $\alpha$ becomes very negative or very positive, which is indeed what is observed in Fig.~\ref{fig:alphabeta}.

 We notice, as expected, that the point $(0,0)$ is always inside the stable region. The tachyonic instability does not appear right away when $\beta<0$ or $\alpha\neq0$, due to the curvature of spacetime. 

\begin{figure}
 \centering
 \includegraphics[width=0.7\textwidth]{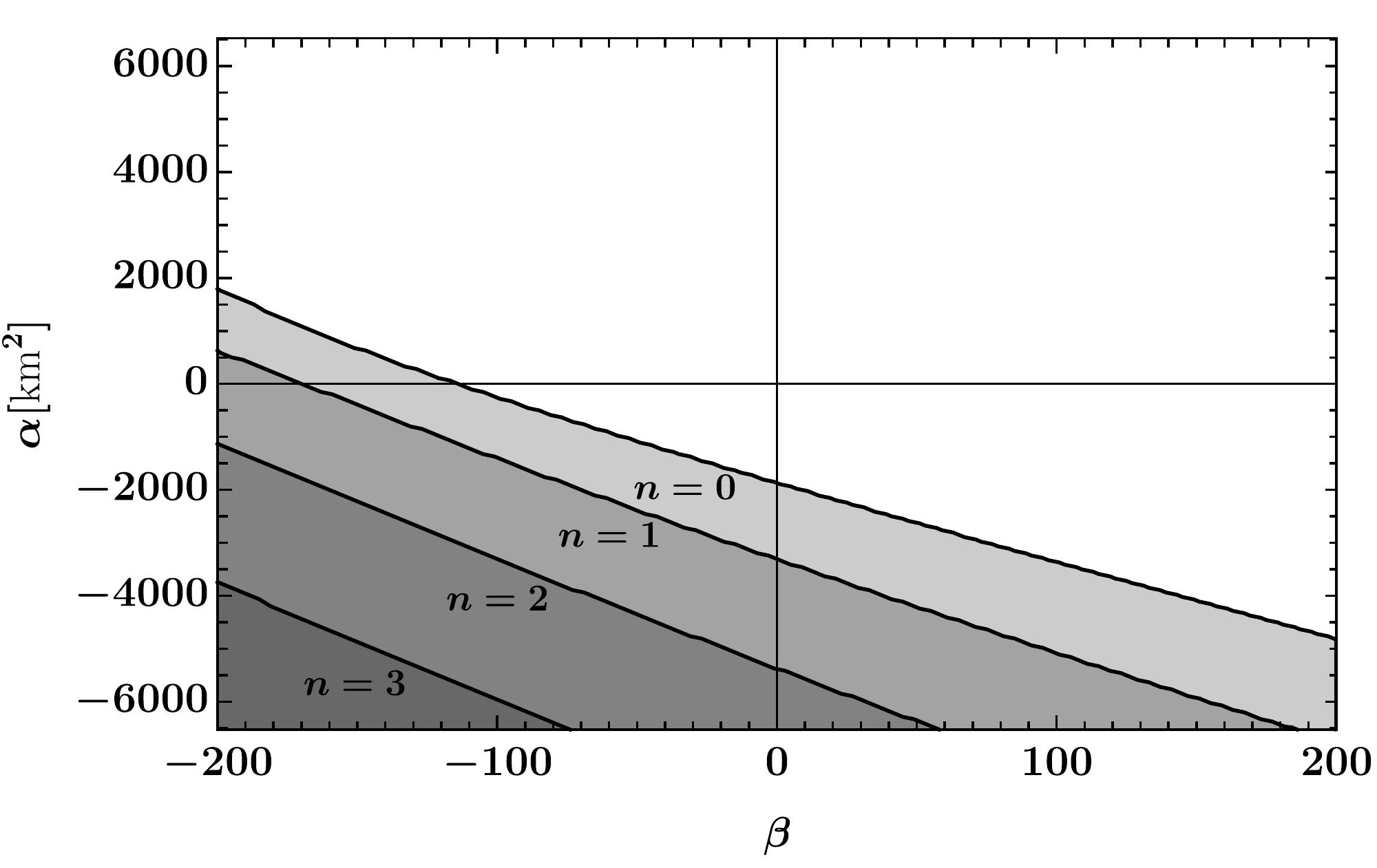}
 \caption[Stable and unstable regions in the parameter space for a light neutron star and a bare scalar field mass]{Stable and unstable regions in the $(\beta,\alpha)$ space for a bare mass of $1.33\times10^{-10}$~eV. As expected, the stable region is enlarged with respect to Fig.~\ref{fig:alphabeta}. Note that the range of the plot for $\beta$ and $\alpha$ is the same as in Fig.~\ref{fig:alphabeta} to facilitate comparison.}
 \label{fig:mphi}
 \end{figure}

 We now consider how the presence of a bare mass affects the results of the previous Section. Fig.~\ref{fig:mphi} shows the region of stability in the $(\beta,\alpha)$ plane when $m_\phi=1$ in the system of units that we used, i.e., scalar particles have a mass of $1.33\times10^{-10}$~eV. The range of parameters in Fig.~\ref{fig:mphi} is the same as in Fig.~\ref{fig:alphabeta} in order to allow comparison. When one zooms out, Fig.~\ref{fig:mphi} looks very similar to Fig.~\ref{fig:alphabeta}  (i.e. the same instability pattern remains valid, but it appears for higher values of the parameters).
 As can be seen from comparing Figs.~\ref{fig:alphabeta} and~\ref{fig:mphi}, the stable region is widened in all directions. As expected, the presence of a bare mass stabilizes the solution (similarly, if the bare mass is tachyonic, the stability region shrinks). Therefore, even a sufficiently large bare mass for the scalar field is able to shield neutron stars from scalarization. This was already noted, {\em e.g.} in~\cite{Ramazanoglu:2016kul}, where the effect of a bare mass in a cosmological setup is also discussed.

\subsection{Changing the properties of the star}\label{Sub:starproperties}

 Let us now examine how changing the background affects the stability. We first consider a more massive star, and then a different equation of state.

\begin{figure}
 \centering
 \includegraphics[width=0.7\textwidth]{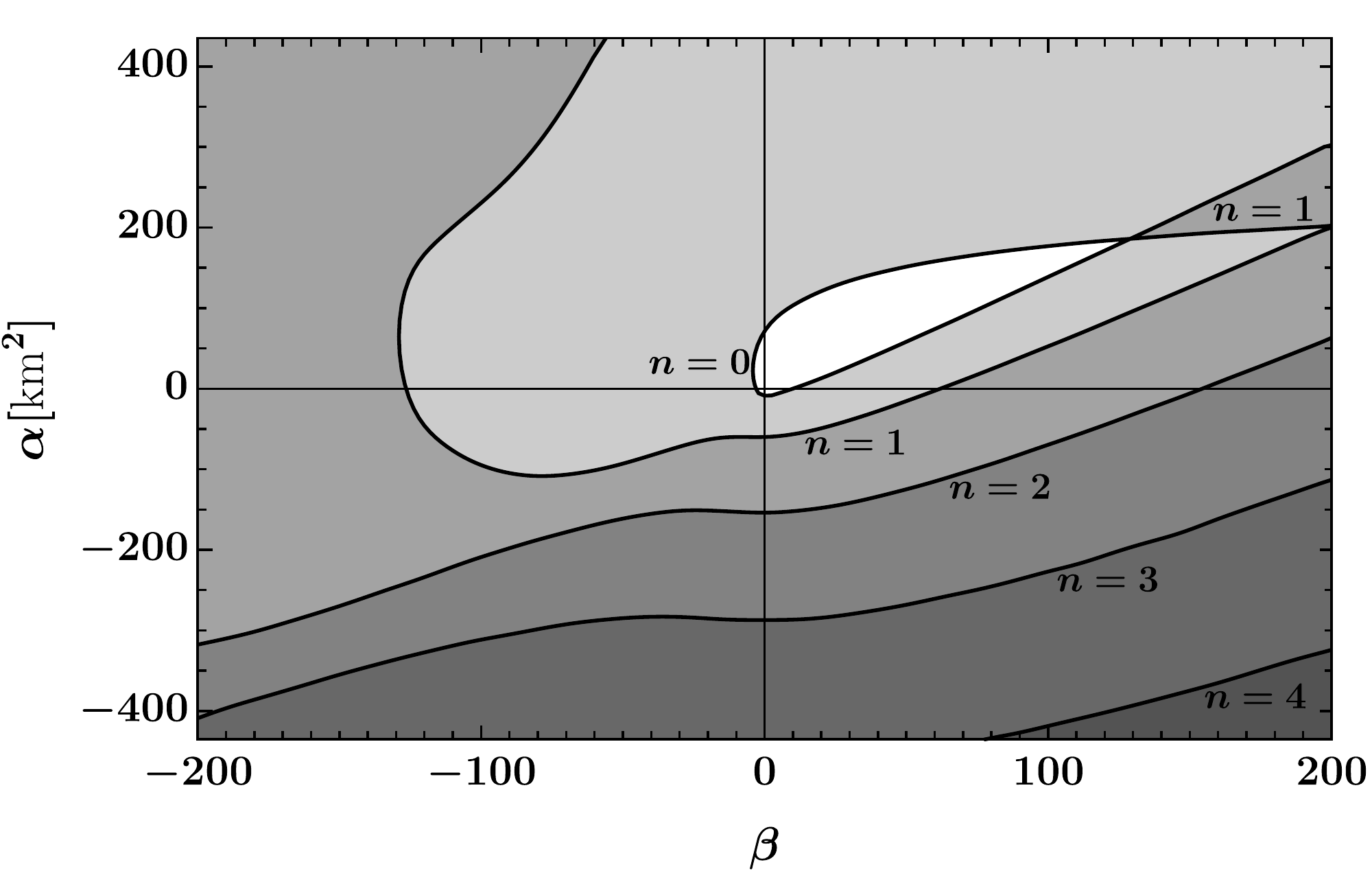}
 \caption[Stable and unstable regions in the parameter space for a heavy neutron star]{Stable and unstable regions in the $(\beta,\alpha)$ space for a heavy neutron star ($M=2.04~M_\odot$, SLy equation of state). The vertical range for $\alpha$ is reduced with respect to Fig.~\ref{fig:alphabeta} for readability. The wedge of stability in Fig.~\ref{fig:alphabeta} has narrowed down to an island around $(\beta,\alpha)=(0,0)$.}
 \label{fig:heavystar}
 \end{figure}

\begin{figure}
 \subfloat[]{%
  \includegraphics[width=0.45\linewidth]{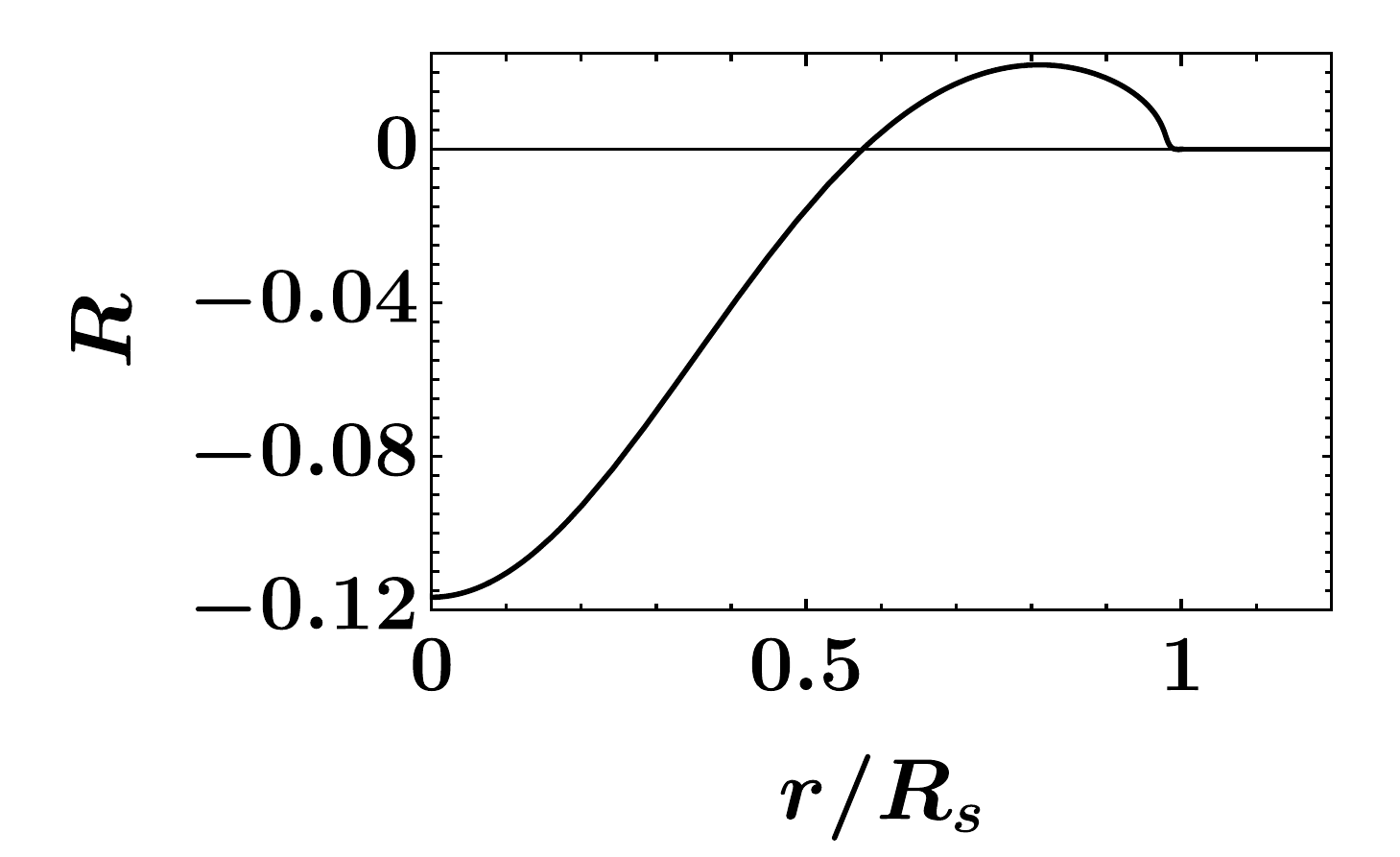}%
 }\hfill
 \subfloat[]{%
  \includegraphics[width=.45\linewidth]{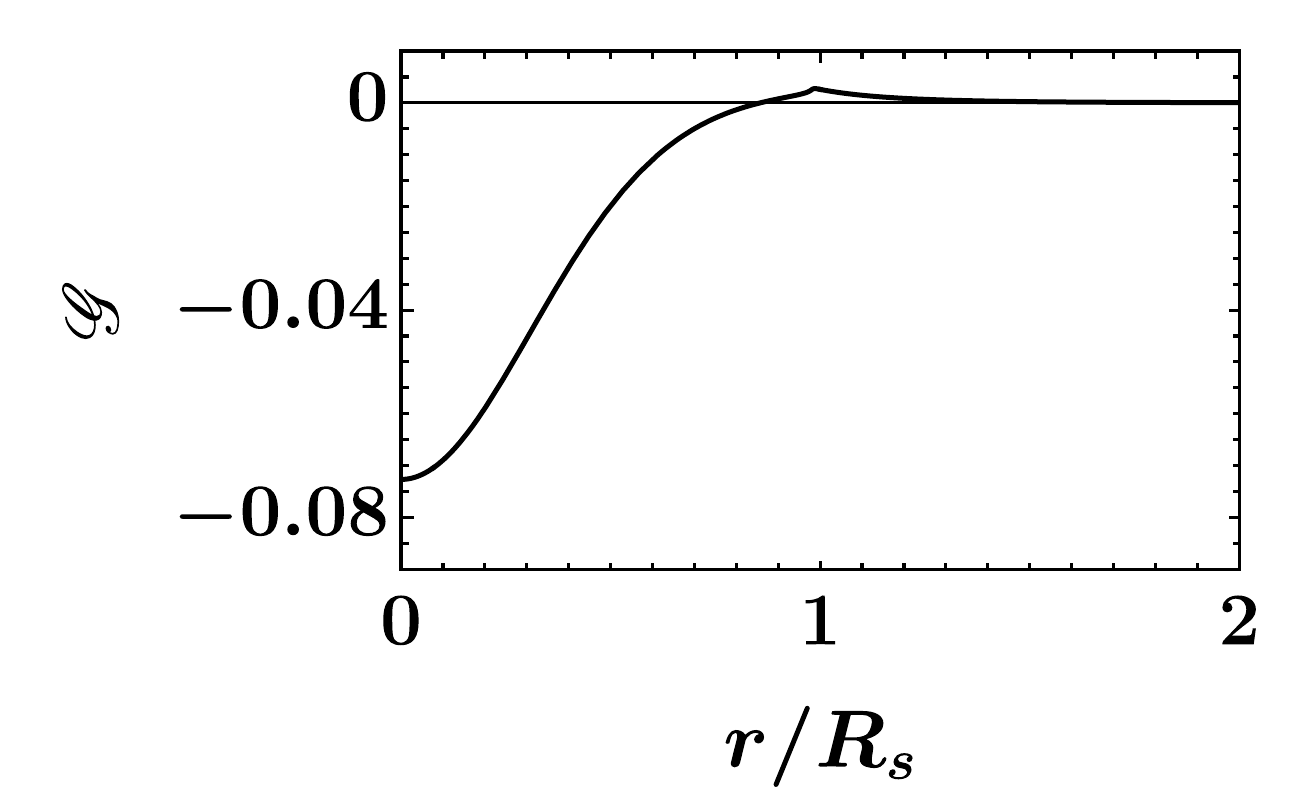}%
 }
 \caption[Ricci scalar and Gauss-Bonnet invariant for a heavy star]{Ricci scalar and Gauss-Bonnet invariant for a heavy star ($M=2.04~M_\odot$, SLy equation of state). Now, the left panel shows that $R$ becomes negative inside the core of the star, allowing instabilities for both signs of $\beta$. The Gauss-Bonnet invariant (right panel) still behaves as in the case of a light star, Fig.~\ref{fig:RGB112}.}
 \label{fig:RGB204}
 \end{figure}

 We first increase the mass of the neutron star (and thus its compactness at the same time). We choose a central density of $\epsilon_0=3.4\times10^{18}$~kg/m$^3$, which corresponds to a mass of $M=2.04~M_\odot$. This is the heaviest spherically symmetric star we can produce with the SLy equation of state; beyond this mass, the solutions become unstable already within general relativity. The results are displayed in Fig.~\ref{fig:heavystar}.
 Since the curvature of the background increased with respect to Fig.~\ref{fig:alphabeta}, it is not surprising that the stable region shrunk. A more unexpected feature is that very positive values of $\beta$ now also lead to an instability. This is due to the fact that the Ricci scalar now becomes negative towards the center of the star, as can be seen in Fig.~\ref{fig:RGB204}.
 In terms of Eq.~\eqref{eq:ricci}, the energy density is no longer dominant with respect to the pressure in the extremely pressurized core of the neutron star, allowing $R<0$.
 This effect was already noted in~\cite{Mendes:2014ufa}, and further studied in~\cite{Palenzuela:2015ima,Mendes:2016fby}.

 We now consider a different stellar model, the MPA1 equation of state~\cite{Gungor:2011vq}. In order to compare with previous results, we keep the same mass as in Section~\ref{Sub:effectivemass}, $M=1.12~M_\odot$. This corresponds to a central density of $\epsilon_0=6.3\times10^{17}$~kg/m$^3$. The stability region is shown in Fig.~\ref{fig:MPA1}.
 \begin{figure}
 \centering
 \includegraphics[width=0.7\textwidth]{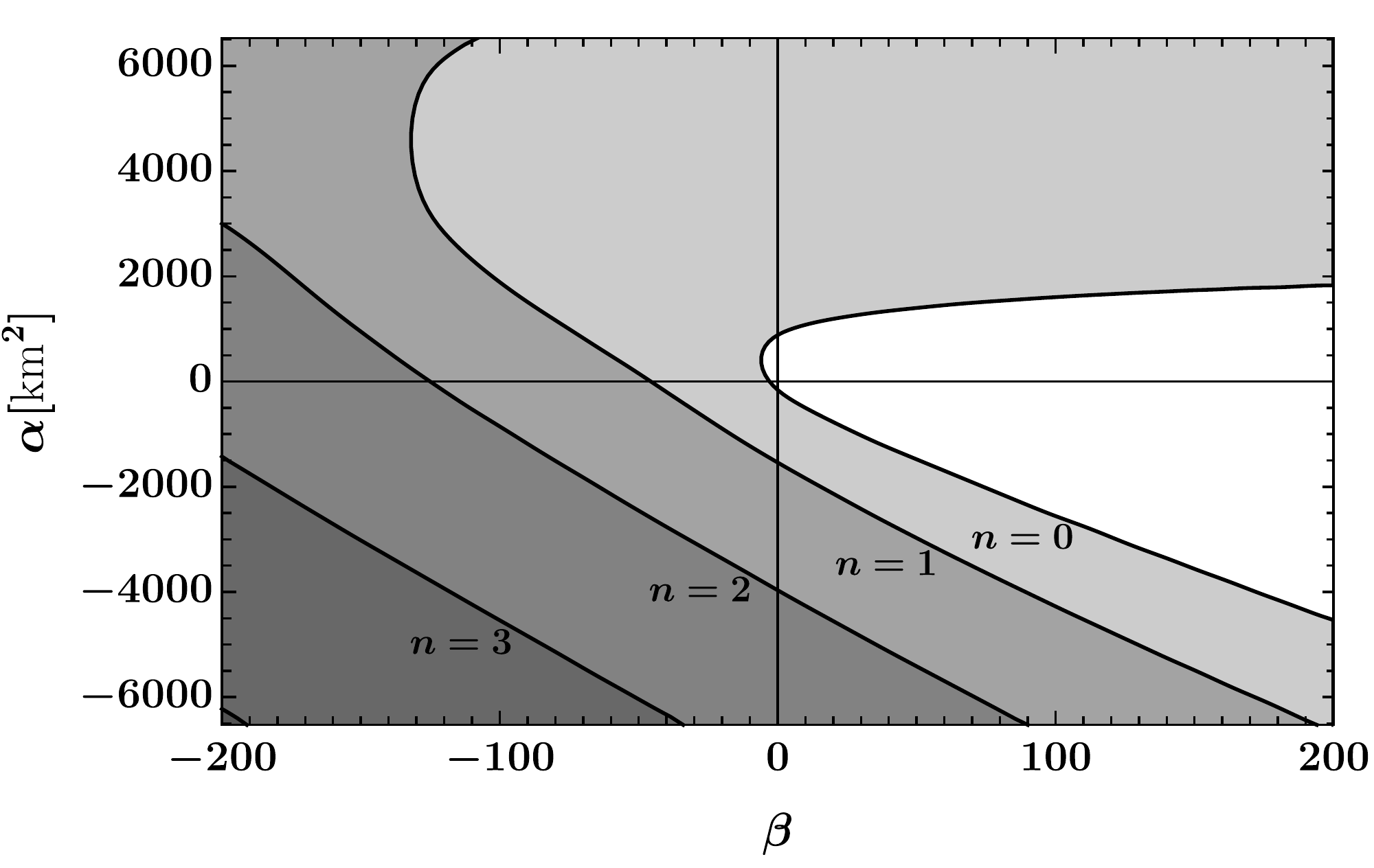}
 \caption[Stable and unstable regions in the parameter space for the MPA1 equation of state]{Stable and unstable regions in the $(\beta,\alpha)$ space for the MPA1 equation of state ($M=1.12~M_\odot$). The range for $\beta$ and $\alpha$ is the same as in Fig.~\ref{fig:alphabeta}. We note that the choice of equation of state does not affect significantly the results.}
 \label{fig:1MPA1}
 \end{figure}
 By comparing Figs.~\ref{fig:alphabeta} and~\ref{fig:1MPA1}, it is clear that the different choice of equation of state does not affect significantly the position of the stable and unstable regions. 

 Although the equation of state influences only mildly stars of similar mass, it can have indirect effects. Indeed, a softer equation of state will lead to a smaller pressure for a given energy density; thus, for a soft equation of state, the Ricci scalar~\eqref{eq:ricci} could remain positive in all configurations, discarding configurations like Fig.~\ref{fig:heavystar}. Similarly, the equation of state can affect the range allowed for the parameter $\gamma$, as we will see in more detail in the next Section.

\subsection{Changing the effective metric}\label{Sub:effectivemetric}

 We now return to the parameter $\gamma$ and examine its role. As mentioned above, the terms controlled by $\gamma$ cannot  generate a tachyonic instability in the absence of the other couplings controlled by $\beta$, $\alpha$ or $m_\phi$. Nonetheless, choosing $\gamma$ beyond a certain range leads to loss of hyperbolicity in the scalar field equation. Also the potential $V_\mathrm{eff}$ does depend on $\gamma$, as can be seen from Eq.~\eqref{eq:Veff}. We analyze these aspects below, before studying numerically the combined effect of $\gamma$ and the other parameters.
 
 The parameter $\gamma$ brings an additional contribution to the effective metric experienced by scalar perturbations, Eq.~\eqref{effMetr}. If this contribution exceeds a certain threshold and becomes dominant, the effective metric becomes elliptic, rendering the background unstable.  We emphasize that this instability is distinct from the usual tachyonic instability associated with scalarization. Depending on the circumstances, it is either a gradient or a ghost instability. It is possible that this instability can be quenched nonlinearly, as in the case for conventional scalarization. Indeed, scalarization through a ghost instability has already been proposed in~\cite{Ramazanoglu:2017yun}. Since here we are not including terms that are nonlinear in the scalar in our analysis, we cannot follow the development and potential quenching.

 One can therefore view the analysis that follows in two different ways. Taking the conservative viewpoint, one may restrict to the well-controlled tachyonic scalarization. In this framework, our results allow to set bounds on the parameter $\gamma$. In a more open-minded perspective, which certainly deserves further investigation in the future, we are setting  bounds beyond which ghost or gradient scalarization can be triggered.

\begin{figure}
 \centering
 \includegraphics[width=0.7\textwidth]{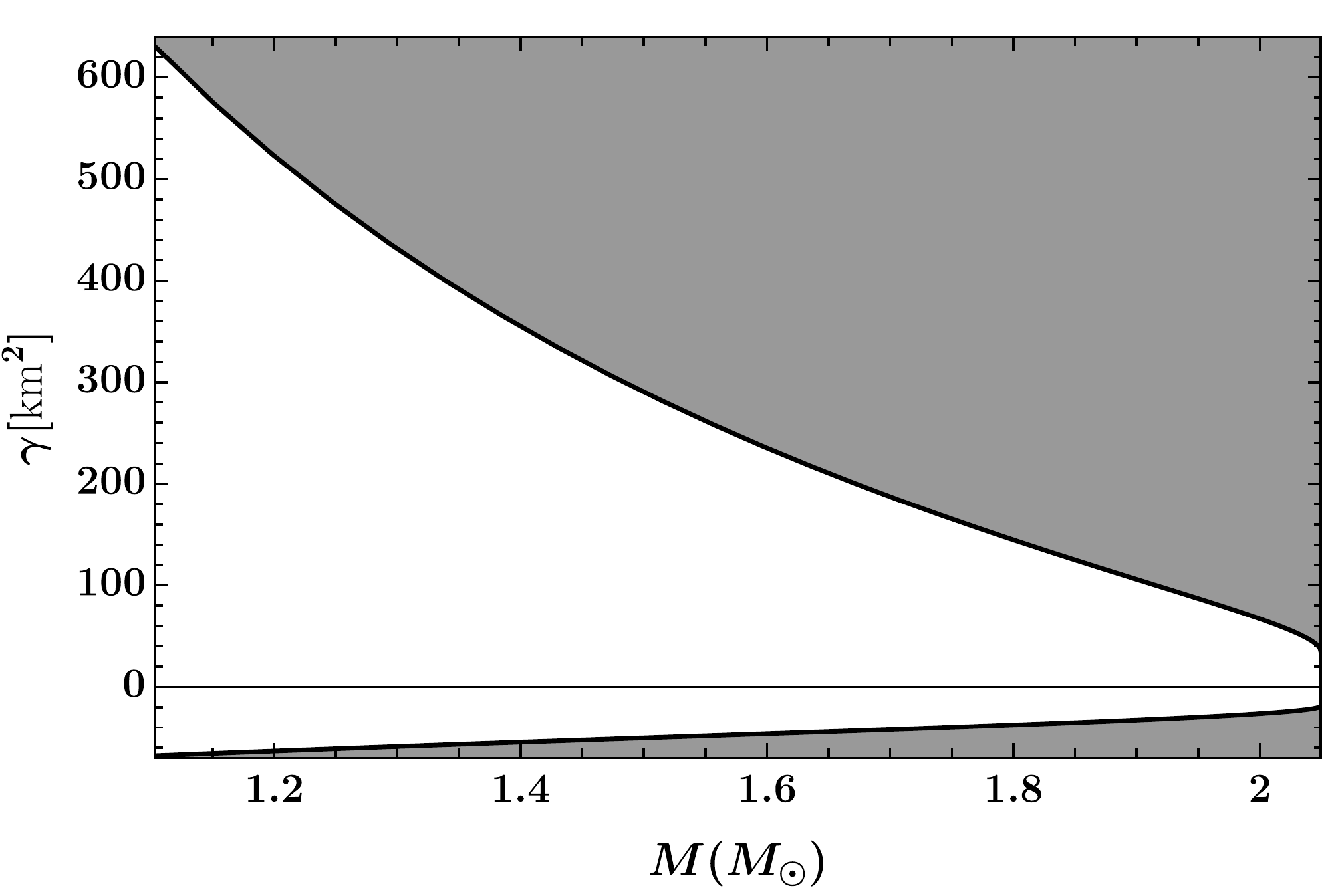}
 \caption[Hyperbolicity of the effective metric]{Hyperbolicity of the effective metric in the $(M,\gamma)$ plane (SLy equation of state). The effective metric is hyperbolic only within the white region. The mass ranges from the putative lowest neutron star mass~\cite{Martinez:2015mya,Suwa:2018uni} to the maximal mass achieved with the SLy equation of state. The range of hyperbolicity narrows down when increasing the mass of the star.}
 \label{fig:Ma41}
 \end{figure}

 Given that we are working on a general relativity background, we can make use of Einstein equations. The inverse of $\tilde g^{\mu\nu}$ for a perfect fluid is then
 \begin{equation}
 \tilde g^{-1}_{\mu\nu} = \dfrac{1}{1-\kappa \gamma p} \left(g_{\mu\nu}-\kappa \gamma\dfrac{\epsilon+p}{1+\kappa \gamma\epsilon}\,u_\mu u_\nu\right),
 \end{equation}
 where $u_\mu$ is the 4-velocity of the fluid. Over the spherically symmetric background that we study, the determinant of the effective metric reads
 \begin{equation}
 \tilde g = g \dfrac{1}{(1-\kappa \gamma p)^3(1+\kappa \gamma\epsilon)}\,.
 \end{equation}
 On a given background, the pressure and energy density are maximal at the center of the star, where we label their value as $p_0$ and $\epsilon_0$. The determinant of the physical metric $g$ is always negative. Thus, the effective metric loses hyperbolicity either when $\kappa \gamma$ becomes larger than $1/p_0$ or when $\kappa \gamma$ becomes more negative than $-1/\epsilon_0$. To summarize, hyperbolicity of the effective metric requires that
 \begin{equation}
 -\dfrac{1}{\kappa \epsilon_0} <\gamma<\dfrac{1}{\kappa p_0}.
 \label{eq:bounda41}
 \end{equation}
 Reference~\cite{Minamitsuji:2016hkk} already noted the existence of the upper bound in a similar context. In the numerical analysis below, we take the conservative approach and restrict to this range. For a given equation of state, the bounds in Eq.~\eqref{eq:bounda41} can be reformulated in terms of the mass of the star, $M$. This is shown in Fig.~\ref{fig:Ma41}.
 The range in which the scalar field equation is hyperbolic closes up around $\gamma=0$ when increasing the mass. The limits presented in Fig.~\ref{fig:Ma41} depend on the equation of state, but only mildly. In the framework of tachyonic scalarization, we can use these limits to put an absolute bound on $\gamma$. For the SLy equation of state,
 \begin{equation}
 -18.7~\mathrm{km}^2 <\gamma<34.9~\mathrm{km}^2.
 \label{eq:absbounda41}
 \end{equation}
 These bounds should only be taken as order of magnitude estimates; they have been established in the decoupling limit and rely on a specific equation of state. However, a more detailed study would allow to put very precise constraints on $\gamma$. We are not aware of previously established bounds on this parameter (except~\cite{Minamitsuji:2016hkk}). Note that the analysis presented in this Section is entirely independent on the other parameters, which do not play a role in the effective metric.
 
 Before moving to the numerical results, we point out an important difference between black holes and stars. In both cases (black hole and neutron star) the scalar field equation can be brought to the form~\eqref{scalarEqPot2}, and we impose that $\mathrm{d}\hat\sigma/\mathrm{d}r_\ast$ vanishes at large $r_\ast$ so that $\int \phi^2$ is finite and the perturbation initially contains a finite amount of energy. The difference between neutron stars and black holes appears on the other side of the $r_\ast$ range. In the case of neutron star, $r_\ast$ goes down to 0, where $\phi=K\hat\sigma\sim\phi_0/r_\ast$ for some constant $\phi_0$, unless $\hat\sigma(0)=0$. We do not want $\phi$ to diverge at the center of the star, so we impose $\hat\sigma(0)=0$. In the black hole case on the other hand, $r_\ast$ goes down to $-\infty$ and nothing particular happens at $r_\ast=0$. The condition for the perturbation to be physical is then that $\mathrm{d}\hat\sigma/\mathrm{d}r_\ast$ vanishes for $r_\ast\to-\infty$, similarly to what happens at large $r_\ast$.

 It is then possible to establish an exact sufficient condition for the presence of an instability in the black hole case. Indeed, the function $\hat\sigma$ then respects the hypotheses of the theorem established in Ref.~\cite{doi:10.1119/1.17935}: if $\int_{-\infty}^{+\infty} V_\mathrm{eff}(r_\ast)\mathrm{d}r_\ast<0$, then $V_\mathrm{eff}$ admits at least one bound state. 

 However, in the case of a star, due to the different boundary conditions, $\int_{0}^{+\infty} V_\mathrm{eff}(r_\ast)\mathrm{d}r_\ast$ can become negative ---~even infinitely negative~--- while $V_\mathrm{eff}$ does not possess any bound state. This is equivalent to the 3D-spherically symmetric quantum well; consider such a well with depth $-V_0$ and width $a$. It admits a bound state only for $V_0a^2>N$ ($N$ is some constant that depends on the mass of the quantum particles), while the integral of the potential is $-V_0 a<0$. Choosing the scaling $V_0=N/2/a^2$, the integral of the potential is then becoming infinitely negative for $a\to0$, while no bound state exists.

\begin{figure}
 \centering
 \includegraphics[width=0.7\textwidth]{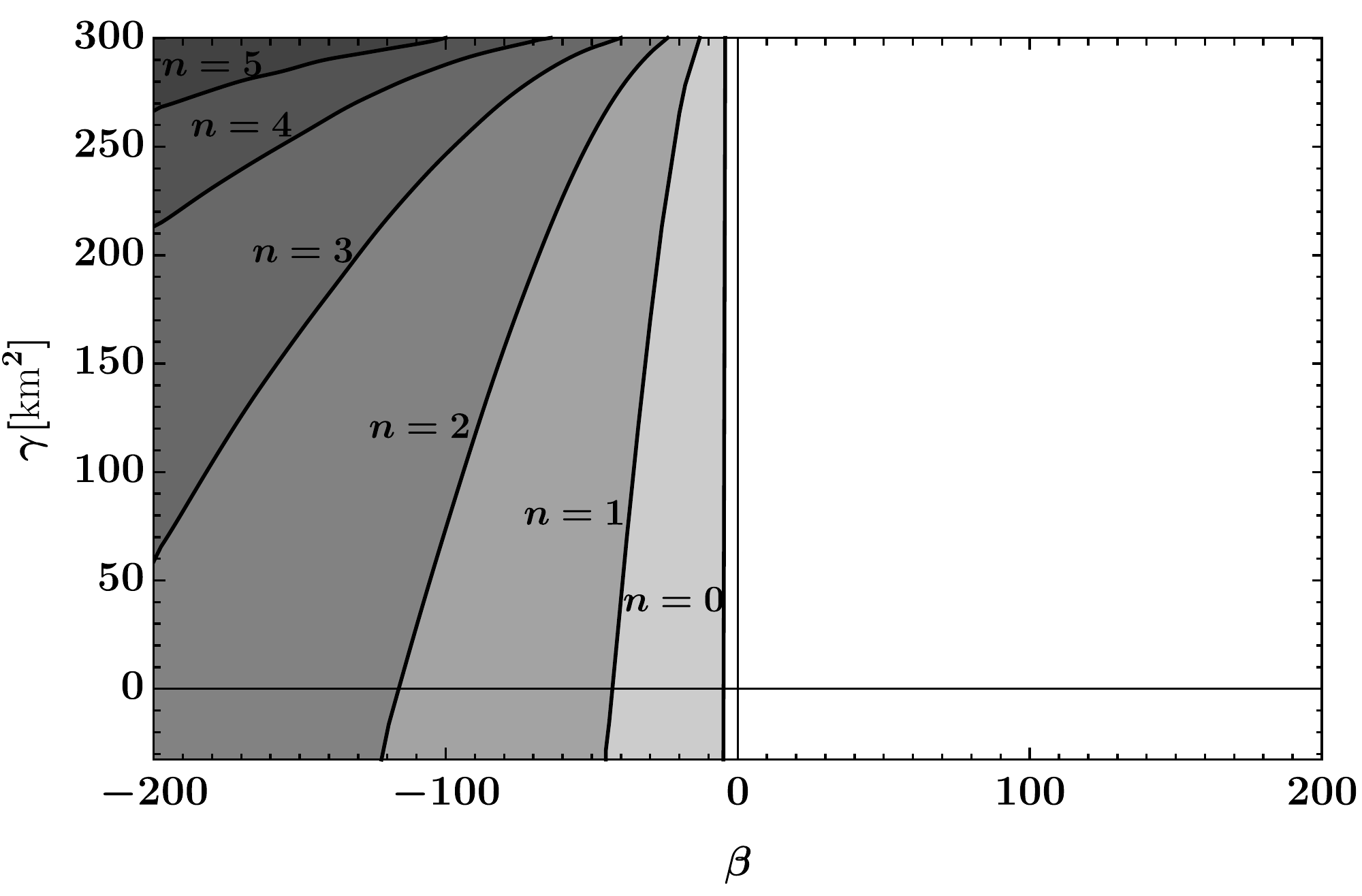}
 \caption[Stable and unstable regions for scalarization with Ricci scalar coupling changing the effective metric]{Stable and unstable regions in the $(\beta,\gamma)$ space ($M=1.12~M_\odot$, SLy equation of state) for $\alpha=0$. The region of stability is rather unaffected by a change in $\gamma$. The bound states pile up close to the upper bound for $\gamma$.}
 \label{fig:alphaa41}
 \end{figure}

 Hereinafter, $\int V_\mathrm{eff}(r_\ast)\mathrm{d}r_\ast$ indeed becomes infinitely negative in the limit where $\gamma$ approaches one of the bounds of Eq.~\eqref{eq:bounda41}. However, this does not necessarily mean that infinitely many bound states develop close to these boundaries. This seems to be true close to the upper bound, but not close to the lower one, as we will see in Fig.~\ref{fig:alphaa41}.

\begin{figure}
 \centering
 \includegraphics[width=0.7\textwidth]{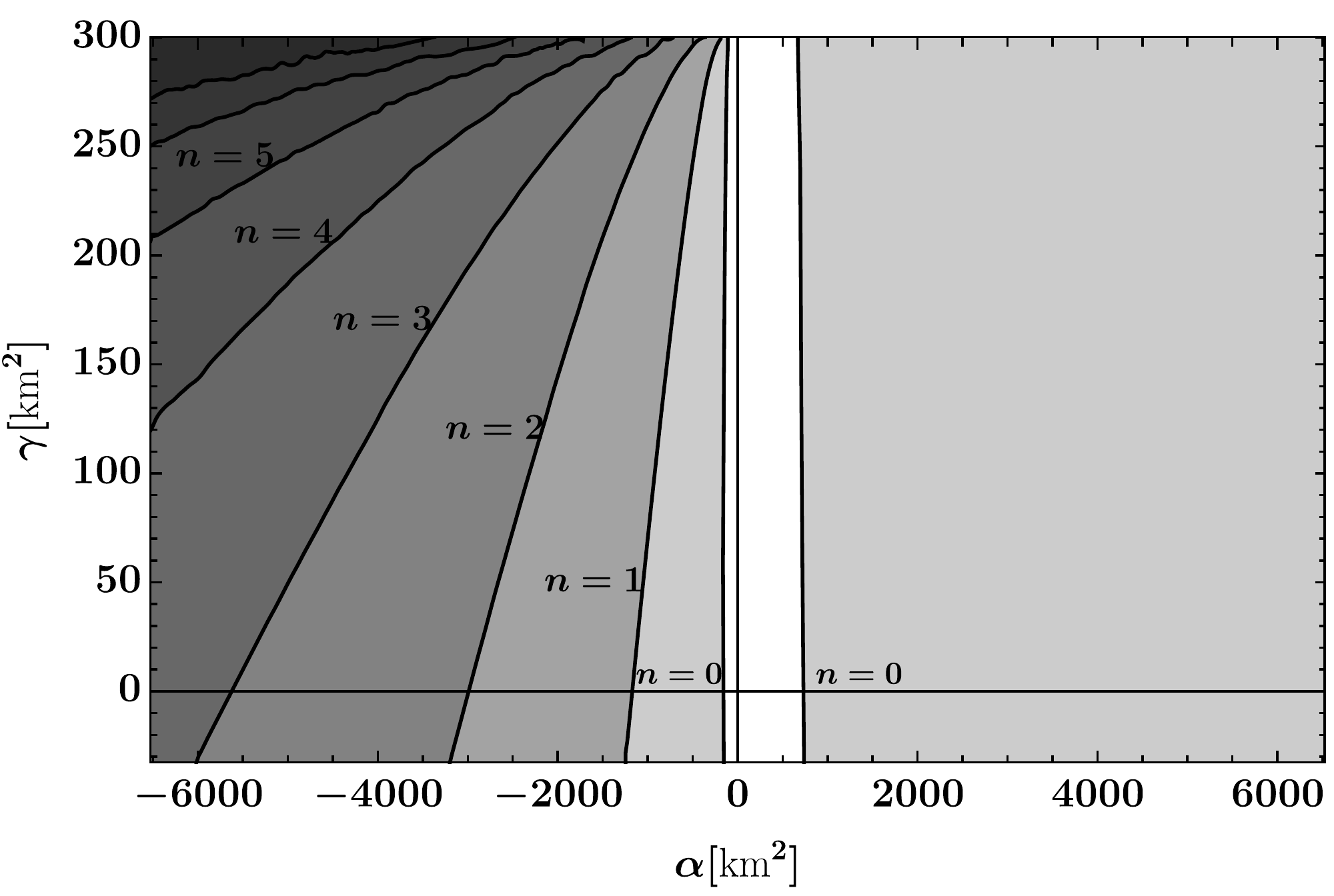}
 \caption[Stable and unstable regions for scalarization with Gauss-Bonnet scalar coupling changing the effective metric]{Stable and unstable regions in the $(\alpha,\gamma)$ space ($M=1.12~M_\odot$, SLy equation of state) for $\beta=0$. The behavior is very similar to what we obtained for $\beta$ in Fig.~\ref{fig:alphaa41}.}
 \label{fig:betaa41}
 \end{figure}

 We now vary $\gamma$ systematically in the range allowed by Eq.~\eqref{eq:bounda41}, and $\beta$ and $\alpha$ in similar ranges as in the previous figures. The results are displayed in Figs.~\ref{fig:alphaa41} and~\ref{fig:betaa41} respectively.
 When $\gamma$ vanishes (cut along the horizontal axis in Figs.~\ref{fig:alphaa41} and~\ref{fig:betaa41}), we retrieve the same stability ranges as in the left and bottom panels of Fig.~\ref{fig:alphabeta} ($\beta>-5.42$ and $722~\mathrm{km}^2>\alpha>-169~\mathrm{km}^2$). It is also natural that the lines $\beta=0$ and $\alpha=0$ (vertical axes in Figs.~\ref{fig:alphaa41} and~\ref{fig:betaa41}) correspond to stable configurations, because $\gamma$ cannot create a tachyonic effective mass on its own. The boundary of the stable region is rather insensitive to the parameter $\gamma$; it does evolve slightly with the value of $\gamma$, but this can be seen only when zooming around smaller values of $\beta$ and $\alpha$ with respect to Figs.~\ref{fig:alphaa41} and~\ref{fig:betaa41}. Close to both bounds of the $\gamma$ range, $\int V_\mathrm{eff}(r_\ast)\mathrm{d}r_\ast$ diverges, but as explained before, this does not necessarily mean that infinitely many bound states should appear (or even that one bound state exists). Indeed, nothing particular happens when approaching the lower bound, while unstable modes pile up when approaching the upper bound; both scenarios are allowed.
 
 To summarize, we have investigated exhaustively the effect of all the terms that play a role in the onset of spontaneous scalarization in action~\eqref{eq:ACI}, in the context of neutron stars. Our analysis has identified the role of each term but has also revealed their combined effects. When taking each terms separately, our results agree with previous results regarding scalarization thresholds. More generally and when all terms are present, we have found that  a very small bare mass suffices to stabilize general relativity solutions, and that the scalarization thresholds are only mildly sensitive to the choice of equation of state.
 \begin{figure}
 \includegraphics[width=0.7\textwidth]{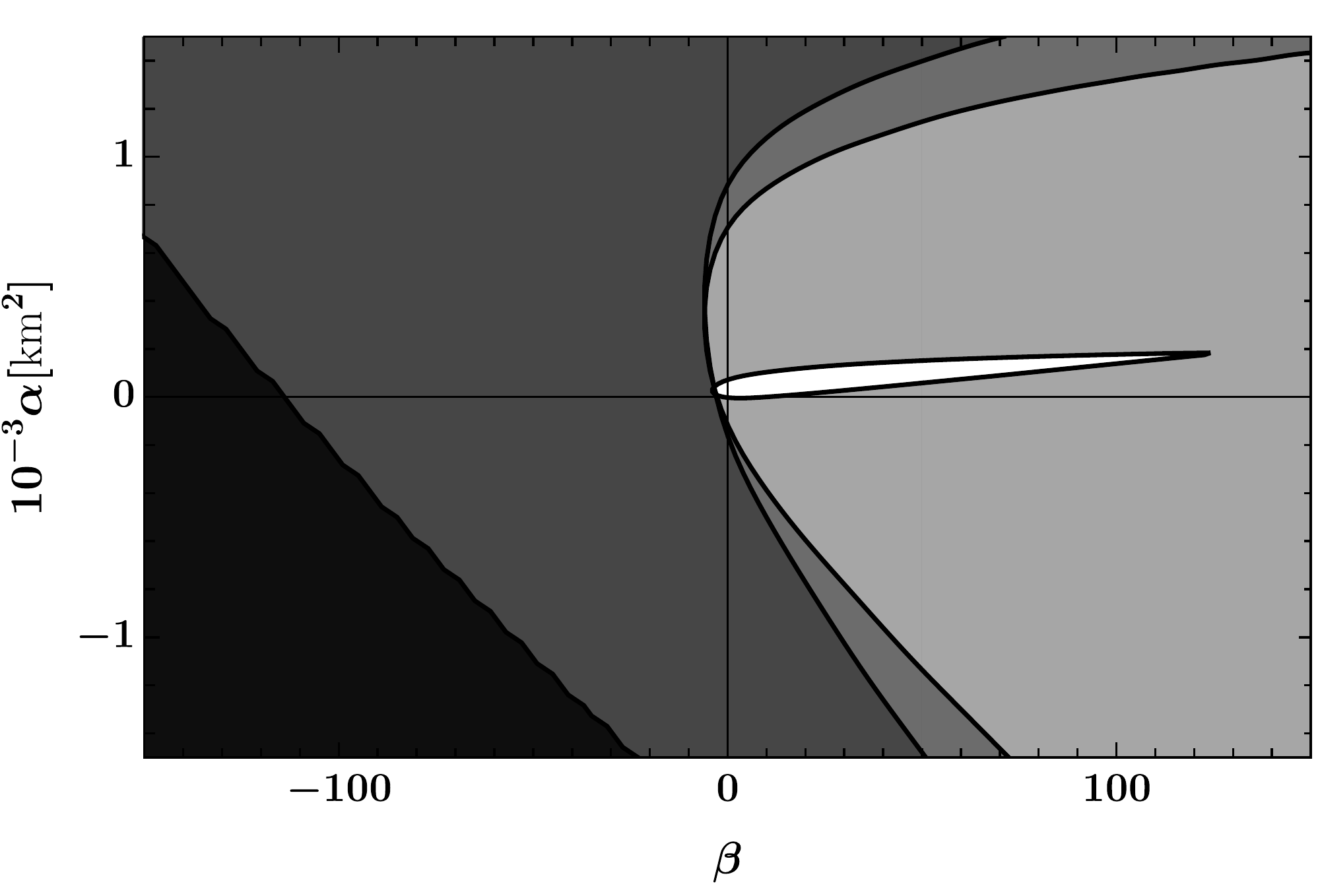}
 \caption[Summary of the stability contour in the parameter space]{Summary of the stability ($n=0$) contours in the $(\beta,\alpha)$ space for comparison. The white region corresponds to the region of stability for a heavy star, $M=2.04~M_\odot$, using the SLy equation of state. The light grey region (together with the white region) is the stability region for $M=1.12~M_\odot$, still with the SLy equation of state. The intermediate grey region (together with the previous paler regions) corresponds to a mass $M=1.12~M_\odot$ and the MPA1 equation of state. The dark grey region (again, with paler regions) corresponds to a mass $M=1.12~M_\odot$ and the SLy equation of state, together with a bare scalar mass of $1.33\times10^{-10}$~eV. Finally, the black region is unstable for all the previous models.}
 \label{fig:thresholds}
 \end{figure}

 Our analysis allowed us to explore, for the first time, the multi-dimensional parameter space and provide scalarization thresholds that depend on more than one coupling. In Fig.~\ref{fig:thresholds} we presented in a single plot a summary of most of the stability contours presented in this Section. 
 One of the striking features revealed by our analysis is the role of the effective metric in which scalar perturbations propagate. It is controlled by a single coupling constant, $\gamma$. There exists a threshold beyond which the effective metric loses hyperbolicity. In the framework of tachyonic scalarization, we interpret this threshold as an absolute bound on the parameter $\gamma$. It is restricted to a rather narrow range (roughly, it should remain small with respect to the characteristic length of curvature). Therefore, it has a very limited effect on the threshold of tachyonic scalarization. On the other hand, the loss of hyperbolicity can be seen as an alternative instability that could lead to scalarization, in line with what was proposed in Ref.~\cite{Ramazanoglu:2017yun} in a more restricted setup.


\chapter[\texorpdfstring{Black hole scalarization with Gauss-Bonnet and Ricci scalar couplings}{Black hole scalarization with Gauss-Bonnet and Ricci scalar couplings}
]{\chaptermark{Black hole scalarization} Black hole scalarization with Gauss-Bonnet and Ricci scalar couplings}
\chaptermark{Black hole scalarization}
\label{Chap:blackholes}

 In the previous Chapter, we investigated the onset of the tachyonic instability that triggers spontaneous scalarization. We have shown that only three terms can in fact contribute to create an effective mass for the scalar field: a bare mass of the scalar field, a coupling between the Gauss-Bonnet invariant and the scalar field, and a coupling between the Ricci scalar and the scalar field. Additionally, there is a fourth term, a coupling between the gravity and the kinetic term, that can affect the threshold of the instability, even though it is not able to trigger it by itself. 

 In the case of black holes the onset of the scalarization is determined entirely by the coupling with the Gauss-Bonnet invariant (and potentially a bare mass of the scalar field). Indeed, general relativity vacuum solution are Ricci flat, therefore the Ricci tensor or scalar cannot contribute to source the instability. However, the properties of the scalarized object depend crucially on nonlinear interactions, as these are the ones that quench the linear instability and determine its endpoint. Non-linearities can originate from scalar self-interactions~\cite{Macedo:2019sem}, from the coupling function to $\mathscr{G}$~\cite{Doneva:2017bvd}, and from the backreaction of the scalar onto the metric. The potential coupling between the scalar field and the Ricci scalar, $R$, has mostly been disregarded in the case of black holes.
 
 As mentioned earlier, this is entirely justified when studying the onset of scalarization, as general relativity black holes have a vanishing $R$. However, it is bound to have an effect on the properties of scalarized objects, as it will contribute to the nonlinear quenching of the tachyonic instability that leads to scalarization. Indeed, as soon as the scalar becomes nontrivial, $R$ will cease to be zero and it will contribute directly to the effective mass of the scalar. From an effective field theory perspective there seems to be no justification to exclude such a coupling. Moreover, it has been shown in Ref.~\cite{Antoniou:2020nax} that this coupling makes general relativity a cosmological attractor and hence reconciles Gauss-Bonnet scalarization  with late-time cosmological observations.

 Motivated by the above, in this Chapter we examine the role a coupling with the Ricci scalar can have on scalarized black holes. We consider the minimal action we employed in Chapter~\ref{Chap:Threshold} to study the threshold of scalarization, i.e. Eq.~\eqref{eq:ACI}, with a standard kinetic term, imposing $\gamma=0$, and in the absence of a bare scalar mass. This choice is justified since if a bare mass is included it needs to be tuned to rather small values else it can prevent scalarization altogether, while $\gamma$ has a very limited effect on the threshold of tachyonic scalarization, as we have shown in Chapter~\ref{Chap:Threshold}. The modified Einstein equation for such theory is
 \begin{equation}\label{eq:fieldeqs}
    G_{\mu\nu}=\kappa T^\text{PF}_{\mu\nu}+T^\phi_{\mu\nu},
 \end{equation}
 where $T^\text{PF}_{\mu\nu}$ is the stress-energy tensor for a perfect fluid and
 \begin{equation}\label{eq:EMtensor}
 \begin{split}
    T^\phi_{\mu\nu}=&-\frac{1}{4}g_{\mu\nu}\nabla_\lambda\phi\nabla^\lambda\phi+\frac{1}{2}\nabla_\mu\phi\nabla_\nu\phi
    +\frac{\beta\phi^2}{4}G_{\mu\nu}\\
    &+\frac{\beta}{4}\left(g_{\mu\nu} \nabla^2 -\nabla_\mu\nabla_\nu \right)\phi^2\\
    &-\frac{\alpha}{2 g}g_{\mu(\rho}g_{\sigma)\nu}\epsilon^{\kappa\rho\alpha\beta}\epsilon^{\sigma\gamma\lambda\tau}R_{\lambda\tau\alpha\beta}\nabla_{\gamma}\nabla_{\kappa}\phi^2
 \end{split}
 \end{equation}
 is the energy momentum tensor contribution that comes from the variation of the $\phi$-dependent part of the action with respect to the metric. In the case of black holes, there is no matter contribution, and Eq.~\eqref{eq:fieldeqs} reduces to
 \begin{equation}\label{eq:grav_eq}
    G_{\mu\nu}=T^\phi_{\mu\nu}.
 \end{equation}
 For the reader's convenience, we report here the scalar field equation,
 \begin{equation}\label{eq:scal_eq}
    \Box \phi =m_\text{eff}^2\phi,
 \end{equation}
 where the effective scalar mass is now given by
 \begin{equation}\label{eq:eff_mass}
    m_\text{eff}^2=\frac{\beta}{2}R-\alpha \gb.
 \end{equation}
 
 We study static, spherically symmetric black holes. We explore the region of existence of scalarized solutions when varying both couplings and the black hole mass. We examine the influence of the Ricci coupling  on the scalar charge of the black holes, which is the quantity that controls the deviations from general relativity in the observation of binaries. We also discuss the role this coupling can play in stability considerations and in rendering black hole scalarization compatible with cosmological observations and strong gravity constraints from neutron stars.
 
 The results presented in this Chapter have been published in Ref.~\cite{Antoniou:2021zoy}.

 
\section{Static, spherically symmetric black holes}\label{Sec:bhspacetime}

 We are interested in a static and spherically symmetric background. Using the line element as defined in Eq.~\eqref{eq:lineElement}, choosing $h(r)=e^{\Gamma(r)}$ and $f(r)=e^{-\Lambda(r)}$ for convenience, the field equations can then be cast as three coupled ordinary differential equations for $\Gamma$, $\Lambda$ and $\phi$, see Appendix~\ref{App:Eqs}. The $(rr)$ component of the metric equations can be solved algebraically with respect to $e^\Lambda$:
 \begin{equation}\label{eq:grr}
    e^\Lambda= \frac{-B+\delta\sqrt{B^2-4A\,C}}{2A},\,\delta=\pm 1,
 \end{equation}
 where
 \begin{align}
    A=&\;4-\beta  \phi^2,\\
    \begin{split}
    B=&\; \beta  \phi ^2+\Gamma ' (\beta  r^2 \phi  \phi '-8 \alpha  \phi\,  \phi '+\beta  r \phi ^2-4 r)\\
    &+r^2 \phi '^2+4 \beta  r \phi\,  \phi '-4,
    \end{split}
    \\
    C=&\; 24 \alpha  \Gamma ' \phi\, \phi '.
 \end{align}
 and a prime denotes differentiation with respect to the radial coordinate. By substituting~\eqref{eq:grr} in the remaining field equations, we end up with a system of two coupled second order differential equations:
 \begin{align}
    \Gamma''=&\, \tilde{\Gamma}(r,\Gamma',\phi,\phi',\alpha,\beta)\label{eq:dif_1},\\
    \phi''=&\, \tilde{\phi}(r,\Gamma',\phi,\phi',\alpha,\beta)\label{eq:dif_2}.
 \end{align}

 In order to search for black hole solutions, we assume the existence of a horizon, where $e^\Gamma\rightarrow 0,\; e^\Lambda\rightarrow \infty$. In line with previous results for different models fashioning a coupling with $\mathscr{G}$ ({\em e.g.}~\cite{Kanti:1995vq,Sotiriou:2014pfa}), only $\delta=+1$ leads to black hole solutions.

\subsection{Near-horizon expansion}\label{Sub:horizon}
 
 Near the horizon, one can perform the following expansion:
 \begin{align}
    e^\Gamma(r\approx r_\text{h})= & \,\gamma_1 (r-r_\text{h})+\gamma_2 (r-r_\text{h})^2+...\label{eq:exp_h1}\\
    e^{-\Lambda}(r\approx r_\text{h})= & \,\lambda_1 (r-r_\text{h})+\lambda_2 (r-r_\text{h})^2+...\label{eq:exp_h2}\\
    \phi(r\approx r_\text{h})= & \,\phi_\text{h} +\phi_1 (r-r_\text{h}) +\phi_2 (r-r_\text{h})^2+...\label{eq:exp_h3}
 \end{align}
 One can substitute these expressions in Eqs.~\eqref{eq:grr},~\eqref{eq:dif_1} and~\eqref{eq:dif_2}, and obtain a near-horizon solution. In particular, $\phi''_\text{h}$ remains finite only provided that
 \begin{equation}\label{eq:phi_h}
    \phi'(r_\text{h})=\phi_1= \big(a+\sqrt{\Delta}\big)/b,
 \end{equation}
 where the expressions for $a,\;b$ and $\Delta$ are as follows:
 \begin{align}
    a &=24 \alpha  \beta  r_h \phi _h^2+r_h^3 \left(-3 \beta ^2 \phi _h^2+\beta  \phi _h^2-4\right),
    \\[4mm]
    \begin{split}
    \Delta &=9216 \alpha ^3 \beta  \phi _h^4+r_h^6 \left(3 \beta ^2 \phi _h^2-\beta  \phi _h^2+4\right)^2\\
    &\quad -192 \alpha ^2 r_h^2 \phi _h^2 \left(9 \beta ^2 \phi _h^2-2 \beta  \phi _h^2+8\right),
    \end{split}
    \\[4mm]
    \begin{split}
    b &= 2 \phi _h \left(8 \alpha -\beta  r_h^2\right) \big[24 \alpha  \beta  \phi _h^2\\
    &\quad +r_h^2 \left(-3 \beta ^2 \phi _h^2+\beta  \phi _h^2-4\right)\big]/(\beta \phi_h^2-4).
    \end{split}
 \end{align}

 Requiring that $\Delta\geq0$  defines a region on the $(r_\text{h},\phi_\text{h})$ space where regular black hole solutions with scalar hair can be found.
 
\subsection{Asymptotic expansion}\label{Sub:bhasymptotic}

 In order to analyze the asymptotic behaviour of the solutions, one can perform a suitable expansion, and solve the equations near spatial infinity imposing that $\phi$ vanishes there. Assuming analyticity, this yields 
 \begin{equation}
 \begin{split}\label{eq:exp_f1}
    g_{tt}(r\gg r_\text{h})= &\; 1-2M\big/{r}+\beta \, Q^2\big/{4\,r^2} +\big(M Q^2-3 \beta  M Q^2\big)\big/{12\, r^3}\\
    &+\big(8 M^2 Q^2-28 \beta  M^2 Q^2-3 \beta ^3 Q^4+5 \beta ^2 Q^4\\
    &-\beta  Q^4\big)\big/{48\, r^4}+\big(288 M^3 Q^2-1040 \beta  M^3 Q^2\\
    & +3072 \alpha  M Q^2-60 \beta ^3 M Q^4+115 \beta ^2 M Q^4\\
    &+10 \beta  M Q^4-9 M Q^4\big)\big/{960r^5}+\mathcal{O}\left(1/r^6\right) \;,
 \end{split}
 \end{equation}
 \begin{equation}
 \begin{split}\label{eq:exp_f2}
    g_{rr}(r\gg r_\text{h})= &\; 1+2M\big/r+\big(16 M^2+2 \beta \, Q^2-Q^2\big)\big/{4\,r^2}+\big(32 M^3\\
    &-5 M Q^2+11 \beta  M Q^2\big)\big/{4\, r^3}+\big(488 \beta  M^2 Q^2\\
    &-208 M^2 Q^2+768 M^4-12 \beta ^3 Q^4+17 \beta ^2 Q^4\\
    &-13 \beta  Q^4+3 Q^4\big)\big/{48\, r^4}+\big(6064 \beta  M^3 Q^2\\
    &-2464 M^3 Q^2+6144 M^5-1536 \alpha  M Q^2\\
    &-348 \beta ^3 M Q^4+589 \beta ^2 M Q^4-442 \beta  M Q^4\\
    &+97
    M Q^4\big)\big/{192\, r^5}+\mathcal{O}\left(1/r^6\right) \;,
 \end{split}
 \end{equation}
 \begin{equation}
 \begin{split}\label{eq:exp_f3}
    \phi(r\gg r_\text{h})= &\; Q\big/r+M Q\big/r^2+\big(32 M^2 Q-3 \beta ^2 Q^3+2 \beta  Q^3\\
    &-Q^3\big)\big/24\, r^3+\big(48 M^3 Q-9 \beta ^2 M Q^3+9 \beta  M Q^3\\
    &-4 M Q^3)\big/24\,r^4+\big(2240 \beta  M^2 Q^3-1680 \beta ^2 M^2 Q^3\\
    &-928 M^2 Q^3-4608 \alpha  M^2 Q+6144 M^4 Q+117 \beta ^4 Q^5\\
    &-144 \beta ^3
    Q^5+86 \beta ^2 Q^5-40 \beta  Q^5+9 Q^5\big)\big/{1920 \, r^5}\\
    &+\mathcal{O}\left(1/r^6\right) \;,
 \end{split}
 \end{equation}
 where $M$ is the ADM mass and $Q$ is the scalar charge. Note that, although $Q$ is not associated to a conservation law, it does determine the decay of the scalar field at large distance.  As one can see from Eqs.~\eqref{eq:exp_f1}-\eqref{eq:exp_f3}, the contribution from the Ricci coupling dominates the asymptotic behaviour of the solutions over the Gauss-Bonnet coupling. Indeed, terms proportional to $\beta$ enter the expansion already at order $r^{-2}$, whereas $\alpha$-dependent terms arise only at order $r^{-5}$.
 
\subsection{Numerical implementation}\label{Sub:bhnumerical}
 
 The system of ordinary differential equations~\eqref{eq:dif_1} and~\eqref{eq:dif_2} can, in principle, be solved by starting from the horizon and integrating towards larger radii. $\alpha$ and $\beta$ are theoretical parameters that are considered fixed. 
 The values of $\Gamma'$, $\Gamma$, $\phi'$, and $\phi$ at $r=r_\text{h}$ appear to be ``initial data''. However, they are not all free to choose. $\Gamma(r_\text{h})$ is fixed by the condition $e^{\Gamma(r_\text{h})}=0$, {\em i.e.}~the fact that $r=r_\text{h}$ is a horizon. $\Gamma'$ has to diverge at $r=r_\text{h}$, else $e^{\Gamma}$ will have a vanishing derivative on the horizon. Finally, $\phi'(r_\text{h})$, and $\phi(r_\text{h})$ are related by the regularity condition~\eqref{eq:phi_h}.  
 One also needs to fix $r_\text{h}$. The field equations are invariant under the global scaling symmetry $r\to\mu r$, $\alpha\to\mu^2\alpha$, where $\mu$ is a free parameter. We can make use of this symmetry to reduce the space of parameters that we have to explore. Practically, we can decide that the horizon is located at $r_\text{h}=1$; solutions with $r_\text{h}\neq1$ can later be obtained by a global scaling.

 Hence, one can treat $\phi(r_\text{h})=\phi_\text{h}$ as the only free parameter. Integrating outwards, one will generically find a solution for arbitrary $\phi_\text{h}$. However, for given $\alpha$ and $\beta$, only one value of $\phi_\text{h}$ has the desired asymptotics, namely $\phi(r\to \infty)=\phi_\infty=0$. Imposing this condition (by a shooting method and to a desired precision) yields a unique solution. The global rescaling mentioned above turns this solution into a one-parameter family, that we can interpret as a family of black holes parametrized by their ADM mass $M$, for fixed couplings $\alpha$ and $\beta$. The scalar charge $Q$ is then determined as a function of $M$, $\alpha$ and $\beta$. 

 A practical complication is that the regularity condition of Eq.~\eqref{eq:phi_h} cannot be imposed numerically with any reasonable accuracy. To circumvent this problem we start the numerical integration at $r\approx r_\text{h}[1 + \mathcal{O}(10^{-4})]$ and use the perturbative expansion in Eqs.~\eqref{eq:exp_h1}--\eqref{eq:exp_h3} to impose the regularity and propagate the data from the horizon to the starting point of the numerical integration. We typically integrate up to distances $r/r_\text{h}\approx 10^4$ and impose that $\phi$ vanishes there to a part in $10^{4}$.
 
 Given a solution, we extract the value of the ADM mass $M$ and the scalar charge $Q$, as defined in the asymptotic expansion~\eqref{eq:exp_f1}-\eqref{eq:exp_f3}. We then have
 \begin{equation}\label{eq:MandQ}
    \begin{split}
    & M = -\left(\frac{1}{2}r^2\Lambda'\,e^{-\Lambda}\right) \bigg|_{r_\text{max}},\\
    & Q = -\left(r^2 \phi'\right)\big|_{r_\text{max}}.
    \end{split}
 \end{equation}

 In the next Section, we use scale-invariant masses and charges, defined as
 \begin{equation}
 \label{eq:scalinvMQ}
    \hat{M}=M/\sqrt{\alpha}\,,\;\; \hat{Q}=Q/\sqrt{\alpha}.
 \end{equation}
 
 To schematically summarize, the algorithm we implement consists of the following steps:
 \begin{enumerate}
    \item We set $r_\text{h}=1$ and we fix $\beta$.
    \item We choose a value for $\alpha$.
    \item Given the parameters $\beta$ and $\alpha$, we find a value for $\phi_\text{h}$ that satisfies the initial conditions, i.e. $\Delta\geq 0$, and gives the right asymptotics, implementing a shooting method. If there is no value of $\phi_\text{h}$ that satisfies these requirements, we treat the parameter set as having no solution.
    \item Given $\alpha$, $\beta$ and $\phi_\text{h}$, we solve the field equations, we extract the value of $M$ and $Q$, and appropriately rescale them as in Eq.~\eqref{eq:scalinvMQ}.
    \item We repeat steps 2-4 for a new value of $\alpha$.
 \end{enumerate}
 
 Note that Eq.~\eqref{eq:scalinvMQ} assumes that $\alpha>0$. 
 Indeed, we will restrict our analysis to positive values of $\alpha$. Evading the no-hair theorem of Ref.~\cite{Silva:2017uqg} requires $\alpha>0$ when $\beta=0$ and $\gb$ is positive, which is the case for a Schwarzschild black hole. Moreover, the Ricci coupling, controlled by $\beta$, does not contribute to linear perturbation theory around general relativity black holes. It is hence unlikely that scalarized spherically symmetric black hole solutions will exist for $\alpha<0$. It should be stressed, however, that the $\alpha<0$ case is particularly interesting when studying rotating black holes~\cite{Dima:2020yac}.
 
\section{Properties of the solutions}\label{Sec:bhproperties}

 \begin{figure*}[h!]
 \begin{center}
    \includegraphics[scale=0.9]{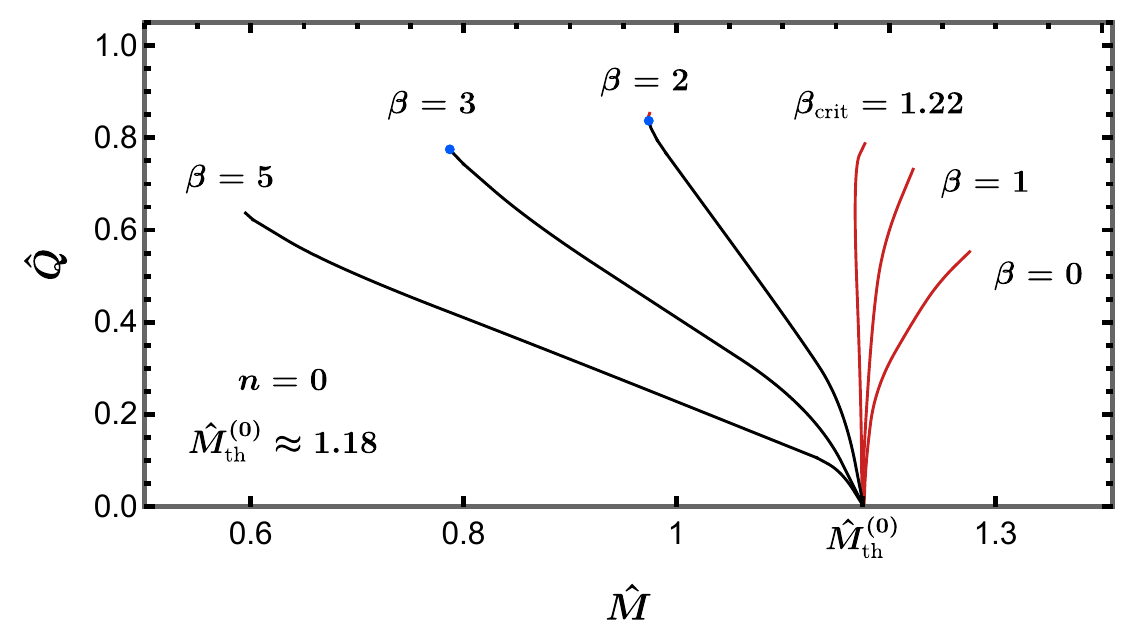} \\
    \includegraphics[scale=0.8]{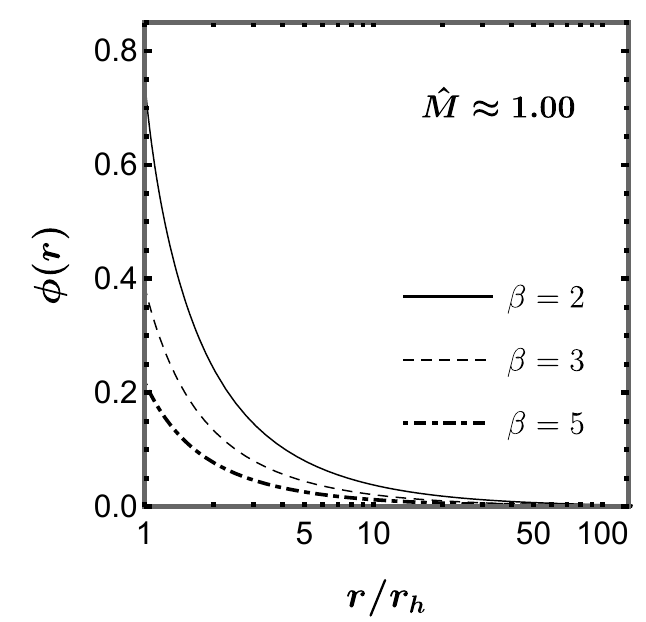} 
    \includegraphics[scale=0.8]{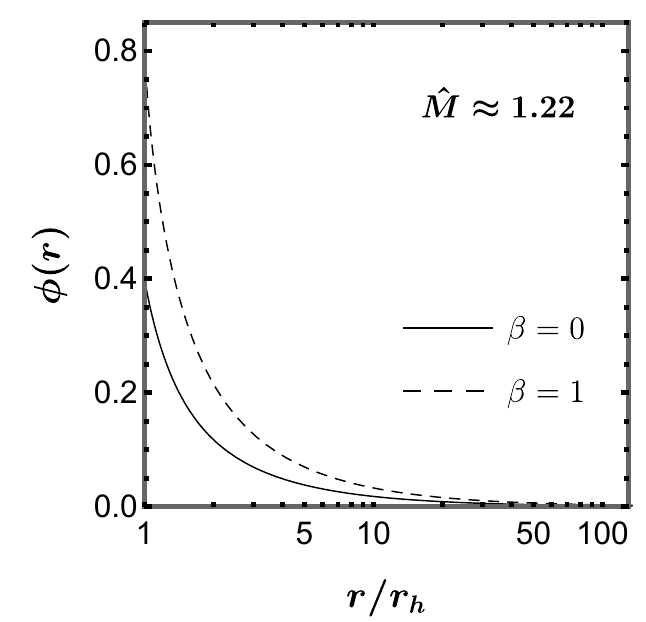}
 \end{center}
 \caption[Scalar charge as a function of the ADM mass for zero nodes solutions and scalar field profiles]{(Top) Normalized scalar charge versus the normalized ADM mass for $n=0$ nodes solutions. The black lines correspond to values of $\beta$ for which all scalarized black holes have masses below the general relativity instability mass threshold, while the red lines mark values of $\beta$ that lead to scalarized black hole masses that are larger than the general relativity mass threshold. The blue dots mark the existence of a turning point, past which solutions are expected to be unstable. (Bottom left) Scalar field profile versus the normalized distance from the horizon for a sample mass $\hat{M}<\hat{M}^{(0)}_\text{th}$, and zero nodes for the scalar field radial profile. (Bottom right) Scalar field profile versus the normalized distance from the horizon for a sample mass $\hat{M}>\hat{M}^{(0)}_\text{th}$.}
 \label{fig:fig_2}
 \end{figure*}

\subsection{Solution with no nodes for the scalar profile}\label{Sub:bhnonodes}

\begin{figure*}[h!]
 \begin{center}
    \includegraphics[width=0.7\textwidth]{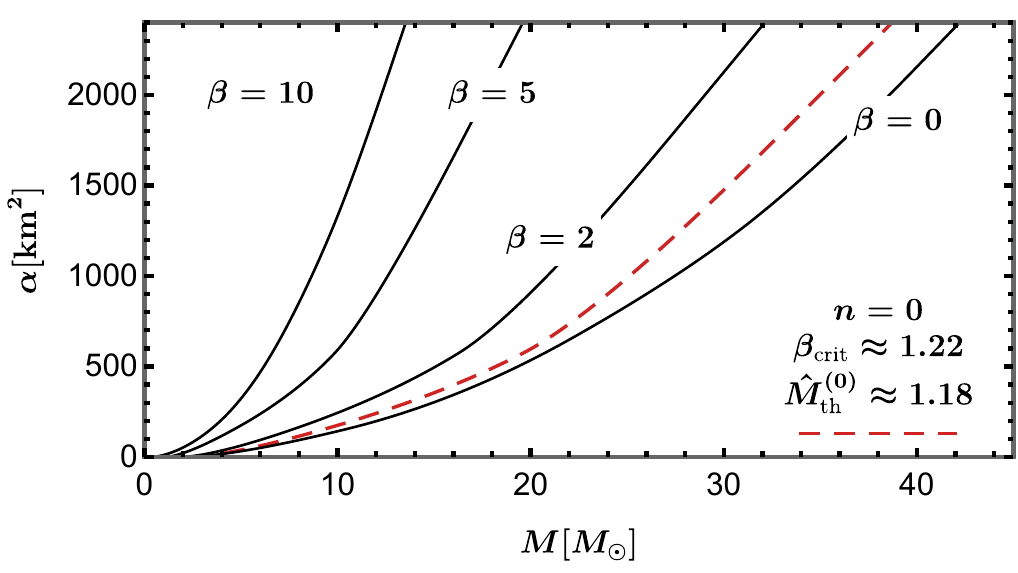}\hspace{1cm}
    \includegraphics[width=0.7\textwidth]{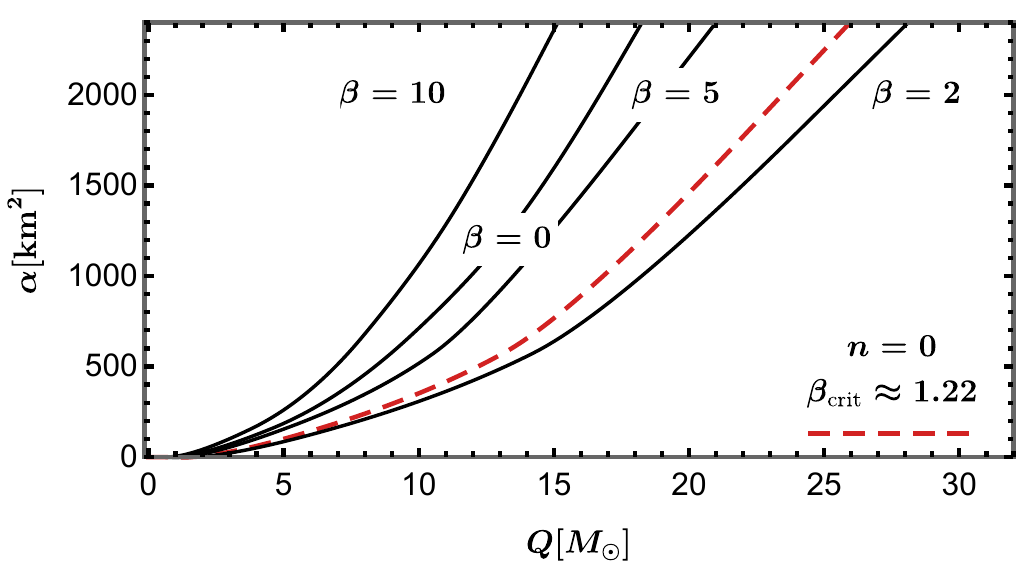}
 \end{center}
 \caption[Domain of existence of zero node scalarized black holes]{(Top) Domain of existence of $n=0$ scalarized black holes on the $\alpha-M$ plane. For a given $\beta$, solutions exist between the corresponding black line, and the dashed, red line. The latter coincides with the line where general relativity solutions of equal mass would become unstable. (Bottom) Same but on the $\alpha-Q$ plane. Both panels can be obtained from an ``unfolding'' of Fig.~\ref{fig:fig_2}.}
 \label{fig:fig_aMQdomain}
 \end{figure*}

 The first scenario we examine is the one where $\beta>0$. This scenario is motivated by the results of Ref.~\cite{Antoniou:2020nax}, where it was shown that positive values of $\beta$ make general relativity a cosmological attractor. We start by exploring the solutions characterized by $n=0$. The results are summarized in Fig.~\ref{fig:fig_2} and  Fig.~\ref{fig:fig_aMQdomain}.
 The top plot of Fig.~\ref{fig:fig_2} shows the dependence of the scalar charge on the black hole ADM mass for different choices of $\beta$. Each line corresponds to a constant value of $\beta$. Each point along a constant $\beta$ line correspond to a specific set of $\alpha$ and $\phi_\text{h}$, given $r_\text{h}=1$.
 When $\beta$  is smaller than some critical value $\beta_{\text{crit}}\approx 1.22$, the charge-mass curve tilts to the right and all scalarized black holes have larger ADM masses than the general relativity mass instability threshold. Such scalarized black holes are unlikely to be produced dynamically. The ADM mass is a measure of energy for the system. The fact that {\em all} scalarized black holes for $\beta<\beta_{\text{crit}}$ have larger mass than {\em all} general relativity black holes that are unstable implies that, if {\em any} scalarized black hole is considered the end point of the tachyonic instability for a general relativity black hole, then this end state would have more energy than the initial state. 

 Based on the argument above, we conjecture that scalarized black holes are unstable for $\beta<\beta_{\text{crit}}$. Conversely, for $\beta>\beta_{\text{crit}}$ the ADM mass for scalarized black holes can be smaller than the general relativity counterparts and hence it is reasonable to expect that scalarized black hole are endpoints of the tachyonic instability. These arguments are consistent with earlier results. In particular, it is already known that for $\beta=0$ scalarized black holes are radially unstable~\cite{Blazquez-Salcedo:2018jnn}. Moreover, the general picture shown in the centre plot of Fig.~\ref{fig:fig_2} is very similar to the one presented in Ref.~\cite{Macedo:2019sem}. In that work, $\beta$ was vanishing and the $\phi^2 R$ term was absent, but a $\phi^4$ self-interaction had a similar effect. Analysis of radial stability did show in that case that stability was associated with whether the curves on the $\hat{Q}-\hat{M}$ plane tilt to the right or the left. 

 These considerations suggest strongly that the coupling between $\phi$ and the Ricci scalar, can have  a very interesting stabilizing effect for scalarized black holes, without having to resort to scalar self-interactions. 

 Note that in some cases, when $\beta>\beta_{\text{crit}}$ and hence the $\hat Q-\hat M$ curve initially leans to the left, this same curve later turns towards the right. The points at which the curves turn right are marked by blue dots in Fig.~\ref{fig:fig_2}. The right-leaning part of these curves is hardly noticeable in Fig.~\ref{fig:fig_2} because it is very short. One expects configurations past the turning point to be unstable, as configurations of the same ADM mass and smaller scalar charge exist. 

 As it is clear from Fig.~\ref{fig:fig_2}, for $\beta>\beta_{\text{crit}}$, the normalized scalar charge $\hat Q$ increases as the normalized ADM mass $\hat M$ decreases, at least in the part of the curves up to the turning point (blue dot), whereas for $\beta<\beta_{\text{crit}}$, the normalized scalar charge $\hat Q$ increases as the normalized ADM mass $\hat M$ increases. Interestingly, the dependence of the curvature near the horizon on the ADM mass turns out to be different in the two cases. For $\beta>\beta_{\text{crit}}$ scalarized black holes tend to have larger curvatures at the horizon when the ADM mass decreases, as is the case in general relativity, whereas for $\beta<\beta_{\text{crit}}$ the curvature on the horizon tends to increase as the mass (and scalar charge) increases. Hence, in both cases, the scalar charge seems to be controlled by the curvature. 

 In Fig.~\ref{fig:fig_aMQdomain}, we show the domain of existence of scalarized black holes on the $\alpha-M$ and $\alpha-Q$ planes. As discussed in Chapter~\ref{Chap:Threshold}, linear analysis showed distinct scalarization thresholds, the first (zero nodes) of which we denote with $\hat{M}^{(0)}_{\text{th}}\approx 1.175$. This threshold  is represented by the dashed, red line in Fig.~\ref{fig:fig_aMQdomain}. Note that $\hat{M}=\text{constant}$ (respectively $\hat{Q}=\text{constant}$) translates to a parabola in the $\alpha-M$ ($\alpha-Q$) plane. The rest of the curves correspond to the existence boundaries for various values of $\beta$. They are related to the horizon condition presented in Eq.~\eqref{eq:phi_h}.
 Solutions for a given $\beta$, then, exist everywhere between the red, dashed general relativity instability line and the corresponding plain, black existence line. Every $\hat{M}=\text{constant}$ ($\hat{Q}=\text{constant}$) parabola included in this existence region corresponds to a point along the respective constant $\beta$ line of Fig.~\ref{fig:fig_2}, so that both panels of Fig.~\ref{fig:fig_aMQdomain} can be obtained from an ``unfolding'' of Fig.~\ref{fig:fig_2}. Note that solutions lying on a horizontal cut in the region of existence, that is points of constant $\alpha$ (and $\beta$), corresponds to solutions with different $\phi_\text{h}$ and $r_\text{h}$, which is no longer restricted to be $r_\text{h}=1$.
 
 Examining the plots reveals something rather interesting: the value of the Ricci coupling $\beta$ can affect the relative position of the existence line with respect to the instability parabola. This should not come as a surprise, based on the results presented in Fig.~\ref{fig:fig_2}, where $\beta$ has a similar effect on the relative position of the curve with respect to the threshold mass $\hat{M}^{(0)}_{\text{th}}$.

 As mentioned earlier, we do not plan to consider the $\beta<0$ case in any detail as positive values appear to be better motivated. However, we can report the following based on a preliminary exploration. There is still a critical value of $\beta$, and for $\beta$ smaller than this value, scalarized black holes have smaller ADM masses than the general relativity instability threshold, together with scalar charges that tend to increase with decreasing mass. For $\beta$ larger than the critical value, the behaviour is reversed. Hence, the equivalent to Fig.~\ref{fig:fig_2} would be qualitatively similar for $\beta<0$.
 
\subsection{Solutions with higher nodes}\label{Sub:bhhighernodes}

\begin{figure*}[h!]
 \begin{center}
    \includegraphics[width=0.7\textwidth]{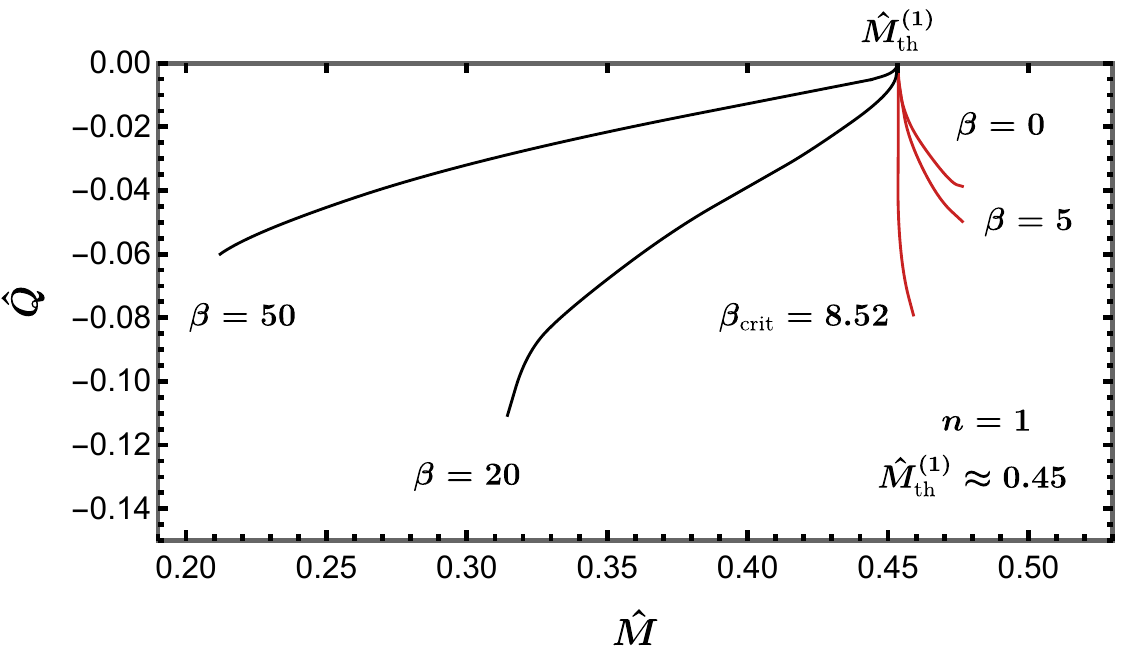}\hspace{1cm}
    \includegraphics[width=0.7\textwidth]{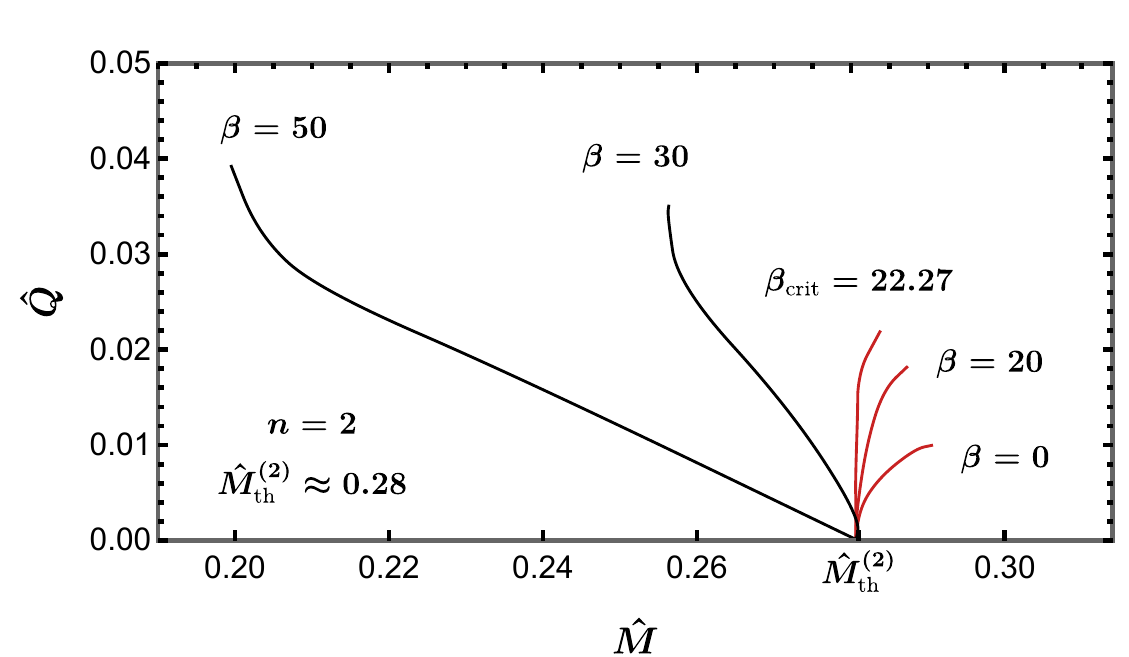}
 \end{center}
 \caption[Scalar charge as a function of ADM mass for one node solutions]{Normalized scalar charge versus normalized ADM mass for the solutions with $n=1$ (top panel) and 2 (bottom panel) nodes.}
    \label{fig:fig_QM1}
 \end{figure*}

 We conclude this Chapter turning to solutions characterized by $n=1$ and $n=2$. For $\beta>0$, the plot of the normalized charge versus the normalized mass is given in Fig.~\ref{fig:fig_QM1}.
 A noticeable pattern is that, for $n=0$, the scalar charge is positive, while it is negative for $n=1$, positive for $n=2$, and so on. This is simply due to the fact that the scalar field has to approach 0 at spatial infinity from a different side, depending on the number of nodes. There is no deep significance in the sign of the charge, since the action~\eqref{eq:ACI} possesses parity symmetry $\phi\to-\phi$, and the signs would have been flipped, had we chosen negative values of $\phi_\text{h}$ as initial conditions. Compared to the $n=0$ case, we see that the order of magnitude of the charge for all different values of $\beta$ is significantly smaller, and the range of masses for which we find scalarized solutions is strongly reduced. Once again there is a critical value of $\beta$ that separates right-leaning curves (likely unstable) from left leaning ones (likely stable). 
 
 We have considered the contribution that a coupling between the scalar field and the Ricci scalar, $R$, can have in black hole spontaneous scalarization. We focused on static, spherically symmetric black holes. The $\beta \phi^2 R$ coupling is known not to affect the threshold of scalarization. However, our results show that it can alter the domain of existence of scalarized black holes, significantly modify their properties, and control their scalar charge. Our results also strongly suggest that the strength of this coupling can have an impact on the stability of scalarized black holes. In particular, having $\beta$ be larger than some critical value, $\beta_\text{crit}$, is expected to resolve the stability problems for models that do not include the $\beta \phi^2 R$ coupling.
 
 We have mostly focused on positive values for $\beta$. We did so for two reasons. First, it has recently been shown that including the $\beta \phi^2 R$ term in black hole scalarization models and selecting a positive $\beta$ makes general relativity a cosmological attractor and allows one to have a consistent cosmological history, at least from the end of the inflationary era~\cite{Antoniou:2020nax}.  The numerical values that we considered here for the couplings are similar to those used in Ref.~\cite{Antoniou:2020nax}.  Second, for positive values of the Ricci coupling (and reasonably small values of the Gauss-Bonnet coupling), neutron stars do not scalarize, as shown in Chapter~\ref{Chap:Threshold}. This allows one to evade the very tight binary pulsar constraints ({\em e.g.}~\cite{Freire:2012mg,Antoniadis:2013pzd,Shao:2017gwu}), related to energy losses due to dipolar emission of gravitational waves, without the need to add a bare mass to the scalar (and tune it appropriately). 

 It is clear that inclusion of the $\beta \phi^2 R$ coupling has multiple benefits in scalarization models. It is worth re-iterating that this coupling has lower mass dimensions than the $\alpha \phi^2 {\cal G}$ coupling, which triggers scalarization at linear level. Moreover, unlike a bare mass term or scalar self-interactions, it allows the scalar to remain massless and free in flat space. Hence, the $\beta \phi^2 R$ coupling can be part of an interesting effective field theory that respects $\phi \to -\phi$ symmetry and in which shift symmetry can be broken only via the coupling to gravity (the complete effective field theory would potentially include more terms, such as  $R\phi^4$ and $G^{\mu\nu}\partial_\mu\phi\partial_\nu\phi$).

 It should be stressed that we only considered the case $\alpha>0$ throughout our analysis, as this is a requirement for having scalarized black holes under the assumptions of staticity and spherical symmetry. However, it has been shown in Ref.~\cite{Dima:2020yac} that, for $\alpha<0$ (and $\beta=0$), scalarization can be triggered by rapid rotation. Indeed, some scalarized black hole have been found in this scenario in Refs.~\cite{Herdeiro:2020wei,Berti:2020kgk}.


\chapter[\texorpdfstring{Neutron star scalarization with Gauss-Bonnet and Ricci scalar couplings}{Neutron star scalarization with Gauss-Bonnet and Ricci scalar couplings}
]{\chaptermark{Neutron star scalarization} Neutron star scalarization with Gauss-Bonnet and Ricci scalar couplings}
\chaptermark{Neutron star scalarization}
\label{Chap:neutronstars}

 The combined results retrieved in Chapter~\ref{Chap:Threshold} and Chapter~\ref{Chap:blackholes}, showed that including the Ricci coupling to a general scalarization theory seems to provide us with several advantages. As we have already discussed, this term has not received much attention in recent literature on spontaneous scalarization. This is mostly due to the fact that, in the black-hole scenario, the onset of scalarization is only controlled by the Gauss-Bonnet invariant, since the Ricci scalar evaluates to zero for general relativity black holes.
 
 However, we determined in Chapter~\ref{Chap:Threshold} that the Ricci term can help in suppressing the scalarization of neutron stars, which would otherwise tend to place significant constraints. Furthermore, in Chapter~\ref{Chap:blackholes}, we showed that this term has very interesting effects on the properties of scalarized black holes. Even though the Ricci coupling does not affect the onset of black hole scalarization, it affects the properties of the scalarized solutions and, consequently, observables. For certain values of the Ricci coupling the presence of this operator is expected to render black holes radially stable, without the need to introduce self-interaction terms. Interestingly, these values of the Ricci coupling happen to be consistent with the ones retrieved in Ref.~\cite{Antoniou:2020nax}, where it was shown that the Ricci term is crucial if one wants to retrieve a late-time attractor to general relativity in a cosmological scenario.

 For the reasons presented above, it is of great interest to examine how the combination of Ricci and Gauss-Bonnet couplings affects neutron star properties. Hence, in this Chapter we focus on solving the complete set of equations for the theory studied in the previous Chapter, i.e. action~\eqref{eq:ACI}, with $\gamma=0$ and $m_\phi=0$, for the case of neutron stars.
 
 We determine over which region of the parameter space scalarized solutions exist, for the three different stellar scenarios already considered in the study of the scalarization threshold in Chapter~\ref{Chap:Threshold}. We examine the properties of the scalarized solutions, in particular their scalar charges and masses. We also investigates in more detail the solutions that always exist near the scalarization thresholds, and we explain how, already at the level of the general relativity solution, a given scalar profile may be favored.
 
 The results presented in this Chapter have been published in Ref.~\cite{Ventagli:2021ubn}.

\section{Static, spherically symmetric neutron stars}\label{Sec:nsspacetime2}
 
 For the purpose of this work, we restrict our analysis to static and spherically symmetric spacetimes and we use the line element as defined in Eq.~\eqref{eq:lineElement}, with $h(r)=e^{\Gamma(r)}$ and $f(r)=e^{-\Lambda(r)}$. As we did in Chapter~\ref{Chap:scalarizationHordneski}, we assume matter to be described by a perfect fluid with $T^\text{PF}_{\mu\nu}$ as a matter stress-energy tensor, where the pressure is  directly related to the energy density through the equation of state. The field equations then take the form of coupled ordinary differential equations~\eqref{eq:fieldeqs} for $\Gamma$, $\Lambda$, $\epsilon$ and $\phi$, see Appendix~\ref{App:Eqs}. We can solve algebraically the $(rr)$ component of the modified Einstein equation for $e^\Lambda$. The result is
 \begin{equation}\label{eq:ExpLambda}
 e^\Lambda=\frac{-B+\delta\sqrt{B^2-4\,A\,C}}{4 A},\,\,\delta=\pm 1
 \end{equation}
 where
 \begin{equation}
 \begin{split}
 & A=1+\kappa\,r^2p-\frac{1}{4}\,\beta\phi^2,\\
 & B=-2+\frac{1}{2}\beta\,\phi^2-2\,r\Gamma'+\frac{1}{2}r\beta\,\phi^2\Gamma'+2\,r\beta\,\phi\phi'\\
 &\qquad -4\,\alpha\,\phi\Gamma'\phi'+\frac{1}{2}r^2\beta\phi\Gamma'\phi'+\frac{1}{2}\,r^2\phi'^2,\\
 & C=24\,\alpha\,\phi\,\Gamma'\phi'.
 \end{split}
 \end{equation}
 For the $\delta=-1$ branch of solutions we do not retrieve general relativity in the limit $\alpha\to 0$ and $\beta \to 0$, henceforth we will assume $\delta=1$. By substituting Eq.~\eqref{eq:ExpLambda} in the remaining differential equations, we can reduce our problem to an integration in three variables: $\Gamma$, $\phi$ and $\epsilon$.

\subsection{Expansion for $r\to 0$}\label{Sub:rto0}

 Close to the center of the star, we can perform an analytic expansion of the form
 \begin{equation}\label{eq:smallr}
 f(r)=\sum_{n=0}^\infty f_n r^n
 \end{equation}
 for the functions $\Gamma$, $\Lambda$, $\epsilon$, $p$ and $\phi$.
 Plugging these expansions in the field equations, we can solve order by order to determine the boundary conditions at the origin. At this point, there are essentially three quantities that one can freely fix: the central density $\epsilon_0$, the value of the scalar field at the center $\phi_0$, and the value of the time component of the metric at the center, determined by $\Gamma_0$. On the other hand, $\Lambda_0$ has to vanish in order to avoid a conical singularity at the center, while $p_0$ is directly related to $\epsilon_0$ by the equation of state. All higher order quantities $\{\Gamma_i, ...,\phi_i\}$, $i\geq1$ can be determined in terms of the three quantities $\{\epsilon_0,\Gamma_0,\phi_0\}$. We will require that spacetime is asymptotically flat, with a trivial scalar field at spatial infinity, which fixes uniquely $\Gamma_0$ and $\phi_0$, or rather restricts $\phi_0$ to a discrete set of values, each corresponding to a different mode; technically, these values are found through a numerical shooting method. Therefore, for given parameters $\alpha$ and $\beta$, a solution is eventually fully determined by the central density $\epsilon_0$. Different choices of $\epsilon_0$ will translate into different masses. 

 We must underline the difference with the black hole case, studied in Chapter~\ref{Chap:blackholes}. For black holes, the equations are scale invariant up to a redefinition of the couplings. Practically, this means that it is enough to explore the full space of parameters $\alpha$ and $\beta$ for a \textit{fixed} mass. One can then deduce all solutions, of arbitrary mass, by an appropriate rescaling. For neutron stars this scaling symmetry is broken by the equation of state that relates $p$ and $\epsilon$. Therefore, one \textit{a priori} has to explore a 3-dimensional space of parameters ($\epsilon_0$, $\alpha$ and $\beta$) in the case of neutron stars. In order to keep this exploration tractable, as it was done in Chapter~\ref{Chap:Threshold}, we focus our study on a selection of central densities and equations of state. We pick these in order to cover very diverse solutions, typically corresponding to the lightest/heaviest observed stars in general relativity. We then explore a wide range of the $(\alpha,\beta)$ parameter space for these fixed densities and equations of state.

 To complete this Section, let us note that solving order by order the field equations for the higher order coefficients in the expansion~\eqref{eq:smallr} does not always yield solutions. All first order coefficients in this expansion have to vanish; one can express $\Gamma_2$, $\epsilon_2$, $p_2$ and $\phi_2$ in terms of $\Lambda_2$; however, $\Lambda_2$ itself is determined by the following equation:
 \begin{equation}\label{eq:Lambda2}
 \begin{split}
 & \Lambda_2^4(256\,\alpha^3\,\phi_0^2-64\,\alpha^3\beta\phi_0^4)+\Lambda_2^3(256\,p_0\alpha^3\kappa\phi_0^2-32\,\alpha^2\beta\phi_0^2\\
 &+8\,\alpha^2\beta^2\phi_0^4)+\Lambda_2^2(3\,\alpha\beta^3\phi_0^4-12\,\alpha\beta^2\phi_0^2-96\,p_0\alpha^2\beta\kappa\phi_0^2)\\
 &+\Lambda_2\left(2\,\beta-\frac{16}{3}\alpha\epsilon_0\kappa-\,\beta^2\,\phi_0^2+\frac{3}{2}\,\beta^3\,\phi_0^2+12\,p_0\alpha\beta^2\kappa\phi_0^2\right.\\
 &\left.+\frac{4}{3}\alpha\beta\epsilon_0\kappa\phi_0^2+\frac{8}{3}\alpha\beta^2\epsilon_0\kappa\phi_0^2+\frac{1}{8}\beta^3\phi_0^4-\frac{3}{8}\beta^4\phi_0^4\right)-\frac{2}{3}\beta\epsilon_0\kappa\\
 &+\frac{16}{9}\alpha\epsilon_0^2\kappa^2-\frac{1}{2}p_0\beta^3\kappa\phi_0^2+\frac{1}{6}\beta^2\epsilon_0\kappa\phi_0^2-\frac{1}{3}\beta^3\epsilon_0\kappa\phi_0^2=0.
 \end{split}
 \end{equation}
 Equation~\eqref{eq:Lambda2} is a fourth order equation in $\Lambda_2$. Such an equation does not necessarily possess real solutions. Therefore, for any choice of parameters $(\alpha,\beta)$ and initial values $(\epsilon_0,\phi_0)$, we need to check that a real solution to Eq.~\eqref{eq:Lambda2} exists. In particular, we need to check this when implementing the shooting method that will allow us to find the values of $\phi_0$ such that the scalar field is trivial at spatial infinity. Such values might actually not exist in the domain where Eq.~\eqref{eq:Lambda2} possesses real solutions.
 In practice, we make sure that each choice of parameters that we consider guarantees not only that Eq.~\eqref{eq:Lambda2} has a positive\footnote{An acceptable solution to Eq.~\eqref{eq:Lambda2} must be positive, otherwise $g_{rr}$ diverges at a finite radius, and consequently the pressure and the energy density diverge as well.} real solution, but that such a solution is connected to the general relativity one. We discard all other parameter combinations that do not respect such criteria.


\subsection{Expansion at spatial infinity}\label{Sub:nsasymptotic}

 We now analyze the asymptotic behaviour of the solutions at spatial infinity. This time, we expand the metric and scalar functions in inverse powers of $r$, and solve order by order.
 We impose that the asymptotic value of the scalar field vanishes, that is $\phi(r\to\infty)\equiv\phi_\infty=0$, and that $\Gamma(r\to\infty)=0$. The asymptotic behavior of the solutions is the same as that of black hole ones, namely Eqs.~\eqref{eq:exp_f1}-\eqref{eq:exp_f3}.
 However, in the case of neutron stars, these expansions are in fact entangled with the boundary conditions at the center of the star, as we already mentioned. For fixed parameters $\alpha$ and $\beta$, the freedom in $M$ directly relates to the freedom in the central density $\epsilon_0$. On the other hand, the fact that only discrete values of $\phi_0$ yield a vanishing scalar field at infinity means that the scalar profile is actually fixed once a central density (or a mass) is chosen. Therefore, $Q$ is fixed as a function of $M$, and does not constitute a free charge; this is sometimes referred to as \textit{secondary hair}.

 The scalar charge constitutes probably the most direct channel to test the theory through observations. Indeed, binaries of compact objects endowed with an asymmetric charge will emit dipolar radiation. This enhances the gravitational wave emission of such systems: in a Post-Newtonian (PN) expansion, dipolar radiation contributes to the energy flux at order -1PN with respect to the usual quadrupolar general relativity flux. Generically, this dipolar emission is controlled by the sensitivities of the compact objects, defined as\footnote{The factor $2$ is added to match the standard definition of the sensitivity in the literature, where a different normalization for the scalar field is generally used.}
 \begin{equation}
    \alpha_I=2 \dfrac{\partial\text{ln}M_I}{\partial\phi_0},
 \end{equation}
 $M_I$ being the mass of the component $I$, and $\phi_0$ the value of the scalar field at infinity. The observation of various binary pulsars, notably the PSR~J1738+0333 system, allows one to set the following constraint:
 \begin{equation}
   |\alpha_A-\alpha_B|\lesssim2\times10^{-3},
    \label{eq:DEFbound}
 \end{equation}
 where $A$ and $B$ label the two components of the system~\cite{Shao:2017gwu,Wex:2020ald}. We can then relate the sensitivity to the scalar charge $Q$, using the generic arguments of~\cite{Damour:1992we}.
 If there is no accidental coincidence in the charge of the two components of the binary, Eq.~\eqref{eq:DEFbound} 
 translate as 
 \begin{equation}
 \left|\dfrac{\tilde{Q}}{M}\right|\lesssim6\times10^{-4}
 \label{eq:boundQ}
 \end{equation}
 for the solutions we consider, where we have defined a new quantity $\tilde{Q}=Q/\sqrt{2\kappa}$, so that $\tilde{Q}/M$ is a dimensionless quantity and we use units where $c=1$, $G=1$ and $M_\odot=1$. Only solutions satisfying this bound on the charge to mass ratio are relevant. It is however a non-trivial task to map this bound onto the parameters of the action~\eqref{eq:ACI}. We will do so by exploring the parameter space in Section~\ref{Sec:nsregions}.


\subsection{Numerical implementation}\label{Sub:nsnumerical}

 We solve the system of three differential equations for the three independent functions $\Gamma$, $\phi$ and $\epsilon$ by starting our integration from $r_0=10^{-5}~\text{km}$. We fix the parameters of the theory $\alpha$ and $\beta$, and the central density $\epsilon_0$, typically to values of order $10^{17}$~kg/m$^3$. Then, we give an initial guess for $\phi_0$, and determine boundary conditions as explained in Section~\ref{Sub:rto0}. The integration will generically give a solution; however, we also demand that the scalar field vanishes at infinity, that is $\phi_\infty=0$. Only a discrete set of $\phi_0$ values will yield $\phi_\infty=0$. Each value corresponds to a different number of nodes of the scalar field in the radial direction. In practice, we integrate up to distances $r_\text{max}=300\, \text{km}$ and we implement a shooting method to select the solutions with $\phi_\text{max}=0$. Generally, we use Mathematica's built-in function FindRoot.

 However, in some cases FindRoot fails to find the right solutions, even if one gives it a limited range $(\phi_{0,\:\text{min}},\phi_{0,\:\text{max}})$ where to look for. When this happens, we resort to  bisection instead. In this latter case, we require that $\phi(r_\text{max})/\phi_0 \leq 10^{-2}$. 

 At each stage of the shooting method, we must check that Eq.~\eqref{eq:Lambda2} gives a real positive solution for $\Lambda_2$ that is connected to the general relativity solution. In some cases, we reach the limit of the region of the parameter space where these criteria are fulfilled before reaching $\phi_\infty=0$. When this is the case, there is no solution associated to the given choice of $\alpha$, $\beta$ and $\epsilon_0$. Note also that, given a set of $\alpha$, $\beta$ and $\epsilon_0$, there is a maximum number of nodes that the solution can have, consequently a maximum number of suitable choices of $\phi_0$ (typically up to three modes in the regions we explore). Solutions with more nodes are encountered only for higher values of the parameters $\alpha$ and $\beta$, or at higher curvatures (that is, at higher $\epsilon_0$).

 Given a solution, we extract the value of the ADM mass $M$ and the scalar charge $Q$, from the asymptotic expansion as in Eq.~\eqref{eq:MandQ}.
 
 Let us now summarize the algorithm we implement:
 \begin{enumerate}
    \item We choose a specific value for the central energy density $\epsilon_0$.
    \item We fix $\beta$.
    \item We choose a value for $\alpha$.
    \item Given $\epsilon_0$, $\beta$ and $\alpha$, we find the value for $\phi_0$ that satisfies the initial conditions on the equation on $\Lambda_2$, i.e. Eq.~\eqref{eq:Lambda2}, and gives the right asymptotics, implementing a shooting method. If there is no value of $\phi_0$ that satisfies these requirements, we treat the parameter set as having no solution.
    \item Given $\alpha$, $\beta$ and $\phi_\text{0}$, we solve the field equations and we extract the value of $M$ and $Q$.
    \item We repeat steps 3-5 for a new value of $\alpha$.
 \end{enumerate}
 
\section{Existence regions of scalarized solutions}\label{Sec:nsregions}

 In this Section, we study the regions where scalarized solutions exist in the $(\alpha,\beta)$ parameter space. We analyze three different neutron star scenarios, which correspond to the three cases studied in Chapter~\ref{Chap:Threshold}.
 
\subsection{Light star with SLy EOS}\label{Sub:lightns}

\begin{figure}[h!]
	\includegraphics[width=0.7\linewidth]{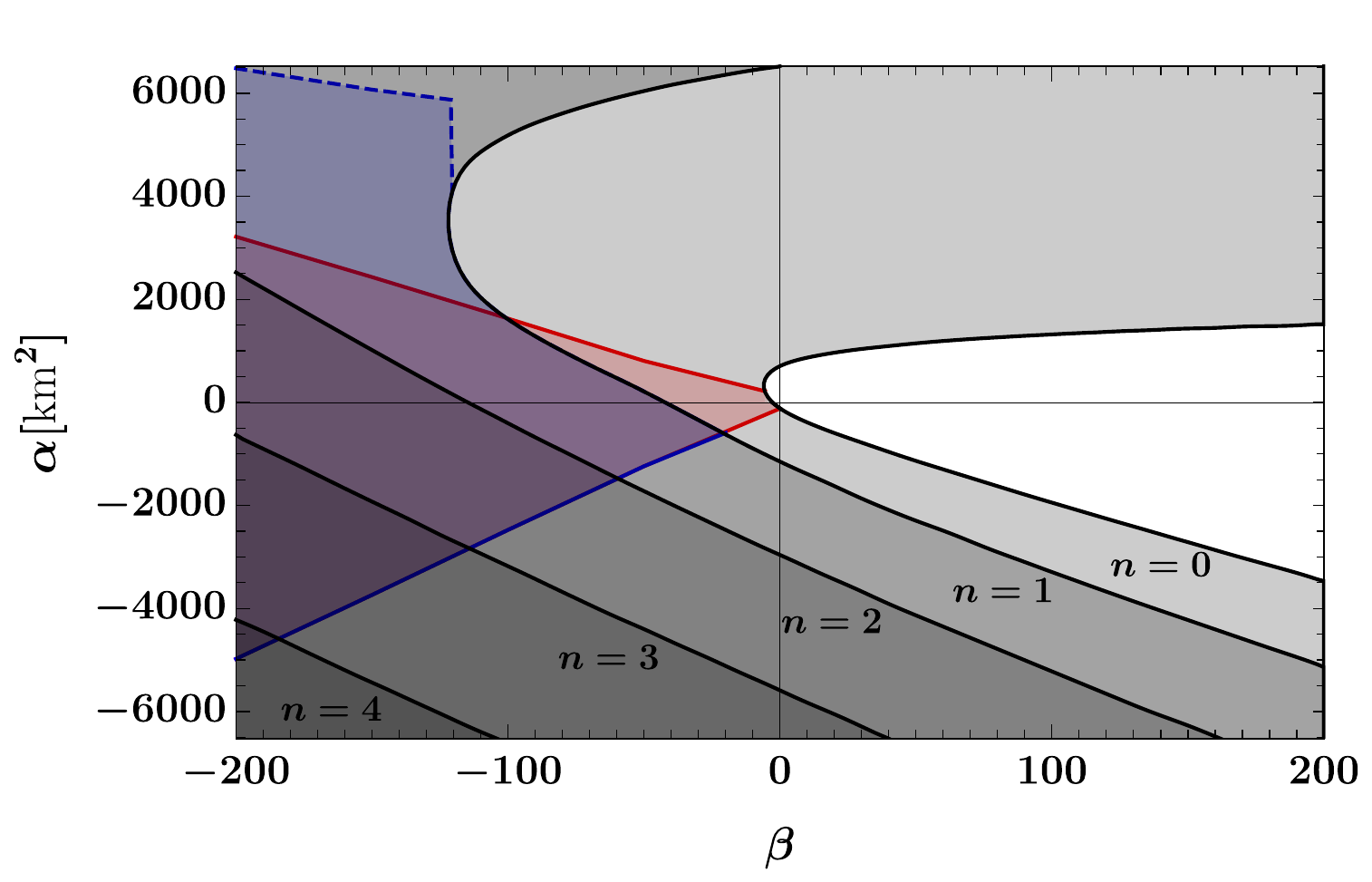}%
	\caption[Regions of existence of scalarized solutions in the parameter space for a light neutron star]{Regions of existence of scalarized solutions in the $(\alpha,\beta)$ space, for the SLy EOS with $\epsilon_0=8.1\times 10^{17}~\text{kg}/\text{m}^3$. The red (respectively blue) region is the region where scalarized solutions with 0 (respectively 1) node exist. We superimposed the grey contours obtained in Chapter~\ref{Chap:Threshold}, which represent the lines beyond which general relativity solutions with the same density are unstable to scalar perturbations with 0, 1, 2, \textit{etc} nodes. We see that the region where there exist scalarized solutions with $n$ nodes is included in the region where general relativity solutions are unstable to scalar perturbations with $n$ nodes, but much smaller. The dashed boundary for the blue region corresponds to a breakdown of the integration inside the star. In general relativity, a star with this choice of $\epsilon_0$ and EOS has a light mass, $M_\text{GR}=1.12 M_\odot$.}
	\label{fig:Sly112}
 \end{figure}

 First, we consider a neutron star described by the SLy equation of state \cite{Haensel:2004nu}, with a central energy density of $\epsilon_0=8.1\times 10^{17}~\text{kg}/\text{m}^3$, so that its gravitational mass in general relativity is $M_{\text{GR}}=1.12 M_\odot$. The results are summarized in Fig.~\ref{fig:Sly112}, where we relate our new results to the previous study of the scalarization thresholds of Chapter~\ref{Chap:Threshold}.
 The white area corresponds to the region of the parameter space where the general relativity solution is stable. When cranking up the parameters $\alpha$ or $\beta$, a new unstable mode appears every time one crosses a black line. The first mode has 0 nodes, the second 1 node, \textit{etc}. We will refer to these black lines as \textit{instability lines}. Any point in the parameter space that lies within some grey region corresponds to a configuration where the general relativity solution is unstable. 
 The red (respectively blue) area corresponds to the region where scalarized solutions with $n=0$ (respectively $n=1$) nodes exist. We do not include the equivalent regions for higher $n$, to not complicate further the analysis. The region where a scalarized solution does exist
is considerably reduced with respect to the region where the general relativity solution is unstable.

 One of our main results is that the parameters $(\alpha,\beta)$ corresponding to the grey areas that are not covered by the colored regions must be excluded. Indeed, there, scalarized solutions do not exist while the general relativity solution itself is unstable. Therefore, neutron stars in these theories, when they reach a critical mass, will be affected by a tachyonic instability, but there does not exist a fixed point (a static scalarized solution) where the growth could halt. This would imply that neutron stars with this mass and EOS do not exist for the corresponding parameters  of the theory~\eqref{eq:ACI}. Considering that the properties of the scalarized star are sensitive to nonlinearities, adding further nonlinear interaction terms to the action, {\em e.g.} self-interactions in a scalar potential, as was proposed in~\cite{Macedo:2019sem}, or non-linear terms in the coupling functions~\cite{Doneva:2017bvd,Silva:2018qhn}, can potentially change this result.

 In Fig.~\ref{fig:Sly112}, the regions where scalarized solutions exist are delimited by \textit{existence lines}, represented by a curve of the respective color. The plain lines correspond to boundaries beyond which it is no longer possible to find a value of $\phi_0$ that allows a suitable solution to Eq.~\eqref{eq:Lambda2}, while providing $\phi_\infty=0$. Beyond dashed lines, on the other hand, nothing special occurs at the center of the star, but the numerical integration breaks down at a finite radius inside the star. We do not know whether, when crossing these dashed lines, our integration is affected by numerical problems, or whether the divergence corresponds to an actual singularity of the solutions. 
 It could be that this singularity emerges as an artifact of the method we employ. Indeed, in our analysis, we keep the central density $\epsilon_0$ fixed while pushing the couplings $\alpha$ and $\beta$ to larger and larger values. However, for each couple of parameters $(\alpha,\beta)$, there probably exists a maximal central density beyond which star solutions do not exist, or equivalently it becomes impossible to sustain such a high central density. The dashed line could correspond to this saturation, where we try to push all the parameters beyond values that can actually be sustained by the model.

 A surprising feature, which is not visible in Fig.~\ref{fig:Sly112}, is that scalarized solutions always exist in a very narrow range along the instability lines. For example, when crossing the black instability line that delimitates the white region where the general relativity solution is stable, from the light-grey region where it is unstable against $n=0$ scalar perturbations, there exists a very narrow band (within the grey region) where scalarized solutions with zero node exist. We observed similar behaviours along each instability line, also in the scenarios discussed in the next paragraphs. We further investigate these particular solutions in Section~\ref{Sub:instabilitylines}.

\subsection{Light star with MPA1 EOS}\label{Sub:mpa1ns}

\begin{figure}[h!]
	\includegraphics[width=0.7\linewidth]{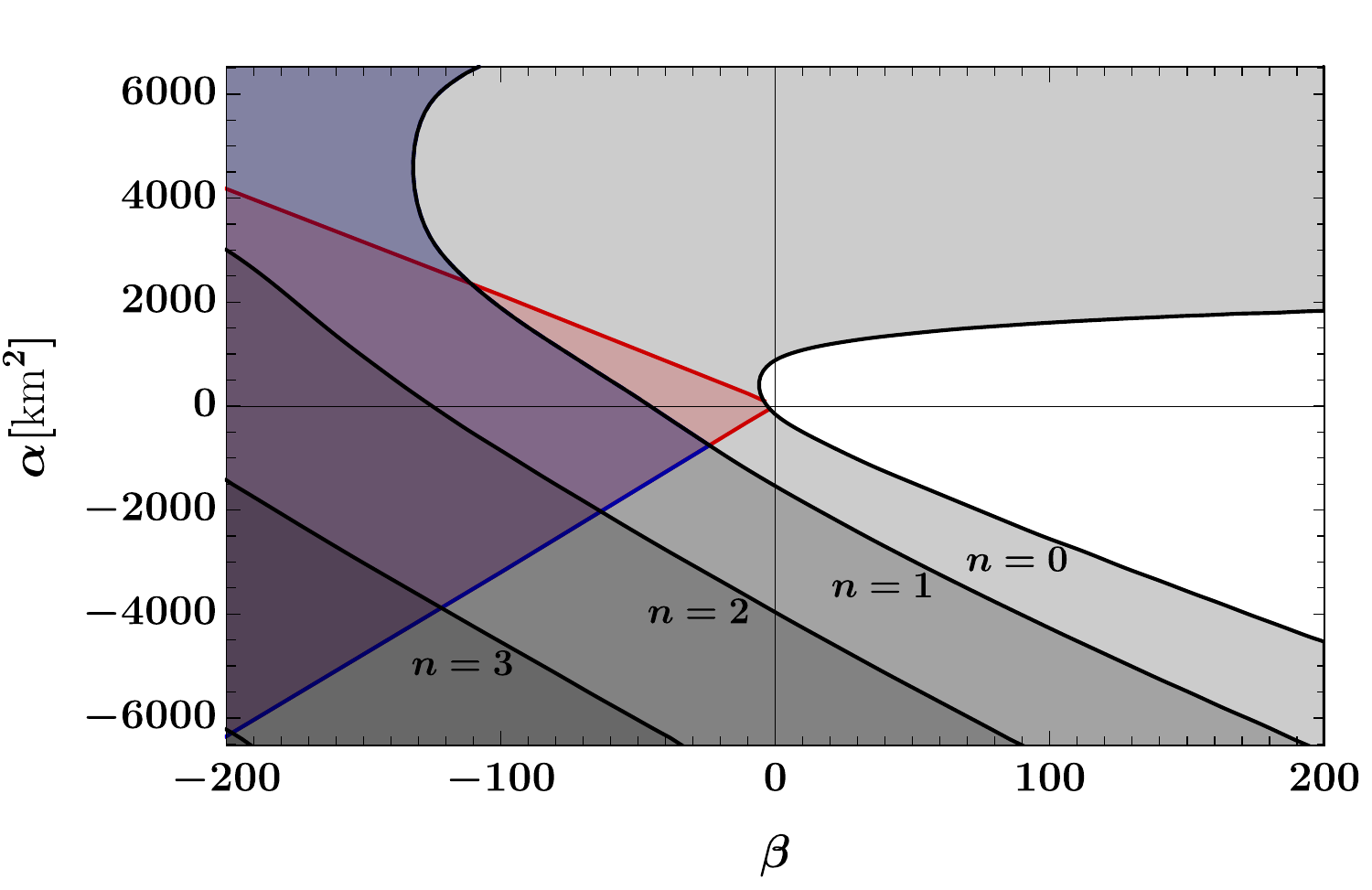}%
	\caption[Regions of existence of scalarized solutions in the parameter space for MPA1 EOS]{Regions of existence of scalarized solutions in the $(\alpha,\beta)$ space, for the MPA1 EOS with $\epsilon_0=6.3\times 10^{17}~\text{kg}/\text{m}^3$. The conventions are the same as in Fig.~\ref{fig:Sly112}.  In general relativity, a star with this choice of $\epsilon_0$ and EOS is again light, with $M_\text{GR}=1.12 M_\odot$.}
	\label{fig:MPA1}
 \end{figure}

 We next consider a stellar model described by the MPA1 equation of state~\cite{Gungor:2011vq}. We choose a central energy density of $\epsilon_0=6.3\times 10^{17}\,\text{kg}/\text{m}^3$, such that it corresponds to the same general relativity mass as in the previous case, that is $M_{\text{GR}}=1.12 M_\odot$. We report the results in Fig.~\ref{fig:MPA1}.
 As one can see, changing the EOS has only mild effects on the region of existence of scalarized solutions. The analysis of the parameter space is qualitatively the same as for the SLy EOS. The main difference is that, for the range of parameters we considered, no numerical divergences (associated with dashed lines) appear with the MPA1 EOS.

\subsection{Heavy star with SLy EOS}\label{Sub:heavyns}

\begin{figure}[h!]
	\subfloat{\includegraphics[width=0.7\linewidth]{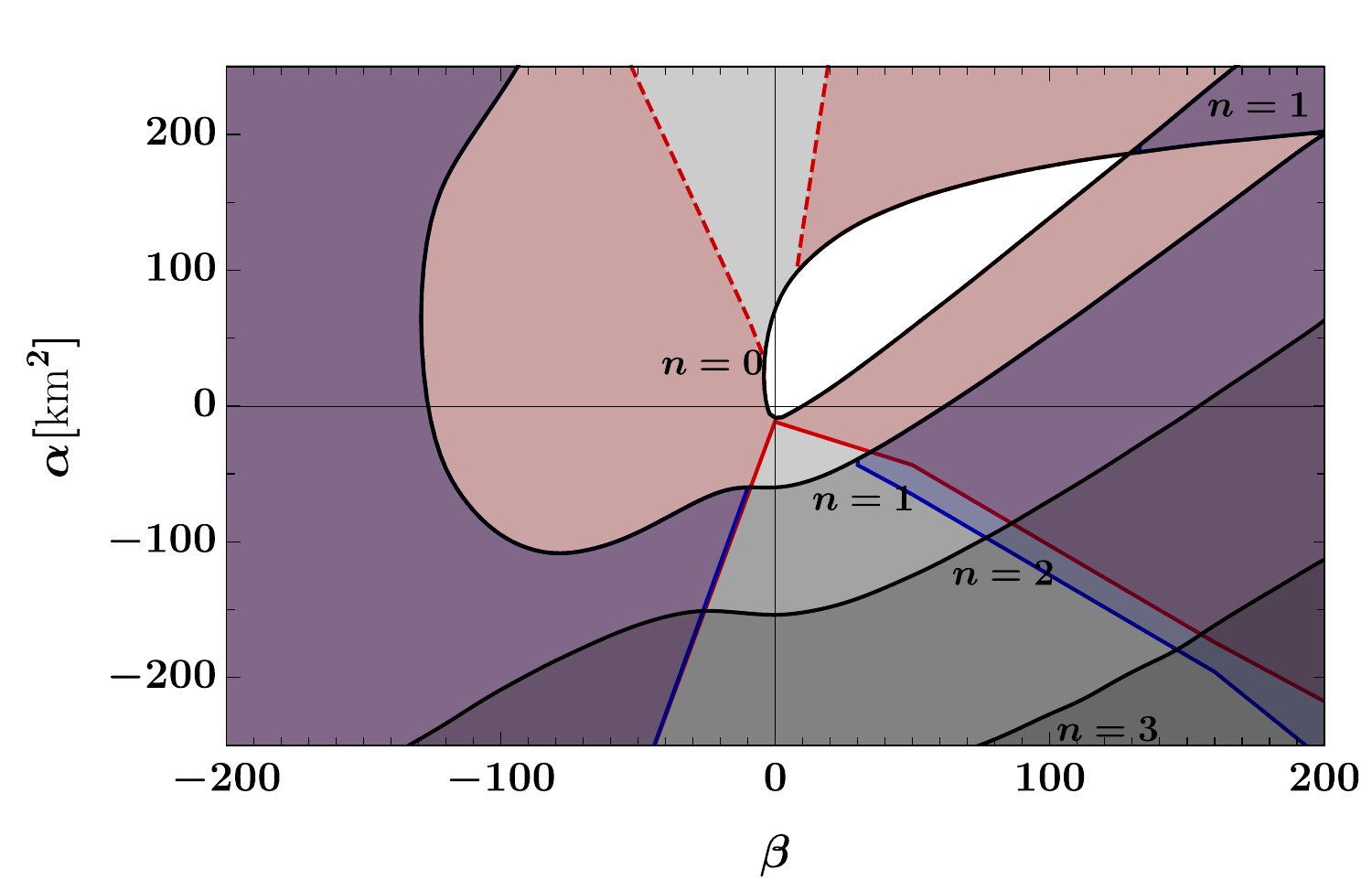}%
	}
	\\
		\subfloat{%
	\includegraphics[width=0.7\linewidth]{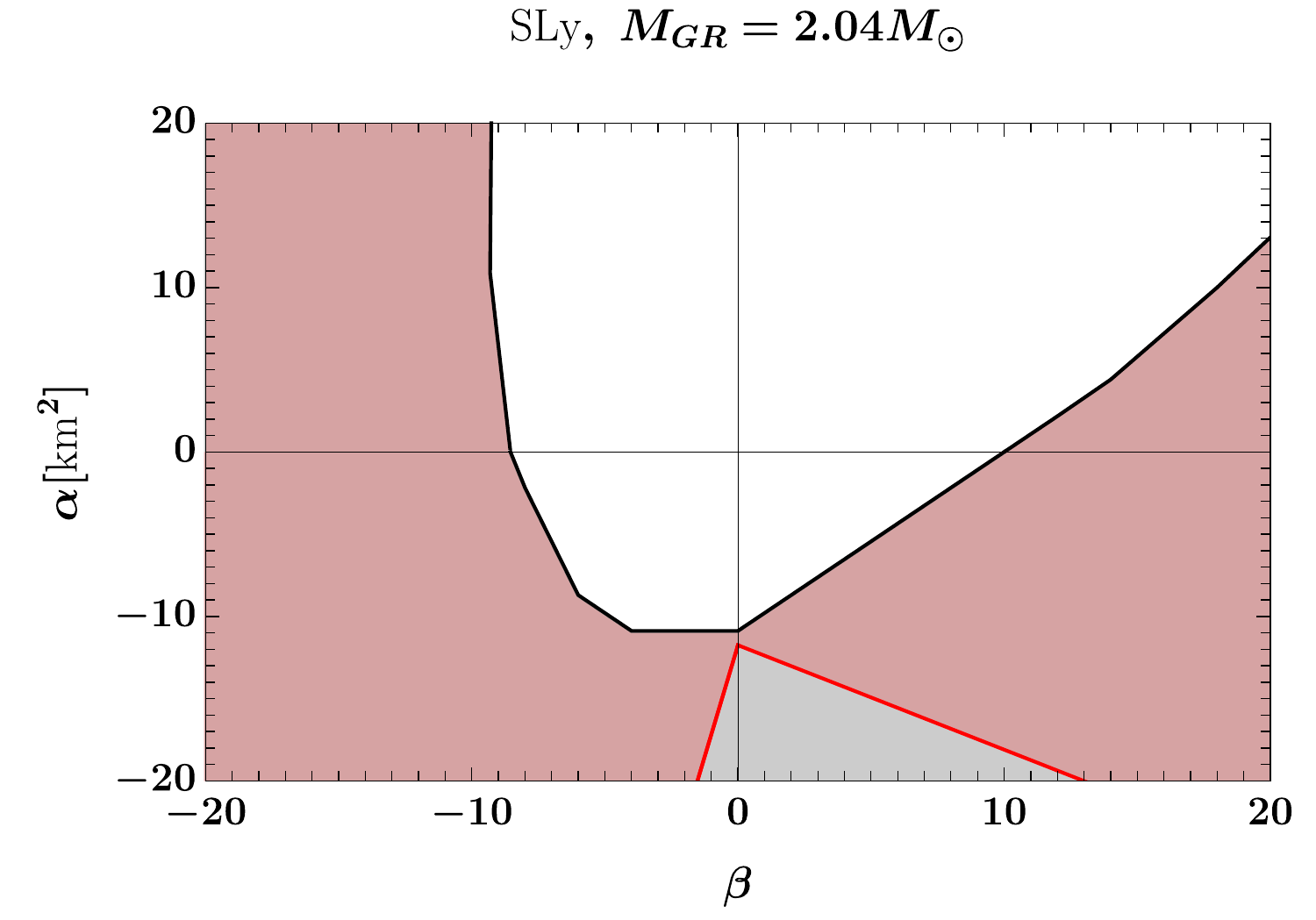}%
	}
	\caption[Regions of existence of scalarized solutions in the parameter space for a heavy neutron star]{Regions of existence of scalarized solutions in the $(\alpha,\beta)$ space, for the SLy EOS with $\epsilon_0=3.4\times 10^{18}~\text{kg}/\text{m}^3$. The conventions are the same as in Fig.~\ref{fig:Sly112}. In general relativity, a star with this choice of $\epsilon_0$ and EOS is the heaviest possible, $M_\text{GR}=2.04 M_\odot$. The bottom panel is simply a zoom of the upper one.}
	\label{fig:SLy204}
 \end{figure}

 Last, we consider a denser neutron star described by the SLy EOS, with $\epsilon_0=3.4\times 10^{18}\,\text{kg}/\text{m}^3$. It corresponds to an increased mass in general relativity of $M_{\text{GR}}=2.04 M_\odot$. The results are shown in Fig.~\ref{fig:SLy204}.
 In this case, positive values of $\beta$ can also lead to scalarized solutions. In Chapter~\ref{Chap:Threshold}, we have already discussed how, in general relativity, dense neutron possess a negative Ricci scalar towards the center, which allows for scalarization to be triggered even when $\beta>0$. As before, a dashed line signals the appearance of divergences, which in this case shows up already for the $n=0$ node.

 In the lower panel of Fig.~\ref{fig:SLy204}, we zoomed on the region of small couplings, in order to understand better what happens for natural values of the Ricci coupling $\beta$. In the absence of the Gauss-Bonnet coupling, scalarization can occur either if $\beta<-8.55$, or $\beta>11.5$. Let us concentrate on the $\beta>0$ scenario, which is motivated by the results of Ref.~\cite{Antoniou:2020nax}, where it was shown that positive values of $\beta$ make general relativity a cosmological attractor. We remind that black hole scalarization (at least for non-rotating black holes) occurs for $\alpha>0$. Hence, we see that there exists an interesting region in the $\alpha>0,~\beta>0$ quadrant where even very compact stars do not scalarize, while black holes do. Such models can therefore \textit{a priori} pass all binary pulsar tests, while being testable with black hole observations. On the other hand, for $\beta\gtrsim11.5$, the red region where general relativity solutions are replaced by scalarized solutions spreads very fast in the $\alpha$ direction, and one has to be careful, when considering black hole scalarization, that such models are not already excluded by neutron star observations.

 To summarize, we have identified the regions of parameter space where solutions exist, considering three different stellar scenarios which correspond to different central densities and EOS. Although we have considered only a limited number of different central densities, we have selected the ones that correspond to the lowest/largest neutron star mass in general relativity, in order to cover very different setups. The regions where scalarized solutions exist are systematically smaller than the ones where the general relativity branch is tachyonically unstable. The complementary regions, where the general relativity solution is unstable while no scalarized solution exists, should be excluded.

\section{Properties of the solutions}\label{Sec:nsproperties}
 
 We now discuss the properties of these solutions, in particular their scalar charge and their mass. We separate this study into two cases: $\beta<0$ (Section~\ref{Sub:negativebeta}) and $\beta>0$ (Section~\ref{Sub:positivebeta}); indeed, these two situations have different motivations and observational interests.
  
\subsection{Mass and scalar charge of the $\beta<0$ solutions}\label{Sub:negativebeta}
 
 We first focus on the scenario where $\beta<0$. This corresponds to the original situation studied by Damour and Esposito-Farèse. Typically, scalarized solutions with $\beta<0$ and $\alpha=0$ are constrained by binary pulsar observations for the massless case~\cite{Freire:2012mg,Antoniadis:2013pzd,Shao:2017gwu}. A particular motivation to study solutions with $\beta<0$ is therefore to determine whether the addition of a non-zero Gauss-Bonnet coupling can improve their properties. We consider three different choices of the Ricci coupling: $\beta=-5.5,-10$ and $-100$. The two first choices are relevant astrophysically: $\beta=-5.5$ is approximately the value where scalarization is triggered for small Gauss-Bonnet couplings, while $\beta=-10$ corresponds to a region where neutron stars are scalarized, but with rather small deviations with respect to general relativity. The third choice, $\beta=-100$, is certainly disfavored observationally, but it allows us to illustrate an interesting behaviour concerning different scalar modes.

\begin{figure*}[h!]
 \begin{center}
	\subfloat{%
	\includegraphics[width=0.5\linewidth]{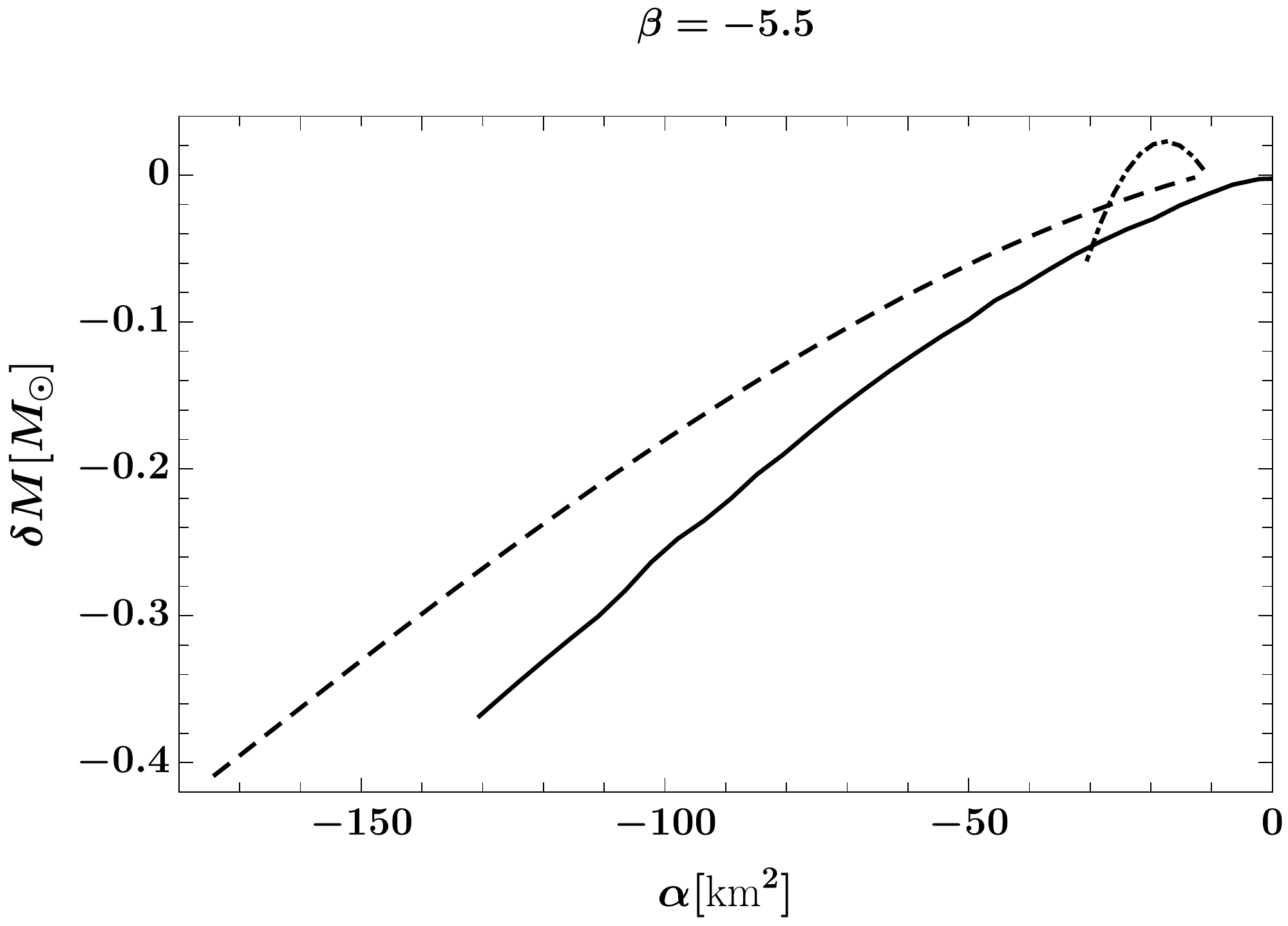}%
	}
	\subfloat{%
	\includegraphics[width=0.5\linewidth]{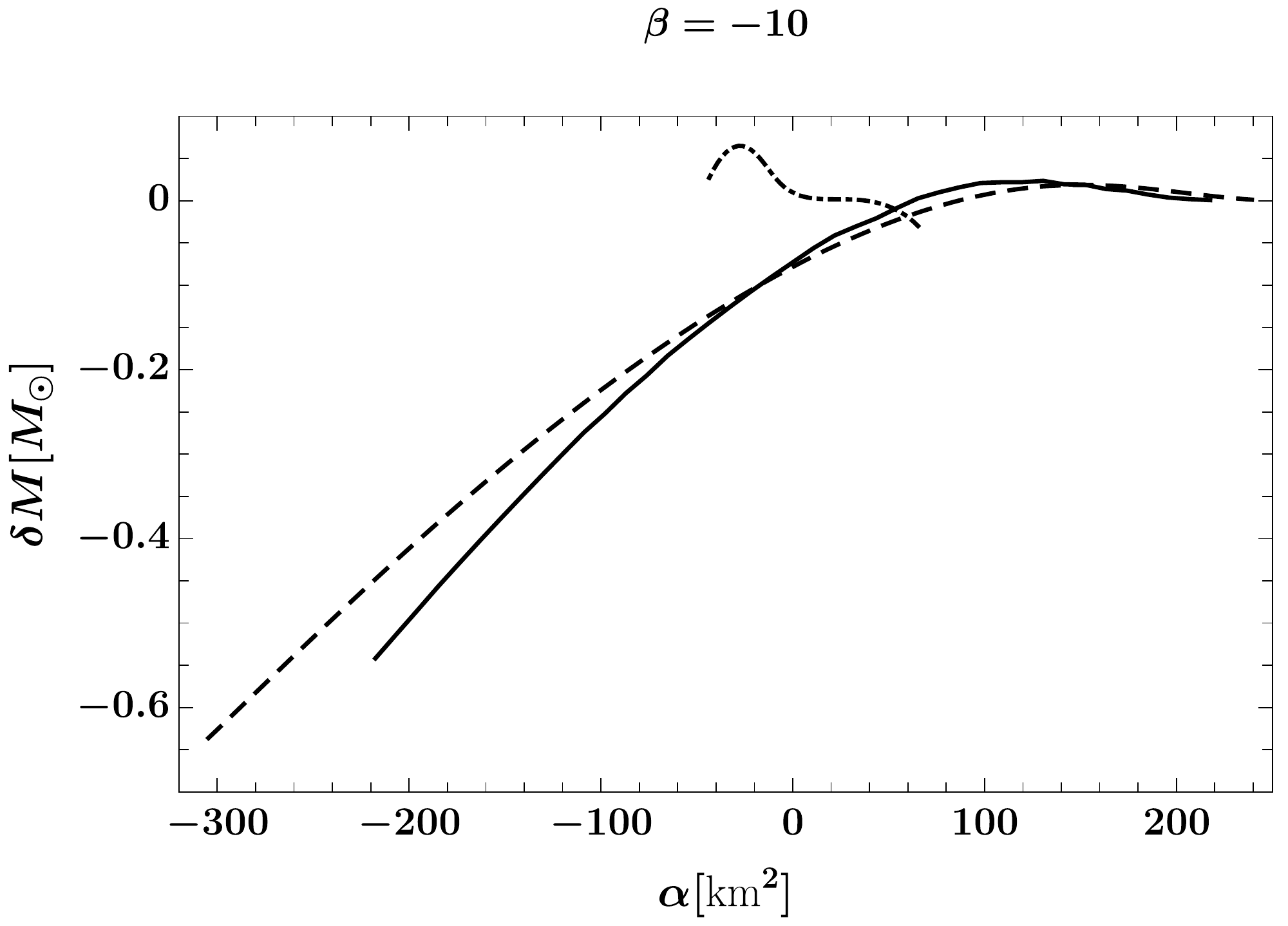}%
	}
	\\
	\subfloat{%
	\includegraphics[width=0.5\linewidth]{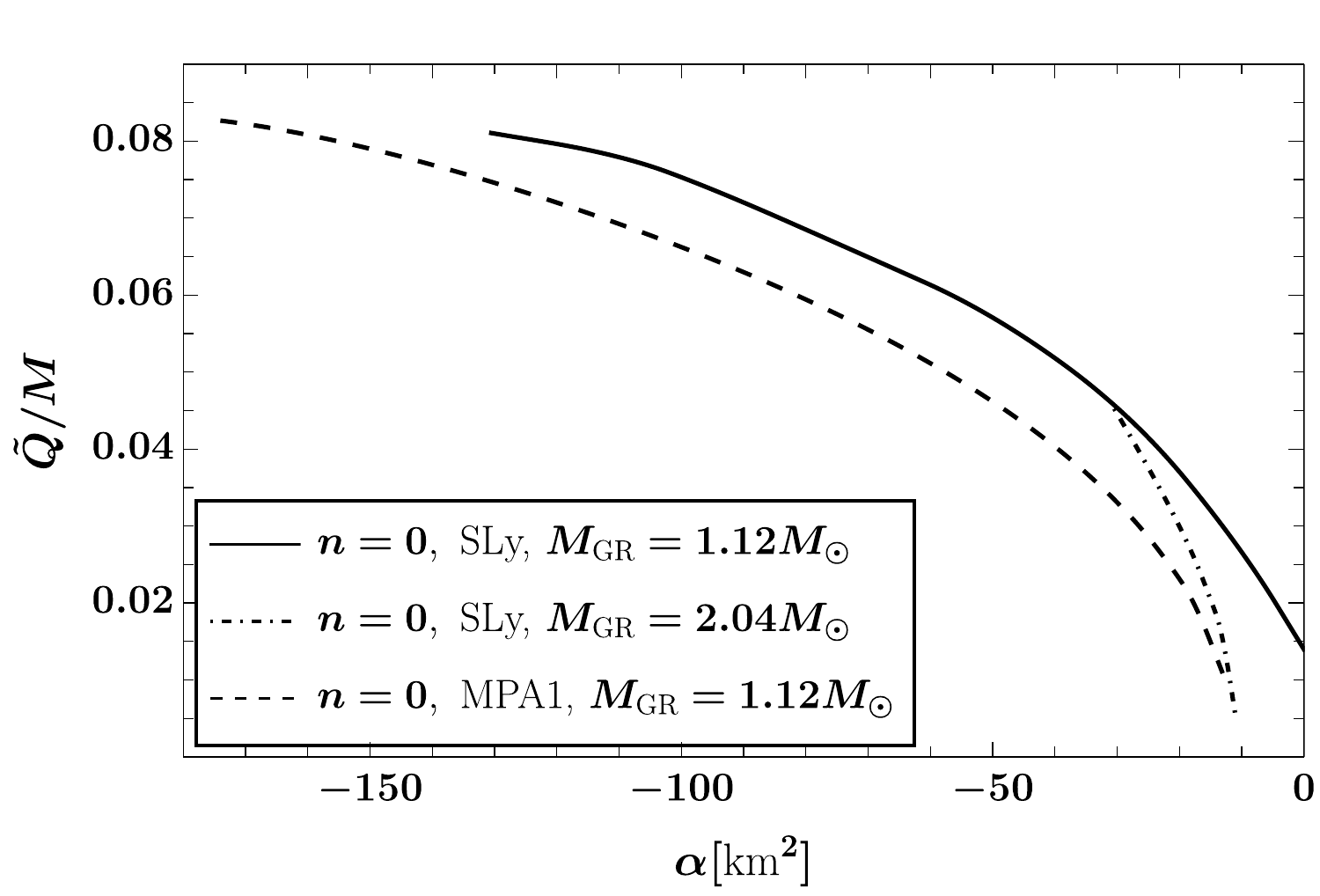}%
	}
	\subfloat{%
	\includegraphics[width=0.5\linewidth]{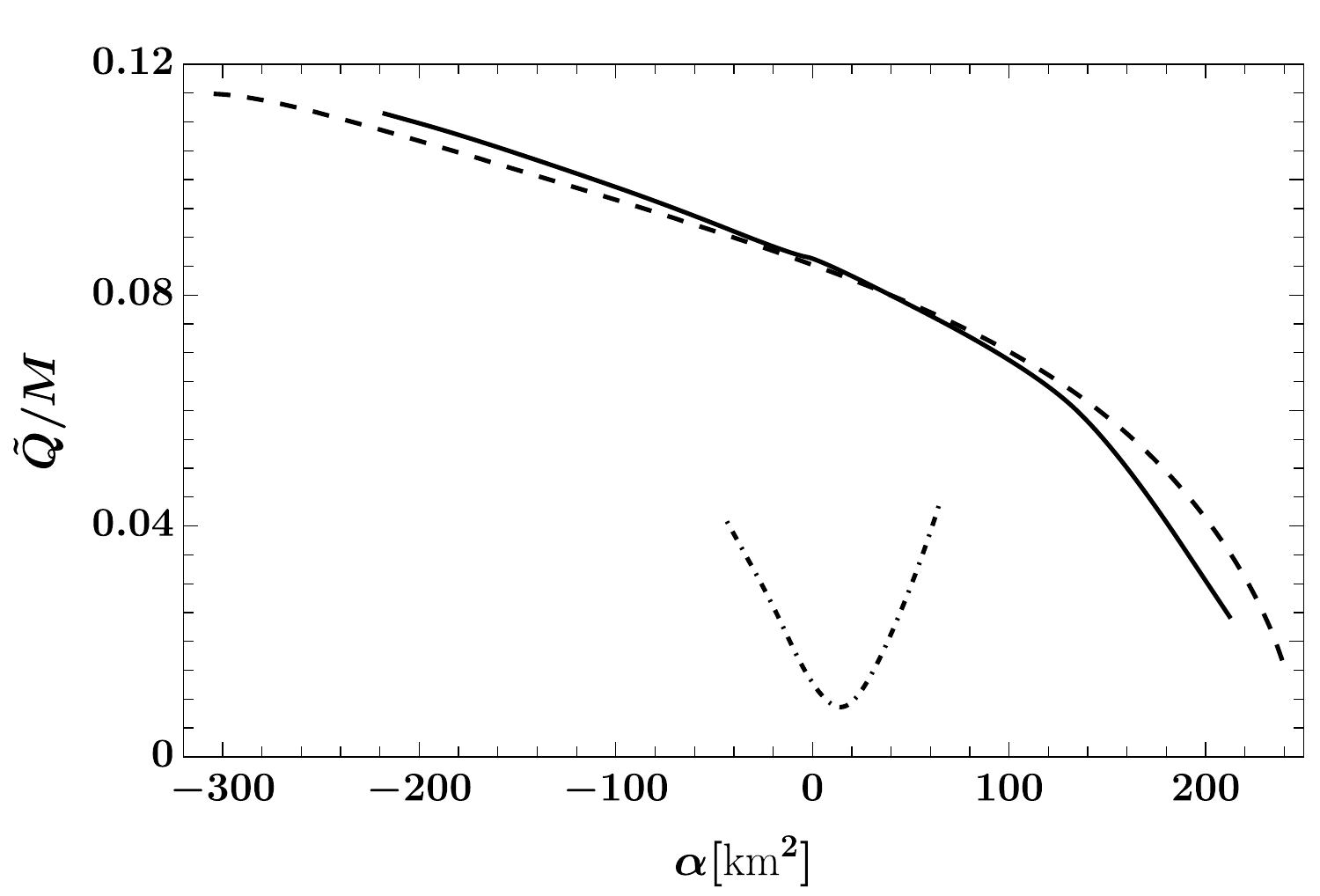}%
	}
	\caption[Mass difference and scalar charge of scalarized solutions for negative Ricci couplings]{Mass difference and scalar charge of scalarized solutions for $\beta<0$. The two left (respectively right) panels show how these quantities evolve when varying $\alpha$ at fixed $\beta=-5.5$ (respectively $-10$). The rescaled scalar charge $\tilde{Q}$ (bottom panels) is normalized to the total mass of the solutions, $M$. For all curves, the mass difference $\delta M$ (upper panels) is computed with respect to a general relativity star with the same central density and EOS. Plain curves correspond to a general relativity mass of $1.12~M_\odot$, using the SLy EOS; dashed curves to the same general relativity mass, but the MPA1 EOS; and dotted-dashed curves to a general relativity mass of $2.04~M_\odot$, using the SLy EOS. In this region of the parameter space, only solutions with 0 nodes for the scalar field exist. A generic feature of lighter stars (plain and dashed curves), is that the charge decreases when $\alpha$ increases, \textit{a priori} offering a way to evade the stringent bound of Eq.~\eqref{eq:boundQ} when increasing $\alpha$. However, it is only for values of $\beta$ very close to the Damour and Esposito-Farèse threshold ($\beta=-5.5$) that we can obtain scalar charges compatible with observations.
	}
	\label{fig:smallCoup}
 \end{center}
 \end{figure*}

 Let us start with the comparison between the cases $\beta=-5.5$ and $-10$. The results are summarized in Fig.~\ref{fig:smallCoup}.
 This figure shows two properties of scalarized stars. First, the mass default (or excess) of scalarized stars with respect to general relativity stars with the same central density and EOS: 
 $\delta M=M-M_{GR}$. 
 Second, the rescaled scalar charge of the scalarized solutions, $\tilde{Q}$. We compare the results for the three different stellar models considered in Section~\ref{Sec:nsregions}, for the two values of $\beta$. 
 All curves extend only over a finite range of $\alpha$. Indeed, passed a certain value of $\alpha$, we exit the red region on the $\beta<0$ side of Figs.~\ref{fig:Sly112}, \ref{fig:MPA1} and \ref{fig:SLy204} (moving vertically, since $\beta$ is fixed to $-5.5$ or $-10$). Scalarized solutions do not exist outside of this region. 

 Figure \ref{fig:smallCoup} shows that the choice of EOS does not affect much the properties of the scalarized solutions.
 However, increasing the density drastically modifies these properties. In particular, at higher densities, there exist solutions with $\delta M>0$. This can appear problematic at first. Indeed, one expects that, in a scalarization process, energy is stored in the scalar field distribution. Hence, the ADM mass, that constitutes a measure of the gravitational energy, should decrease in the process. 
 However, we stress that we are not studying a dynamical process. Indeed, the stars for which we are computing the mass difference $\delta M$ have, by construction, the same central energy density $\epsilon_0$. In the scalarization process of a general relativity neutron star, the central energy density will not remain fixed. Hence, our results do not necessarily mean that a star will gain mass when undergoing scalarization.

 Perhaps more interestingly for observations, Fig.~\ref{fig:smallCoup} also shows the behaviour of the scalar charge. For the light neutron stars, the scalar charge always decreases when $\alpha$ increases. Therefore, the constraint on the scalar charge, Eq.~\eqref{eq:boundQ}, disfavors the solutions with $\alpha<0$ with respect to standard Damour and Esposito-Farèse ($\alpha=0$) solutions. On the contrary, one could hope that a positive Gauss-Bonnet coupling could help evade these constraints even for $\beta<-5.5$, by quenching the charge. Effectively, there will be a direction in the $\alpha>0$ and $\beta<0$ quadrant where the effects of the two operators, Ricci and Gauss-Bonnet, combine to yield a small scalar charge.
 This interesting possibility is moderated by what happens in the case of denser stars (dotted-dashed line in Fig.~\ref{fig:smallCoup}). For large negative values of the Ricci coupling ($\beta=-10$), the scalar charge does not have a monotonic behaviour with $\alpha$. In particular, as shown in the bottom-right panel of Fig.~\ref{fig:smallCoup}, $\tilde{Q}$ starts increasing for positive values of $\alpha$. Even at the point where $\tilde{Q}$ is minimal, its value ($\tilde{Q}/M\simeq8\times10^{-3}$) already exceeds the bound of Eq.~\eqref{eq:boundQ}. Therefore, it is only for values of $\beta$ that are very close to the Damour and Esposito-Farèse threshold $\beta\simeq-5.5$, that the addition of the Gauss-Bonnet coupling can help to reduce the scalar charge, and to pass the stringent binary pulsar tests.

\begin{figure}[h!]
	\includegraphics[width=0.7\linewidth]{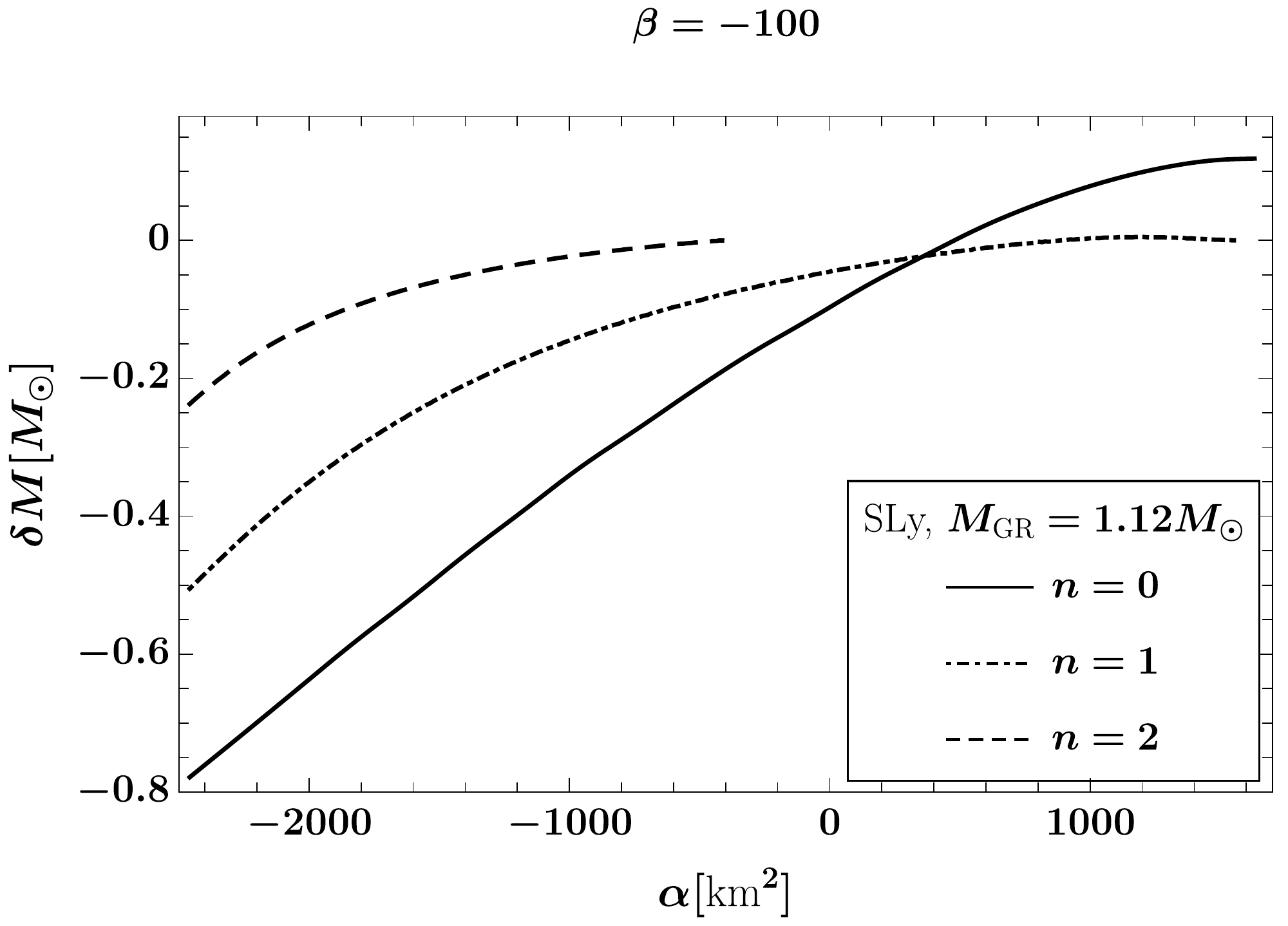}
	\caption[Mass difference for a large negative value of the Ricci coupling]{Mass difference $\delta M$ vs $\alpha$ at $\beta=-100$. The EOS considered here is the SLy one, with $\epsilon_0=8.1\times 10^{17}\,\text{kg}/\text{m}^3$, which in general relativity corresponds to $M_\text{GR}=1.12~M_\odot$. The color and dashing conventions is the same as in Fig.~\ref{fig:smallCoup}. We have more modes in this region of parameter space, that we represent as dotted-dashed (for $n=1$ node) and dashed (for $n=2$ nodes) curves. For $\alpha\gtrsim350\, \text{km}^2$, solutions with 1 node start having a smaller mass than solutions with 0 node, which can indicate that solutions with 1 node are more energetically favored.}
	\label{fig:betaNeg100}
 \end{figure}

 To conclude the study of the $\beta<0$ region, we consider a significantly more negative Ricci coupling, namely $\beta=-100$. To illustrate what happens at these large negative values of $\beta$, it is enough to consider one scenario, for example the one of lighter neutron stars with the SLy EOS. For such negative values of $\beta$, there exist several scalarized solutions, with different number of nodes. We can then compare the mass difference of these solutions between each other. Figure \ref{fig:betaNeg100} shows that, for $\alpha>\alpha_\text{c}\simeq350\, \text{km}^2$, scalarized solutions with 1 node become lighter than scalarized solutions with 0 node.
 This is a hint that, for $\alpha>\alpha_c$, the one node solution will be preferred energetically to the zero node solution. We cannot conclude definitively on this issue, as the ADM mass does not take into account the energy stored in the scalar distribution (which is non-zero for the two scalarized solutions). However, in the regime where this inversion happens, the mass difference with respect to general relativity, $\delta M$, is rather small. If our interpretation in terms of energetic preference is correct, the transition from a preferred solution with zero node to a solution with one node is interesting. Indeed, the scalarized solution with zero node is associated with the fundamental mode of the general relativity background instability. At the perturbative level, all the other modes of instability have higher energies. It would then be natural to expect that, at the non-linear level of scalarized solutions, this energy hierarchy is respected. This is the case up to $\alpha=\alpha_\text{c}$, but not anymore beyond. In Section~\ref{Sec:scalarprofile}, we provide a putative explanation for this inversion: that for $\alpha>\alpha_\text{c}$, the profile of the effective mass over the general relativity background tends to favor the growth of scalar field solutions with one node, rather than zero.
 
\subsection{Mass and scalar charge of the $\beta>0$ solutions}\label{Sub:positivebeta}

\begin{figure}[h!]
	\subfloat{\includegraphics[width=0.7\linewidth]{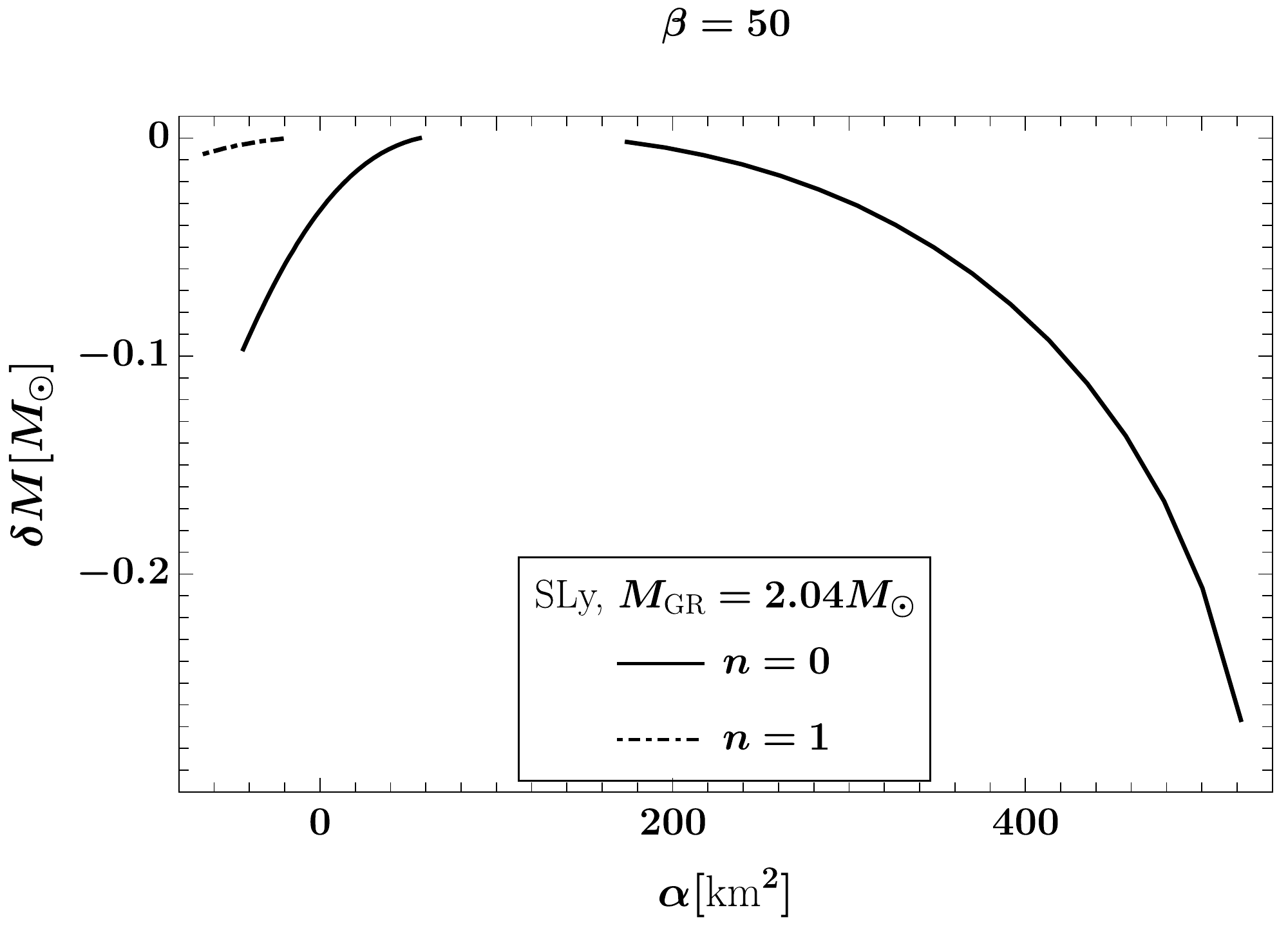}}
	\\
	\subfloat{\includegraphics[width=0.7\linewidth]{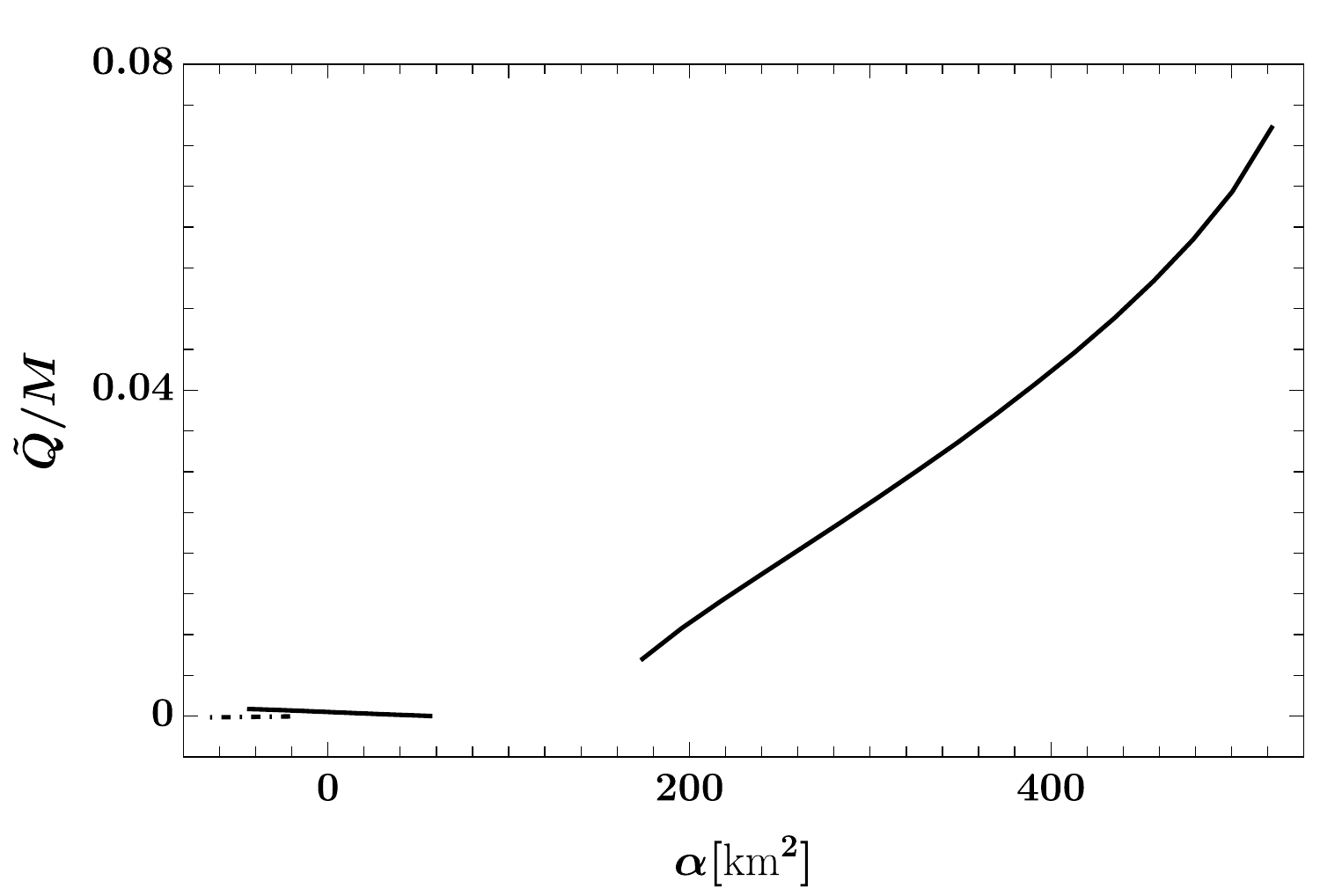}}      
	\caption[Mass difference and scalar charge of scalarized solutions for positive Ricci couplings]{Mass difference and rescaled scalar charge of scalarized solutions for $\beta>0$ ($\beta=50$ here). Among the three neutron star scenarios that we considered throughout the Chapter, only the heavier star ($\epsilon_0=5.51 \times 10^{-3}$~kg/m$^3$, $M_\text{GR}=2.04~M_\odot$, SLy EOS) possesses some scalarized solutions in this region. The dashing convention is the same as in Fig.~\ref{fig:betaNeg100}. Solutions that correspond to the interval of $\alpha$ centered on 0 are interesting observationally, as they yield very small scalar charges, compatible with Eq.~\eqref{eq:boundQ}.}
	\label{fig:50M204}
 \end{figure}

 We now consider the case of positive $\beta$. Such solutions are less constrained by observations than their $\beta<0$ counterparts. They are also very interesting from a cosmological perspective, where $\beta>0$ allows a consistent history throughout different epochs~\cite{Antoniou:2020nax}. We have seen in Section~\ref{Sec:nsregions} that, among the three different possible neutron star configurations we focus on, only the denser one leads to scalarized solutions for $\beta>0$. In Fig.~\ref{fig:50M204}, we show the mass difference $\delta M$ and rescaled scalar charge $\tilde{Q}$ as functions of $\alpha$ when $\beta=50$.
 Note that scalarized solutions with zero node exist over two disconnected ranges of $\alpha$ ($-44~\text{km}^2<\alpha<57~\text{km}^2$ and $174~\text{km}^2<\alpha<522~\text{km}^2$). In the gap, general relativity solutions are stable and no scalarized solutions exist. This is obvious from Fig.~\ref{fig:SLy204}, taking a cut along the vertical line $\beta=50$. 
    
 Over the first interval, $\alpha$ is rather small and the scalarization process is dominated by the negative Ricci scalar. For strictly vanishing $\alpha$, the scalarization phenomenon with $\beta>0$ has already been examined in \cite{Mendes:2014ufa,Palenzuela:2015ima,Mendes:2016fby}. Here, we find that, in the interval of small values of $\alpha$, the scalar charges of the $n=0$ solutions (as well as of the $n=1$ solutions) are very small. Typically, $\tilde{Q}/M \simeq 10^{-4}-10^{-5}$, compatible with Eq.~\eqref{eq:boundQ}. Hence, all solutions  with $\beta>0$ and rather small values of $\alpha$ are interesting observationally: they display either no scalarization effects for neutron stars (for $\beta\lesssim 11.51$) or very mild scalar charges (for $\beta\gtrsim 11.51$). At the same time, they allow for a consistent cosmological history; finally, together with positive values of $\alpha$, they will generically give rise to black hole scalarization, as studied in detail in Chapter~\ref{Chap:blackholes}. In this region of the parameter space, we can therefore hope to discover scalarization effects in the future gravitational wave signals of binary black holes, that are either absent or suppressed in the case of neutron stars.
    
 Over the second interval ($174~\text{km}^2<\alpha<522~\text{km}^2$), the contribution of the Gauss-Bonnet invariant tends to dominate, and the scalar charges are more significant, as one can immediately notice in Fig.~\ref{fig:SLy204}. Such setups are not compatible with Eq.~\eqref{eq:boundQ}, and therefore less interesting phenomenologically.

\subsection{Scalarized solutions along the instability lines}\label{Sub:instabilitylines}

\begin{figure}[h!]
	\subfloat{\includegraphics[width=0.7\linewidth]{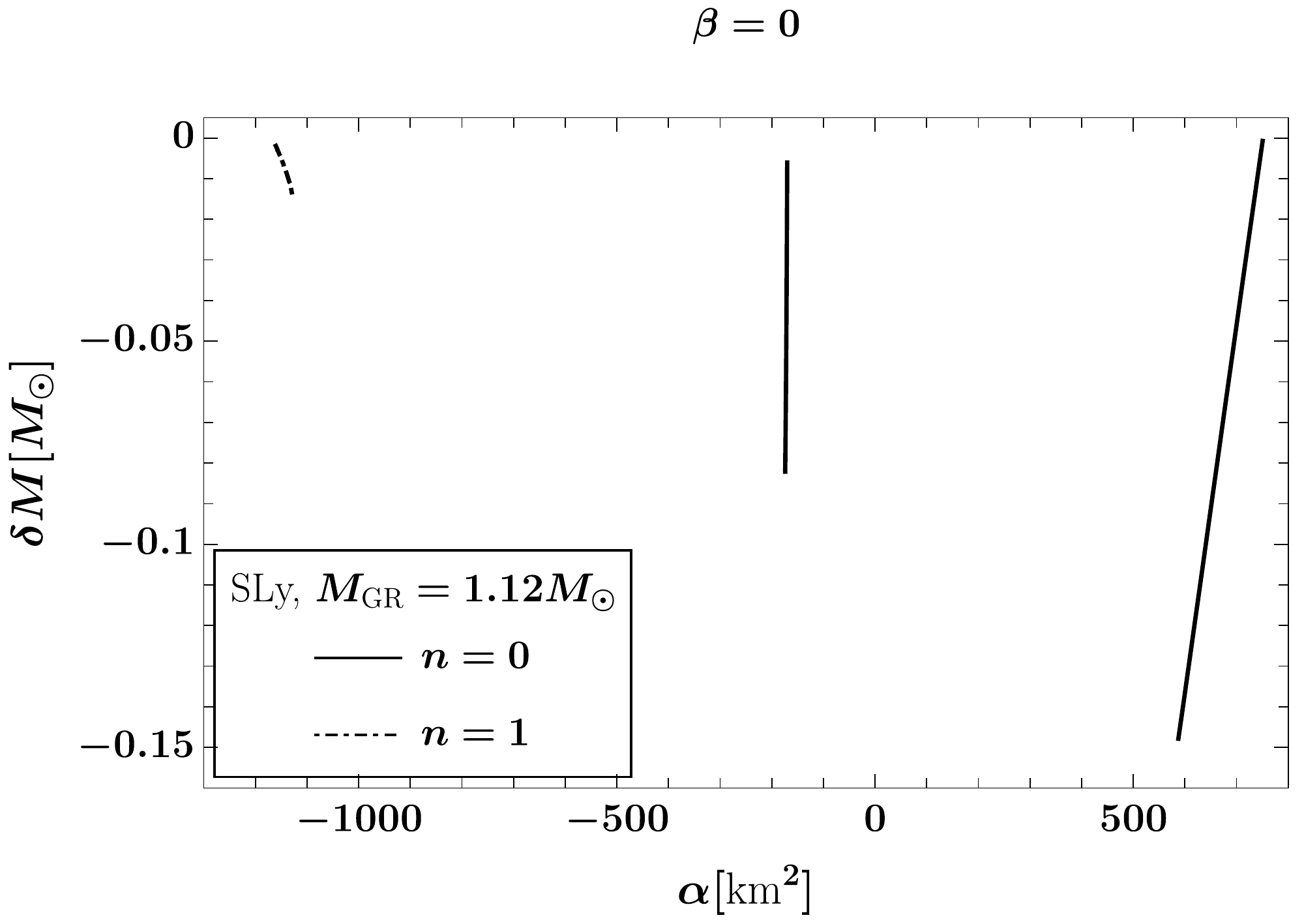}}
	\\
	\subfloat{\includegraphics[width=0.7\linewidth]{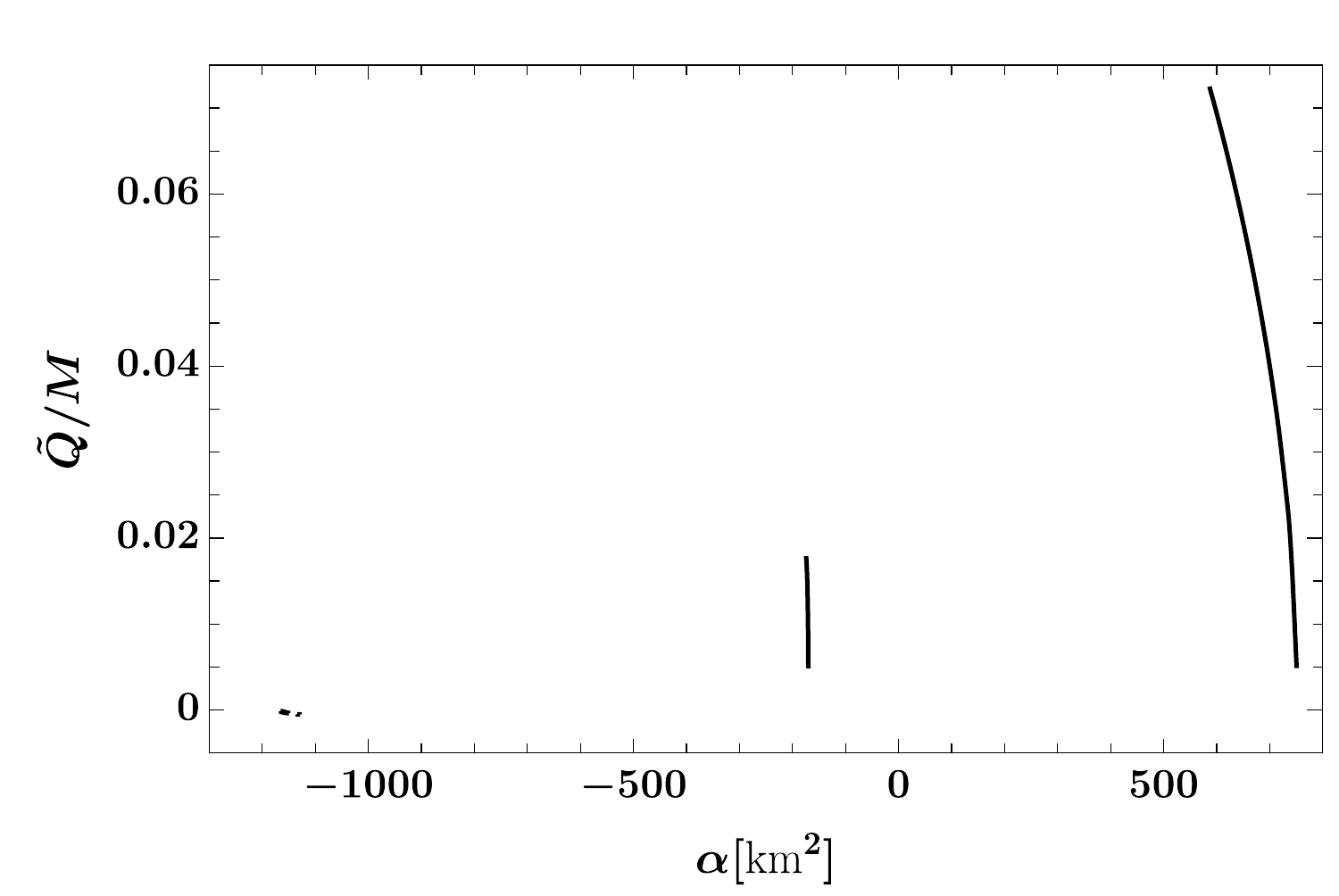}}      
	\caption[Mass difference and rescaled scalar charge of the scalarized solutions along the instability lines]{Mass difference and scalar charge of the scalarized solutions along the instability lines, for $\beta=0$. The scenario considered here corresponds to $\epsilon_0=8.1\times 10^{17}~\text{kg}/\text{m}^3$ ($M_\text{GR}=1.12M_\odot$) together with the SLy EOS. Solutions with zero node acquire a significant charge and mass difference, and are apparently disconnected from general relativity when they appear while increasing $\alpha$ towards positive values. Solutions with $n=1$ nodes are very close to general relativity, with a small charge and mass difference. Since they extend only over a small range of $\tilde{Q}$ and $\delta M$, they are difficult to spot. They lie at the upper left (respectively lower left) of the top (respectively bottom) panel.
	}
	\label{fig:beta0}
 \end{figure}

 As we mentioned at the end of Section~\ref{Sec:nsregions}, a generic feature that is not observable in Figs.~\ref{fig:Sly112}, \ref{fig:MPA1} and \ref{fig:SLy204}, is that scalarized solutions are present in a tiny band close to each instability line.
 Let us illustrate this with the light star model (with SLy EOS), that is the one which corresponds to Fig.~\ref{fig:Sly112}. For simplicity, we also restrict our study to solutions with $\beta=0$ (\textit{i.e.}, we take a cut along the vertical axis in Fig.~\ref{fig:Sly112}). The characteristics of the solutions are shown in Fig.~\ref{fig:beta0}.
 Scalarized solutions with zero nodes (the ones lying close to the $n=0$ instability line of the general relativity solution) have a characteristic mass difference and scalar charge which is not particularly small. It is of the same order as for the solutions we previously examined (Figs.~\ref{fig:smallCoup}--\ref{fig:50M204}). They also exhibit a surprising behaviour: when increasing $\alpha$ progressively from 0 towards positive values, the mass and scalar charge suddenly deviate from general relativity, instead of being smoothly connected; further increasing $\alpha$, $\delta M$ and $\tilde{Q}$ then tend to decrease. This behaviour is significantly different from what we could observe in Figs.~\ref{fig:smallCoup}--\ref{fig:50M204}.
 
 Solutions with more nodes ($n=1$, 2, 3...) exhibit a clear feature: they deviate very slightly from general relativity in terms of mass, and acquire only a small scalar charge (typically $\delta M < 10^{-2}$ and $\tilde{Q}/M < 10^{-4}$). We verified this behaviour for all higher nodes admitted; however, for simplicity, in Fig.~\ref{fig:beta0} we show only the case $n=1$. This feature can be understood as follows; close to some instability line (on the unstable side), an unstable mode of the effective potential associated with the general relativity solution has just appeared. A very small deformation of the potential can therefore easily restore the equilibrium. This deformation can be caused by the back-reaction of the scalar onto the metric: the instability is triggered, the scalar field starts growing, but it immediately back-reacts on the potential, making it shallower and suppressing the instability. Clearly, such a behaviour can only happen close to instability lines, where a specific mode is on the edge of stability.
 
 We conclude this Section with the outline of our results. We investigated in detail the physical characteristics of the scalarized solutions. In general, large parameters ($|\beta|\gg1$ or $|\alpha|\gg L^2$, where $L\simeq10$~km is the typical curvature scale) lead to scalar charges that would be in conflict with binary pulsar constraints. However, it is interesting to notice that solutions with $\beta>0$ and reasonably small $\alpha$ (typically $|\alpha|\lesssim 50$~km$^2$) lead either to stable general relativity configurations, or to scalarized stars with small charges. Remarkably, this is the region of the $(\alpha,\beta)$ parameter space for which general relativity is a cosmological attractor \cite{Antoniou:2020nax} and black holes scalarization can take place \cite{Antoniou:2021zoy}. Therefore, it is possible to construct scalarization models that are consistent with current observations, while still having interesting strong field phenomenology.

 We have also discovered that scalarized solutions systematically exist near the thresholds that delimit the stability of the general relativity solutions, and provided a putative explanation for this.

\section[\texorpdfstring{Predicting the scalar profile of scalarized stars from general relativity solutions}{Predicting the scalar profile of scalarized stars from general relativity solutions}
]{\sectionmark{Predicting the scalar profile of scalarized stars} Predicting the scalar profile of scalarized stars from general relativity solutions}
\sectionmark{Predicting the scalar profile of scalarized stars}
\label{Sec:scalarprofile}

 We conclude this Chapter by arguing that, already at the perturbative level of the general relativity solution, we can identify an influence on the profile of the scalar field in the fully scalarized solution. To this end, let us focus on the effective mass given in Eq.~\eqref{eq:eff_mass}, that is $ m_\text{eff}^2=\beta R/2-\alpha \mathscr{G}$. This is a radially dependent quantity, and the scalar field is most likely to grow at radii where $m_\text{eff}^2$ is most negative. In particular, it is natural to expect that, if $m_\text{eff}^2$ has a minimum at $r=0$, this will favor a monotonic profile for the scalar field, and hence an $n=0$ type of solution. On the contrary, if $m_\text{eff}^2$ has a minimum at $r>0$, this favors a peaked profile for the scalar field, which is more common in $n\geq1$ solutions. Let us illustrate this with a concrete example. We will
 consider the scenario that corresponds to $M_\text{GR}=1.12 M_\odot$, together with the SLy EOS, and two choices of $\beta$: $\beta=-10$ and $\beta=-100$. In the first case, only solutions with 0 node exist; in the second case, we can construct solutions with 0 or 1 node.

\begin{figure}[h!]
	\subfloat{\includegraphics[width=0.6\linewidth]{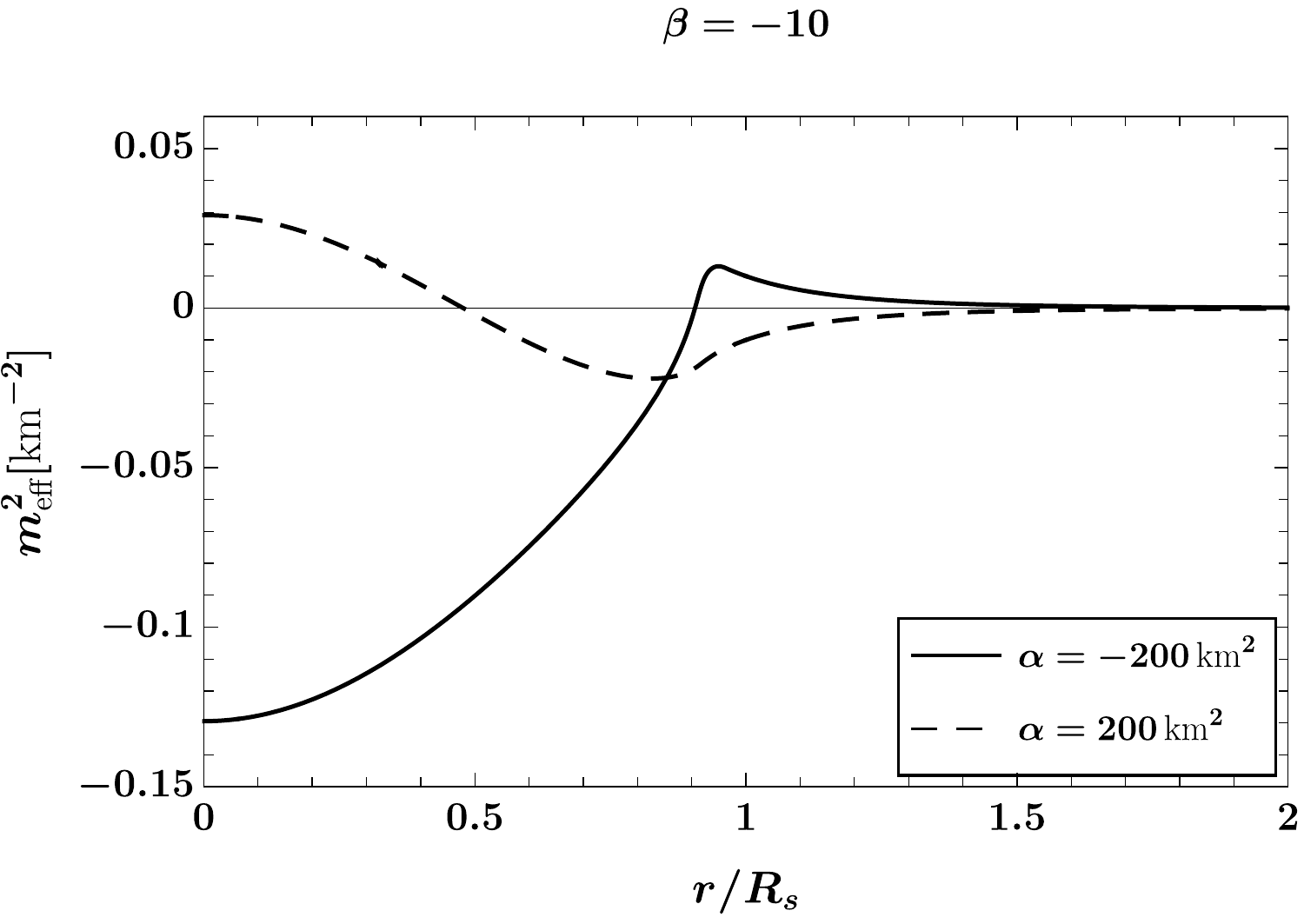}}
	\\%
	\subfloat{\includegraphics[width=0.6\linewidth]{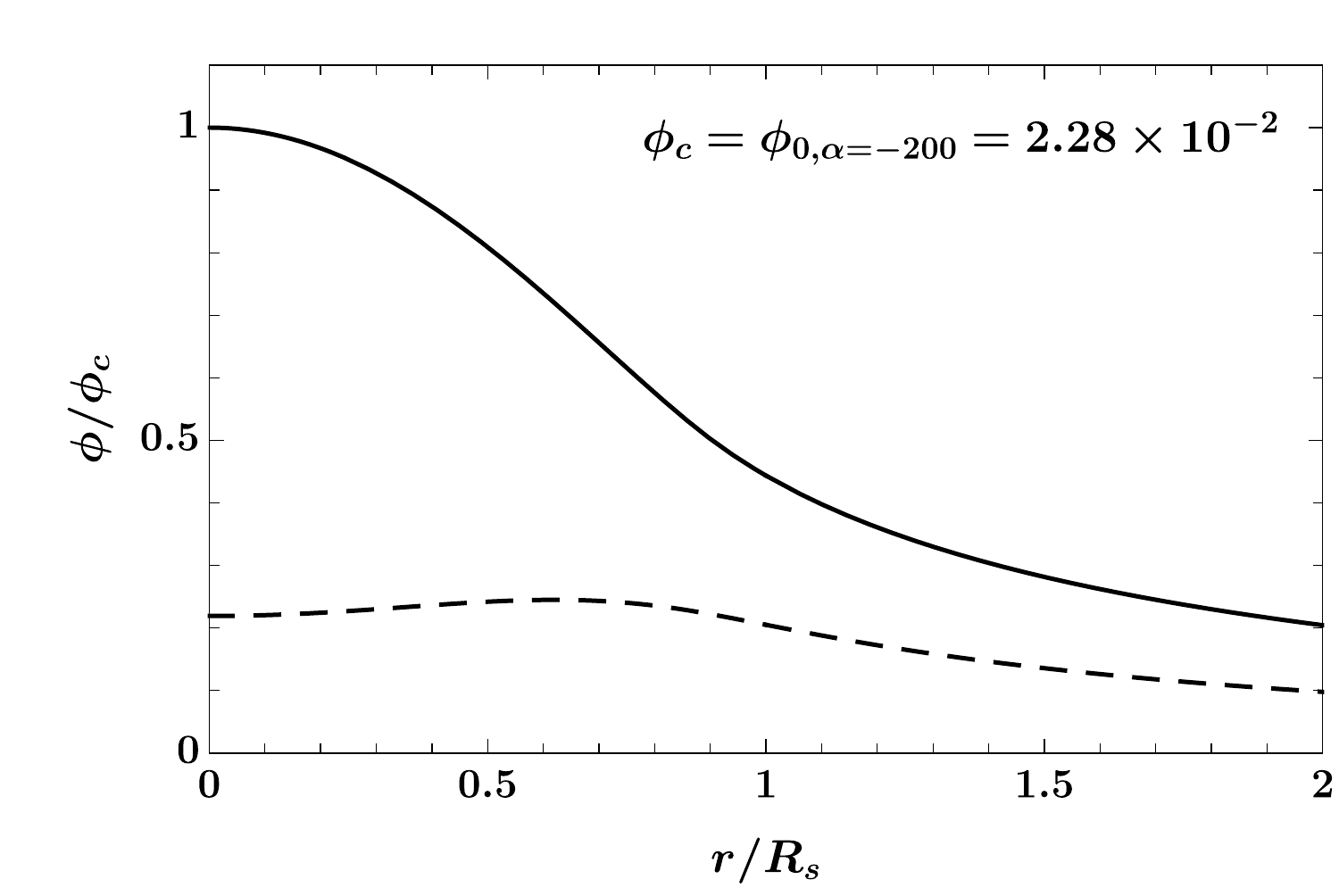}}
	\caption[Scalar field profile and effective mass squared for small negative value of Ricci coupling]{Upper panel: radial profile of the effective mass squared over the general relativity background, using the SLy EOS and a central density $\epsilon_0=8.1\times 10^{17}\,\text{kg}/\text{m}^3$ (yielding $M_{\text{GR}}=1.12 M_{\odot}$), for $\beta=-10$ and $\alpha=\pm200$~km$^2$; Lower panel: radial profile of the scalar field, this time in the fully scalarized solution with the same EOS, central density, and Lagrangian parameters. The radial coordinate is normalized by $R_\text{s}$, the radius of the star surface. In the lower panel, the scalar field is normalized to its central value for $\alpha=-200\, \text{km}^2$. When the minimum of $m_\text{eff}^2$ is shifted to $r>0$, so is the peak of $\phi$.}
	\label{fig:EffMass2}
 \end{figure}

 We first focus on the case $\beta=-10$. The Ricci scalar is everywhere positive over the background we consider, with a maximum at $r=0$; hence, $\beta R$ contributes negatively to the squared mass, favouring the growth of the scalar field close to the center. The Gauss-Bonnet scalar, on the other hand, is negative in the central region of the star, and becomes positive towards the surface. Therefore, $-\alpha\mathscr{G}$ reinforces the effect of $\beta R$ if $\alpha<0$, while counterbalancing it if $\alpha>0$. This is illustrated in the top panel of Fig.~\ref{fig:EffMass2}.
 The bottom panel shows the scalar profile of the fully scalarized solutions associated with the same parameters. In this range of parameters, only solutions with 0 node are allowed (as one can check in Fig.~\ref{fig:Sly112}); hence, pushing the minimum of $m_\text{eff}^2$ away from the center cannot favour $n=1$ solutions, which do not exist. Still, we notice that positive $\alpha$ values, which have the effect of displacing the minimum of $m_\text{eff}^2$ to $r>0$, also displace the peak of the scalar field to $r>0$. The peak of the scalar field is located approximately at the minimum of $m_\text{eff}^2$. Again, one must be careful in the comparison of the two panels, as one of them corresponds to a general relativity star while the other one corresponds to a scalarized star. However, our analysis seems to capture what happens during the transition from the general relativity to the scalarized branch.

\begin{figure}[h!]
	\subfloat{\includegraphics[width=0.6\linewidth]{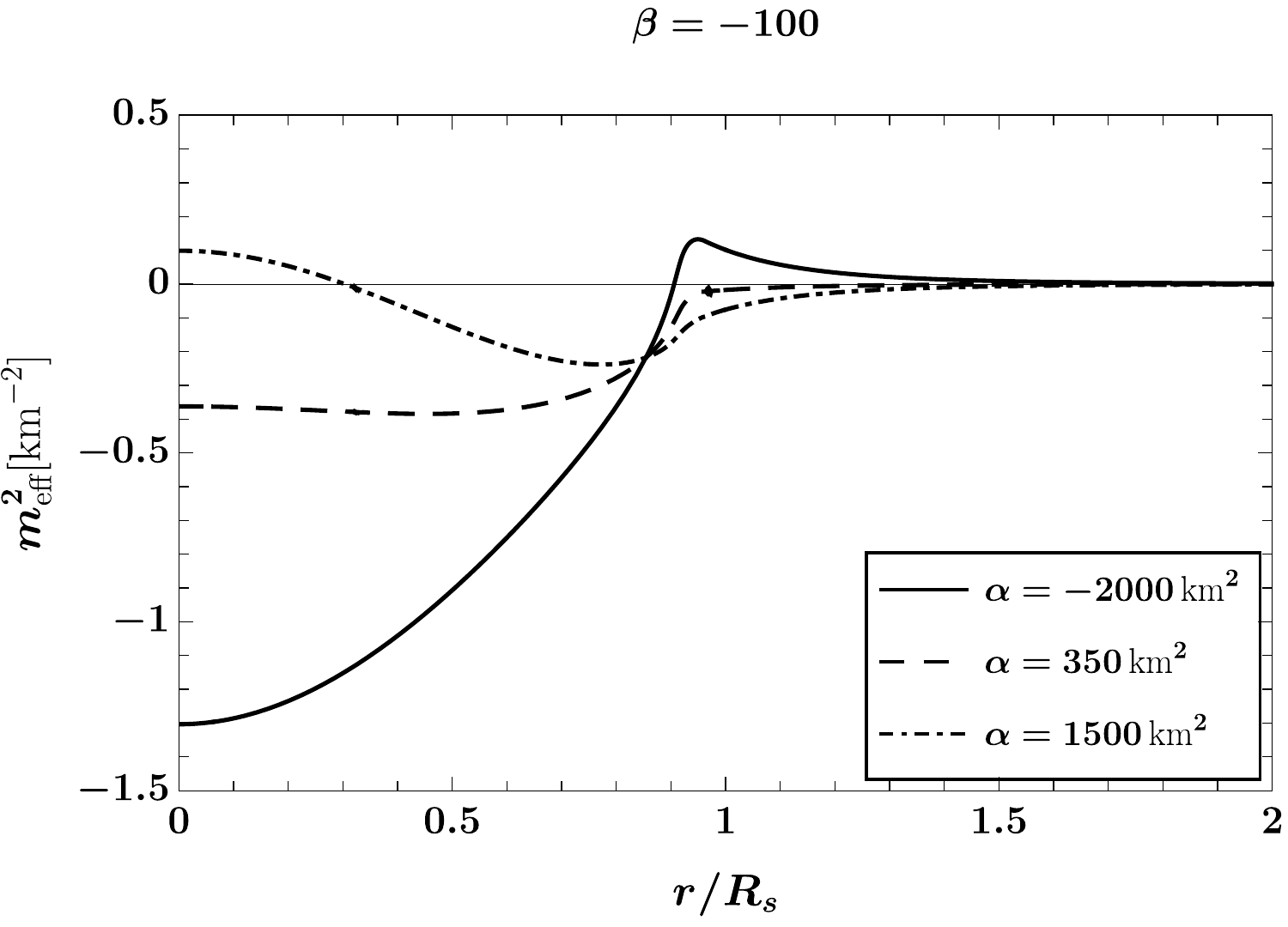}}
		\\%
	\subfloat{\includegraphics[width=0.6\linewidth]{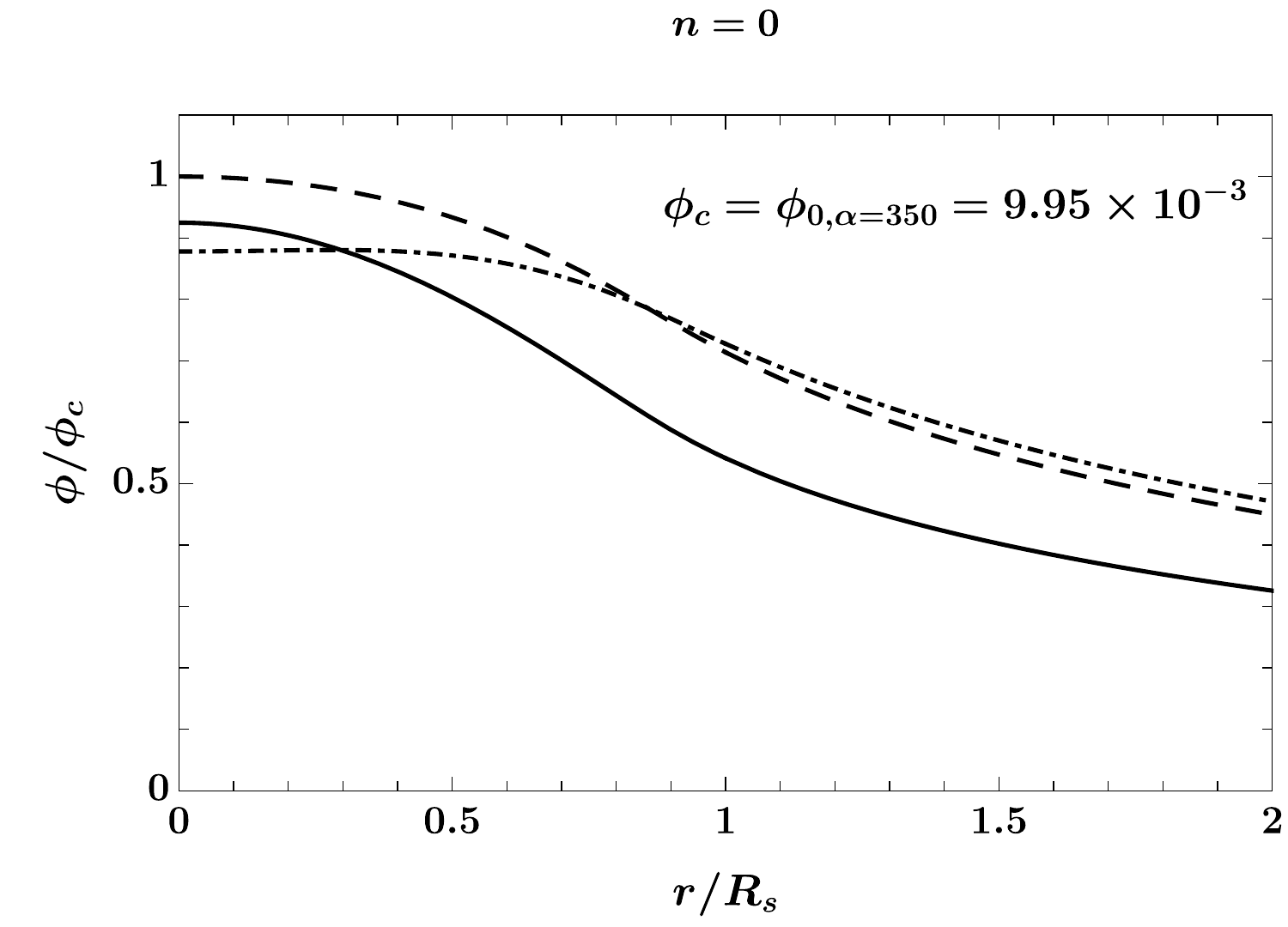}}
	    \\%
	\subfloat{\includegraphics[width=0.6\linewidth]{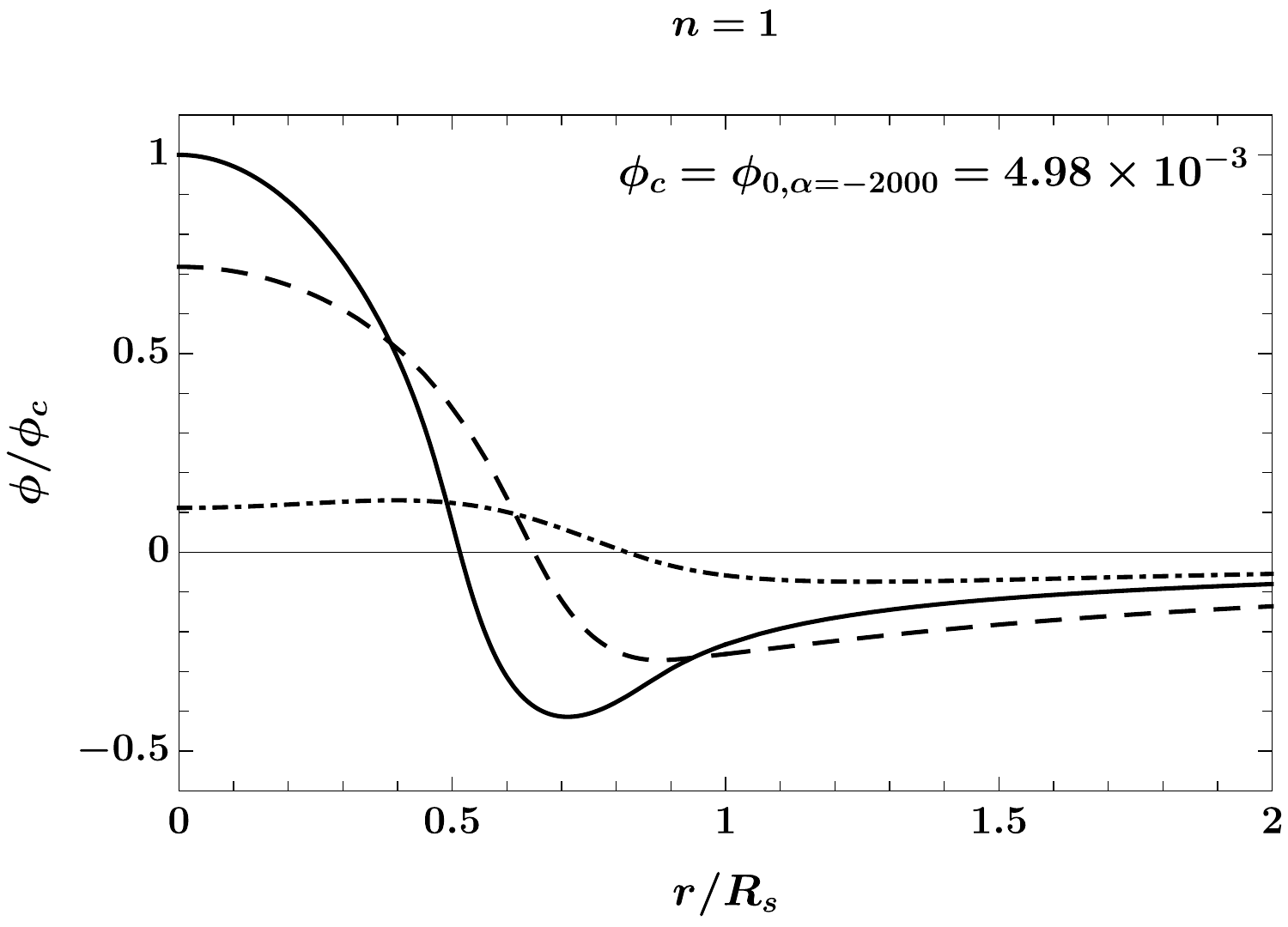}}
	\caption[Scalar field profile and effective mass squared for large negative value of Ricci coupling]{Upper panel: radial profile of the effective mass squared over the general relativity background, using the SLy EOS and a central density $\epsilon_0=8.1\times 10^{17}\,\text{kg}/\text{m}^3$ (yielding $M_{\text{GR}}=1.12 M_{\odot}$), for $\beta=-100$ and $\alpha=-200$, 350 or 1500~km$^2$; Center (respectively lower) panel: radial profile of the scalar field solution with 0 (respectively 1) node in the fully scalarized solution with the same EOS, central density, and Lagrangian parameters. The normalization is similar to the one of Fig.~\ref{fig:EffMass2}. When increasing $\alpha$, the minimum of $m_\text{eff}^2$ is progressively shifted from $r=0$ to a finite radius, alternatively favoring the growth of $n=0$ and $n=1$ solutions.}
	\label{fig:EffMass3}
 \end{figure}

 To illustrate better the transition between $n=0$ and $n=1$ solutions, let us now consider the case $\beta=-100$. 
 The qualitative discussion about the effect of $\beta R$ and $-\alpha\mathscr{G}$ over the effective mass is exactly the same as in the previous case. We will therefore consider again a large negative and a large positive value of $\alpha$, as well as an intermediate one: $\alpha=-2000, \,350$ and 1500~km$^2$. Note that the intermediate value corresponds to $\alpha_\text{c}$ in Section~\ref{Sub:negativebeta}, the critical value at which scalarized stars with $n=0$ node become more massive (and hence probably less stable) than those with $n=1$ node. We show the results in Fig.~\ref{fig:EffMass3}.
 The top panel shows the profile of the effective scalar mass. It behaves exactly as in the case $\beta=-10$, with a minimum at $r=0$ for negative values of $\alpha$, which is progressively shifted to larger radii when we increase $\alpha$. For the parameters we chose, this time, both solutions with zero and one node exist. In the center (respectively bottom)  panel of Fig.~\ref{fig:EffMass3}, we show the $n=0$ (respectively $n=1$) solutions. In Section~\ref{Sub:negativebeta}, we stated that for $\alpha<\alpha_c$ we expected that the zero node solution will be energetically preferred over the one node solution, and vice-versa for $\alpha>\alpha_c$. The profiles of the effective mass squared give a complementary argument that strengthens this expectation. Indeed, for $\alpha=-2000\,\text{km}^2\ll\alpha_c$ the shape of $m_\text{eff}^2$ favours a scalar solution with a maximum at the center of the star, which decays monotonically with $r$, \textit{i.e.} a $n=0$ solution. For $\alpha=1500\,\text{km}^2\gg\alpha_c$, the tachyonic instability is still triggered inside the star, but away from the center. Thus, we expect that a solution with one node will be favoured. The transition between a minimum at $r=0$ and $r>0$ indeed seems to occur around $\alpha_\text{c}$.
 
 We have shown that the profile of the effective mass at the general relativity level can foster the growth of certain modes with respect to others. We focused on the two cases $\beta=-10,-100$. Even though, the former case only admits zero node solutions, one can still relate the peak of the scalar field profile to the minimum of the effective mass square. In the case $\beta=-100$, the profile of $m_\text{eff}^2$ seems to favour the one node solutions for values of $\alpha$ above a critical threshold $\alpha_c$. This result appears to confirm our analysis in Section~\ref{Sec:nsproperties}. We stress that one must be careful in comparing the profiles of the scalar field and the effective mass square, since the latter corresponds to a general relativity star. Nevertheless, this simplistic analysis seems to capture the transition between the general relativity to the scalarized branch.


\chapter{Conclusions}\label{Chap:conclusion}

\section{Summary}

 The new era of gravitational-wave observations allows us to explore the gravitational interaction at the strongest regime ever probed so far. Hence, there is hope that new physics can be detected. Testing gravity in this new limit could give some answers to the big open questions in gravitational and particle physics, such as quantum gravity, the dark energy or the dark matter problem.
 
 Searching for new physics means looking for new fundamental fields. Scalar fields are particularly important since they are ubiquitous in both extensions of the Standard Model and in alternative theories of gravity. Moreover, they are the simplest fields to work with. The work of this thesis is solely focused on generalized scalar-tensor theories. Thus, in Chapter~\ref{Chap:sttheories}, we presented a review of scalar-tensor theories and their generalization to the Horndeski class.
 
 Scalar fields can act as basic probes for alternative theories of gravity. Hence, a lot of attention has been drawn to understand if these fields leave an imprint on compact objects. However, in all tests performed so far in the weak-field regime, scalars have not been detected. Thus, if we indeed expect them to leave an imprint in strong gravity phenomenology, we need to be able to explain why these fields do not manifest at regimes we have tested so far. The screening mechanism is a useful tool able to explain such behavior. In this thesis we focused on the specific phenomenon of spontaneous scalarization. It consists of a mechanism that triggers a phase transition from general relativity solutions to scalarized ones. As a consequence, the scalar field is in fact screened far away from compact objects, whereas it can leave imprints around black holes and inside or close to the surface of neutron stars.

 Such mechanism was first formulated by Damour and Esposito-Far\`{e}se, in the context of scalar-tensor theories. This model, however, is covered by a no hair theorem, thus only neutron stars are able to scalarize, whereas black holes are still described by general relativity solutions. It was recently shown that another theory, scalar Gauss-Bonnet gravity exhibits the same mechanism for both neutron stars and black holes. In Chapter~\ref{Chap:spontaneousscalarization}, we first provided a description of this mechanism as a tachyonic instability, at linear level, which is later quenched by nonlinearities included in the system. We then reviewed spontaneous scalarization in the literature. 
 
 The recent formulation of spontaneous scalarization in scalar Gauss-Bonnet gravity has revealed that the Damour and Esposito-Far\`{e}se model is not uniquely affected by this phenomenon. It is then interesting to investigate if there are other theories that present the same mechanism. In the work of this thesis, we address this issue, focusing on the study of spontaneous scalarization in the context of Horndeski theory.
 
 First, we investigated the possibility of identifying classes of theories within the Horndeski action that exhibit spontaneous scalarization triggered by a tachyonic instability. In Chapter~\ref{Chap:scalarizationHordneski}, we first determined the conditions that need to be satisfied so that solutions of general relativity are admissible. We probed whether or not there will be a tachyonic instability by calculating the effective mass of scalar perturbation on a fixed spacetime background that is a solution of Einstein’s equations. Though this approximation neglects backreaction, it is adequate for the purpose of simply identifying scalarization models. We were able to determine a minimal action that contains all of the terms that contribute to the effective mass at linearized level. Such action contains four distinct terms that contribute to scalarization. Through suitable field redefinitions, one of them can be directly linked to the known Damour and Esposito-Far\`{e}se model and another to the scalar Gauss-Bonnet scalarization models. The third term can be thought of as a disformal coupling to matter and relates to a model studied in Ref.~\cite{Minamitsuji:2016hkk}. The fourth term comes from a potential for a scalar and, although it cannot trigger spontaneous scalarization on its own, it affects the onset of the tachyonic instability in all other models. One can start from our minimal action, supplement it with terms that contribute only nonlinearly to the scalar equation, and construct scalarization models. The onset of the tachyonic instability that will kickstart scalarization will be determined by the minimal action, while the end state depends on the choice of the extra terms that contribute nonlinearly. This is because scalarization is triggered by a linear tachyonic instability and later quenched by nonlinear effects.

 We then proceed to investigate exhaustively the effect of all these terms in Chapter~\ref{Chap:Threshold}.
 We first identified the role of each term separately, and then studied their combined effects. In the former case, our results agree with previous results regarding scalarization thresholds. In the latter, more general case, we have found that a very small bare mass suffices to stabilize general relativity solutions, and that the scalarization thresholds are only mildly sensitive to the choice of equation of state.
 
 We were also able to explore, for the first time, the multi-dimensional parameter space and provide scalarization thresholds that depend on more than one coupling. 
 Our analysis revealed a particular striking features: the role of the effective metric in which scalar perturbations propagate, which is controlled by a single coupling constant. There exists a threshold beyond which the effective metric loses hyperbolicity. This threshold can be interpreted as an absolute bound on the coupling parameter on which the effective metric depends, in the framework of tachyonic scalarization. As a consequence, the value of this coupling constant is restricted to a rather narrow range and thus it has a very limited effect on the threshold of tachyonic scalarization. We stress that the loss of hyperbolicity could alternatively be seen as a source of instability leading to scalarization, in line with what was proposed in Ref.~\cite{Ramazanoglu:2017yun} in a more restricted setup. It would be interesting to understand if such an instability can indeed be controlled and give rise to a sensible scalarization process. 
 
 The analysis done in Chapter~\ref{Chap:scalarizationHordneski} and~\ref{Chap:Threshold} were performed assuming a general relativity background. In the following two Chapters, we then proceed to solve the full system of field equations, thus including the backreaction of the metric on the scalar field equation, for a theory that included both the Ricci scalar and Gauss-Bonnet couplings.
 
 We first focused on the case of static spherically symmetric black holes in Chapter~\ref{Chap:blackholes}. 
 The coupling with the Ricci scalar is known not to affect the threshold of scalarization. Nonetheless, we showed that such term can in fact influence the domain of existence of scalarized black holes, significantly modify their properties, and control their scalar charge. Our analysis appears to indicate that the strength of this coupling potentially has an impact on the stability of scalarized black holes. In particular, we expect that having a coupling parameter larger than some critical value can resolve the stability problems for models that do not include such coupling. We mainly focused on positive values for the Ricci coupling, since this choice makes general relativity a cosmological attractor, allowing to have a consistent cosmological history from the end of the inflationary era as shown in Ref.~\cite{Antoniou:2020nax}. Moreover, the analysis done in Chapter~\ref{Chap:Threshold} shows that, for positive values of the Ricci coupling and reasonably small value of the Gauss-Bonnet coupling, neutron stars do not scalarize, allowing us to evade tight binary pulsar constraints ({\em e.g.}~\cite{Freire:2012mg,Antoniadis:2013pzd,Shao:2017gwu}).
 
 Lastly, in Chapter~\ref{Chap:neutronstars}, we focused on the case of neutron stars. We considered three different stellar scenarios which correspond to different central energy densities and EOS. The regions in the parameter space where scalarized solutions exist are smaller than those where the general relativity branch is affected by a tachyonic instability. One of the main result of our work is that the complementary regions, where general relativity solutions are unstable while no scalarized solutions exists, must be excluded. We then proceed to investigate the physical characteristics of the scalarized solutions. Generally, large values for both the Ricci and Gauss-Bonnet couplings lead to scalar charges in conflict with binary pulsar constraints. At the same time, solutions with positive Ricci coupling and small Gauss-Bonnet coupling parameter either lead to stable general relativity configurations or scalarized stars with small scalar charges. Notably, this region coincides with that where black hole scalarization can take place. As a consequence, one can construct models consistent with current observations, while having interesting strong field phenomenology. It is worth noting that future gravitational-wave observations, such as for instance the observations of extreme mass ratio inspirals by LISA~\cite{Maselli:2020zgv, Maselli:2021men}, will reach the precision to measure small scalar charges for neutron stars and black holes. We have also found that scalarized solutions systematically exist near the thresholds that delimit the stability of the general relativity solutions, providing a putative explanation for this behavior. Finally, we have shown that the profile of the effective mass at the general relativity level can foster the growth of certain modes with respect to others.


\section{Future perspectives}

 The work of this thesis was aimed at developing a theoretical understanding of spontaneous scalarization in generalized scalar-tensor theories.
 A natural next step would be to further connect these results to observations.

 For example, gravitational-wave observation of binaries that contain black holes would still be able to measure or constrain the Ricci and Gauss-Bonnet coupling parameters. A detailed post-Newtonian analysis of the inspiral phase would be sufficient to provide some first constraints. In this regard, some preliminary work has been done in Refs.~\cite{Julie:2019sab,Julie:2022huo} for the study of black hole sensitivities in the context of scalar Gauss-Bonnet gravity. One could also extend the results retrieved in Ref.~\cite{Perkins:2021mhb}, where they obtained gravitational-wave constraints on the coupling parameter of Einstein-dilaton-Gauss-Bonnet gravity, to scalarization models. At the same time, scalarization theories in which the scalar charge is non-zero only below a mass threshold are also expected to be severely constrained by extreme mass ratio inspirals observations by LISA: the supermassive black hole would be described by the Kerr metric, whereas the small black hole can carry a scalar charge. This is the ideal scenario to apply the considerations of Ref.~\cite{Maselli:2020zgv}.
 
 Furthermore, when studying black holes scalarization we only considered positive values for the Gauss-Bonnet coupling, as this is a requirement for having scalarized black holes under the assumptions of staticity and spherical symmetry. However, it has been shown in Ref.~\cite{Dima:2020yac} that, for negative values of this coupling, in the absence of a Ricci coupling, scalarization can be triggered by rapid rotation. Indeed, some scalarized black holes have been found in this scenario in Refs.~\cite{Herdeiro:2020wei,Berti:2020kgk}. It would thus be very interesting to consider the effect of the Ricci scalar coupling in this scenario of black holes scalarization induced by rotation. At the same time, the effect of rotation on neutron star scalarization was investigated in the framework of the Damour and Esposito-Far\`{e}se model in Ref.~\cite{Doneva:2013qva}. It would be worth exploring an extension of this analysis to coupled Ricci and Gauss-Bonnet couplings, or pure Gauss-Bonnet ones.

 It will also be interesting to combine the bounds coming from neutron star and black hole observations with the theoretical constraints that relate to the requirement that scalarization models have a well-posed initial value problem~\cite{Ripley:2020vpk}. So far, the combined theory with both Ricci and Gauss-Bonnet couplings has not been studied in detail from the initial value problem perspective.
 Finally, another possible aspect worth investigating would be the stability analysis of the scalarized solutions, both the neutron stars and the black holes.
 
 Lastly, there are other mechanisms, not addressed in this thesis, that can trigger a similar phase transition to that of spontaneous scalarization. It is the case of \textit{dynamical scalarization}, for instance, where binaries can undergo a process of scalarization during the evolution of the systems, even when initially none of the two companions carried any scalar charge. This mechanisms was first discovered for neutron stars binaries in the Damour and Esposito-Far\`{e}se model~\cite{Barausse:2012da,Palenzuela:2013hsa}, and recently studied in the case of binary black holes in the context of scalar Gauss-Bonnet gravity~\cite{Silva:2020omi}. At the same time, it is also possible to extend the concepts of scalarization to different field contents, as it was shown in Refs. \cite{Herdeiro:2018wub,Ramazanoglu:2017xbl,Ramazanoglu:2018hwk}.
 
 For all these reasons, there are still several directions worth pursuing in the broader framework of scalarization. On one hand, future observations will provide better insights on this phenomenon and tighter constraints on the theories that are affected by it. On the other, there are still open theoretical challenges that are waiting to be explored.

\appendix
\appendixpage
\noappendicestocpagenum
\addappheadtotoc
\renewcommand{\thechapter}{\Alph{chapter}}


\chapter{Conformal transformation rules}\label{App:conformal}

 In this Appendix we provide some useful rules of conformal transformation for the most used geometric quantities as well as for the matter stress-energy tensor. For a more complete review on conformal transformation see Ref.~\cite{Dabrowski:2008kx}. We will work in the general case of $D$ dimensions, and the case $D=4$ can then be easily retrieved. Given the conformal transformation
 \begin{equation}
 \tilde{g}_{\mu\nu}=\Omega^2 g_{\mu\nu},
 \end{equation}
 it follows that the determinant of the metric and the inverse of the metric respectively transform as
 \begin{subequations}
 \begin{equation}
 \sqrt{-\tilde{g}}=\Omega^D\sqrt{-g},
 \end{equation}
 \begin{equation}\label{eq:contrMetr}
 \tilde{g}^{\mu\nu}=\Omega^{-2}g^{\mu\nu}.
 \end{equation}
 \end{subequations}
 Using these rules to transform the Christoffel symbols yields
 \begin{subequations}
 \begin{equation}\label{eq:ChristTilde}
 \tilde{\Gamma}^{\rho}_{\mu\nu}=\Gamma^{\rho}_{\mu\nu}+\frac{1}{\Omega}(\delta^\rho_\mu\,\Omega_{,\nu}+\delta^\rho_\nu\,\Omega_{,\mu}-g_{\mu\nu}g^{\rho\sigma}\Omega_{,\sigma})\, , \qquad \tilde{\Gamma}^{\rho}_{\mu\rho}=\Gamma^{\rho}_{\mu\rho}+D\frac{\Omega_{,\mu}}{\Omega}
 \end{equation}
 \begin{equation}\label{eq:Christ}
 \Gamma^{\rho}_{\mu\nu}=\tilde{\Gamma}^{\rho}_{\mu\nu}-\frac{1}{\Omega}(\delta^\rho_\mu\,\Omega_{,\nu}+\delta^\rho_\nu\,\Omega_{,\mu}-\tilde{g}_{\mu\nu}\tilde{g}^{\rho\sigma}\Omega_{,\sigma})\, , \qquad \Gamma^{\rho}_{\mu\rho}=\tilde{\Gamma}^{\rho}_{\mu\rho}-D\frac{\Omega_{,\mu}}{\Omega}.
 \end{equation}
 \end{subequations}
 Eqs.~\eqref{eq:ChristTilde}-\eqref{eq:Christ} can in turn be used to write the transformation rule for the Riemann tensor
 \begin{subequations}
 \begin{equation}
 \begin{split}
 \tensor{\tilde{R}}{^\rho_\mu_\nu_\sigma} & =\tensor{R}{^\rho_\mu_\nu_\sigma}+\frac{2}{\Omega^2}(\delta^{\rho}_{\nu}\Omega_{,\mu}\Omega_{,\sigma}-\delta^{\rho}_{\sigma}\Omega_{,\mu}\Omega_{\nu}+g_{\mu\sigma}\Omega^{,\rho}\Omega_{,\nu}-g_{\mu\nu}\Omega^{,\rho}\Omega_{,\sigma})\\
    & +\frac{1}{\Omega}(\delta^{\rho}_{\sigma}\Omega_{;\mu\nu}-\delta^{\rho}_{\nu}\Omega_{;\mu\sigma}+g_{\mu\nu}\tensor{\Omega}{^{;\rho}_{;\sigma}}-g_{\mu\sigma}\tensor{\Omega}{^{;\rho}_{;\nu}})\\
     &+\frac{1}{\Omega^2}(\delta^{\rho}_{\sigma}g_{\mu\nu}-\delta^{\rho}_{\nu}g_{\mu\sigma})\Omega^{\kappa}\Omega_{\kappa},
 \end{split}
 \end{equation} 
 \begin{equation}
 \begin{split}
 \tensor{R}{^\rho_\mu_\nu_\sigma} & =\tensor{\tilde{R}}{^\rho_\mu_\nu_\sigma}-\frac{1}{\Omega}(\tilde{\delta}^{\rho}_{\sigma}\Omega_{\tilde{;}\mu\nu}-\tilde{\delta}^{\rho}_{\nu}\Omega_{\tilde{;}\mu\sigma}+\tilde{g}_{\mu\nu}\tensor{\Omega}{^{\tilde{;}\rho}_{\tilde{;}\sigma}}-\tilde{g}_{\mu\sigma}\tensor{\Omega}{^{\tilde{;}\rho}_{\tilde{;}\nu}})\\
                        & +\frac{1}{\Omega^2}(\tilde{\delta}^{\rho}_{\sigma}\tilde{g}_{\mu\nu}-\tilde{\delta}^{\rho}_{\nu}\tilde{g}_{\mu\sigma})\tilde{g}_{\kappa\lambda}\Omega^{\kappa}\Omega^{\lambda\kappa},
 \end{split}
 \end{equation}
 \end{subequations}
 where $\tilde{;}$ refers to the covariant derivative in the frame described by the metric $\tilde{g}_{\mu\nu}$. Given the rules for the Riemann tensor, it is easy to show that the following transformations hold 
 \begin{subequations}
 \begin{equation}
 \begin{split}
 \tilde{R}_{\mu\nu}&=R_{\mu\nu}+\frac{1}{\Omega^2}[2(D-2)\Omega_{,\mu}\Omega_{,\nu}-(D-3)g_{\mu\nu}\Omega_{\rho}\Omega^\rho]\\
     &-\frac{1}{\Omega}[(D-2)\Omega_{;\mu\nu}+g_{\mu\nu}\Box\Omega],
 \end{split}
 \end{equation}
 \begin{equation}
 R_{\mu\nu}=\tilde{R}_{\mu\nu}-\frac{1}{\Omega^2}(D-1)\tilde{g}_{\mu\nu}\Omega_{\rho}\Omega^\rho+\frac{1}{\Omega}[(D-2)\Omega_{\tilde{;}\mu\nu}+\tilde{g}_{\mu\nu}\tilde{\Box}\Omega],
 \end{equation}
 \begin{equation}
 \tilde{R}=\Omega^{-2}\left[ R-2(D-1)\frac{\Box\Omega}{\Omega}-(D-1)(D-4)g^{\mu\nu}\frac{\Omega_{,\mu}\Omega_{,\nu}}{\Omega^2} \right],
 \end{equation}
 \begin{equation}
 R=\Omega^2\left[ \tilde{R}+2(D-1)\frac{\tilde{\Box}\Omega}{\Omega}-D(D-1)\tilde{g}^{\mu\nu}\frac{\Omega_{,\mu}\Omega_{,\nu}}{\Omega^2} \right].
 \end{equation}
 One can further employ Eqs.~\eqref{eq:contrMetr},~\eqref{eq:ChristTilde} and~\eqref{eq:Christ} to recover the transformation rule for the d'Alembertian operator. The result is
 \begin{equation}
 \tilde{\Box}\phi=\Omega^{-2} \left[ \Box\phi+(D-2)g^{\mu\nu}\frac{\Omega_{,\mu}\phi_{,\nu}}{\Omega} \right],
 \end{equation}
 \begin{equation}
 \Box\phi=\Omega^{2} \left[ \tilde{\Box}\phi-(D-2)\tilde{g}^{\mu\nu}\frac{\Omega_{,\mu}\phi_{,\nu}}{\Omega} \right],
 \end{equation}
 \end{subequations}
 where, as for the covariant derivative, $\tilde{\Box}$ is the d'Alembertian in the frame with the metric $\tilde{g}_{\mu\nu}$. 

 Lastly, let us focus on how conformal transformations affect the matter stress-energy tensor. In order to do so, we first consider the matter action
 \begin{equation}
 \tilde{S}_M=\int d^Dx \sqrt{-\tilde{g}}\tilde{\mathcal{L}}_M=\int d^Dx\sqrt{-g}\mathcal{L}_M=S_M,
 \end{equation}
 where the Lagrangians in the two conformal frames are connected by the following equation
 \begin{equation}
 \tilde{\mathcal{L}}_M=\Omega^{-D}\mathcal{L}_M.
 \end{equation}
 Deriving the stress-energy tensor in the usual way, for the frame described by the metric $\tilde{g}_{\mu\nu}$, we find
 \begin{equation}
 \tilde{T}_{\mu\nu}=\frac{-2}{\sqrt{-\tilde{g}}}\frac{\delta}{\delta \tilde{g}^{\mu\nu}}(\sqrt{-\tilde{g}}\,\tilde{\mathcal{L}}_M)=\Omega^{-D}\frac{-2}{\sqrt{-g}}\frac{\partial g^{\rho\sigma}}{\partial \tilde{g}^{\mu\nu}}\frac{\delta}{\delta g^{\rho\sigma}}(\sqrt{-g}\,\mathcal{L}_M).
 \end{equation}
 We are thus able to define the relations between the stress-energy tensors in the two frames
 \begin{equation}
 \begin{split}
 &\tilde{T}_{\mu\nu}=\Omega^{-D+2}T_{\mu\nu},\quad\tilde{T}^{\mu\nu}=\Omega^{-D-2}T^{\mu\nu},\\
 &\tensor{\tilde{T}}{^{\mu}_{\nu}}=\Omega^{-D}\tensor{T}{^{\mu}_{\nu}},\quad\,\,\,\tilde{T}=\Omega^{-D}T.
 \end{split}
 \end{equation}

\chapter{Horndeski equations of motion}\label{App:Horndeski}
 We give here the explicit expressions for the terms in the field equations presented in Section~\ref{Sec:Horndeski}. Throughout this Appendix we use the notation $\phi_\mu\equiv \nabla_\mu \phi$ and $\phi_{\mu\nu}\equiv \nabla_\mu\nabla_\nu\phi$. The $\mathcal{G}^{i}_{\mu\nu}$ functions appearing in the modified Einstein equations are
 \begin{subequations}
 \begin{equation}
 \label{G2}
 \mathcal{G}^2_{\mu\nu}=-\frac{1}{2}G_{2X}\phi_\mu\phi_\nu-\frac{1}{2}G_2 g_{\mu\nu},
 \end{equation}
 \begin{equation}
 \label{G3}
 \mathcal{G}^3_{\mu\nu}=\frac{1}{2}G_{3X}\DAlembert\phi\phi_\mu\phi_\nu+\nabla_{(\mu} G_3\phi_{\nu)}-\frac{1}{2} g_{\mu\nu}\nabla_\lambda G_3\phi^\lambda,
 \end{equation}
 \begin{equation}
 \label{G4}
 \begin{split}
 \mathcal{G}^4_{\mu\nu}&= G_4G_{\mu\nu}-\frac{1}{2}G_{4X}R\phi_\mu\phi_\nu -\frac{1}{2}G_{4XX}\left[(\DAlembert\phi)^2-(\phi_{\alpha\beta})^2 \right]\phi_\mu\phi_\nu\\
 &-G_{4X}\DAlembert\phi\phi_{\mu\nu} +  G_{4X}\phi_{\mu\lambda}\phi^\lambda_{\;\nu}+2\nabla_\lambda G_{4X}\phi^\lambda_{\;(\mu}\phi_{\nu)} -\nabla_\lambda G_{4X}\phi^\lambda\phi_{\mu\nu}\\
 & +g_{\mu\nu}(G_{4\phi}\DAlembert\phi-2XG_{4\phi\phi})+ g_{\mu\nu}\big\lbrace -2G_{4\phi X}\phi_{\alpha\beta}\phi^\alpha\phi^\beta\\
 & +G_{4XX}\phi_{\alpha\lambda}\phi^\lambda_{\;\beta}\phi^\alpha\phi^\beta +\frac{1}{2}G_{4X}\left[ (\DAlembert\phi)^2-(\phi_{\alpha\beta})^2 \right]  \big\rbrace\\
 & + 2\big[ G_{4X}R_{\lambda (\mu}\phi_{\nu )}\phi^\lambda -\nabla_{(\mu}G_{4X}\phi_{\nu)}\DAlembert\phi\big]-g_{\mu\nu}\left[ G_{4X}R^{\alpha\beta}\phi_\alpha\phi_\beta \right.\\
 & \left. -\nabla_\lambda G_{4X}\phi^\lambda\DAlembert\phi\right] + G_{4X}R_{\mu\alpha\nu\beta}\phi^\alpha\phi^\beta -G_{4\phi}\phi_{\mu\nu}-G_{4\phi\phi}\phi_\mu\phi_\nu \\
 &+2G_{4\phi X}\phi^\lambda\phi_{\lambda(\mu}\phi_{\nu )} - G_{4XX}\phi^\alpha\phi_{\alpha\mu}\phi^\beta\phi_{\beta\nu},
 \end{split}
 \end{equation}
 \begin{equation}
 \label{G5}
 \begin{split}
 \mathcal{G}^5_{\mu\nu}&=  G_{5X}R_{\alpha\beta}\phi^\alpha\phi^\beta_{\;(\mu}\phi_{\nu)} -G_{5X}R_{\alpha(\mu}\phi_{\nu)}\phi^\alpha\DAlembert\phi  -\frac{1}{2}G_{5X}R_{\alpha\beta}\phi^\alpha\phi^\beta\phi_{\mu\nu}\\
 & -\frac{1}{2}G_{5X}R_{\mu\alpha\nu\beta}\phi^\alpha\phi^\beta\DAlembert\phi+ G_{5X}R_{\alpha\lambda\beta(\mu}\phi_{\nu)}\phi^\lambda\phi^{\alpha\beta} \\
 &+G_{5X}R_{\alpha\lambda\beta(\mu}\phi_{\nu)}^\lambda\phi^\alpha\phi^\beta -\frac{1}{2}\left\{\nabla_{(\mu}[G_{5X}\phi^\alpha]\phi_{\alpha\nu)} \right.\\
 &\left.-\nabla_{(\mu}[G_{5X}\phi_{\nu)}]\right\}\DAlembert\phi- \nabla_\lambda[G_{5\phi}\phi_{(\mu}]\phi_{\nu)}^{\;\lambda} +\frac{1}{2}\left[ \nabla_\lambda(G_{5\phi}\phi^\lambda) \right.\\
 &\left.-\nabla_\alpha(G_{5X}\phi_\beta)\phi^{\alpha\beta} \right]\phi_{\mu\nu}
 +\nabla^\alpha G_5\phi^\beta R_{\alpha(\mu\nu)\beta} -\nabla_{(\mu}G_5G_{\nu)\lambda}\phi^\lambda \\
 &+ \frac{1}{2}\nabla_{(\mu}G_{5X}\phi_{\nu)}\left[ (\DAlembert\phi)^2-(\phi_{\alpha\beta})^2 \right]-\nabla^\lambda G_5 R_{\lambda(\mu}\phi_{\nu)} \\
 &+ \nabla_\alpha[G_{5X}\phi_\beta]\phi^\alpha_{\;(\mu}\phi_{\nu)}^{\;\beta} -\frac{1}{2}G_{5X}G_{\alpha\beta}\phi^{\alpha\beta}\phi_\mu\phi_\nu \\
 &- \nabla_\beta G_{5X}\left[ \DAlembert\phi\phi^\beta_{\;(\mu}-\phi^{\alpha\beta}\phi_{\alpha(\mu} \right] \phi_{\nu)}  +\frac{1}{2}\phi^\alpha\nabla_\alpha G_{5X} \left[ \DAlembert\phi\phi_{\mu\nu}\right.\\
 &\left.-\phi_{\beta\mu}\phi^\beta_{\;\nu} \right] -\frac{1}{2}G_{5X}\DAlembert\phi\phi_{\alpha\mu}\phi^\alpha_{\;\nu}+ \frac{1}{2}G_{5X}(\DAlembert\phi)^2\phi_{\mu\nu}\\
 &+\frac{1}{12}G_{5XX}\left[(\DAlembert\phi)^3 -3\DAlembert\phi(\phi_{\alpha\beta})^2  +2(\phi_{\alpha\beta})^3\right]\phi_\mu\phi_\nu \\
 &+\frac{1}{2}\nabla_\lambda G_5 G_{\mu\nu}\phi^\lambda+ g_{\mu\nu} \Biggl\{ -\frac{1}{6} G_{5X} \left[ (\DAlembert\phi)^3 -3\DAlembert\phi(\phi_{\alpha\beta})^2\right.\\
 &\left.+2(\phi_{\alpha\beta})^3 \right] + \nabla_\alpha G_5 R^{\alpha\beta}\phi_\beta  -\frac{1}{2}\nabla_\alpha(G_{5\phi}\phi^\alpha)\DAlembert\phi\\
 & + \frac{1}{2}\nabla_\alpha(G_{5\phi}\phi_\beta)\phi^{\alpha\beta}-\frac{1}{2}\nabla_\alpha G_{5X}\nabla^\alpha X\DAlembert\phi  +\frac{1}{2} \nabla_\alpha G_{5X}\nabla_\beta X \phi^{\alpha\beta}\\
 &-\frac{1}{4}\nabla^\lambda G_{5X}\phi_\lambda \left[ (\DAlembert\phi)^2-(\phi_{\alpha\beta})^2 \right]+ \frac{1}{2}G_{5X}R_{\alpha\beta}\phi^\alpha\phi^\beta\DAlembert\phi\\
 & -\frac{1}{2}G_{5X}R_{\alpha\lambda\beta\rho}\phi^{\alpha\beta}\phi^\lambda\phi^\rho \Biggr\}. \\
 \end{split}
 \end{equation}
 \end{subequations}
 The function $P^i_\phi$ appearing in the scalar field equations are
 \begin{align}
 \label{P2}
 P^2_\phi & =G_{2\phi}, \\
 \label{P3}
 P^3_\phi & =\nabla_\mu G_{3\phi}\phi^\mu, \\
 \label{P4}
 P^4_\phi & =G_{4\phi}R+G_{4\phi X}\left[ (\DAlembert\phi)^2-(\phi_{\alpha\beta})^2 \right], \\
 \label{P5}
 P^5_\phi & =-\nabla_\mu G_{5\phi}G^{\mu\nu}\phi_\nu-\frac{1}{6}G_{5\phi X}\left[ (\DAlembert\phi)^3 -3\DAlembert\phi(\phi_{\alpha\beta})^2+2(\phi_{\alpha\beta})^3 \right],
 \end{align}
 whereas the $J^i_\mu$ functions are
 \begin{align}
 \label{J2}
 J^2_\mu = & \,-\mathcal{L}_{2X}\phi_\mu, \\
 \label{J3}
 J^3_\mu = & \,-\mathcal{L}_{3X}\phi_\mu+G_{3X}\nabla_\mu X+2G_{3\phi}\phi_\mu, \\
 \notag
 J^4_\mu = & \,-\mathcal{L}_{4X}\phi_\mu+2G_{4X}R_{\mu\nu}\phi^\nu-2G_{4XX}(\DAlembert\phi\nabla_\mu X-\nabla^\nu X \phi_{\mu\nu})\\
 \label{J4}
 & \, -2G_{4\phi X}(\DAlembert\phi\phi_\mu+\nabla_\mu X), \\
 \notag
 J^5_\mu = & \, -\mathcal{L}_{5X}\phi_\mu-2G_{5\phi}G_{\mu\nu}\phi^\nu \\
 \notag
 & \, -G_{5X}\left[ G_{\mu\nu}\nabla^\nu X +R_{\mu\nu}\DAlembert\phi\phi^\nu-R_{\nu\lambda}\phi^\nu\phi^\lambda_{\;\mu} -R_{\alpha\mu\beta\nu}\phi^\nu\phi^{\alpha\beta}\right] \\
 \notag
 & \, +G_{5XX}\Biggl\{\frac{1}{2}\nabla_\mu X\left[ (\DAlembert\phi)^2-(\phi_{\alpha\beta})^2 \right]\\
 & \,-\nabla_\nu X (\DAlembert\phi \phi_\mu^{\;\nu}-\phi_{\alpha\mu}\phi^{\alpha\nu}) \Biggl\}\\
 \label{J5}
 & \, +G_{5\phi X} \left\{ \frac{1}{2}\phi_\mu\left[ (\DAlembert\phi)^2-(\phi_{\alpha\beta})^2 \right]+\DAlembert\phi\nabla_\mu X-\nabla^\nu X \phi_{\mu\nu} \right\}.
 \end{align}
 The explicit expression for the scalar field equation is
 \begin{equation}
 \begin{split}
 &  -G_{2\phi} -G_{2X}\DAlembert\phi -G_{2\phi X} \phi^\mu \phi_\mu +G_{2XX}\phi^\mu\phi^\nu\phi_{\mu\nu}
 +2G_{3\phi}\DAlembert\phi \\
 & +G_{3X}\left[ (\DAlembert\phi)^2-R_{\mu\nu}\phi^\mu\phi^\nu - (\phi_{\mu\nu})^2 \right] + G_{3\phi\phi}\phi_\mu\phi^\mu \\
 &  + G_{3\phi X}\phi^\mu \left( \phi_\mu \DAlembert\phi -2 \phi^\nu \phi_{\mu\nu} \right)+ G_{3XX}\phi^\mu\phi^\nu \left( \tensor{\phi}{^\lambda_\mu}\phi_{\lambda\nu}-\phi_{\mu\nu}\DAlembert\phi \right)\\
 &-G_{4\phi}R +G_{4X}G_{\mu\nu}\phi^{\mu\nu}    +G_{4\phi X} \left[ 4 R_{\mu\nu}\phi^\mu\phi^\nu-R\phi^\mu\phi_\mu-3 (\DAlembert\phi)^2\right.\\
 &\left.+3(\phi_{\mu\nu})^2 \right] + G_{4XX} \big\{ \DAlembert\phi \left[ 3 (\phi_{\lambda\sigma})^2-(\DAlembert\phi)^2 \right] -2(\phi^{\mu\nu})^3 \\
 &+ \phi^\mu\phi^\nu \big( R\phi_{\mu\nu}-4R_{\mu\lambda}\tensor{\phi}{^\lambda_\nu} +2R_{\mu\nu}\DAlembert\phi-2R_{\mu\lambda\nu\sigma}\phi^{\lambda\sigma} \big) \big\}\\
 & +2 G_{4\phi\phi X}\phi^\mu \left( \phi^\nu \phi_{\mu\nu}-\phi_\mu \DAlembert\phi \right) + G_{4\phi XX}\phi^\mu \big\{ 4\phi^\nu \left( \phi_{\mu\nu}\DAlembert\phi-\phi_{\lambda\nu}\tensor{\phi}{^\lambda_\mu} \right)\\
 &-\phi_\mu \big[ (\DAlembert\phi)^2-(\phi_{\lambda\sigma})^2 \big] \big\} +G_{4XXX}\phi^\mu\phi^\nu \big\{ 2\tensor{\phi}{^\lambda_\mu}\left( \phi_{\lambda\sigma}\tensor{\phi}{_\nu^\sigma}- \phi_{\lambda\nu}\DAlembert\phi \right)\\
 &  + \phi_{\mu\nu}\big[ (\DAlembert\phi)^2-(\phi_{\lambda\sigma})^2 \big] \big\}
 -2 G_{5\phi}G_{\mu\nu}\phi^{\mu\nu} +\frac{1}{2}G_{5X} \big[ R (\DAlembert\phi)^2\\
 &+2R^{\mu\lambda}R_{\nu\lambda}\phi_\mu\phi^\nu -R_{\mu\nu}R\phi^\mu\phi^\nu+2R^{\lambda\sigma}R_{\mu\lambda\nu\sigma}\phi^\mu\phi^\nu \\
 &-R^{\mu\lambda\sigma\rho}R_{\nu\lambda\sigma\rho}\phi_\mu\phi^\nu -R(\phi_{\mu\nu})^2 -4R^{\mu\nu}\phi_{\mu\nu}\DAlembert\phi  + 4R^{\mu\nu}\phi_{\lambda\nu}\tensor{\phi}{^\lambda_\mu}\\
 &+2R_{\mu\lambda\nu\sigma}\phi^{\mu\nu}\phi^{\sigma\lambda} \big] - G_{5\phi\phi}G_{\mu\nu}\phi^\mu\phi^\nu
 + G_{5\phi X} \big\{ \phi^\mu \phi^\nu \big[ 4R_{\mu\lambda}\tensor{\phi}{^\lambda_\nu}\\
 &-2R_{\mu\nu}\DAlembert\phi -R\phi_{\mu\nu} +2R_{\mu\lambda\nu\sigma}\phi^{\lambda\sigma} \big] + \frac{2}{3} \big[ 2(\phi^{\mu\nu})^3+\DAlembert\phi \big( (\DAlembert\phi)^2\\
 &-3(\phi_{\mu\nu})^2 \big)\big] -G_{\mu\nu}\phi^{\mu\nu}\phi^\lambda\phi_\lambda \big\} +\frac{1}{6}G_{5XX}\big\{ 3\phi^{\mu\nu}\phi_{\lambda\sigma}\big( \phi_{\mu\nu}\phi^{\lambda\sigma} \\
 &-2\tensor{\phi}{^\lambda_\mu}\tensor{\phi}{^\sigma_\nu}\big) +\DAlembert\phi \big[ 8(\phi^{\mu\nu})^3 +\DAlembert\phi \left( (\DAlembert\phi)^2-6(\phi_{\mu\nu})^2 \right) \big]\\
 & -3\phi^\mu\phi^\nu \big[ 2R^{\lambda\sigma}\phi_{\lambda\mu}\phi_{\sigma\nu}-2G_{\lambda\sigma}\phi^{\lambda\sigma}\phi_{\mu\nu} +R_{\mu\nu}(\DAlembert\phi)^2 \\
 & -\tensor{\phi}{^\lambda_\nu}\left( R\phi_{\lambda\mu}+4R_{\lambda\mu}\DAlembert\phi\right)+ 4 R_{\mu\lambda}\phi^{\lambda\sigma}\phi_{\sigma\nu}-R_{\mu\nu}(\phi_{\sigma\lambda})^2\\
 &+2R_{\mu\sigma\nu\rho}\phi^{\sigma\lambda}\tensor{\phi}{^\rho_\lambda}-2R_{\mu\sigma\nu\rho}\phi^{\rho\sigma}\DAlembert\phi\\
 & + 4R_{\nu\sigma\lambda\rho}\tensor{\phi}{^\lambda_\mu}\phi^{\rho\sigma} \big] \big\} + \frac{1}{2} G_{5\phi\phi X}\phi^\mu \big\{ 2\phi^\nu \left( \phi_{\lambda\nu}\tensor{\phi}{^\lambda_\mu}-\phi_{\mu\nu}\DAlembert\phi \right)\\
 &+ \phi_\mu \left[ (\DAlembert\phi)^2 - (\phi_{\lambda\sigma})^2 \right] \big\}  + \frac{1}{6} G_{5\phi XX} \phi^\mu \big\{ \phi_\mu \big[ 2(\phi^{\lambda\nu})^3+\DAlembert\phi \left( (\DAlembert\phi)^2\right.\\
 &\left.-3(\phi_{\lambda\sigma})^2 \right) \big]+ 6\phi^\nu \big[ 2\tensor{\phi}{^\lambda_\mu}\big( \phi_{\lambda\nu}\DAlembert\phi-\phi_{\lambda\sigma}\tensor{\phi}{^\sigma_\nu} \big)  \\
 &- \phi_{\mu\nu} \left( (\DAlembert\phi)^2 - (\phi_{\lambda\sigma})^2  \right) \big] \big\}-\frac{1}{6} G_{5XXX}\phi^\mu \phi^\nu \big\{ \phi_{\mu\nu} \big[ 2 (\phi^{\lambda\sigma})^3\\
 &+\DAlembert\phi \left( (\DAlembert\phi)^2-3(\phi_{\lambda\sigma})^2 \right) \big] +3\tensor{\phi}{^\lambda_\mu} \big[ 2\tensor{\phi}{^\sigma_\nu} \big(\phi_{\lambda\sigma}\DAlembert\phi  -\phi_{\rho\sigma} \tensor{\phi}{_\lambda^\rho} \big)\\
 & -\phi_{\lambda\nu}\left( (\DAlembert\phi)^2-(\phi_{\rho\sigma})^2 \right) \big] \big\}=0.\\
 \end{split}
 \end{equation}

\chapter[\texorpdfstring{Disformal invariance of the Horndeski Lagrangian}{Disformal invariance of the Horndeski lagrangian}
]{\chaptermark{Disformal invariance in Horndeski} Disformal invariance of the Horndeski lagrangian}
\chaptermark{Disformal invariance in Horndeski}
\label{App:Disftransf}
  
 The Horndeski Lagrangian~\eqref{Horndeski} is formally invariant under the transformation~\eqref{eq:disformal}~\cite{Bettoni:2013diz}. We derived independently these transformations and we found a mismatch with the results in Ref.~\cite{Bettoni:2013diz} which cannot be explained with differences in notation. Formal invariance means that the Lagrangian maintains the same structure, upon redefinition of the free functions $G_i(\phi,X)$. For completeness, we report these transformations. Written with respect to the metric $\bar{g}_{\mu\nu}$, the Lagrangian reads
 \begin{equation}\label{HorndeskiBar}
 \bar{S}=\frac{1}{2\kappa}\sum_{i=2}^{5}\int \dd^4x\,\sqrt{-\bar{g}}\bar{\mathcal{L}}_i,
 \end{equation}
 where we have defined
 \begin{align}
 \label{L2bar}
 \bar{\mathcal{L}}_2 = & \,  \bar{G}_2(\phi,\bar{X}),\\
 \label{L3bar}
 \bar{\mathcal{L}}_3 = & \, -\bar{G}_3(\phi,\bar{X})\bar{\DAlembert}\phi,\\
 \label{L4bar}
 \bar{\mathcal{L}}_4 = & \, \bar{G}_4(\phi,\bar{X})\bar{R} +\bar{G}_{4\bar{X}}[(\bar{\DAlembert}\phi)^2-(\bar{\nabla}_\mu\bar{\nabla}_\nu\phi)^2],\\
 \notag
 \bar{\mathcal{L}}_5 = & \, \bar{G}_5(\phi,\bar{X})\bar{G}_{\mu\nu}\bar{\nabla}^\mu\bar{\nabla}^\nu\phi \\
 \label{L5bar}
 - & \frac{\bar{G}_{5\bar{X}}}{6}\left[\left(\bar{\DAlembert}\phi\right)^3 -3\DAlembert\phi(\bar{\nabla}_\mu\bar{\nabla}_\nu\phi)^2+2(\bar{\nabla}_\mu\bar{\nabla}_\nu\phi)^3\right],
 \end{align}
 where the barred quantities are evaluated with the metric $\bar{g}_{\mu\nu}$. We can now define a new metric $g_{\mu\nu}$ which is related to $\bar{g}_{\mu\nu}$ through a disformal transformation
 \begin{equation}\label{eq:disformal2}
 \bar{g}_{\mu\nu} \equiv C(\phi)\left[g_{\mu\nu}+D(\phi)\nabla_\mu\phi\nabla_\nu\phi\right].
 \end{equation}
 As anticipated, under this transformation, Lagrangian~\eqref{HorndeskiBar} becomes Lagrangian~\eqref{Horndeski}, defined as in Eqs.~\eqref{L2}--\eqref{L5}. We can map the functions $G_i(\phi,X)$ in term of the barred functions $\bar{G}_i(\phi,\bar{X})$,
 \begin{subequations}
 \begin{equation}\label{eq:G2transf}
 \begin{split}
 G_2(\phi,X) & = C^2\sqrt{1-2DX}\bar{G}_2(\phi,\bar{X}) + \frac{2X\bar{G}_3(\phi,\bar{X})}{\sqrt{1-2DX}}\bigg( C'\\ 
 & +\frac{CD'X}{1-2DX}\bigg)+2XI_{3\phi} +\frac{3X\bar{G}_4(\phi,\bar{X})}{\sqrt{1-2DX}}\left(-\frac{C'^2}{C}+2C''\right.\\ 
 & \left.+\frac{2XC'D'}{1-2DX} \right)-4X\bigg[\bar{G}_4(\phi,\bar{X})\left(\frac{1+2D^2X^2}{\sqrt{1-2DX}}C'\right.\\
 &\left.-CD'X\sqrt{1-2DX} \right) \bigg]_\phi + \frac{12X^3C'D'\bar{G}_{4\bar{X}}(\phi,\bar{X})}{C(1-2DX)^{5/2}} +2XI_{4\phi}\\
 &+\frac{3X^2C'\bar{G}_5(\phi,\bar{X})}{C^2(1-2DX)^{3/2}}\left(-\frac{2C'^2}{C}+2C''+\frac{3XC'D'}{1-2DX} \right) \\
 &+ \frac{2X^3C'^2\bar{G}_{5\bar{X}}(\phi,\bar{X})}{C^3(1-2DX)^{5/2}}\left(-\frac{C'}{C} +\frac{3XD'}{1-2DX}\right)\\
 &-2X\bigg[\frac{X\bar{G}_5(\phi,\bar{X})}{\sqrt{1-2DX}}\left( \frac{(1+DX)C'^2}{(1-2DX)C^2}+\frac{C'D'X}{C}\right.\\
 &\left.-\frac{2D'^2X^2}{1-2DX}\right)\bigg]_\phi  + 2XI_{5\phi},
 \end{split}
 \end{equation}

 \begin{equation}\label{eq:G3transf}
 \begin{split}
 G_3(\phi,X) & = \frac{C\bar{G}_3(\phi,\bar{X})}{\sqrt{1-2DX}} + I_3 -\frac{\bar{G}_4(\phi,\bar{X})}{\sqrt{1-2DX}}\big[ 4CD'X(1-2DX)\\
 &-C'(5-4DX+4D^2X^2) \big]\\
 &
 + \frac{2X\bar{G}_{4\bar{X}}}{(1-2DX)^{3/2}}\left[(1+2DX)\frac{C'}{C}+2D'X\right]\\
 & +\frac{4CDX\bar{G}_{4\phi}(\phi,\bar{X})}{\sqrt{1-2DX}} + I_4 
 -\frac{X\bar{G}_5}{\sqrt{1-2DX}}\bigg[ -\frac{C'^2}{2C^2}\\
 &+\frac{XC'D'}{C}-\frac{4X^2D'^2(2-DX)}{(1-2DX)^2}-\frac{2XD''}{1-2DX} \bigg]\\
 &-\frac{X^2\bar{G}_{5\bar{X}}}{(1-2DX)^{5/2}}\left( -\frac{C'^2}{C^2}+\frac{2XC'D'}{C}-\frac{4X^2D'^2}{1-2DX} \right) \\
 &-\frac{2X\bar{G}_{5\phi}}{(1-2DX)^{3/2}}\left( \frac{C'}{C}-XD' \right)
 \end{split}
 \end{equation}
 
 \begin{equation}\label{eq:G4transf}
 G_4(\phi,X) = C\sqrt{1-2DX}\bar{G}_4(\phi,\bar{X})+\frac{D'X^2\bar{G}_5(\phi,\bar{X})}{(1-2DX)^{3/2}} +XK_{5\phi},
 \end{equation}

 \begin{equation}\label{eq:G5transf}
 G_5(\phi,X)=\frac{\bar{G}_5(\phi,\bar{X})}{\sqrt{1-2DX}}+K_5,
 \end{equation}
 \end{subequations}
 where a prime or a subscript $\phi$ denotes a derivative with respect to $\phi$, a subscript $\bar{X}$ denotes a derivative with respect to $\bar{X}$, defined as
 \begin{equation}
 \bar{X}=-\frac{1}{2}\bar{g}^{\mu\nu}\partial_\mu\phi\partial_\nu\phi = \frac{X}{C(1-2DX)},
 \end{equation}
 and
 \begin{gather}\label{eq:IntegralDisf}
 I_3 = -CD\int\dd X\frac{\bar{G}_3(\phi,\bar{X})}{(1-2DX)^{3/2}}, \\
 \begin{multlined}
  I_4 = -\int\dd X \bigg[3\bar{G}_4(\phi,\bar{X})\sqrt{1-2DX}(CD)' \\+2\bar{G}_{4\phi}(\phi,\bar{X})\frac{CD}{\sqrt{1-2DX}}\bigg], 
  \end{multlined}\\
 \begin{multlined}
 I_5=-\int\dd X\bigg\{\frac{\bar{G}_5(\phi,\bar{X})}{(1-2DX)^{3/2}}\bigg[\frac{ (1-DX)C'^2}{2C^2}\\
 -\frac{(2-3DX)C'D'X}{C}+3D'^2X^2-D''X \bigg] \\
 + \frac{C'-CD'X}{C(1-2DX)^{3/2}}\bar{G}_{5\bar{X}}(\phi,\bar{X})-K_{5\phi\phi} \bigg\},
 \end{multlined} \\
 K_{5}=-D\int\dd X \frac{\bar{G}_5(\phi,\bar{X})}{(1-2DX)^{3/2}}.
 \end{gather}
 Our results of Eqs.~\eqref{eq:G2transf} and~\eqref{eq:G3transf} do not coincide with those of Eqs.~(C7) and~(C8) of Appendix C of Ref.~\cite{Bettoni:2013diz}.

\chapter{Background equations}\label{App:TOV}

 In this Appendix, we derive the background field equations studied in Chapter~\ref{Chap:Threshold}.
 Matter is described as a perfect fluid with stress-energy tensor
 \begin{equation}
 T^\text{PF}_{\mu\nu}=(\epsilon+p)u_\mu u_\nu+p g_{\mu\nu},
 \end{equation}
 where $\epsilon$ is the energy density of the fluid, $p$ its pressure and $u_\mu$ its 4-velocity. The system of coordinates~\eqref{eq:lineElement} is chosen so that the fluid is at rest. Therefore,
 \begin{equation}
 u_\mu=(-c\sqrt{h},0,0,0).
 \end{equation}
 The local mass density is defined as $\rho=\epsilon/c^2$. In this setup, Einstein's field equations take the form
 \begin{align}
 0&=(r f)'- 1 +\kappa \epsilon r^2,
 \\
 0&=\dfrac{f}{h}(rh)'- 1 -\kappa p r^2 .
 \end{align}
 Additionally, the conservation equation $\nabla_\mu T^{\mu\nu}_\text{PF}=0$ can be put in the form
 \begin{equation}
 0=-\dfrac{1}{2 r f}[(-1 + f - \kappa p r^2) (p + \epsilon)] + p'.
 \end{equation}
 Together with an equation of state $p(\epsilon)$, these equations allow to solve for the background geometry and the matter distribution. Note that the equation of states we used are 23 parameters fits of the actual equations of state~\cite{Gungor:2011vq}.

 
\chapter{The effective potential}\label{App:Veff}

 In terms of the background functions $f$ and $h$ and the parameters $m_\phi$, $\beta$, $\alpha$ and $\gamma$, the effective potential presented in Chapter~\ref{Chap:Threshold} reads
 \begingroup
 \allowdisplaybreaks
 \begin{align} \notag
 V_\text{eff}(r) & = \bigg\{ h \{4 \gamma  f^2 h'^2 \{[-2 \beta  r^2+r^2-2 \gamma  f' r+16 \alpha +2 \gamma \\\notag
 &-8 (2 \alpha +\gamma ) f] h'-2 r \gamma  f h''\} r^3+f h \{\{-16 \gamma  (4 \alpha +5 \gamma ) f^2\\\notag
 &+4 \{8 \alpha  (r^2+4 \gamma )+\gamma [(6 \beta +1) r^2+14 \gamma ]+4 r \gamma  [(12 \alpha +5 \gamma ) f'\\\notag
 &+r \gamma  f'']\} f +[(4 \beta -3) r^2-32 \alpha -6 \gamma ] (r^2+2 \gamma )\\\notag
 &+4 r \gamma  f' [(2 \beta +1) r^2+\gamma  f' r+2 (\gamma -8 \alpha )]\} h'^2+8 r \gamma  f [(2 \beta  r^2\\\notag
 &+4 \gamma  f' r-16 \alpha+16 \alpha  f+5 \gamma  f) h''+2 r \gamma  f h^{(3)}] h'\\\notag
 &-4 r^2 \gamma ^2 f^2 h''^2\} r^2 +2 h^2 \{8 \gamma  \{r [4 (2 \alpha +\gamma ) h''+r \gamma  h^{(3)}]\\\label{eq:Veff}
 &-5 \gamma  h'\} f^3+4 f^2 \{\gamma  h' [8 \beta  r^2+4 \gamma  f'' r^2+3 (8 \alpha +5 \gamma ) f' r+10 \gamma ]\\\notag
 &+r \{-2 [(4 \alpha -\beta  \gamma +2 \gamma ) r^2-2 \gamma ^2 f' r+\gamma  (16 \alpha +5 \gamma )] h''\\\notag
 &-r \gamma  (r^2+2 \gamma ) h^{(3)}\}\}  +2 r \{(r^2+2 \gamma ) [(1-2 \beta ) r^2-5 \gamma  f' r\\\notag
 &+2 (8 \alpha +\gamma )] h''-h' \{2 (2 \beta -1) r^3+4 (2 m_\phi^2 r^2+4 \beta -1) \gamma  r \\\notag
 &+2 \gamma  (r^2+2 \gamma ) f'' r+f' \{-2 r f' \gamma ^2+5 [(3-2 \beta ) r^2+6 \gamma ] \gamma\\\notag
 &+8 \alpha  (3 r^2+8 \gamma )\}\}\} f +r (r^2+2 \gamma ) f' [(1-2 \beta ) r^2-2 \gamma  f' r\\\notag
 &+2 (8 \alpha +\gamma )] h'\} r+4 h^3 \{-12 \gamma ^2 f^3+4 \gamma  [2 \beta  r^2+r^2\\\notag
 &+\gamma  (2 f'+r f'') r+6 \gamma ] f^2+\{-4 \beta  (r^2+4 \gamma ) r^2\\\notag
 &+\gamma  \{f' [(8 \beta -6) r^2+\gamma  f' r-12 \gamma ]-2 r (r^2+2 \gamma ) f''\} r\\\notag
 &-4 \gamma  (2 m_\phi^2 r^4+r^2+3 \gamma )\} f+r (r^2+2 \gamma ) \{-r \gamma  f'^2\\\notag
 &+2 [(1-2 \beta ) r^2+\gamma ] f'+4 r (m_\phi^2 r^2+\beta )\}\}\} [r^2+2 \gamma\\\notag
 & -2 \gamma  (f+r f')]^2-5 f [(r^2+2 \gamma -2 \gamma f) h-2 r \gamma  f h']^2 \{r [r^2\\\notag
 &+2 \gamma-2 \gamma  (f+r f')] h'+2 \gamma  h (f'' r^2-2 f+2)\}^2-2 [-r^2\\\notag
 &-2 \gamma +2 \gamma  (f+r f')] [(-r^2-2 \gamma +2 \gamma  f) h \\\notag
 &+2 r \gamma  f h'] \{-4 \gamma ^2 [16 h^3-4 r (r h''-6 h') h^2-3 r^2 h' (r h''\\\notag
 &-5 h') h+5 r^3 h'^3] f^3 +2 \gamma  \{4 \{3 r^2+\gamma  [4 f'+r (2 f''-r f^{(3)})] r\\\notag
 &+16 \gamma \} h^3-2 r \{r [2 r^2+\gamma  (r f''-2 f') r+6 \gamma ] h''+h' \{-5 r^2 \\\notag
 &+\gamma  [r (2 r f^{(3)}-7 f'')-8 f'] r-34 \gamma \}\} h^2+r^2 h' \{h' [9 r^2\\\notag
 &+2 \gamma  (5 r f''-3 f') r+38 \gamma ] -3 r (r^2-2 \gamma  f' r+2 \gamma ) h''\} h\\\notag
 &+5 r^3 (r^2-2 \gamma  f' r+2 \gamma ) h'^3\} f^2-2 h \{2 \gamma  \{[(r^2+4 \gamma ) f''\\\notag
 &-r (r^2+2 \gamma ) f^{(3)}] r^2 +6 r^2+16 \gamma +f' (2 \gamma  f'' r^3+3 r^3\\\notag
 &+10 \gamma  r)\} h^2+r \{\gamma  h' \{10 (r^2+2 \gamma )+r [f' (3 r^2-4 \gamma  f' r+10 \gamma )\\\notag
 &+4 r (r^2+\gamma  f' r+2 \gamma ) f'']\}-r (r^2+2 \gamma ) (r^2-2 \gamma  f' r\\\notag
 &+2 \gamma ) h''\} h +2 r^2 (r^2-2 \gamma  f' r+2 \gamma ) (r^2+\gamma  f' r+2 \gamma ) h'^2\} f\\\notag
 &+r (r^2+2 \gamma ) h^2 f' [r (r^2-2 \gamma  f' r+2 \gamma ) h' +2 \gamma  h (f'' r^2\\\notag
 &+2)]\} \bigg\} / \bigg\{ 16 r^2 h^2 [2 \gamma-2 \gamma  (r f'+f)+r^2]^3 [h (2 \gamma -2 \gamma  f\\\notag
 &+r^2)-2 \gamma  r f h'] \bigg\}.
 \end{align}
 \endgroup
 Numerical calculations are more accurate in terms of Eq.~\eqref{LinEq} formulated in terms of $r$, rather than in terms of $r_\ast$ for two reasons. First, obtaining $V_\mathrm{eff}(r_\ast)$ requires to reconstruct the coordinate $r_\ast$ numerically for every change in the background or in the choice of $\gamma$, as can be seen from Eq.~\eqref{eq:drast}. Second, $V_\mathrm{eff}$ contains up to third order derivatives of the background functions. Numerically obtaining these quantities introduces errors, which can be avoided when working in terms of $r$.


\chapter{Field equations for the minimal theory}\label{App:Eqs}

 We report here the field equations for action~\eqref{eq:ACI}, where we set $\gamma=0$ and $m_\phi=0$, for a static and spherically symmetric spacetime and with matter described as a perfect fluid:
 \begin{subequations}
 \begin{equation}
 \begin{split}
    \underline{\boldsymbol{tt}}:\quad &e^{2 \Lambda}(\beta    \phi ^2+4 \kappa  r^2 \epsilon -4)
    +e^{\Lambda}(-8 \alpha    \phi  \Lambda ' \phi '+16 \alpha    \phi '^2\\
    &+16 \alpha    \phi  \phi ''-\beta   \phi ^2+\beta    r^2 \phi  \Lambda ' \phi '-2 \beta   r^2 \phi '^2 -2 \beta    r^2 \phi  \phi ''\\
    &+  r^2 \phi '^2+\beta   r \phi ^2 \Lambda '-4 \beta   r \phi  \phi '-4 r \Lambda '+4)+ 24 \alpha  \phi  \Lambda ' \phi '\\
    &-16 \alpha   \phi '^2-16 \alpha  \phi  \phi ''=0,
 \end{split}
 \end{equation}
 \begin{equation}
 \begin{split}
    \underline{\boldsymbol{rr}}:\quad &e^{2 \Lambda} (\beta   \phi ^2-4 \kappa  p r^2-4)
    +e^\Lambda (8 \alpha    \phi  \Gamma' \phi '-\beta   r^2 \phi  \Gamma' \phi '\\&-\beta    r \phi ^2 \Gamma'+4 r \Gamma'-\beta  \phi ^2-  r^2 \phi '^2-4 \beta   r \phi  \phi '+4)\\&-24 \alpha   \phi  \Gamma' \phi '=0,
 \end{split}
 \end{equation}
 \begin{equation}
 \begin{split}
    \underline{\textbf{Scalar}}:\quad & 4 \beta   \phi\,e^{2 \Lambda } +e^{\Lambda }(-8 \alpha   \phi  \Gamma ' \Lambda '+8 \alpha   \phi  \Gamma '^2+16 \alpha   \phi  \Gamma ''-4 \beta  \phi
   \\&+\beta   r^2 \phi  \Gamma ' \Lambda '-\beta   r^2 \phi  \Gamma '^2
   -2 \beta   r^2 \phi  \Gamma ''-2   r^2 \Gamma ' \phi '+2  r^2 \Lambda ' \phi '\\&-4   r^2 \phi ''-4 \beta   r \phi  \Gamma '+4 \beta   r \phi  \Lambda '-8  r \phi ')+24 \alpha   \phi  \Gamma ' \Lambda '\\&-8 \alpha   \phi (\Gamma '^2+16 \alpha  \phi  \Gamma '')=0,
 \end{split}
 \end{equation}
 \begin{equation}
 \begin{split}
    \underline{\boldsymbol{T_{(\text{PF}),\mu}^{\mu\nu}}}:\quad &2p'+(\epsilon+p)\Gamma'=0.
 \end{split}
 \end{equation}
 \end{subequations}
 Clearly, writing the above equations in the vacuum case, that is by setting to zero all contributions for matter, one retrieves the set of three coupled differential equations describing black holes.

\addcontentsline{toc}{chapter}{Bibliography}
\bibliographystyle{ieeetr}
\bibliography{bibnote}

\begin{thebibliography}{100}

\bibitem{Misner:1973prb}
C.~W. Misner, K.~S. Thorne, and J.~A. Wheeler, {\em {Gravitation}}.
\newblock San Francisco: W. H. Freeman, 1973.

\bibitem{Wald:1984rg}
R.~M. Wald, {\em {General Relativity}}.
\newblock Chicago, USA: Chicago Univ. Pr., 1984.

\bibitem{LIGOScientific:2016aoc}
B.~P. Abbott {\em et~al.}, ``{Observation of Gravitational Waves from a Binary
  Black Hole Merger},'' {\em Phys. Rev. Lett.}, vol.~116, no.~6, p.~061102,
  2016.

\bibitem{LIGOScientific:2016sjg}
B.~P. Abbott {\em et~al.}, ``{GW151226: Observation of Gravitational Waves from
  a 22-Solar-Mass Binary Black Hole Coalescence},'' {\em Phys. Rev. Lett.},
  vol.~116, no.~24, p.~241103, 2016.

\bibitem{LIGOScientific:2018jsj}
B.~P. Abbott {\em et~al.}, ``{Binary Black Hole Population Properties Inferred
  from the First and Second Observing Runs of Advanced LIGO and Advanced
  Virgo},'' {\em Astrophys. J. Lett.}, vol.~882, no.~2, p.~L24, 2019.

\bibitem{LIGOScientific:2020ibl}
R.~Abbott {\em et~al.}, ``{GWTC-2: Compact Binary Coalescences Observed by LIGO
  and Virgo During the First Half of the Third Observing Run},'' {\em Phys.
  Rev. X}, vol.~11, p.~021053, 2021.

\bibitem{LIGOScientific:2021djp}
R.~Abbott {\em et~al.}, ``{GWTC-3: Compact Binary Coalescences Observed by LIGO
  and Virgo During the Second Part of the Third Observing Run},'' 11 2021.

\bibitem{LIGOScientific:2017vwq}
B.~P. Abbott {\em et~al.}, ``{GW170817: Observation of Gravitational Waves from
  a Binary Neutron Star Inspiral},'' {\em Phys. Rev. Lett.}, vol.~119, no.~16,
  p.~161101, 2017.

\bibitem{LIGOScientific:2020aai}
B.~P. Abbott {\em et~al.}, ``{GW190425: Observation of a Compact Binary
  Coalescence with Total Mass $\sim 3.4 M_{\odot}$},'' {\em Astrophys. J.
  Lett.}, vol.~892, no.~1, p.~L3, 2020.

\bibitem{LIGOScientific:2021qlt}
R.~Abbott {\em et~al.}, ``{Observation of Gravitational Waves from Two Neutron
  Star\textendash{}Black Hole Coalescences},'' {\em Astrophys. J. Lett.},
  vol.~915, no.~1, p.~L5, 2021.

\bibitem{Yunes:2013dva}
N.~Yunes and X.~Siemens, ``{Gravitational-Wave Tests of General Relativity with
  Ground-Based Detectors and Pulsar Timing-Arrays},'' {\em Living Rev. Rel.},
  vol.~16, p.~9, 2013.

\bibitem{Stelle:1976gc}
K.~S. Stelle, ``{Renormalization of Higher Derivative Quantum Gravity},'' {\em
  Phys. Rev.}, vol.~D16, pp.~953--969, 1977.

\bibitem{Penrose:1969pc}
R.~Penrose, ``{Gravitational collapse: The role of general relativity},'' {\em
  Riv. Nuovo Cim.}, vol.~1, pp.~252--276, 1969.

\bibitem{Mathur:2009hf}
S.~D. Mathur, ``{The Information paradox: A Pedagogical introduction},'' {\em
  Class. Quant. Grav.}, vol.~26, p.~224001, 2009.

\bibitem{Hawking:1974rv}
S.~W. Hawking, ``{Black hole explosions},'' {\em Nature}, vol.~248, pp.~30--31,
  1974.

\bibitem{Zwicky:1933gu}
F.~Zwicky, ``{Die Rotverschiebung von extragalaktischen Nebeln},'' {\em Helv.
  Phys. Acta}, vol.~6, pp.~110--127, 1933.

\bibitem{Bertone:2004pz}
G.~Bertone, D.~Hooper, and J.~Silk, ``{Particle dark matter: Evidence,
  candidates and constraints},'' {\em Phys. Rept.}, vol.~405, pp.~279--390,
  2005.

\bibitem{Turner:1999kz}
M.~S. Turner, ``{Dark matter and energy in the universe},'' {\em Phys. Scripta
  T}, vol.~85, pp.~210--220, 2000.

\bibitem{SDSS:2005xqv}
D.~J. Eisenstein {\em et~al.}, ``{Detection of the Baryon Acoustic Peak in the
  Large-Scale Correlation Function of SDSS Luminous Red Galaxies},'' {\em
  Astrophys. J.}, vol.~633, pp.~560--574, 2005.

\bibitem{SupernovaSearchTeam:2004lze}
A.~G. Riess {\em et~al.}, ``{Type Ia supernova discoveries at z \ensuremath{>}
  1 from the Hubble Space Telescope: Evidence for past deceleration and
  constraints on dark energy evolution},'' {\em Astrophys. J.}, vol.~607,
  pp.~665--687, 2004.

\bibitem{WMAP:2006bqn}
D.~N. Spergel {\em et~al.}, ``{Wilkinson Microwave Anisotropy Probe (WMAP)
  three year results: implications for cosmology},'' {\em Astrophys. J.
  Suppl.}, vol.~170, p.~377, 2007.

\bibitem{Carroll:2000fy}
S.~M. Carroll, ``{The Cosmological constant},'' {\em Living Rev. Rel.}, vol.~4,
  p.~1, 2001.

\bibitem{Maltoni:2003da}
M.~Maltoni, T.~Schwetz, M.~A. Tortola, and J.~W.~F. Valle, ``{Status of three
  neutrino oscillations after the SNO salt data},'' {\em Phys. Rev. D},
  vol.~68, p.~113010, 2003.

\bibitem{deGouvea:2014xba}
A.~de~Gouvea, D.~Hernandez, and T.~M.~P. Tait, ``{Criteria for Natural
  Hierarchies},'' {\em Phys. Rev. D}, vol.~89, no.~11, p.~115005, 2014.

\bibitem{Lovelock:1971yv}
D.~Lovelock, ``{The Einstein tensor and its generalizations},'' {\em J. Math.
  Phys.}, vol.~12, pp.~498--501, 1971.

\bibitem{Lovelock:1972vz}
D.~Lovelock, ``{The four-dimensionality of space and the einstein tensor},''
  {\em J. Math. Phys.}, vol.~13, pp.~874--876, 1972.

\bibitem{Sotiriou:2014yhm}
T.~P. Sotiriou, ``{Gravity and Scalar Fields},'' {\em Lect. Notes Phys.},
  vol.~892, pp.~3--24, 2015.

\bibitem{Jordan}
P.~Jordan, {\em {Schwerkraft und Weltall}}.
\newblock Friedrich Vieweg und Sohn, Braunschweig, 1955.

\bibitem{fujii_maeda_2003}
Y.~Fujii and K.-i. Maeda, {\em The Scalar-Tensor Theory of Gravitation}.
\newblock Cambridge Monographs on Mathematical Physics, Cambridge University
  Press, 2003.

\bibitem{Cho:1975sf}
Y.~M. Cho and P.~G.~O. Freund, ``{Nonabelian Gauge Fields in Nambu-Goldstone
  Fields},'' {\em Phys. Rev. D}, vol.~12, p.~1711, 1975.

\bibitem{Madore:1998hu}
J.~Madore, ``{Introduction to noncommutative geometry},'' {\em PoS},
  vol.~corfu98, p.~016, 1998.

\bibitem{Kokado:1996ua}
A.~Kokado, G.~Konisi, T.~Saito, and K.~Uehara, ``{Brans-Dicke theory on M(4) x
  Z(2) geometry},'' {\em Prog. Theor. Phys.}, vol.~96, pp.~1291--1300, 1996.

\bibitem{Kokado:1998uw}
A.~Kokado, G.~Konisi, T.~Saito, and Y.~Tada, ``{Brans-Dicke theory from M(4) x
  Z(2) geometry},'' {\em Prog. Theor. Phys.}, vol.~99, pp.~293--303, 1998.

\bibitem{Peccei:1977hh}
R.~D. Peccei and H.~R. Quinn, ``{CP Conservation in the Presence of
  Instantons},'' {\em Phys. Rev. Lett.}, vol.~38, pp.~1440--1443, 1977.

\bibitem{Peccei:1977ur}
R.~D. Peccei and H.~R. Quinn, ``{Constraints Imposed by CP Conservation in the
  Presence of Instantons},'' {\em Phys. Rev. D}, vol.~16, pp.~1791--1797, 1977.

\bibitem{Ringwald:2014vqa}
A.~Ringwald, ``{Axions and Axion-Like Particles},'' in {\em {49th Rencontres de
  Moriond on Electroweak Interactions and Unified Theories}}, pp.~223--230,
  2014.

\bibitem{Barrow:1977}
J.~D. Barrow and R.~A. Matzner, ``{The homogeneity and isotropy of the
  Universe},'' {\em Monthly Notices of the Royal Astronomical Society},
  vol.~181, p.~719–727, 1977.

\bibitem{Guth:1980zm}
A.~H. Guth, ``{The Inflationary Universe: A Possible Solution to the Horizon
  and Flatness Problems},'' {\em Phys. Rev. D}, vol.~23, pp.~347--356, 1981.

\bibitem{Lidsey:1995np}
J.~E. Lidsey, A.~R. Liddle, E.~W. Kolb, E.~J. Copeland, T.~Barreiro, and
  M.~Abney, ``{Reconstructing the inflation potential : An overview},'' {\em
  Rev. Mod. Phys.}, vol.~69, pp.~373--410, 1997.

\bibitem{Bassett:2005xm}
B.~A. Bassett, S.~Tsujikawa, and D.~Wands, ``{Inflation dynamics and
  reheating},'' {\em Rev. Mod. Phys.}, vol.~78, pp.~537--589, 2006.

\bibitem{Pajer:2013fsa}
E.~Pajer and M.~Peloso, ``{A review of Axion Inflation in the era of Planck},''
  {\em Class. Quant. Grav.}, vol.~30, p.~214002, 2013.

\bibitem{Amin:2014eta}
M.~A. Amin, M.~P. Hertzberg, D.~I. Kaiser, and J.~Karouby, ``{Nonperturbative
  Dynamics Of Reheating After Inflation: A Review},'' {\em Int. J. Mod. Phys.
  D}, vol.~24, p.~1530003, 2014.

\bibitem{Nicolis:2008in}
A.~Nicolis, R.~Rattazzi, and E.~Trincherini, ``{The Galileon as a local
  modification of gravity},'' {\em Phys. Rev. D}, vol.~79, p.~064036, 2009.

\bibitem{Chow:2009fm}
N.~Chow and J.~Khoury, ``{Galileon Cosmology},'' {\em Phys. Rev. D}, vol.~80,
  p.~024037, 2009.

\bibitem{Brax:2011sv}
P.~Brax, C.~Burrage, and A.-C. Davis, ``{Laboratory Tests of the Galileon},''
  {\em JCAP}, vol.~09, p.~020, 2011.

\bibitem{Barreira:2014zza}
A.~Barreira, B.~Li, W.~A. Hellwing, L.~Lombriser, C.~M. Baugh, and S.~Pascoli,
  ``{Halo model and halo properties in Galileon gravity cosmologies},'' {\em
  JCAP}, vol.~04, p.~029, 2014.

\bibitem{Khoury:2003aq}
J.~Khoury and A.~Weltman, ``{Chameleon fields: Awaiting surprises for tests of
  gravity in space},'' {\em Phys. Rev. Lett.}, vol.~93, p.~171104, 2004.

\bibitem{Khoury:2003rn}
J.~Khoury and A.~Weltman, ``{Chameleon cosmology},'' {\em Phys. Rev. D},
  vol.~69, p.~044026, 2004.

\bibitem{Damour:1993hw}
T.~Damour and G.~Esposito-Far\`ese, ``{Nonperturbative strong field effects in
  tensor - scalar theories of gravitation},'' {\em Phys. Rev. Lett.}, vol.~70,
  pp.~2220--2223, 1993.

\bibitem{Silva:2017uqg}
H.~O. Silva, J.~Sakstein, L.~Gualtieri, T.~P. Sotiriou, and E.~Berti,
  ``{Spontaneous scalarization of black holes and compact stars from a
  Gauss-Bonnet coupling},'' {\em Phys. Rev. Lett.}, vol.~120, no.~13,
  p.~131104, 2018.

\bibitem{Doneva:2017bvd}
D.~D. Doneva and S.~S. Yazadjiev, ``{New Gauss-Bonnet Black Holes with
  Curvature-Induced Scalarization in Extended Scalar-Tensor Theories},'' {\em
  Phys. Rev. Lett.}, vol.~120, no.~13, p.~131103, 2018.

\bibitem{Antoniou:2017acq}
G.~Antoniou, A.~Bakopoulos, and P.~Kanti, ``{Evasion of No-Hair Theorems and
  Novel Black-Hole Solutions in Gauss-Bonnet Theories},'' {\em Phys. Rev.
  Lett.}, vol.~120, no.~13, p.~131102, 2018.

\bibitem{Herdeiro:2018wub}
C.~A.~R. Herdeiro, E.~Radu, N.~Sanchis-Gual, and J.~A. Font, ``{Spontaneous
  Scalarization of Charged Black Holes},'' {\em Phys. Rev. Lett.}, vol.~121,
  no.~10, p.~101102, 2018.

\bibitem{Ramazanoglu:2017xbl}
F.~M. Ramazano\u{g}lu, ``{Spontaneous growth of vector fields in gravity},''
  {\em Phys. Rev. D}, vol.~96, no.~6, p.~064009, 2017.

\bibitem{Ramazanoglu:2018hwk}
F.~M. Ramazano\u{g}lu, ``{Spontaneous growth of spinor fields in gravity},''
  {\em Phys. Rev. D}, vol.~98, no.~4, p.~044011, 2018.
\newblock [Erratum: Phys.Rev.D 100, 029903 (2019)].

\bibitem{Brans:1961sx}
C.~Brans and R.~H. Dicke, ``{Mach's principle and a relativistic theory of
  gravitation},'' {\em Phys. Rev.}, vol.~124, pp.~925--935, 1961.

\bibitem{Schlamminger:2007ht}
S.~Schlamminger, K.~Y. Choi, T.~A. Wagner, J.~H. Gundlach, and E.~G.
  Adelberger, ``{Test of the equivalence principle using a rotating torsion
  balance},'' {\em Phys. Rev. Lett.}, vol.~100, p.~041101, 2008.

\bibitem{Berti:2015itd}
E.~Berti {\em et~al.}, ``{Testing General Relativity with Present and Future
  Astrophysical Observations},'' {\em Class. Quant. Grav.}, vol.~32, p.~243001,
  2015.

\bibitem{Scharer:2014kya}
A.~Sch\"arer, R.~Ang\'elil, R.~Bondarescu, P.~Jetzer, and A.~Lundgren,
  ``{Testing scalar-tensor theories and parametrized post-Newtonian parameters
  in Earth orbit},'' {\em Phys. Rev. D}, vol.~90, no.~12, p.~123005, 2014.

\bibitem{Palenzuela:2013hsa}
C.~Palenzuela, E.~Barausse, M.~Ponce, and L.~Lehner, ``{Dynamical scalarization
  of neutron stars in scalar-tensor gravity theories},'' {\em Phys. Rev. D},
  vol.~89, no.~4, p.~044024, 2014.

\bibitem{Flanagan:2004bz}
E.~E. Flanagan, ``{The Conformal frame freedom in theories of gravitation},''
  {\em Class. Quant. Grav.}, vol.~21, p.~3817, 2004.

\bibitem{Sotiriou:2007zu}
T.~P. Sotiriou, V.~Faraoni, and S.~Liberati, ``{Theory of gravitation theories:
  A No-progress report},'' {\em Int. J. Mod. Phys. D}, vol.~17, pp.~399--423,
  2008.

\bibitem{Deffayet:2011gz}
C.~Deffayet, X.~Gao, D.~A. Steer, and G.~Zahariade, ``{From k-essence to
  generalised Galileons},'' {\em Phys. Rev. D}, vol.~84, p.~064039, 2011.

\bibitem{Kobayashi:2011nu}
T.~Kobayashi, M.~Yamaguchi, and J.~Yokoyama, ``{Generalized G-inflation:
  Inflation with the most general second-order field equations},'' {\em Prog.
  Theor. Phys.}, vol.~126, pp.~511--529, 2011.

\bibitem{Horndeski:1974wa}
G.~W. Horndeski, ``{Second-order scalar-tensor field equations in a
  four-dimensional space},'' {\em Int. J. Theor. Phys.}, vol.~10, pp.~363--384,
  1974.

\bibitem{Bettoni:2013diz}
D.~Bettoni and S.~Liberati, ``{Disformal invariance of second order
  scalar-tensor theories: Framing the Horndeski action},'' {\em Phys. Rev.},
  vol.~D88, p.~084020, 2013.

\bibitem{Bekenstein:1992pj}
J.~D. Bekenstein, ``{The Relation between physical and gravitational
  geometry},'' {\em Phys. Rev. D}, vol.~48, pp.~3641--3647, 1993.

\bibitem{Zumalacarregui:2013pma}
M.~Zumalacárregui and J.~García-Bellido, ``{Transforming gravity: from
  derivative couplings to matter to second-order scalar-tensor theories beyond
  the Horndeski Lagrangian},'' {\em Phys. Rev.}, vol.~D89, p.~064046, 2014.

\bibitem{BenAchour:2016cay}
J.~Ben~Achour, D.~Langlois, and K.~Noui, ``{Degenerate higher order
  scalar-tensor theories beyond Horndeski and disformal transformations},''
  {\em Phys. Rev. D}, vol.~93, no.~12, p.~124005, 2016.

\bibitem{Langlois:2015cwa}
D.~Langlois and K.~Noui, ``{Degenerate higher derivative theories beyond
  Horndeski: evading the Ostrogradski instability},'' {\em JCAP}, vol.~02,
  p.~034, 2016.

\bibitem{Ostrogradsky:1850fid}
M.~Ostrogradsky, ``{M\'emoires sur les \'equations diff\'erentielles, relatives
  au probl\`eme des isop\'erim\`etres},'' {\em Mem. Acad. St. Petersbourg},
  vol.~6, no.~4, pp.~385--517, 1850.

\bibitem{Woodard:2006nt}
R.~P. Woodard, ``{Avoiding dark energy with 1/r modifications of gravity},''
  {\em Lect. Notes Phys.}, vol.~720, pp.~403--433, 2007.

\bibitem{Woodard:2015zca}
R.~P. Woodard, ``{Ostrogradsky's theorem on Hamiltonian instability},'' {\em
  Scholarpedia}, vol.~10, no.~8, p.~32243, 2015.

\bibitem{BenAchour:2016fzp}
J.~Ben~Achour, M.~Crisostomi, K.~Koyama, D.~Langlois, K.~Noui, and G.~Tasinato,
  ``{Degenerate higher order scalar-tensor theories beyond Horndeski up to
  cubic order},'' {\em JHEP}, vol.~12, p.~100, 2016.

\bibitem{Campbell:1991kz}
B.~A. Campbell, N.~Kaloper, and K.~A. Olive, ``{Classical hair for Kerr-Newman
  black holes in string gravity},'' {\em Phys. Lett. B}, vol.~285,
  pp.~199--205, 1992.

\bibitem{Boulware:1985wk}
D.~G. Boulware and S.~Deser, ``{String Generated Gravity Models},'' {\em Phys.
  Rev. Lett.}, vol.~55, p.~2656, 1985.

\bibitem{Green:2012oqa}
M.~B. Green, J.~H. Schwarz, and E.~Witten, {\em {Superstring Theory Vol. 1}:
  {25th Anniversary Edition}}.
\newblock Cambridge Monographs on Mathematical Physics, Cambridge University
  Press, 11 2012.

\bibitem{Green:2012pqa}
M.~B. Green, J.~H. Schwarz, and E.~Witten, {\em {Superstring Theory Vol. 2}:
  {25th Anniversary Edition}}.
\newblock Cambridge Monographs on Mathematical Physics, Cambridge University
  Press, 11 2012.

\bibitem{Burgess:2003jk}
C.~P. Burgess, ``{Quantum gravity in everyday life: General relativity as an
  effective field theory},'' {\em Living Rev. Rel.}, vol.~7, pp.~5--56, 2004.

\bibitem{Mignemi:1992nt}
S.~Mignemi and N.~R. Stewart, ``{Charged black holes in effective string
  theory},'' {\em Phys. Rev. D}, vol.~47, pp.~5259--5269, 1993.

\bibitem{Kanti:1995vq}
P.~Kanti, N.~E. Mavromatos, J.~Rizos, K.~Tamvakis, and E.~Winstanley,
  ``{Dilatonic black holes in higher curvature string gravity},'' {\em Phys.
  Rev. D}, vol.~54, pp.~5049--5058, 1996.

\bibitem{Sotiriou:2013qea}
T.~P. Sotiriou and S.-Y. Zhou, ``{Black hole hair in generalized scalar-tensor
  gravity},'' {\em Phys. Rev. Lett.}, vol.~112, p.~251102, 2014.

\bibitem{Macedo:2019sem}
C.~F.~B. Macedo, J.~Sakstein, E.~Berti, L.~Gualtieri, H.~O. Silva, and T.~P.
  Sotiriou, ``{Self-interactions and Spontaneous Black Hole Scalarization},''
  {\em Phys. Rev. D}, vol.~99, no.~10, p.~104041, 2019.

\bibitem{Silva:2018qhn}
H.~O. Silva, C.~F.~B. Macedo, T.~P. Sotiriou, L.~Gualtieri, J.~Sakstein, and
  E.~Berti, ``{Stability of scalarized black hole solutions in
  scalar-Gauss-Bonnet gravity},'' {\em Phys. Rev. D}, vol.~99, no.~6,
  p.~064011, 2019.

\bibitem{Minamitsuji:2018xde}
M.~Minamitsuji and T.~Ikeda, ``{Scalarized black holes in the presence of the
  coupling to Gauss-Bonnet gravity},'' {\em Phys. Rev. D}, vol.~99, no.~4,
  p.~044017, 2019.

\bibitem{Hawking:1972qk}
S.~W. Hawking, ``{Black holes in the Brans-Dicke theory of gravitation},'' {\em
  Commun. Math. Phys.}, vol.~25, pp.~167--171, 1972.

\bibitem{Bekenstein:1971hc}
J.~D. Bekenstein, ``{Nonexistence of baryon number for static black holes},''
  {\em Phys. Rev.}, vol.~D5, pp.~1239--1246, 1972.

\bibitem{Sotiriou:2011dz}
T.~P. Sotiriou and V.~Faraoni, ``{Black holes in scalar-tensor gravity},'' {\em
  Phys. Rev. Lett.}, vol.~108, p.~081103, 2012.

\bibitem{Harada:1997mr}
T.~Harada, ``{Stability analysis of spherically symmetric star in scalar -
  tensor theories of gravity},'' {\em Prog. Theor. Phys.}, vol.~98,
  pp.~359--379, 1997.

\bibitem{Andreou:2019ikc}
N.~Andreou, N.~Franchini, G.~Ventagli, and T.~P. Sotiriou, ``{Spontaneous
  scalarization in generalized scalar-tensor theory},'' {\em Phys. Rev.},
  vol.~D99, no.~12, p.~124022, 2019.

\bibitem{Saravani:2019xwx}
M.~Saravani and T.~P. Sotiriou, ``{Classification of shift-symmetric Horndeski
  theories and hairy black holes},'' {\em Phys. Rev. D}, vol.~99, no.~12,
  p.~124004, 2019.

\bibitem{McManus:2016kxu}
R.~McManus, L.~Lombriser, and J.~Peñarrubia, ``{Finding Horndeski theories
  with Einstein gravity limits},'' {\em JCAP}, vol.~1611, no.~11, p.~006, 2016.

\bibitem{Motohashi:2018wdq}
H.~Motohashi and M.~Minamitsuji, ``{General Relativity solutions in modified
  gravity},'' {\em Phys. Lett.}, vol.~B781, pp.~728--734, 2018.

\bibitem{Sperhake:2017itk}
U.~Sperhake, C.~J. Moore, R.~Rosca, M.~Agathos, D.~Gerosa, and C.~D. Ott,
  ``{Long-lived inverse chirp signals from core collapse in massive
  scalar-tensor gravity},'' {\em Phys. Rev. Lett.}, vol.~119, no.~20,
  p.~201103, 2017.

\bibitem{Rosca-Mead:2019seq}
R.~Rosca-Mead, C.~J. Moore, M.~Agathos, and U.~Sperhake, ``{Inverse-chirp
  signals and spontaneous scalarisation with self-interacting potentials in
  stellar collapse},'' {\em Class. Quant. Grav.}, vol.~36, no.~13, p.~134003,
  2019.

\bibitem{Rosca-Mead:2020ehn}
R.~Rosca-Mead, U.~Sperhake, C.~J. Moore, M.~Agathos, D.~Gerosa, and C.~D. Ott,
  ``{Core collapse in massive scalar-tensor gravity},'' {\em Phys. Rev. D},
  vol.~102, no.~4, p.~044010, 2020.

\bibitem{Minamitsuji:2016hkk}
M.~Minamitsuji and H.~O. Silva, ``{Relativistic stars in scalar-tensor theories
  with disformal coupling},'' {\em Phys. Rev.}, vol.~D93, no.~12, p.~124041,
  2016.

\bibitem{Ventagli:2020rnx}
G.~Ventagli, A.~Leh\'ebel, and T.~P. Sotiriou, ``{Onset of spontaneous
  scalarization in generalized scalar-tensor theories},'' {\em Phys. Rev. D},
  vol.~102, no.~2, p.~024050, 2020.

\bibitem{Tolman:1939jz}
R.~C. Tolman, ``{Static solutions of Einstein's field equations for spheres of
  fluid},'' {\em Phys. Rev.}, vol.~55, pp.~364--373, 1939.

\bibitem{Oppenheimer:1939ne}
J.~R. Oppenheimer and G.~M. Volkoff, ``{On Massive neutron cores},'' {\em Phys.
  Rev.}, vol.~55, pp.~374--381, 1939.

\bibitem{Gungor:2011vq}
C.~Gungor and K.~Y. Eksi, ``{Analytical Representation for Equations of State
  of Dense Matter},'' 2011.

\bibitem{TheLIGOScientific:2017qsa}
B.~P. Abbott {\em et~al.}, ``{GW170817: Observation of Gravitational Waves from
  a Binary Neutron Star Inspiral},'' {\em Phys. Rev. Lett.}, vol.~119, no.~16,
  p.~161101, 2017.

\bibitem{schiff1955quantum}
L.~Schiff, {\em Quantum Mechanics}.
\newblock International series in pure and applied physics, McGraw-Hill, 1955.

\bibitem{Blazquez-Salcedo:2018jnn}
J.~L. Blázquez-Salcedo, D.~D. Doneva, J.~Kunz, and S.~S. Yazadjiev, ``{Radial
  perturbations of the scalarized Einstein-Gauss-Bonnet black holes},'' {\em
  Phys. Rev. D}, vol.~98, no.~8, p.~084011, 2018.

\bibitem{Doneva:2019vuh}
D.~D. Doneva, K.~V. Staykov, and S.~S. Yazadjiev, ``{Gauss-Bonnet black holes
  with a massive scalar field},'' {\em Phys. Rev. D}, vol.~99, no.~10,
  p.~104045, 2019.

\bibitem{Martinez:2015mya}
J.~G. Martinez, K.~Stovall, P.~C.~C. Freire, J.~S. Deneva, F.~A. Jenet, M.~A.
  McLaughlin, M.~Bagchi, S.~D. Bates, and A.~Ridolfi, ``{Pulsar J0453+1559: A
  Double Neutron Star System with a Large Mass Asymmetry},'' {\em Astrophys.
  J.}, vol.~812, no.~2, p.~143, 2015.

\bibitem{Suwa:2018uni}
Y.~Suwa, T.~Yoshida, M.~Shibata, H.~Umeda, and K.~Takahashi, ``{On the minimum
  mass of neutron stars},'' {\em Mon. Not. Roy. Astron. Soc.}, vol.~481, no.~3,
  pp.~3305--3312, 2018.

\bibitem{Ramazanoglu:2016kul}
F.~M. Ramazanoğlu and F.~Pretorius, ``{Spontaneous Scalarization with Massive
  Fields},'' {\em Phys. Rev.}, vol.~D93, no.~6, p.~064005, 2016.

\bibitem{Mendes:2014ufa}
R.~F.~P. Mendes, ``{Possibility of setting a new constraint to scalar-tensor
  theories},'' {\em Phys. Rev. D}, vol.~91, no.~6, p.~064024, 2015.

\bibitem{Palenzuela:2015ima}
C.~Palenzuela and S.~L. Liebling, ``{Constraining scalar-tensor theories of
  gravity from the most massive neutron stars},'' {\em Phys. Rev. D}, vol.~93,
  no.~4, p.~044009, 2016.

\bibitem{Mendes:2016fby}
R.~F.~P. Mendes and N.~Ortiz, ``{Highly compact neutron stars in scalar-tensor
  theories of gravity: Spontaneous scalarization versus gravitational
  collapse},'' {\em Phys. Rev. D}, vol.~93, no.~12, p.~124035, 2016.

\bibitem{Ramazanoglu:2017yun}
F.~M. Ramazanoğlu, ``{Regularization of instabilities in gravity theories},''
  {\em Phys. Rev. D}, vol.~97, no.~2, p.~024008, 2018.
\newblock [Erratum: Phys.Rev.D 99, 069905 (2019)].

\bibitem{doi:10.1119/1.17935}
W.~F. Buell and B.~A. Shadwick, ``Potentials and bound states,'' {\em American
  Journal of Physics}, vol.~63, no.~3, pp.~256--258, 1995.

\bibitem{Antoniou:2020nax}
G.~Antoniou, L.~Bordin, and T.~P. Sotiriou, ``{Compact object scalarization
  with general relativity as a cosmic attractor},'' {\em Phys. Rev. D},
  vol.~103, no.~2, p.~024012, 2021.

\bibitem{Antoniou:2021zoy}
G.~Antoniou, A.~Leh\'ebel, G.~Ventagli, and T.~P. Sotiriou, ``{Black hole
  scalarization with Gauss-Bonnet and Ricci scalar couplings},'' {\em Phys.
  Rev. D}, vol.~104, no.~4, p.~044002, 2021.

\bibitem{Sotiriou:2014pfa}
T.~P. Sotiriou and S.-Y. Zhou, ``{Black hole hair in generalized scalar-tensor
  gravity: An explicit example},'' {\em Phys. Rev.}, vol.~D90, p.~124063, 2014.

\bibitem{Dima:2020yac}
A.~Dima, E.~Barausse, N.~Franchini, and T.~P. Sotiriou, ``{Spin-induced black
  hole spontaneous scalarization},'' {\em Phys. Rev. Lett.}, vol.~125, no.~23,
  p.~231101, 2020.

\bibitem{Freire:2012mg}
P.~C. Freire, N.~Wex, G.~Esposito-Farese, J.~P. Verbiest, M.~Bailes, B.~A.
  Jacoby, M.~Kramer, I.~H. Stairs, J.~Antoniadis, and G.~H. Janssen, ``{The
  relativistic pulsar-white dwarf binary PSR J1738+0333 II. The most stringent
  test of scalar-tensor gravity},'' {\em Mon. Not. Roy. Astron. Soc.},
  vol.~423, p.~3328, 2012.

\bibitem{Antoniadis:2013pzd}
J.~Antoniadis {\em et~al.}, ``{A Massive Pulsar in a Compact Relativistic
  Binary},'' {\em Science}, vol.~340, p.~6131, 2013.

\bibitem{Shao:2017gwu}
L.~Shao, N.~Sennett, A.~Buonanno, M.~Kramer, and N.~Wex, ``{Constraining
  nonperturbative strong-field effects in scalar-tensor gravity by combining
  pulsar timing and laser-interferometer gravitational-wave detectors},'' {\em
  Phys. Rev. X}, vol.~7, no.~4, p.~041025, 2017.

\bibitem{Herdeiro:2020wei}
C.~A.~R. Herdeiro, E.~Radu, H.~O. Silva, T.~P. Sotiriou, and N.~Yunes,
  ``{Spin-induced scalarized black holes},'' {\em Phys. Rev. Lett.}, vol.~126,
  no.~1, p.~011103, 2021.

\bibitem{Berti:2020kgk}
E.~Berti, L.~G. Collodel, B.~Kleihaus, and J.~Kunz, ``{Spin-induced black-hole
  scalarization in Einstein-scalar-Gauss-Bonnet theory},'' {\em Phys. Rev.
  Lett.}, vol.~126, no.~1, p.~011104, 2021.

\bibitem{Ventagli:2021ubn}
G.~Ventagli, G.~Antoniou, A.~Leh\'ebel, and T.~P. Sotiriou, ``{Neutron star
  scalarization with Gauss-Bonnet and Ricci scalar couplings},'' {\em Phys.
  Rev. D}, vol.~104, no.~12, p.~124078, 2021.

\bibitem{Wex:2020ald}
N.~Wex and M.~Kramer, ``{Gravity Tests with Radio Pulsars},'' {\em Universe},
  vol.~6, no.~9, p.~156, 2020.

\bibitem{Damour:1992we}
T.~Damour and G.~Esposito-Farese, ``{Tensor multiscalar theories of
  gravitation},'' {\em Class. Quant. Grav.}, vol.~9, pp.~2093--2176, 1992.

\bibitem{Haensel:2004nu}
P.~Haensel and A.~Y. Potekhin, ``{Analytical representations of unified
  equations of state of neutron-star matter},'' {\em Astron. Astrophys.},
  vol.~428, pp.~191--197, 2004.

\bibitem{Maselli:2020zgv}
A.~Maselli, N.~Franchini, L.~Gualtieri, and T.~P. Sotiriou, ``{Detecting scalar
  fields with Extreme Mass Ratio Inspirals},'' {\em Phys. Rev. Lett.},
  vol.~125, no.~14, p.~141101, 2020.

\bibitem{Maselli:2021men}
A.~Maselli, N.~Franchini, L.~Gualtieri, T.~P. Sotiriou, S.~Barsanti, and
  P.~Pani, ``{Detecting new fundamental fields with LISA},'' 6 2021.

\bibitem{Julie:2019sab}
F.-L. Juli\'e and E.~Berti, ``{Post-Newtonian dynamics and black hole
  thermodynamics in Einstein-scalar-Gauss-Bonnet gravity},'' {\em Phys. Rev.
  D}, vol.~100, no.~10, p.~104061, 2019.

\bibitem{Julie:2022huo}
F.-L. Juli\'e, H.~O. Silva, E.~Berti, and N.~Yunes, ``{Black hole sensitivities
  in Einstein-scalar-Gauss-Bonnet gravity},'' 2 2022.

\bibitem{Perkins:2021mhb}
S.~E. Perkins, R.~Nair, H.~O. Silva, and N.~Yunes, ``{Improved
  gravitational-wave constraints on higher-order curvature theories of
  gravity},'' {\em Phys. Rev. D}, vol.~104, no.~2, p.~024060, 2021.

\bibitem{Doneva:2013qva}
D.~D. Doneva, S.~S. Yazadjiev, N.~Stergioulas, and K.~D. Kokkotas, ``{Rapidly
  rotating neutron stars in scalar-tensor theories of gravity},'' {\em Phys.
  Rev. D}, vol.~88, no.~8, p.~084060, 2013.

\bibitem{Ripley:2020vpk}
J.~L. Ripley and F.~Pretorius, ``{Dynamics of a $\mathbb Z_2$ symmetric EdGB
  gravity in spherical symmetry},'' {\em Class. Quant. Grav.}, vol.~37, no.~15,
  p.~155003, 2020.

\bibitem{Barausse:2012da}
E.~Barausse, C.~Palenzuela, M.~Ponce, and L.~Lehner, ``{Neutron-star mergers in
  scalar-tensor theories of gravity},'' {\em Phys. Rev. D}, vol.~87, p.~081506,
  2013.

\bibitem{Silva:2020omi}
H.~O. Silva, H.~Witek, M.~Elley, and N.~Yunes, ``{Dynamical Descalarization in
  Binary Black Hole Mergers},'' {\em Phys. Rev. Lett.}, vol.~127, no.~3,
  p.~031101, 2021.

\bibitem{Dabrowski:2008kx}
M.~P. Dabrowski, J.~Garecki, and D.~B. Blaschke, ``{Conformal transformations
  and conformal invariance in gravitation},'' {\em Annalen Phys.}, vol.~18,
  pp.~13--32, 2009.

\end{thebibliography}

\end{document}